\documentclass[11pt]{article}
\usepackage[english]{babel}
\usepackage[utf8]{inputenc}
\usepackage{lmodern}
\usepackage[T1]{fontenc}
\usepackage{marginnote}
\usepackage{amsmath,amssymb,lmodern,amsthm}
\usepackage{mathtools}
\usepackage{graphicx}
\usepackage{mathrsfs}
\usepackage{bm}
\usepackage[breaklinks]{hyperref}
\usepackage{url}
\usepackage{breakurl}
\usepackage{array}
\usepackage{tensor}
\usepackage{pifont}
\usepackage[toc,page]{appendix}
\usepackage[left=2.5cm,right=2.5cm,top=3cm,bottom=3cm]{geometry}
\usepackage{tikz}
\usetikzlibrary{arrows,positioning}
\usepackage{enumitem}
\usepackage{empheq}
\usepackage{cleveref}
\usepackage{cases}
\usepackage{csquotes}
\usepackage{accents}
\usepackage{setspace}

\theoremstyle{plain}
\usepackage{xargs}
\usepackage{xcolor}
\usepackage{subcaption}
\usepackage[justification=raggedright]{caption}
\usepackage[affil-it]{authblk}
\usepackage{tocloft}

\hypersetup{
	colorlinks=true,
	urlcolor=blue,
	linkcolor={black},
	citecolor={green!40!black}
}
\usepackage{thmtools}
\usepackage[backend=bibtex,style=numeric-comp,maxnames=4,sorting=none,url=false,hyperref=true,date=year,eprint=true,doi=true]{biblatex}

\newlist{properties}{enumerate}{10}
\setlist[properties]{label*=\roman*)}
\crefname{propertiesi}{\text{property}}{\text{properties}}
\Crefname{propertiesi}{\text{Property}}{\text{Properties}}
\newlist{points}{enumerate}{10}
\setlist[points]{label*=\arabic*)}
\crefname{pointsi}{\text{point}}{\text{points}}
\Crefname{pointsi}{\text{Point}}{\text{Points}}

\newcommand{\ubar}[1]{\underaccent{\bar}{#1}}
\newcommand{\lmultieqref}[1]{(\ref{#1},~}
\newcommand{\rmultieqref}[1]{~\ref{#1})}
\newcommand{\eqfref}[1]{eq.~(\ref{#1})}

\numberwithin{equation}{section}
\newtheorem{thm}{Theorem}[]
\theoremstyle{definition}
\newtheorem{prop}{Proposition}[section]
\newtheorem{corollary}{Corollary}[section]
\newtheorem{deff}{Definition}[section]
\newtheorem*{crit*}{Criterion}
\newtheorem{crit}{Criterion}[]

\newtheorem{lemma}{Lemma}[section]
\newtheorem{remark}{Remark}[section]
\renewcommand{\iff}{\Longleftrightarrow}

\newcommand{\lied}{\pounds}
\newcommand{\defeq}{\vcentcolon=}
\newcommand{\defeqs}{\stackrel{\scri}{\vcentcolon=}}
\newcommand{\defeqc}{\stackrel{\Sc}{\vcentcolon=}}
\newcommand{\defeqcn}{\stackrel{\Scn}{\vcentcolon=}}
\newcommand{\defeqI}{\stackrel{\I}{\vcentcolon=}}
\newcommand{\scri}{\mathscr{J}}

\newcommand{\pt}[2]{\tensor{\hat{#1}}{#2}}
\newcommand{\ptrd}[3]{\tensor[_{{\ }_{\scriptscriptstyle#1}\!}]{\hat{#2}}{#3}}

\newcommand{\ctru}[3]{\tensor[^{{\,}^{\scriptscriptstyle\!#1\!}\!}]{#2}{#3}}
\newcommand{\ctrd}[3]{\tensor[_{{\ }_{\scriptscriptstyle#1}\!}]{#2}{#3}}

\newcommand{\csrd}[2]{\tensor[_{{\ }_{\scriptscriptstyle\!#1\!}\!}]{#2}{}}

\newcommand{\ctrdupm}[3]{\tensor[_{{\ }_{\scriptscriptstyle#1}\!\!\!\!}^{{\,}^{\scriptscriptstyle\!\pm\!}\!}]{#2}{#3}}

\newcommand{\ct}[2]{\tensor{{#1}}{#2}}
\newcommand{\ctp}[2]{\tensor[^{{\,}^{\scriptscriptstyle\!+\!}\!}]{{#1}}{#2}}
\newcommand{\ctl}[2]{\tensor[^{{\,}^{\scriptscriptstyle\!\ell\!}\!}]{{#1}}{#2}}
\newcommand{\ctps}[2]{\tensor[^{{\,}^{\scriptscriptstyle\!+\!}\!}]{\underline{#1}}{#2}}
\newcommand{\ctpms}[2]{\tensor[^{{\,}^{\scriptscriptstyle\!\pm\!}\!}]{\underline{#1}}{#2}}
\newcommand{\ctm}[2]{\tensor[^{{\,}^{\scriptscriptstyle\!-\!}\!}]{{#1}}{#2}}
\newcommand{\ctms}[2]{\tensor[^{{\,}^{\scriptscriptstyle\!-\!}\!}]{\underline{#1}}{#2}}
\newcommand{\ctcmp}[2]{\tensor[^{{\,}^{\scriptscriptstyle\!-+\!}\!}]{{\mathring{#1}}}{#2}}
\newcommand{\ctcpm}[2]{\tensor[^{{\,}^{\scriptscriptstyle\!+-\!}\!}]{{\mathring{#1}}}{#2}}
\newcommand{\ctcupm}[2]{\tensor[^{{\,}^{\scriptscriptstyle\!\pm\!}\!}]{{\mathring{#1}}}{#2}}
\newcommand{\ctcump}[2]{\tensor[^{{\,}^{\scriptscriptstyle\!\mp\!}\!}]{{\mathring{#1}}}{#2}}
\newcommand{\ctpm}[2]{\tensor[^{{\,}^{\scriptscriptstyle\!\pm\!}\!}]{{#1}}{#2}}
\newcommand{\ctmp}[2]{\tensor[^{{\,}^{\scriptscriptstyle\!\mp\!}\!}]{{#1}}{#2}}
\newcommand{\ctrmp}[2]{\tensor[^{{\,}^{\scriptscriptstyle\!-+\!}\!}]{{#1}}{#2}}
\newcommand{\ctrpm}[2]{\tensor[^{{\,}^{\scriptscriptstyle\!+-\!}\!}]{{#1}}{#2}}
\newcommand{\ctc}[2]{\tensor{\mathring{#1}}{#2}}
\newcommand{\ctt}[2]{\tensor{\grave{#1}}{#2}}
\newcommand{\cttcn}[2]{\tensor{\grave{\ubar{#1}}}{#2}}
\newcommand{\cts}[2]{\tensor{\overline{#1}}{#2}}
\newcommand{\ctcp}[2]{\tensor[^{{\,}^{\scriptscriptstyle\!+\!}\!}]{\mathring{#1}}{#2}}

\newcommand{\ctcm}[2]{\tensor[^{{\,}^{\scriptscriptstyle\!-\!}\!}]{\mathring{#1}}{#2}}

\newcommand{\ctsl}[2]{\tensor[^{{\,}^{\scriptscriptstyle\!\ell\!}\!}]{\underline{#1}}{#2}}
\newcommand{\ctd}[2]{\tensor{\dot{#1}}{#2}}

\newcommand{\ctcnp}[2]{\tensor[^{{\,}^{\scriptscriptstyle\!+\!}\!}]{\ubar{{#1}}}{#2}}
\newcommand{\ctcnm}[2]{\tensor[^{{\,}^{\scriptscriptstyle\!-\!}\!}]{\ubar{{#1}}}{#2}}
\newcommand{\ctcnmp}[2]{\tensor[^{{\,}^{\scriptscriptstyle\!-+\!}\!}]{{\ubar{{#1}}}}{#2}}
\newcommand{\ctcnpm}[2]{\tensor[^{{\,}^{\scriptscriptstyle\!+-\!}\!}]{{\ubar{{#1}}}}{#2}}
\newcommand{\ctcnupm}[2]{\tensor[^{{\,}^{\scriptscriptstyle\!\pm\!}\!}]{{\ubar{{#1}}}}{#2}}

\newcommand{\cscn}[1]{\ubar{{#1}}}
\newcommand{\ctcn}[2]{\tensor{\ubar{{#1}}}{#2}}
\newcommand{\cdcn}[1]{\tensor{\ubar{\D}}{#1}}
\newcommand{\cdcng}[1]{\tensor{\ubar{{\tilde{\D}}}}{#1}}
\newcommand{\ctcng}[2]{\tensor{\tilde{\ubar{#1}}}{#2}}
\newcommand{\ctg}[2]{\tensor{\tilde{#1}}{#2}}
\newcommand{\ctcg}[2]{\tensor{\tilde{\mathring{#1}}}{#2}}
\newcommand{\ctsg}[2]{\tensor{\tilde{\overline{#1}}}{#2}}
\newcommand{\csg}[1]{\tilde{#1}}
\newcommand{\cs}[1]{#1}
\newcommand{\cst}[1]{\grave{#1}}
\newcommand{\csp}[1]{\tensor[^{{\,}^{\scriptscriptstyle\!+\!}\!}]{#1}{}}
\newcommand{\csm}[1]{\tensor[^{{\,}^{\scriptscriptstyle\!-\!}\!}]{#1}{}}
\newcommand{\cscnp}[1]{\tensor[^{{\,}^{\scriptscriptstyle\!+\!}\!}]{\ubar{#1}}{}}
\newcommand{\cscnm}[1]{\tensor[^{{\,}^{\scriptscriptstyle\!-\!}\!}]{\ubar{#1}}{}}
\newcommand{\csl}[1]{\tensor[^{{\,}^{\scriptscriptstyle\!\ell\!}\!}]{#1}{}}
\newcommand{\cspm}[1]{\tensor[^{{\,}^{\scriptscriptstyle\!\pm\!}\!}]{#1}{}}
\newcommand{\csmp}[1]{\tensor[^{{\,}^{\scriptscriptstyle\!\mp\!}\!}]{#1}{}}
\newcommand{\cscnpm}[1]{\tensor[^{{\,}^{\scriptscriptstyle\!\pm\!}\!}]{\ubar{#1}}{}}

\newcommand{\csC}[1]{\mathring{#1}}
\newcommand{\csS}[1]{\overline{#1}}
\newcommand{\csSg}[1]{\tilde{\overline{#1}}}
\newcommand{\csCg}[1]{\tilde{\mathring{#1}}}
\newcommand{\ps}[1]{\hat{#1}}

\newcommand{\bzeta}{{\bar{\zeta}}}

\newcommand{\df}[1]{\text{d}#1}

\newcommand{\fblue}[1]{\textcolor{blue}{#1}}

\newcommand{\prn}[1]{\left(#1\right)}
\newcommand{\bprn}[1]{\Big(#1\Big)}
\newcommand{\brkt}[1]{\left[#1\right]}
\newcommand{\bbrkt}[1]{\Big[#1\Big]}
\newcommand{\cbrkt}[1]{\left\lbrace#1\right\rbrace}

\newcommand{\Bcbrkt}[1]{\Bigg\lbrace#1\Bigg\rbrace}
\newcommand{\abs}[1]{\left|#1\right|}
\newcommand{\evalat}[1]{\Big|_{#1}}
\newcommand{\eqs}{\stackrel{\scri}{=}}
\newcommand{\eqsflat}{\stackrel{\scri_0}{=}}
\newcommand{\eqsopen}{\stackrel{\Delta}{=}}
\newcommand{\neqsopen}{\stackrel{\Delta}{\neq}}
\newcommand{\eqc}{\stackrel{\Sc}{=}}
\newcommand{\eqcn}{\stackrel{\Scn}{=}}
\newcommand{\eqSv}[1]{\stackrel{\Sc_{#1}}{=}}
\newcommand{\eqI}{\stackrel{\I}{=}}

\newcommand{\neqs}{\stackrel{\scri}{\neq}}

\newcommand{\cd}[1]{\tensor{\nabla}{#1}}

\newcommand{\cds}[1]{\tensor{\overline{\nabla}}{#1}}
\newcommand{\cdc}[1]{\tensor{\D}{#1}}
\newcommand{\pd}[1]{\tensor{\hat{\nabla}}{#1}}
\newcommand{\cdg}[1]{\tensor{\tilde{\nabla}}{#1}}

\newcommand{\cdcg}[1]{\tensor{\tilde{\D}}{#1}}
\newcommand{\commute}[2]{\left[#1,#2\right]}
\newcommand{\dpart}[1]{\partial_{#1}}

\newcommand{\spacef}{\ }

\newcommand{\Q}{\mathcal{Q}}
\newcommand{\T}{\mathcal{T}}
\newcommand{\D}{\mathcal{D}}
\newcommand{\W}{\mathcal{W}}
\newcommand{\Z}{\mathcal{Z}}
\newcommand{\V}{\mathcal{V}}
\newcommand{\Sc}{\mathcal{S}}
\newcommand{\Pc}{\mathcal{P}}
\renewcommand{\P}{\mathcal{P}}
\newcommand{\I}{\mathcal{I}}
\newcommand{\J}{\mathcal{J}}
\newcommand{\R}{\mathcal{R}}
\newcommand{\B}{\mathcal{B}}
\newcommand{\Y}{\mathcal{Y}}
\newcommand{\C}{\mathcal{C}}

\newcommand{\fX}{\mathfrak{X}}

\newcommand{\form}[1]{\bm{#1}}
\newcommand{\fB}{\mathsf{B}}
\newcommand{\fT}{\mathsf{T}}
\newcommand{\fCS}{\mathsf{CS}}
\newcommand{\fb}{\mathfrak{b}}
\newcommand{\ft}{\mathfrak{t}}
\newcommand{\fcs}{\mathfrak{cs}}
\newcommand{\Scn}{\mathbf{S}_2}
\newcommand{\ms}[1]{\ct{h}{#1}}

\newcommand{\msg}[1]{\ct{\tilde{h}}{#1}}
\newcommand{\mc}[1]{\ct{q}{#1}}
\newcommand{\mcg}[1]{\ct{\tilde{q}}{#1}}
\newcommand{\mcn}[1]{\ct{\ubar{q}}{#1}}
\newcommand{\mcng}[1]{\ct{\ubar{{\tilde{q}}}}{#1}}

 \renewcommand{\thesubsection}{\thesection.\Alph{subsection}}

\newcolumntype{M}[1]{>{\centering\arraybackslash}m{#1}}
\newcolumntype{N}{@{}m{0pt}@{}}

\def\be{\begin{equation}}
\def\ee{\end{equation}}
\def\bea{\begin{eqnarray}}
\def\eea{\end{eqnarray}}
\def\bean{\begin{eqnarray*}}
	\def\eean{\end{eqnarray*}}

\newcounter{marginnotecount}[section]

\newcommand{\limitl}{\lim_{\Lambda\rightarrow0}}

\usepackage{subfiles}
\graphicspath{{./figures/}}

	\title{\Large \textbf{Asymptotic Structure}\\ \textbf{with a positive cosmological constant}}
	\author[]{Francisco Fernández-Álvarez\thanks{francisco.fernandez@ehu.eus}\ }
	\author[]{\ José M. M. Senovilla\thanks{josemm.senovilla@ehu.eus}} 
	\affil[]{Departamento de Física\\ Universidad del País Vasco UPV/EHU\\ Apartado 644, 48080 Bilbao, Spain }
	\date{\today{}}
\begin{document}
	
	\maketitle

	\begin{abstract}
		This is the second of two papers \cite{Fernandez-Alvarez_Senovilla-afs} that study the asymptotic structure of space-times with a non-negative cosmological constant $\Lambda$. This paper deals with the case $\Lambda >0$. Our approach is founded on
the `tidal energies' built with the Weyl curvature and, specifically, we use the asymptotic
super-Poynting vector computed from the rescaled Bel-Robinson tensor at infinity to provide a covariant, gauge-invariant, criterion for the existence, or absence, of gravitational radiation at infinity. The fundamental idea we put forward is that the physical asymptotic properties are encoded in $(\scri,\ms{_{ab}},\ct{D}{_{ab}})$, where the first element of the triplet is a 3-dimensional manifold, the second is a representative of a conformal class of Riemannian metrics on $\scri$, and the third element is a traceless symmetric tensor field on $\scri$. The full set of physically relevant properties of the space-time cannot be characterised at infinity without taking $\ct{D}{_{ab}}$ into consideration, and our radiation criterion takes this fully into account. We similarly propose a no-incoming radiation criterion based also on the triplet $(\scri,\ms{_{ab}},\ct{D}{_{ab}})$ and on radiant supermomenta deduced from the rescaled Bel-Robinson tensor too.
		We search for news tensors encoding the two degrees of freedom of gravitational radiation and argue that any news-like object must be associated to, and depends on, 2-dimensional cross-sections of $\scri$. We identify one component of news for every such cross-section and present a general strategy to find the second component, which depends on the particular physical situation.
		We put in connection the radiation condition and the news-like tensors with the directional structure of the gravitational field at infinity and the criterion of no-incoming radiation. We also introduce the concept of equipped $\scri$ by endowing the conformal boundary with a selected congruence of curves which may be determined by the algebraic structure of the asymptotic Weyl tensor. We also define a group of asymptotic symmetries preserving the new structures. We consider the limit $\Lambda\rightarrow 0$, and apply  all our results to selected exact solutions of Einstein Field Equations in order to illustrate their validity.
	\end{abstract}

	\newpage
	\tableofcontents
	\listoffigures
	\newpage
	\section{Introduction}\label{sec:introduction}

	This is the second of a couple of papers that we present simultaneously and complement each other. The subject of the papers is the asymptotic structure of space-times that admit a conformal completion {\em \`a la} Penrose with a non-negative cosmological constant $\Lambda \geq 0$. The present paper deals with $\Lambda >0$, while the companion \cite{Fernandez-Alvarez_Senovilla-afs} considers the better understood case with vanishing cosmological constant. Notation, nomenclature and some important ideas can be obtained from the Introduction of that first paper. Some of the notation used is summarised in \cref{tab:tablenotation} though.
	
	The geometry and physics present at infinity with a vanishing cosmological constant differ notably from the ones with a positive $ \Lambda $. These differences are discussed thoroughly in this paper. The fundamental distinctions emerge as a consequence of the change in the causal character of the conformal boundary, no matter \emph{how tiny} the cosmological constant is \cite{Ashtekar2017}. 
	Attention to the complications arising when $\Lambda>0$ was revived in \cite{Penrose2011} and later the problematic was exposed accurately in \cite{Ashtekar2014} from the point of view of an analysis of the topology and geometry of the conformal boundary, called $\scri$. As explained in \cite{Fernandez-Alvarez_Senovilla-afs}, such conformal boundary carries a universal structure when $\Lambda=0$ which is simply missing, in general, in the case with $\Lambda>0$. All possible 3-dimensional Riemannian manifolds are good enough, in principle, for $\scri$, and intrinsic asymptotic symmetries can be simply absent. One may try to restrict the asymptotic structure but, if the constraints are too strong one may lose too much information. 
	Recall that all conformally related metrics on $\scri$ describe the same physical situation, leading to the well-known conformal gauge freedom. Physically relevant results must be gauge independent.
	
		During the last years some understanding of the situation with $\Lambda >0$ has been accumulating and different approaches explored, which include a quadrupole formula in the linear regime \cite{Ashtekar2015,Hoque2019}, linear physical fields in de Sitter (dS) universe \cite{Ashtekar2015-b,Lewandowski2020}, radiation of black-holes in dS backgrounds \cite{Podolsky2000,Podolsky2009}, spinorial definitions of mass \cite{Szabados2013,Szabados2015}, spin-coefficient/coordinate-based approaches \cite{Saw2016,Saw2017,Saw2017ziu,Saw2018,Saw2018i,Poole2019}, gauge-fixing methods and foliations \cite{Compere2019,Compere2020}, power emitted by a binary system \cite{Bonga2017} and others \cite{He2016,Chrusciel2016,Chrusciel2013,Virmani2019,Dolan2019,Poole2019,Hoque2021} -- see \cite{Szabados2019} for a review. Despite these advances, the absence of a universal structure and of a general group of asymptotic symmetries together with the problem of the directional-dependence as one approaches infinity \cite{Krtous2004} makes the whole picture rather incomplete. This work tries to resolve many of the open problems, in particular concerning the characterisation of radiation at infinity with a positive cosmological constant, the existence of news-like tensors, the question of in- and out-going radiation, and the definition of asymptotic symmetries. For that, we study the asymptotic structure from the ground up, reviewing some basic ideas and giving the expressions of the relevant fields on the conformal boundary. 
		
		To that purposes we have changed completely the perspective and resort to considering the {\em tidal} nature of the gravitational field and the tidal energies associated to the Weyl curvature, represented by the Bel-Robinson tensor \cite{Bel1958,Senovilla2000}. This change of perspective allows us to avoid the (many times misleading) analogies based on the $\Lambda=0$ case, opening novel and independent routes and new aspects of the problem. In particular, as a fundamental result, a geometric and covariant characterisation of gravitational radiation at infinity in the presence of a positive cosmological constant was presented in \cite{Fernandez-Alvarez_Senovilla20b} and is thorougly discussed in the present paper. It is formulated in terms of the {\em super-Poynting vector} built from the rescaled Bel-Robinson tensor and relative to the preferred observer determined by the timelike normal to $\scri$. A crucial point is that our radiation criterion has been successfully tested in the asymptotically flat case \cite{Fernandez-Alvarez_Senovilla20} (in more detail in the companion paper \cite{Fernandez-Alvarez_Senovilla-afs}), where it is shown to be equivalent to the classical characterisation based on the News tensor.  
		
		The key point we would like to bring up is that the asymptotic physical properties of the space-times cannot be encoded on the (conformal) Riemannian manifold describing $\scri$ {\em exclusively}, one also needs to take into consideration a symmetric and trace-free tensor field $\ct{D}{_{ab}}$ defined on $\scri$ {\em but not determined by it}. In $\Lambda$-vacuum this tensor is also divergenceless, and thus a  `TT'-tensor. This tensor field is {\em inherited} from the ambient manifold: it represents the (rescaled) electric part of the Weyl tensor with respect to the timelike normal to $\scri$. Thus, the physical properties are appropriately encoded at infinity in the triplet $(\scri,\ms{_{ab}},\ct{D}{_{ab}})$, where the first two provide the manifold (with its topology) and the conformal class of Riemannian metrics, and the third leg is the essential inherited element which completes the picture. This perspective can be robustly founded in 
		\begin{itemize}
		\item the analysis performed in \cite{Starobinsky1983} of the Einstein field equations with a positive $\Lambda$, where an appropriate expansion of the metric shows that the first three terms are a spatial metric, its curvature, and a traceless symmetric tensor $\ct{D}{_{ab}}$ and that these three terms {\em determine the entire expansion}. 
		\item the important result proved in \cite{Friedrich1986a} stating that a solution of $\Lambda>0$-vacuum field equations is totally determined by data at infinity consisting on a conformal class of Riemannian metrics $\ms{_{ab}}$ on a 3-dimensional manifold $\scri$ together with a TT-tensor $\ct{D}{_{ab}}$.
		\end{itemize} 
		 Furthermore, this compares quite well with the $\Lambda=0$ case in which one also needs a third object, the inherited connection or its curvature, {\em not determined by the universal structure on $\scri$}, to completely determine the physical properties of the space-time and in particular to define the news tensor, see the companion paper for detailed explanations \cite{Fernandez-Alvarez_Senovilla-afs}. In summary, the entire information of the physical space-time is encoded in $(\scri,\ms{_{ab}},\ct{D}{_{ab}})$ and one cannot aspire to provide a characterisation of any relevant physical asymptotic property without taking $\ct{D}{_{ab}}$ into account --- including gravitational radiation at $\scri$, incoming/outcoming radiation, news, etc. Our radiation criterion takes this fully into account.
		
		We explore the possibility of defining news-like tensor fields {\em associated to 2-dimensional cross-sections of $\scri$}. This is an intricate task, mainly due to the possible mixing of incoming and outgoing radiation at infinity when $\Lambda >0$. To that end, we first generalise a result by Geroch \cite{Geroch1977} on the existence of a tensor called $\ct{\rho}{_{ab}}$ which was determinant in his covariant definition of the news tensor when $\Lambda=0$. We provide a neat geometric interpretation of this tensor as an object of 2-dimensional (conformal) Riemannian manifolds and clarify its role in the definition of news-like tensors. Thereby, we identify a first component of the news tensor for any cross-section of $\scri$ and provide a scheme of how to find the second component involving $\ct{D}{_{ab}}$. In particular, we introduce two news-like objects that we call {\em radiant news} and their relationship with the radiation criterion is discussed at large. 
		
		As a next step forward, we introduce the concept of {\em equipped} $\scri$, and their strict and strong versions. This basically requires the identification of a vector field intrinsic to $\scri$ which we would like to use as `evolution' direction. Such vector field defines a congruence of curves on $\scri$, and we argue that in some important cases it can be appropriately chosen by taking into account the principal null directions of the rescaled Weyl tensor at $\scri$, which determine what we call an {\em orientation} on $\scri$. If the congruence of curves is orthogonal to a family of 2-surfaces foliating $\scri$ then we say that the equipment is strict, see also \cite{Compere2019,Compere2020}; if in addition the leaves of the foliation are umbilical we speak of strong equipment. This latter case can be put in correspondence with properties of the lightlike normals to the leaves via a Goldberg-Sachs-type theorem that we establish.
		
		We then connect the radiation condition and the news-like tensors with the directional structure of the gravitational field at infinity, formulating a criterion of no-incoming radiation at equipped $\scri$, see also \cite{Ashtekar2019}. Once again, this criterion is based on the triplet $(\scri,\ms{_{ab}},\ct{D}{_{ab}})$, as it must, and on the radiant supermomenta built with the Bel-Robinson tensor of the rescaled Weyl tensor at $\scri$. It actually distinguishes a class of space-times worth exploring in which $\ct{D}{_{ab}}$ is determined by the intrinsic geometry of $(\scri,\ms{_{ab}})$ {\em except} for one single component related to the Coulomb part of the gravitational field. The combination of the radiation and the no-incoming radiation criteria leads to illuminating results. 
		
		The definition of asymptotic symmetries preserving the new structures is also considered. We introduce the concept of {\em basic asymptotic symmetry}, characterised by leaving invariant the fundamental structure $(\scri,\ms{_{ab}},\ct{D}{_{ab}})$, and shown to arise as the extensions to $\scri$ of Killing vector fields of the physical space-time. A second kind of asymptotic symmetries, relevant for equipped $\scri$, is also introduced and its structure analysed in detailed. We prove that they may form infinite-dimensional Lie algebras in some circumstances, and how they can also be obtained as the extension to $\scri$ of approximate symmetries in the physical space-time. Charges and balance laws are briefly considered using all these infinitesimal symmetries.
		
		
		Finally, we put all our results to test by discussing some exact solutions of Einstein's Field Equations in order to illustrate their validity. We consider de Sitter, Kerr-de Sitter, C-metric, and Robinson-Trautman space-times. 
		
		To end this long introduction, we would like to stress that what we provide is a general framework to study the asymptotic properties of the space-times with a positive cosmological constant and that, in particular, the question of whether or not gravitational radiation arrives to (or departs from) $\scri$ can be easily confirmed, with very elementary calculations (and easily implemented on a computer), using our radiation criterion.
		
		\subsubsection*{Notation}
				\begin{table}[ht!]
								\centering
								\begin{tabular}{ | M{2cm}| M{3cm} | M{3cm} |M{3cm}| N }
									\hline
									\quad& Notation for projected objects & Projector &  First appearance  & \\ \hline 
									\textbf{Spacelike hypersurfaces $ \I $, 3+1 decomposition}& $ \cts{T}{} $ & $ \ct{P}{^\alpha_{\beta}}$ & \Cref{eq:notation-3-1} on page \pageref{eq:notation-3-1}
									(see section \ref{app:spatial-hypersurface} of  \cref{app:spatial-hypersurfaces})  &\\[1cm] \hline 
									\textbf{Surfaces $ \Sc $, 2+1 decomposition}&	$\ctc{T}{}$& $ \ctc{P}{^\alpha_{\beta}} $  & \Cref{eq:notation-2-1} on page \pageref{eq:notation-2-1} (see section \ref{ssec:foliationAndCuts} of  \cref{app:spatial-hypersurfaces}) & \\[1cm]\hline
									\textbf{Projected surface $ \Scn  $}& $ 
									\ctcn{T}{}$	&$ \ctcn{P}{^\alpha_{\beta}} $	&	\Cref{eq:decomp-Schouten-m} and subsequent equations on page \pageref{eq:decomp-Schouten-m} (see section \ref{app:congruences} of \cref{app:spatial-hypersurfaces})& \\[1cm]\hline					
								\end{tabular}
								\caption[Additional notation]{The rest of the notation is summarised in \cite{Fernandez-Alvarez_Senovilla-afs}.}
								\label{tab:tablenotation}
							\end{table}
							

		
	\section{Superenergy}\label{sec:supernergy}
			
			The impossibility of defining a local energy-momentum tensor for the gravitational field does not rule out the possibility of describing the field's strength. Apart from the asymptotic definitions of energy mentioned in \cref{sec:introduction}, a local notion of \emph{energy density divided by area} exists. This particular weighting of the energy is naturally associated to tidal forces and is described by a rank-four tensor, quadratic in the Weyl curvature, called the Bel-Robinson tensor \cite{Bel1958}. Importantly, this object is defined locally and shows lots of analogies with the energy-momentum tensor of the electromagnetic field \cite{Maartens1998,Senovilla97}. As a matter of fact, it can not have physical units of energy density and this led to assign the name \emph{supernergy quantities} to objects defined with this tensor. Since its formulation, the superenergy turned out to be a useful tool in a variety of studies of gravity, such as the causal propagation of gravity \cite{Bonilla_Senovilla97}, the algebraic characterisation of the Weyl tensor \cite{Senovilla2011}, the formation of black holes \cite{Christodoulou2009} or the global non-linear stability of Minkowski space-time \cite{Christodoulou-Klainerman1993}. In addition, it emerges in the quasilocal formulations of energy \cite{Szabados2004} and can be formulated for fields other than gravity \cite{Chevreton64,Teyssandier1999,Senovilla2000} exhibiting interchanges of superenergy quantities and conserved mixed currents of different fields \cite{Senovilla2000,Senovilla2000b,Lazkoz2003}. Superenergy is one of the key ingredients in this work.\\
			
			There exists a general definition of superenergy tensor \cite{Senovilla2000} and, within that broader class, the Bel-Robinson tensor is the \emph{basic superenergy tensor} constructed with the Weyl tensor. Next subsections apply to any basic superenergy tensor built with a Weyl candidate, i. e., a traceless tensor $ \ct{W}{_{\alpha\beta\gamma}^\delta} $ sharing all the algebraic symmetries of the Weyl tensor. 
			 
		\subsection{Basic superenergy tensors}\label{ssec:basic-superenergy}
			The basic superenergy tensor $ \ct{\T}{_{\alpha\beta\gamma\delta}}\{\ct{W}{}\} $ constructed with a Weyl candidate $ \ct{W}{_{\alpha\beta\gamma\delta}} $ is defined in 4 dimensions as
				\begin{equation}\label{eq:def-BR-tensor}
					\ct{\T}{_{\alpha\beta\gamma\delta}}\defeq \ct{W}{_{\alpha\mu\gamma}^\nu}\ct{W}{_{\delta\nu\beta}^\mu} + \ctru{*}{W}{_{\alpha\mu\gamma}^\nu}\ctru{*}{W}{_{\delta\nu\beta}^\mu}=\ct{W}{_{\alpha\mu\gamma\nu}}\ct{W}{_\beta^\mu_\delta^\nu}+\ct{W}{_{\alpha\mu\delta\nu}}\ct{W}{_\beta^\mu_\gamma^\nu}-\frac{1}{8}\ct{g}{_{\alpha\beta}}\ct{g}{_{\gamma\delta}}\ct{W}{_{\mu\nu\rho\sigma}}\ct{W}{^{\mu\nu\rho\sigma}}.
				\end{equation}	
			Observe that, being quadratic in the Weyl candidate, its geometrised dimensions are $ L^{-4} $. However, the physical dimensions are energy-density per area, $ ML^{-3}T^{-2} $ \cite{Capella1961,Senovilla2000,Szabados2004,Horowitz82,Teyssandier1999}, and we presented yet another simple proof in \cite{Fernandez-Alvarez_Senovilla20}. The properties of this tensor include:
				\begin{properties}[label=\roman*)]
					\item it is completely symmetric $ \ct{\T}{_{(\alpha\beta\gamma\delta)}}= \ct{\T}{_{\alpha\beta\gamma\delta}} $,
					\item it is traceless $ \ct{\T}{^\mu_{\beta\gamma\mu}}=0 $,
					\item obeys a dominant superenergy property, $ \ct{\T}{_{\mu\nu\rho\sigma}}\ct{v}{^\mu}\ct{w}{^\nu}\ct{u}{^\rho}\ct{q}{^\sigma}\geq 0 $, where the four vectors are causal and future oriented. In particular,\label{it:dominant-se-condition}
					\item  $ \ct{\T}{^\alpha_{\nu\rho\sigma}}{k}{^\nu}{\ell}{^\rho}{z}{^\sigma}$ is future pointing and lightlike, if $ {k}{^\nu}$, ${\ell}{^\rho}$, ${z}{^\sigma} $ are lightlike and future oriented.  \cite{Penrose1986,Bergqvist2004}.\label{it:dominant-lightlike-se-condition}
				\end{properties}
			In addition to these algebraic properties, let us include a differential one,
				\begin{equation}\label{eq:div-superenergy-tensor}
					\cd{_\mu}\ct{\T}{_{\alpha\beta\gamma}^\mu}=2\ct{W}{_{\mu\gamma\nu\alpha}}\ct{Y}{_{\beta}^{\nu\mu}}+2\ct{W}{_{\mu\gamma\nu\beta}}\ct{Y}{_{\alpha}^{\nu\mu}}+ \ct{g}{_{\alpha\beta}}\ct{W}{^{\mu\nu\rho}_\gamma}\ct{Y}{_{\mu\nu\rho}}\spacef, 
				\end{equation}
			where
				\begin{equation}\label{eq:currentCandidate}
					 \ct{Y}{_{\alpha\beta\gamma}} \defeq -\cd{_\mu} \ct{W}{_{\alpha\beta\gamma}^\mu},
				\end{equation}
			and point out the following case:
				\begin{equation}
					 \ct{Y}{_{\alpha\beta\gamma}}=0 \implies \cd{_\mu}\ct{\T}{_{\alpha\beta\gamma}^\mu}=0\spacef. 
				\end{equation}
		 At this point, it is worth mentioning that for the particular case with $ \ct{\T}{_{\alpha\beta\gamma\delta}} $ constructed with the Weyl tensor, $ \ct{Y}{_{\alpha\beta\gamma}} $ is the Cotton tensor and, if Einstein's field equations hold, the absence of matter fields implies $ \ct{Y}{_{\alpha\beta\gamma}} =0$  and $\ct{\T}{_{\alpha\beta\gamma\mu}}$ is divergence-free.\\
		
		 Later, we will be interested in relating the algebraic classification of $ \ct{W}{_{\alpha\beta\gamma\delta}} $ with $  \ct{\T}{_{\alpha\beta\gamma\delta}} $ in a precise way. A characterisation of the Petrov type of $ \ct{W}{_{\alpha\beta\gamma}^\delta} $ and its repeated (or degenerated) principal null directions (PND) is  \cite{Bergqvist98,Senovilla2011}  
				\begin{itemize}\label{it:petrov-class-BR}
					\item $\ct{\T}{_{\alpha\beta\gamma\mu}}\ct{k}{^\mu} =0$ and $ \ct{\T}{_{\alpha\beta\gamma\delta}}\neq 0 $ $\Longleftrightarrow$ Petrov type N and $ \ct{k}{^\alpha} $ is a quadruple PND.
					\item $\ct{\T}{_{\alpha\beta\nu\mu}}\ct{k}{^\nu}\ct{k}{^\mu} =0$ and $\ct{\T}{_{\alpha\beta\gamma\mu}}\ct{k}{^\mu} \neq 0$ $\Longleftrightarrow$ Petrov type III and $ \ct{k}{^\alpha} $ is a triple PND.
					\item $\ct{\T}{_{\alpha\rho\nu\mu}}\ct{k}{^\rho}\ct{k}{^\nu}\ct{k}{^\mu} =0$ and $\ct{\T}{_{\alpha\beta\nu\mu}}\ct{k}{^\nu}\ct{k}{^\mu} \neq 0$ $\Longleftrightarrow$ Petrov type II or D and $ \ct{k}{^\alpha} $ is a double PND.
					\item 	$\ct{\T}{_{\sigma\rho\nu\mu}}\ct{k}{^\sigma}\ct{k}{^\rho}\ct{k}{^\nu}\ct{k}{^\mu} =0$  and $\ct{\T}{_{\alpha\rho\nu\mu}}\ct{\ell}{^\rho}\ct{\ell}{^\nu}\ct{\ell}{^\mu}\neq 0\quad\forall \text{ lightlike } \ct{\ell}{^\alpha}$  $\Longleftrightarrow$ Petrov type I and $ \ct{k}{^\alpha} $ is a non-degenerate PND.
					\item  $\ct{\T}{_{\alpha\rho\nu\mu}}\ct{k}{^\rho}\ct{k}{^\nu}\ct{k}{^\mu} =0$, $\ct{\T}{_{\alpha\rho\nu\mu}}\ct{\ell}{^\rho}\ct{\ell}{^\nu}\ct{\ell}{^\mu} =0$ and $ \ct{\T}{_{\alpha\beta\gamma\delta}}\neq 0 $  $ \Longleftrightarrow $ Petrov type D and $ \ct{k}{^\alpha} $, $ \ct{\ell}{^\alpha} $ are double PND.
				\end{itemize}
			 For a detailed description of these and more general properties, see \cite{Senovilla2000,Senovilla2011} and references therein.

			\subsubsection{Orthogonal decomposition}\label{sssec:orthogonal-decomposition}
	 			Choose a unit timelike future-pointing vector field $ \ct{u}{^\alpha} $. At each point, introduce a basis $ \{\ct{e}{^\alpha_a} \}$ spanning the vector space whose elements are all the vectors orthogonal to $ \ct{u}{^\alpha} $ at that point. Also, define a basis $ \{\ct{\omega}{_\alpha^a}\}$ for the dual space of one-forms such that they are orthogonal to $ \ct{u}{^\alpha} $. We call these objects orthogonal to $ \ct{u}{^\alpha} $ `spatial with respect to $ \ct{u}{^\alpha} $' or, for simplicity, spatial ---Latin indices denote spatial components and run from 1 to 3. Let $ \ct{q}{^\alpha}=\ct{e}{^\alpha_a}\ct{q}{^a} $ be any spatial vector,  $ \ct{q}{^\alpha}\ct{u}{_\alpha}=0 $. A projector can be defined,
	 				\begin{equation}
	 					\ct{P}{^\alpha_\beta}\defeq\ct{e}{^\alpha_i}\ct{\omega}{_\beta^i}= \ct{\delta}{^\alpha_\beta}+\ct{u}{^\alpha}\ct{u}{_\beta},\quad\ct{P}{^\alpha_\mu}\ct{u}{^\mu}=0,\quad\ct{P}{^\alpha_\mu}\ct{q}{^\mu}=\ct{q}{^\alpha}\spacef.
	 				\end{equation} 
	 			In this way, any space-time vector $ \ct{w}{^\alpha} $ can be decomposed as a spatial part with respect to $ \ct{u}{^\alpha} $ together with another one tangent to $ \ct{u}{^\alpha} $,
	 				\begin{equation}\label{eq:notation-3-1}
	 					\ct{w}{^\alpha}=-\ct{u}{^\alpha}\ct{u}{_\mu}\ct{w}{^\mu}+\cts{w}{^\alpha},\text{ with }\cts{w}{^\alpha}=\cts{w}{^p}\ct{e}{^\alpha_p}\spacef,
	 				\end{equation}
	 			such that 
	 				\begin{equation}
	 					\ct{P}{^\alpha_\mu}\ct{w}{^\mu}=\cts{w}{^\alpha}\spacef.
	 				\end{equation}
	 			This is generalised to higher-rank tensors in an obvious way. Latin indices are raised and lowered with 
	 				\begin{equation}
	 					\ms{_{ab}}\defeq\ct{e}{^\alpha_a}\ct{e}{^\beta_b}\ct{g}{_{\alpha\beta}},\quad\ms{^{ab}}\defeq\ct{\omega}{_\alpha^a}\ct{\omega}{_\beta^b}\ct{g}{^{\alpha\beta}}\spacef.
	 				\end{equation}
	 			\\
	 			
	 			 With this, it is possible to decompose the Weyl candidate and $ \ct{\T}{^\alpha_{\beta\gamma\delta}} $ into tangent and spatial parts with respect to $ \ct{u}{^\alpha} $. The former is fully determined by its \emph{electric} and \emph{magnetic} parts with respect to $ \ct{u}{^\alpha} $
	 				\begin{align}
		 				\ct{D}{_{\alpha\beta}}&\defeq \ct{u}{^\mu}\ct{u}{^\nu}\ct{W}{_{\mu\alpha\nu\beta}}\spacef,\\
		 				\ct{C}{_{\alpha\beta}}&\defeq \ct{u}{^\mu}\ct{u}{^\nu}\ctru{*}{W}{_{\mu\alpha\nu\beta}}\spacef.
	 				\end{align}
	 			which are symmetric, traceless, spatial fields, 
	 				\begin{align}
		 				\ct{D}{_{\alpha\beta}}&= \ct{D}{_{ab}}\ct{\omega}{_\alpha^a}\ct{\omega}{_\beta^b}\spacef,\\
		 				\ct{C}{_{\alpha\beta}}&= \ct{C}{_{ab}}\ct{\omega}{_\alpha^a}\ct{\omega}{_\beta^b}\spacef.
	 				\end{align}
	 			The full splitting of $ \ct{\T}{_{\alpha\beta\gamma\delta}} $ is 
	 				\begin{equation}\label{eq:BR-orthogonaldecomp}
	 					\ct{\T}{_{\alpha\beta\gamma\delta}}=  \cs{\W}\ct{u}{_\alpha}\ct{u}{_\beta}\ct{u}{_\gamma}\ct{u}{_\delta}+ 4\cts{\P}{_{(\alpha}}\ct{u}{_{\beta}}\ct{u}{_{\gamma}}\ct{u}{_{\delta)}}+4\ct{u}{_{(\alpha}}\ct{Q}{_{\beta\gamma\delta)}}+6\ct{t}{_{(\alpha\beta}}\ct{u}{_{\gamma}}\ct{u}{_{\delta)}}+\ct{t}{_{\alpha\beta\gamma\delta}}\spacef.
	 				\end{equation}
	 			We are interested in the first three terms on the right-hand side. They have obvious definitions in terms of the projector and $ \ct{u}{^\alpha} $. In particular, $ \cts{\P}{^\alpha} $ and $ \ct{Q}{_{\alpha\beta\gamma}} $ are spatial, i. e., $ \cts{\P}{^\alpha} =\ct{e}{^\alpha_a}\cts{\P}{^a}$, $ \ct{Q}{_{\alpha\beta\gamma}}=\ct{\omega}{_\alpha^a}\ct{\omega}{_\beta^b}\ct{\omega}{_\gamma^c}\ct{Q}{_{abc}} $. The first two are called the \emph{superenergy density} and \emph{super-Poynting} vector field, and have the following expressions in terms of the electric and magnetic parts of the Weyl candidate (\cite{Bel1962,Maartens1998}):
	 				\begin{align}
		 				\W&=  \ct{D}{_{ab}} \ct{D}{^{ab}}+  \ct{C}{_{ab}} \ct{C}{^{ab}}\spacef,\label{eq:superenergy}\\
		 				\cts{\P}{^a}&= \commute{C}{D}_{rs}\ct{\epsilon}{^{rsa}}= 2\ct{C}{_r^t}\ct{D}{_ {ts}}\ct{\epsilon}{^{rsa}}\spacef,\label{eq:commutator-se}
	 				\end{align}
	 			where  we have defined an alternating 3-dimensional tensor 
	 				\begin{equation}
	 					 -\ct{u}{_\alpha}\ct{\epsilon}{_{abc}} \defeq \ct{\eta}{_{\alpha\beta\gamma\delta}}\ct{e}{^\beta_a}\ct{e}{^\gamma_b}\ct{e}{^\delta_c}
	 				\end{equation}
	 			Also, the third one can be expressed as \cite{Alfonso2008}
	 				\begin{equation}
	 					\ct{Q}{_{\alpha\beta\gamma}}=\ct{P}{_{\alpha\beta}}\cts{\P}{_\gamma}-2\prn{\ct{D}{_{\alpha\mu}}\ct{C}{_{\beta\nu}}+\ct{D}{_{\beta\mu}}\ct{C}{_{\alpha\nu}}}\ct{u}{^\rho}\ct{\eta}{_{\rho\gamma}^{\nu\mu}}\spacef,\quad\ct{Q}{^\mu_{\mu\gamma}}=\cts{\P}{_\gamma}\spacef.\label{eq:Qalfonso}
	 				\end{equation}
	 			The superenergy density and the super-Poynting vector field inherit their name from the analogy with electromagnetism when $ \ct{W}{_{\alpha\beta\gamma}^\delta} $ is the Weyl tensor: the former represents energy density per unit area and the latter, the spatial direction of propagation of supernergy with respect to $ \ct{u}{^\alpha} $. Following this analogy, one can also define a supermomentum
	 				\begin{equation}\label{eq:supermomentum}
 						\ct{\P}{^\alpha} \defeq -\ct{u}{^\mu}\ct{u}{^\nu}\ct{u}{^\rho}\ct{\T}{^\alpha_{\mu\nu\rho}} = \cs{\W}\ct{u}{^\alpha} + \cts{\P}{^\alpha}\spacef,
	 				\end{equation}
	 			which is non-spacelike and future pointing. One important feature is that
	 				\begin{equation}\label{eq:noWnoT}
	 					\ct{W}{_{\alpha\beta\gamma}^\delta}=0\iff\ct{\T}{_{\alpha\beta\gamma\delta}}=0\iff \cs{\W}=0\spacef.
	 				\end{equation}
	 			Particularly inspiring for us is the following definition by Bel \cite{Bel1962}:
	 			\begin{deff}
	 				There is a state of intrinsic gravitational radiation at a point $ p $ when $ \cts{\P}{^\alpha}|_p \neq 0$ for all unit timelike $\ct{u}{^\alpha}  $.\label{Bel}
	 			\end{deff}
			 This classical definition agrees with the discussion by Pirani \cite{Pirani57} and is based on the analogy with null electromagnetic fields which are precisely the fields with a non-zero Poynting vector for all possible observers. More recently, another similar characterization was put forward in \cite{Alfonso2008} in the following terms
			\begin{deff}
	 				There is a superenergy state of intrinsic gravitational radiation at a point $ p $ when $ \ct{Q}{_{\alpha\beta\gamma}}|_p \neq 0$ for all unit timelike $\ct{u}{^\alpha}  $.\label{Alfonso}
	 			\end{deff}
				Using (\ref{eq:Qalfonso}) it is easy to check that every state of intrinsic gravitational radiation is also a superenergy such state, but there are more cases of the latter in general.
 			
		\subsection{Lightlike projections}\label{ssec:lightlike-projections}
		 	Let $ \ct{u}{^\alpha} $ be the unit timelike vector field, as defined in previous subsection, and $ \ct{r}{^\alpha} $ a unit, vector (field) --non necessarily defined everywhere-- spatial with respect to $ \ct{u}{^\alpha} $, i. e., $ \ct{r}{_\alpha}\ct{u}{^\alpha}= 0 $, $ \ct{r}{^\alpha}=\ct{r}{^a}\ct{e}{^\alpha_a} $. There are two (up to a boost) independent lightlike directions coplanar with $ \ct{u}{^\alpha} $ and $ \ct{r}{^\alpha} $:
				\begin{align}
					\ctp{k}{^\alpha}&\defeq\frac{1}{\sqrt{2}}\left(\ct{u}{^\alpha}+\ct{r}{^\alpha}\right)\spacef,\label{eq:kp-def}\\
					\ctm{k}{^\alpha}&\defeq\frac{1}{\sqrt{2}}\left(\ct{u}{^\alpha}-\ct{r}{^\alpha}\right)\quad\label{eq:km-def}
				\end{align}
			such that $ \ctp{k}{^\alpha}\ctm{k}{_\alpha}=-1 $. At each point, introduce a basis for the vector space constituted by all vectors orthogonal to $ \ctp{k}{_\alpha} $, $ \{\ctp{e}{^\alpha_{\hat{a}}}\}=\{\ctp{k}{^\alpha},\ct{E}{^\alpha_A}\}$, and do the same with respect to $ \ctm{k}{_\alpha} $, $ \{\ctm{e}{^\alpha_{\tilde{a}}}\}=\{\ctm{k}{^\alpha},\ct{E}{^\alpha_A}\}$. Notice that, as these vector fields are lightlike, $ \ctm{k}{^\alpha}=\ctm{k}{^{\tilde{1}}}\ctm{e}{^\alpha_{\tilde{1}}} $ and $ \ctp{k}{^\alpha}=\ctp{k}{^{\hat{1}}}\ctp{e}{^\alpha_{\hat{1}}} $. One can introduce dual bases $ \{\ctp{\omega}{_\alpha^{\hat{a}}}\}=\{-\ctm{k}{_\alpha},\ct{W}{_\alpha^A}\}$, $ \{\ctm{\omega}{_\alpha^{\tilde{a}}}\}=\{-\ctp{k}{_\alpha},\ct{W}{_\alpha^A}\} $, such that $ \ctp{k}{^\alpha}\ctm{\omega}{_\alpha^{\tilde{a}}} =0$, $ \ctm{k}{^\alpha}\ctp{\omega}{_\alpha^{\hat{a}}} = 0$, $  \ctm{k}{^\alpha}\ctm{\omega}{_\alpha^{\tilde{a}}} = \ctm{k}{^{\tilde{a}}} $, $  \ctp{k}{^\alpha}\ctp{\omega}{_\alpha^{\hat{a}}} = \ctp{k}{^{\hat{a}}} $. Here, $ \{\ct{E}{^\alpha_A}\} $, $ \{\ct{W}{_\alpha^A}\} $ are bases spanning the two-dimensional vector space orthogonal to $ \ct{r}{^\alpha} $ and $ \ct{u}{^\alpha} $ --equivalently, orthogonal to $ \ctpm{k}{^\alpha} $. 
			Any space-time vector $ \ct{w}{^\alpha} $ decomposes into  a part tangent to $ \ct{u}{^\alpha} $ and a spatial part, $ \cts{w}{^\alpha} $, which splits into a part tangent to $ \ct{r}{^\alpha} $ and another one that is orthogonal to both vector fields $ \ct{u}{^\alpha} $ and $ \ct{r}{^\alpha} $,  $ \ctc{w}{^\alpha}=\ctc{w}{^A}\ct{E}{^\alpha_A} $, 
				\begin{equation}\label{eq:notation-2-1}
					\ct{w}{^\alpha}=-\ct{w}{_\mu}\ct{u}{^\mu}\ct{u}{^\alpha}+\cts{w}{^\alpha}=-\ct{w}{_\mu}\ct{u}{^\mu}\ct{u}{^\alpha}+\ct{w}{_\mu}\ct{r}{^\mu}\ct{r}{^\alpha}+\ctc{w}{^\alpha}\spacef.
				\end{equation}
			The object
				\begin{equation}
				\ctc{P}{^\alpha_\beta} = \ct{\delta}{^\alpha_\beta}+\ct{u}{^\alpha}\ct{u}{_\beta}-\ct{r}{^\alpha}\ct{r}{_\beta}=\ct{\delta}{^\alpha_\beta}+\ctm{k}{^\alpha}\ctp{k}{_\beta}+\ctp{k}{^\alpha}\ctm{k}{_\beta}\quad
				\end{equation}
			is the projector orthogonal to $ \ct{n}{^\alpha} $ and $ \ct{r}{^\alpha} $
				\begin{equation}
				\ctc{P}{^\alpha_\beta}\ct{w}{^\beta}=\ctc{w}{^\beta},\quad\ctc{P}{^\alpha_\beta}\ct{r}{_\alpha}=\ctc{P}{^\alpha_\beta}\ct{u}{_\alpha}=0\spacef.
				\end{equation}
			Again, all this is generalised to higher-rank tensors in a natural way.\\
			
		 	There are some useful lightlike projections of the Weyl candidate we are interested in. Some of them are expressed in terms of $ \{\ctp{e}{^\alpha_{\hat{a}}}\} $ and $ \{\ctp{\omega}{_\alpha^{\hat{a}}}\} $ only, and are orthogonal to $ \ctp{k}{_\alpha} $ in their contravariant indices and to $ \ctm{k}{^\alpha} $ in the covariant ones:
				\begin{align}
				\ctp{D}{^{\alpha\beta}}&\defeq\ctp{k}{^\mu}\ctp{k}{^\nu}\ct{W}{_\mu^\alpha_\nu^\beta}=\ctp{D}{^{\hat{a}\hat{b}}}\ctp{e}{^\alpha_{\hat{a}}}\ctp{e}{^\beta_{\hat{b}}}\spacef,\label{eq:DpDef}\\					\ctp{C}{^{\alpha\beta}}&\defeq\ctp{k}{^\mu}\ctp{k}{^\nu}\ctru{*}{W}{_\mu^\alpha_\nu^\beta}=\ctp{C}{^{\hat{a}\hat{b}}}\ctp{e}{^\alpha_{\hat{a}}}\ctp{e}{^\beta_{\hat{b}}}\spacef,\label{eq:CpDef}\\					\ctrmp{D}{_\alpha^\beta}&\defeq\ctm{k}{^\mu}\ctp{k}{^\nu}\ct{W}{_{\mu\alpha\nu}^\beta}=\ctrmp{D}{_{\hat{a}}^{\hat{b}}}\ctp{\omega}{_\alpha^{\hat{a}}}\ctp{e}{^\beta_{\hat{b}}}\spacef,\\
				\ctrmp{C}{_\alpha^\beta}&\defeq\ctm{k}{^\mu}\ctp{k}{^\nu}\ctru{*}{W}{_{\mu\alpha\nu}^\beta}=\ctrmp{C}{_{\hat{a}}^{\hat{b}}}\ctp{\omega}{_\alpha^{\hat{a}}}\ctp{e}{^\beta_{\hat{b}}}\spacef,
				\end{align}
			whereas others are written in terms of $ \{\ctm{e}{^\alpha_{\tilde{a}}}\} $ and $ \{\ctm{\omega}{_\alpha^{\tilde{a}}}\}$, and are orthogonal to $ \ctm{k}{_\alpha} $ in their contravariant indices and to $ \ctp{k}{^\alpha} $ in the convariant ones:
				\begin{align}
				\ctm{D}{^{\alpha\beta}}&\defeq\ctm{k}{^\mu}\ctm{k}{^\nu}\ct{W}{_\mu^\alpha_\nu^\beta}=\ctm{D}{^{\tilde{k}\tilde{l}}}\ctm{e}{^\alpha_{\tilde{k}}}\ctm{e}{^\beta_{\tilde{l}}}\spacef,\\					\ctm{C}{^{\alpha\beta}}&\defeq\ctm{k}{^\mu}\ctm{k}{^\nu}\ctru{*}{W}{_\mu^\alpha_\nu^\beta}=\ctm{C}{^{\tilde{k}\tilde{l}}}\ctm{e}{^\alpha_{\tilde{k}}}\ctm{e}{^\beta_{\tilde{l}}}\spacef,\\					\ctrpm{D}{_\alpha^\beta}&\defeq\ctp{k}{^\mu}\ctm{k}{^\nu}\ct{W}{_{\mu\alpha\nu}^\beta}	=\ctrpm{D}{_{\tilde{k}}^{\tilde{l}}}\ctm{\omega}{_\alpha^{\tilde{k}}}\ctm{e}{^\beta_{\tilde{l}}}\spacef,\\
				\ctrpm{C}{_\alpha^\beta}&\defeq\ctp{k}{^\mu}\ctm{k}{^\nu}\ctru{*}{W}{_{\mu\alpha\nu}^\beta}=\ctrpm{C}{_{\tilde{k}}^{\tilde{l}}}\ctm{\omega}{_\alpha^{\tilde{k}}}\ctm{e}{^\beta_{\tilde{l}}}\spacef.
				\end{align}
			Notice that these quantities are not completely independent from each other but all the information of $ \ct{W}{_{\alpha\beta\gamma}^\delta} $ is contained in the first pair of the upper set plus the first pair of the lower one. In terms of the Weyl scalars, the first set of equations contains $ \ct{\phi}{_{2,3,4}} $; the second, $ \ct{\phi}{_{2,0,1}} $ --- see \cref{app:NPformulation}. \\
			
			It will turn out to be very practical to introduce the following notation for any symmetric tensor $ \ct{B}{_{\mu\nu}}$:
				\begin{equation}
				\ct{B}{_{\alpha\beta}}=\ct{u}{^{\mu}}\ct{u}{^{\nu}}\ct{B}{_{\mu\nu}}\ct{u}{_{\alpha}}\ct{u}{_{\beta}}+\ct{u}{^{\mu}}\ctc{P}{^{\nu}_{(\alpha}}\ct{u}{_{\beta)}} \ct{B}{_{\mu\nu}}+2\ct{B}{_{\mu}}\ct{u}{^\mu}\ct{r}{_{(\alpha}}\ct{u}{_{\beta)}}+\ct{r}{_\alpha}\ct{r}{_\beta}\cs{B}+2\ctc{B}{_{(\alpha}}\ct{r}{_{\beta)}}+\ctc{B}{_{\alpha\beta}}\spacef,
				\end{equation}
			and 
				\begin{equation}
					\ctt{B}{_{\alpha\beta}}\defeq \ctc{B}{_{\alpha\beta}}-\frac{1}{2}\ctc{P}{_{\alpha\beta}}\ctc{P}{^{\mu\nu}}\ctc{B}{_{\mu\nu}}\spacef, \quad  \ctt{B}{^{\rho}_{\rho}}=0
				\end{equation}
			where
				\begin{equation}\label{eq:notation-symmetric-tensor}
					\ctc{B}{_{\alpha\beta}}\defeq \ctc{P}{^\mu_\alpha}\ctc{P}{^\nu_\beta}\ct{B}{_{\mu\nu}},\quad\ct{B}{_{\alpha}}\defeq \ct{r}{^\mu}\ct{B}{_{\mu\alpha}},\quad	\ctc{B}{_{\alpha}}\defeq \ctc{P}{^\nu_\alpha}\ct{r}{^\mu}\ct{B}{_{\mu\nu}},\quad\cs{B}\defeq \ct{r}{^\mu}\ct{r}{^\nu}\ct{B}{_{\mu\nu}}.
				\end{equation}
			Obviously, $ 	\ctc{B}{_{\alpha\beta}} $ and $ \ctc{B}{_\alpha} $ are orthogonal to $ \ct{r}{^\alpha} $ and $ \ct{u}{^\alpha} $. The same notation will be used with uppercase indices. Define 
			$$ \mc{_{AB}}\defeq \ct{E}{^\alpha_A}\ct{E}{^\beta_B}\ctc{P}{_{\alpha\beta}} = \ct{E}{^\alpha_A}\ct{E}{^\beta_B}\ct{g}{_{\alpha\beta}} $$ 
			and 
			$$ \mc{^{AB}}\defeq \ct{W}{_\alpha^A}\ct{W}{_\beta^B}\ctc{P}{^{\alpha\beta}} = \ct{W}{_\alpha^A}\ct{W}{_\beta^B}\ct{g}{^{\alpha\beta}}$$
			 to lower and raise capital Latin indices. Then,
				\begin{equation}\label{eq:notation-symmetric-tensorLatin}
				\ctc{B}{_{AB}}\defeq \ct{E}{^\alpha_A}\ct{E}{^\beta_B}\ctc{B}{_{\alpha\beta}} ,\quad	\ctt{B}{_{AB}}\defeq \ctc{B}{_{AB}}-\frac{1}{2}\mc{_{AB}}\ctc{B}{^C_{C}},\quad \ctc{B}{_{A}}\defeq \ct{E}{^A_\alpha}\ct{r}{^\mu}\ct{B}{_{\mu\alpha}},\quad  \ctt{B}{^{A}_{A}}=0.
				\end{equation}
			
			 An alternating two-dimensional tensor can be defined by
				\begin{equation}
					\ct{r}{_m}\ctc{\epsilon}{_{AB}}= \ct{\epsilon}{_{mab}}\ct{E}{^a_A}\ct{E}{^b_B}\spacef.
				\end{equation}
			 A list of properties of these quantities has been placed in \cref{app:lightlike-projections}. \\
			 
			 The electric and magnetic parts of the rescaled Weyl tensor read:
			 \begin{align}
				\ct{D}{_{ab}}=&D \prn{\ct{r}{_a}\ct{r}{_b}-\frac{1}{2}\ctc{P}{_{ab}}}+2\ct{r}{_{(a}}\ct{W}{_{b)}^B} \ctc{D}{_B}+\ct{W}{_a^A}\ct{W}{_b^B}\ctt{D}{_{AB}}\spacef,\label{eq:Dab-decomposition0}\\
				\ct{C}{_{ab}}=&C \prn{\ct{r}{_a}\ct{r}{_b}-\frac{1}{2}\ctc{P}{_{ab}}}+2\ct{r}{_{(a}}\ct{W}{_{b)}^B} \ctc{C}{_B}+\ct{W}{_a^A}\ct{W}{_b^B}\ctt{C}{_{AB}}\quad \label{eq:Cab-decomposition0}
				\end{align}
			 and equivalently, in terms of the lightlike components that we have just presented,
				\begin{align}
				\ct{D}{_{ab}}=& \ctmp{k}{^\mu}\ctmp{k}{^\nu}\ctpm{D}{_{_{\mu\nu}}}\ct{r}{_a}\ct{r}{_b}+2\ct{r}{_{(a}}\ct{W}{_{b)}^B}\prn{\ctcp{D}{_B}+\ctcm{D}{_B}}+\nonumber\\
				&\frac{1}{2}\ct{W}{_a^A}\ct{W}{_b^B}\prn{\ctcp{D}{_{AB}}+\ctcm{D}{_{AB}}}-\frac{1}{2}\ctmp{k}{^\mu}\ctmp{k}{^\nu}\ctpm{D}{_{\mu\nu}}\ctc{P}{_{ab}}\spacef,\label{eq:Dab-decomposition}\\
				\ct{C}{_{ab}}=& \ctmp{k}{^\mu}\ctmp{k}{^\nu}\ctpm{C}{_{_{\mu\nu}}}\ct{r}{_a}\ct{r}{_b}+2\ct{r}{_{(a}}\ct{W}{_{b)}^B}\prn{\ctcp{C}{_B}+\ctcm{C}{_B}}+\nonumber\\
				&\frac{1}{2}\ct{W}{_a^A}\ct{W}{_b^B}\prn{\ctcp{C}{_{AB}}+\ctcm{C}{_{AB}}}-\frac{1}{2}\ctmp{k}{^\mu}\ctmp{k}{^\nu}\ctpm{C}{_{\mu\nu}}\ctc{P}{_{ab}}\spacef.\label{eq:Cab-decomposition}
				\end{align}
			Another quantity that will appear later on is:
				\begin{equation}\label{eq:tABC}
				\ctcupm{t}{_{ABC}}\defeq \ct{E}{^\alpha_A}\ct{E}{^\beta_B}\ct{E}{^\gamma_C}\ctpm{k}{^\mu}\ct{W}{_{\alpha\beta\gamma\mu}}=\pm 2\sqrt{2}\mc{_{C[A}}\ctcupm{D}{_{B]}}\spacef.\\
				\end{equation}

			\subsection{Radiant superenergy}\label{ssec:radiant-superenergy}
			Now, we introduce a new kind of superenergy quantities associated to lightlike directions. The use of these objects helps in identifying the radiative sectors of the superenergy tensor. For such reason, we refer to them as \emph{radiant superenergy} quantities. Given a future-oriented lightlike vector field $ \ct{\ell}{^\alpha} $ we define its associated \emph{radiant supermomentum},
				\begin{equation}\label{eq:radiant-sm-def}
					\ctl{\Q}{^\alpha}\defeq -\ct{\ell}{^\mu}\ct{\ell}{^\nu}\ct{\ell}{^\sigma}\ct{\T}{^\alpha_{\mu\nu\sigma}}\spacef.
				\end{equation}
			Given any lightlike vector field $ \ct{k}{^\alpha} $ such that $ \ct{k}{_\alpha}\ct{\ell}{^\alpha}=-1 $, $ \ct{\Q}{^\alpha} $ decomposes as\footnote{The underlining used here should not be confused with the short bar placed under quantities associated to a congruence (compare to the notation of \cref{app:congruences}).}
				\begin{equation}
					\ctl{\Q}{^\alpha}=\csl{\W}\ct{k}{^\alpha}+\ctsl{\Q}{^\alpha}=\csl{\W}\ct{k}{^\alpha}+\csl{\Z}\ct{\ell}{^\alpha} + \prn{\delta^\alpha_\mu+\ct{\ell}{^\alpha}\ct{k}{_\mu}+\ct{k}{^\alpha}\ct{\ell}{_\mu}}\ctsl{\Q}{^\mu}\spacef,\label{eq:radiant-sm-decomp}
				\end{equation}
			where $ \ctsl{\Q}{^\alpha} $ is the \emph{radiant super-Poynting vector} and $ \csl{\W} $ and $ \csl{\Z} $ the corresponding transverse and longitudinal\footnote{For the Bel-Robinson tensor, the `transverse' and `longitudinal'  modes of the gravitational radiation  determine completely $ \csl{\W} $ and $ \csl{\Z} $, respectively. This can be easily inferred from the expressions of these quantities in terms of the Weyl candidate scalars in \cref{app:NPformulation}.} \emph{radiant superenergy densities}. Note that $ \csl{\Z} $ and $ \prn{\delta^\alpha_\mu+\ct{\ell}{^\alpha}\ct{k}{_\mu}+\ct{k}{^\alpha}\ct{\ell}{_\mu}}\ctsl{\Q}{^\mu} $ depend on the choice of $ \ct{k}{^\alpha} $. In particular, for the previously defined $ \ctpm{k}{^\alpha} $, the supermomenta read
				\begin{align}
				\ctp{\Q}{^\alpha}&\defeq -\ctp{k}{^\mu}\ctp{k}{^\nu}\ctp{k}{^\rho}\ct{\T}{^\alpha_{\mu\nu\rho}}=\csp{\W}\ctm{k}{^\alpha}+\ctps{\Q}{^\alpha}=\csp{\W}\ctm{k}{^\alpha}+\ctps{\Q}{^a}\ctp{e}{^\alpha_a}\spacef,\label{eq:Qpdef}\\
				\ctm{\Q}{^\alpha}&\defeq -\ctm{k}{^\mu}\ctm{k}{^\nu}\ctm{k}{^\rho}\ct{\T}{^\alpha_{\mu\nu\rho}}=\csm{\W}\ctp{k}{^\alpha}+\ctms{\Q}{^\alpha}=\csm{\W}\ctp{k}{^\alpha}+\ctms{\Q}{^k}\ctm{e}{^\alpha_k}\spacef.\label{eq:Qmdef}
				\end{align} 
			where
				\begin{align}
				\ctps{\Q}{^a}&= \csp{\Z}\ctp{k}{^a} + \ctps{\Q}{^A}\ct{E}{^a_A}\spacef,\label{eq:Qbarpdef}\\
				\ctms{\Q}{^k}&= \csm{\Z}\ctm{k}{^k} + \ctms{\Q}{^A}\ct{E}{^k_A}\spacef,\label{eq:Qbarmdef}
				\end{align}
			Also, the following formulae hold\footnote{This can be done for any lightlike vector fields $ \ct{\ell}{^\alpha} $ and $ \ct{k}{^\alpha} $ as defined above. Just for convenience, we present them for $ \ctpm{k}{^\alpha} $ .}	
			\begin{align}
							\csp{\W}&\defeq -\ctp{k}{_\mu}\ctp{\Q}{^\mu}=2\ctp{C}{_{\mu\nu}}\ctp{C}{^{\mu\nu}}=2\ctp{D}{_{\mu\nu}}\ctp{D}{^{\mu\nu}}= 2\ctcp{C}{_{AB}}\ctcp{C}{^{AB}}=2\ctcp{D}{_{AB}}\ctcp{D}{^{AB}}\geq 0 \label{eq:Wpdef},\\
							\csm{\W}&\defeq -\ctm{k}{_\mu}\ctm{\Q}{^\mu}=2\ctm{C}{_{\mu\nu}}\ctm{C}{^{\mu\nu}}=2\ctm{D}{_{\mu\nu}}\ctm{D}{^{\mu\nu}}= 2\ctcm{C}{_{AB}}\ctcm{C}{^{AB}}=2\ctcm{D}{_{AB}}\ctcm{D}{^{AB}}\geq 0\label{eq:Wmdef},\\
							\csp{\Z}&\defeq -\ctm{k}{_\mu}\ctp{\Q}{^\mu}=2\ctrmp{C}{_{\mu\nu}}\ctp{C}{^{\mu\nu}}= 2\ctrmp{D}{_{\mu\nu}}\ctp{D}{^{\mu\nu}}=4\ctcp{C}{_A}\ctcp{C}{^A}\geq 0 \label{eq:Zpdef},\\
							\csm{\Z}&\defeq -\ctp{k}{_\mu}\ctm{\Q}{^\mu}=2\ctrpm{C}{_{\mu\nu}}\ctm{C}{^{\mu\nu}}= 2\ctrpm{D}{_{\mu\nu}}\ctm{D}{^{\mu\nu}}=4\ctcm{C}{_A}\ctcm{C}{^A}\geq 0\label{eq:Zmdef},\\
							\ctp{\Q}{^A}&= 4\sqrt{2}\ctcp{C}{_P}\ctcp{C}{^{AP}},\\
							\ctm{\Q}{^A}&= -4\sqrt{2}\ctcm{C}{_P}\ctcm{C}{^{AP}}\spacef.
							\end{align}
			The expressions on the right-hand side can be derived by direct computation, using \cref{it:nullDecompProp1,it:nullDecompProp4,it:nullDecompProp5,it:nullDecompProp11,it:alternativeRelationC,it:alternativeRelationF} on page \pageref{it:nullDecompProp1} . In addition, we define the \emph{Coulomb superenergy density} as\footnote{For the Bel-Robinson tensor, this is completely determined by the Coulomb part of the gravitational field, in the sense of its expression in terms of the Weyl scalars --see \cref{app:NPformulation}.}:
				\begin{equation}
				\cs{\V}\defeq  \ctp{k}{^\mu}\ctm{k}{^\nu}\ctp{k}{^\rho}\ctm{k}{^\sigma}\ct{\T}{_{\mu\nu\rho\sigma}}=\ctrpm{C}{^{AB}}\ctrpm{C}{_{AB}}+\ctrpm{D}{^{AB}}\ctrpm{D}{_{AB}}=\cs{C}^2+\cs{D}^2 \geq 0.\label{eq:Vdef}
				\end{equation}
			Observe the non-negativity of \cref{eq:Zmdef,eq:Zpdef,eq:Wmdef,eq:Wpdef,eq:Vdef}. Contracting  \cref{eq:BR-orthogonaldecomp} four times with $ \ct{u}{^\alpha} $ in the form
				\begin{equation}\label{eq:u-kp-km}
					\ct{u}{^\alpha}=\frac{1}{\sqrt{2}}\prn{\ctp{k}{^\alpha}+\ctm{k}{^\alpha}},
				\end{equation}
			one gets the relation
				\begin{equation}
				\cs{\W} = \frac{1}{4}\brkt{\csp{\W}+4\csp{\Z}+6\cs{\V}+4\csm{\Z}+\csm{\W}}\spacef.\label{eq:superneregySum}
				\end{equation}
			Indeed, it is easy to generalise this formula for any kind of coplanarity and to obtain the following lemma
				\begin{lemma}\label{thm:noRadCulSE-noSE}
				Let $ \cs{\W} $ be the superenergy density associated to a unit timelike vector field $ \ct{u}{^\alpha} $, and  $ \cspm{\W} $, $ \cspm{\Z} $, $ \cs{\V} $, the superenergy densities associated to a couple of lightlike vector fields $ \ctpm{k}{^\alpha} $ such that
					\begin{equation}
						\ct{u}{^\alpha}= \prn{a\ctm{k}{^\alpha}+b\ctp{k}{^\alpha}},\quad\text{with }ab=\frac{1}{2}\spacef.
					\end{equation}
				Then,
					\begin{equation}
						\cs{\W} = \brkt{b^4\csp{\W}+4b^3a\csp{\Z}+6a^2b^2\cs{\V}+4ba^3\csm{\Z}+a^4\csm{\W}}\spacef,\label{eq:superneregySumgeneralCoplanarity}
					\end{equation}
					\begin{equation}
					\cs{\W} = 0 \iff \cbrkt{\cs{\V}=0,\quad\cspm{\W}=0,\quad\cspm{\Z}=0}\spacef.
					\end{equation}
				\end{lemma}
		
			
		Any radiant supermomentum $ \ctl{\Q}{^\alpha} $ constructed with a future-pointing lightlike vector field as in \cref{eq:radiant-sm-def} has some basic properties,
				\begin{properties}
					\item  $  \ctl{\Q}{^\alpha}$ is lightlike, $  \ctl{\Q}{^\mu}\ctl{\Q}{_\mu} = 0  $, and future pointing. This follows by the dominant superenergy condition in the version of \cref{it:dominant-lightlike-se-condition} on page \pageref{it:dominant-lightlike-se-condition}.\label{it:radiantNullProperty}
					\item  $\prn{\delta^\nu_\mu+\ct{\ell}{^\nu}\ct{k}{_\mu}+\ct{k}{^\nu}\ct{\ell}{_\mu}}\ctsl{\Q}{^\mu}\ctsl{\Q}{_\nu} = 2\csl{\Z}\csl{\W}$, which can be shown applying \cref{it:radiantNullProperty} to $ \ctl{\Q}{^\alpha} $ and using \cref{eq:radiant-sm-decomp}\label{it:normQpmA}. From these same equations, it follows that
					\item $ \ctsl{\Q}{^\alpha}= 0 \iff \csl{\Z}= 0$. And, also, $ \ctl{\Q}{^\alpha}= 0 \iff \csl{\W}= 0 = \csl{\Z} $.\label{it:noQnoWnoZ}
				\end{properties}
		If we contract the radiant supermomenta with $ \ct{P}{^\alpha_\beta} $, we obtain their parts   orthogonal to $ \ct{u}{^\alpha} $,
				\begin{align}
				\ct{P}{^\alpha_\mu}\ctp{\Q}{^\mu}&= \frac{1}{\sqrt{2}}\prn{\csp{\Z}-\csp{\W}}\ct{r}{^\alpha}+\ctps{\Q}{^A}\ct{E}{^\alpha_A}=\ctps{\Q}{^\alpha}\spacef,\label{eq:QpProjected}\\
				\ct{P}{^\alpha_\mu}\ctm{\Q}{^\mu}&= -\frac{1}{\sqrt{2}}\prn{\csm{\Z}-\csm{\W}}\ct{r}{^\alpha}+\ctms{\Q}{^A}\ct{E}{^\alpha_A}=\ctms{\Q}{^\alpha}\spacef.\label{eq:QmProjected}
				\end{align}
		Using \cref{it:nullDecompProp12,it:nullDecompProp11,it:nullDecompProp14,it:nullDecompProp15} on page \pageref{it:nullDecompProp12}, we can write the radiant decomposition in terms of the electric and magnetic parts of the Weyl candidate
				\begin{align}
				\csp{\Z}&=\prn{\ctc{D}{_A}+\ctc{\epsilon}{_A^E}\ctc{C}{_E}}\prn{\ctc{D}{^A}+\ctc{\epsilon}{^A^D}\ctc{C}{_D}}\quad\label{eq:Zp},\\
				\csm{\Z}&=\prn{\ctc{D}{_A}-\ctc{\epsilon}{_A^E}\ctc{C}{_E}}\prn{\ctc{D}{^A}-\ctc{\epsilon}{^A^D}\ctc{C}{_D}}\quad\label{eq:Zm},\\
				\csp{\W}&= 2\prn{\ctt{C}{_{AB}} -\ctt{D}{^T_{(B}}\ctc{\epsilon}{_{A)T}}}\prn{\ctt{C}{^{AB}} -\ctt{D}{_M^{(B}}\ctc{\epsilon}{^{A)M}}}\quad\label{eq:Wp},\\
				\csm{\W}&= 2\prn{\ctt{C}{_{AB}} +\ctt{D}{^T_{(B}}\ctc{\epsilon}{_{A)T}}}\prn{\ctt{C}{^{AB}}+ \ctt{D}{_M^{(B}}\ctc{\epsilon}{^{A)M}}}\quad\label{eq:Wm},\\
				\ctp{\Q}{^A}&= 	2\sqrt{2} \prn{\ctt{C}{^{AB}} -\ctt{D}{_M^{(B}}\ctc{\epsilon}{^{A)M}}}\prn{\ctc{C}{_B}-\ctc{\epsilon}{_B^E}\ctc{D}{_E}}\quad\label{eq:Qp},\\
				\ctm{\Q}{^A}&= -	2\sqrt{2} \prn{\ctt{C}{^{AB}} +\ctt{D}{_M^{(B}}\ctc{\epsilon}{^{A)M}}}\prn{\ctc{C}{_B}+\ctc{\epsilon}{_B^E}\ctc{D}{_E}}\quad\label{eq:Qm}.
				\end{align}
		 And with these relations it is straightforward to compute 
				\begin{align}
				\csp{\Z}-\csm{\Z}&= 4\ctc{D}{^A}\ctc{\epsilon}{_{AB}}\ctc{C}{^B}\quad\label{eq:z-z},\\
				\csp{\Z}+\csm{\Z}&= 2\prn{\ctc{C}{_A}\ctc{C}{^A}+\ctc{D}{_A}\ctc{D}{^A}}\quad\label{eq:z+z},\\
				\csp{\W}-\csm{\W}&= 8\ctt{D}{^{TB}}\ctc{\epsilon}{_{TA}}\ctt{C}{^A_B}\spacef,\label{eq:w+w}\\
				\csp{\W}+\csm{\W}&= 4\prn{\ctt{D}{_{AB}}\ctt{D}{^{AB}}+\ctt{C}{_{AB}}\ctt{C}{^{AB}}}\spacef,\label{eq:w-w}\\
				\sqrt{2}\prn{\ctp{\Q}{^A}-\ctm{\Q}{^A}}&= 8\prn{\ctc{C}{_P}\ctt{C}{^{PA}}+\ctc{D}{_E}\ctt{D}{^{EA}}}\spacef,\label{eq:QA+QA}\\
				\sqrt{2}\prn{\ctp{\Q}{^A}+\ctm{\Q}{^A}}&=- 8\ctc{\epsilon}{_T^A}\prn{\ctc{D}{_P}\ctt{C}{^{PT}}-\ctc{C}{_E}\ctt{D}{^{ET}}}\spacef.\label{eq:QA-QA}		
				\end{align} 

		We would like to have a complete relation between the radiant and the standard supermomentum, and we already have the relation between superenergy densitites \cref{eq:superneregySum}. Thus, it only remains to find an expression for the super-Poynting vector in terms of the new quantities. To that purpose, substitute $ \ctpm{k}{^\alpha} $ in terms of $ \ct{u}{^\alpha} $ and $ \ct{r}{^\alpha} $ in \cref{eq:Qmdef,eq:Qpdef},
			\begin{align}
			\ctp{\Q}{^\alpha}= \frac{1}{2\sqrt{2}}\prn{\ct{\Q}{^\alpha}-3\ct{\T}{^\alpha_{\mu\nu\rho}}\ct{u}{^\mu}\ct{r}{^\nu}\ct{r}{^\rho}-3\ct{\T}{^\alpha_{\mu\nu\rho}}\ct{r}{^\mu}\ct{u}{^\nu}\ct{u}{^\rho}-\ct{\T}{^\alpha_{\mu\nu\rho}}\ct{r}{^\mu}\ct{r}{^\nu}\ct{r}{^\rho}}\spacef,\\
			\ctm{\Q}{^\alpha}= \frac{1}{2\sqrt{2}}\prn{\ct{\Q}{^\alpha}-3\ct{\T}{^\alpha_{\mu\nu\rho}}\ct{u}{^\mu}\ct{r}{^\nu}\ct{r}{^\rho}+3\ct{\T}{^\alpha_{\mu\nu\rho}}\ct{r}{^\mu}\ct{u}{^\nu}\ct{u}{^\rho}+\ct{\T}{^\alpha_{\mu\nu\rho}}\ct{r}{^\mu}\ct{r}{^\nu}\ct{r}{^\rho}}\spacef.
			\end{align}
		Thus,
			\begin{equation}\label{eq:+Q+-Q}
			\ctp{\Q}{^\alpha}+\ctm{\Q}{^\alpha}=\frac{1}{\sqrt{2}}\prn{\ct{\Q}{^\alpha}-3\ct{\D}{^\alpha_{\mu\nu\rho}}\ct{u}{^\mu}\ct{r}{^\nu}\ct{r}{^\rho}}\spacef.
			\end{equation}
		If we contract this equation with $ \ct{P}{^\alpha_\beta} $,  the right-hand side is determined by \cref{eq:Qalfonso},
			\begin{equation}
			\ct{P}{^\alpha_\beta}\ct{\T}{_{\alpha\mu\nu\rho}}\ct{u}{^\mu}\ct{r}{^\nu}\ct{r}{^\rho}=-\ct{Q}{_{\beta\nu\rho}}\ct{r}{^\nu}\ct{r}{^\rho}= -\cts{\P}{_\beta} + 4\ct{r}{_\beta}\ctc{\epsilon}{_{AB}}\ct{C}{^A}\ct{D}{^B}+4\ct{W}{_\beta^E}\ctc{\epsilon}{_{EA}}\prn{\ctc{C}{^A}\cs{D}-\ctc{D}{^A}\cs{C}}\spacef,
			\end{equation}
		which, after inserting \cref{eq:z-z}, becomes
			\begin{equation}
			\ct{P}{^\alpha_\beta}\ct{\T}{_{\alpha\mu\nu\rho}}\ct{u}{^\mu}\ct{r}{^\nu}\ct{r}{^\rho}=-\cts{\P}{_\beta}-\prn{\csp{\Z}-\csm{\Z}}\ct{r}{_\beta}+4\ct{W}{_\beta^E}\ctc{\epsilon}{_{EA}}\prn{\ctc{C}{^A}\cs{D}-\ctc{D}{^A}\cs{C}}\spacef.
			\end{equation}
		The left-hand side is known by \cref{eq:QpProjected,eq:QmProjected}, therefore
			\begin{align}
			\frac{1}{\sqrt{2}}\prn{\csp{\Z}-\csm{\Z}-\csp{\W}+\csm{\W}}\ct{r}{^\alpha}+\prn{\ctps{\Q}{^A}+\ctms{\Q}{^A}}\ct{E}{^\alpha_A}\nonumber\\
			=\frac{1}{\sqrt{2}}\brkt{4\cts{\P}{^\alpha}+3\prn{\csp{\Z}-\csm{\Z}}\ct{r}{^\alpha}-12\ct{E}{^\alpha_E}\ctc{\epsilon}{^E_A}\prn{\ctc{C}{^A}\cs{D}-\ctc{D}{^A}\cs{C}}}\spacef.
			\end{align}
		After recombining the terms, one finds the following relation:
			\begin{equation}\label{eq:poynting-radiant}
			4\cts{\P}{^a}= \prn{2\csm{\Z}-2\csp{\Z}-\csp{\W}+\csm{\W}}\ct{r}{^a}+\brkt{\sqrt{2}\prn{\ctps{\Q}{^A}+\ctms{\Q}{^A}}+12\ctc{\epsilon}{^A_E}\prn{\ctc{C}{^E}\cs{D}-\ctc{D}{^E}\cs{C}}}\ct{E}{^a_A}\spacef.
			\end{equation}
		The whole supermomentum is determined by \cref{eq:poynting-radiant,eq:superneregySum},
			\begin{align}\label{eq:supermomentum-radiant}
			4\ct{\Q}{^\alpha}=&\brkt{\csp{\W}+4\csp{\Z}+6\cs{\V}+4\csm{\Z}+\csm{\W}}\ct{u}{^\alpha}+\prn{2\csm{\Z}-2\csp{\Z}-\csp{\W}+\csm{\W}}\ct{r}{^\alpha}+\\
			+&\brkt{\sqrt{2}\prn{\ctps{\Q}{^A}+\ctms{\Q}{^A}}+12\ctc{\epsilon}{^A_E}\prn{\ctc{C}{^E}\cs{D}-\ctc{D}{^E}\cs{C}}}\ct{E}{^\alpha_A}\spacef.
			\end{align}
			
		The radiant components contain no information about the traces $ \ct{D}{^A_A} $ and $ \ct{C}{^A_A} $, whose squares determine the Coulomb superenergy density $ \V $. Besides, the longitudinal radiant superenergy densities $ \cspm{\Z} $ are controlled by $ \ctc{C}{_A} $, $ \ctc{D}{_A} $. Note that \Cref{eq:supermomentum-radiant,eq:poynting-radiant} contain a `mixed' term
			\begin{equation}\label{eq:dA}
			\ct{d}{^{A}}\defeq \ctc{\epsilon}{^A_E}\prn{\ctc{C}{^E}\cs{D}-\ctc{D}{^E}\cs{C}}\spacef.
			\end{equation} 
	
		Let us finish this section with some results. The first one follows from the Petrov classification on page \pageref{it:petrov-class-BR}
			\begin{lemma}
				A radiant supermomentum, $ \ctl{\Q}{^\alpha} $, constructed with a lightlike vector $ \ct{\ell}{^\alpha} $ vanishes if and only if $ \ct{\ell}{^\alpha} $ is a repeated PND of the corresponding Weyl candidate.
			\end{lemma}
			\begin{lemma}\label{thm:oneQvanishes}
				Consider the lightlike projections of the Weyl candidate tensor for a couple of lighlike vectors $ \ctpm{k}{^\alpha} $ as in \cref{eq:u-kp-km}. Let one of the associated radiant supermomenta vanish, $ \ctpm{\Q}{^\alpha}=0 $. Then,
				\begin{enumerate}[label=\roman*)]
					\item $ \ctcupm{D}{_A}=0=\ctcupm{C}{_A} \quad(\iff \ctc{D}{_A}\ctc{\epsilon}{^{AB}}=\mp\ctc{C}{^B})$ and then $ \ctc{C}{_A}=\ctcump{C}{_A} $, $ \ctc{D}{_A}=\ctcump{D}{_A} $,
					\item $ \ctt{D}{_{T(B}}\ctc{\epsilon}{_{A)}^T}=\pm\ctt{C}{_{AB}} $,
					\item \begin{align}
					\csmp{\Z}&= 4\ctc{C}{_A}\ctc{C}{^A}=4\ctc{D}{_A}\ctc{D}{^A}\spacef,\\
					\csmp{\W}&=\mp 16\ctt{D}{^{TB}}\ctc{\epsilon}{_{TA}}\ctt{C}{^A_B}= 16\ctt{C}{_{AB}}\ctt{C}{^{AB}}=16\ctt{D}{_{AB}}\ctt{D}{^{AB}}\spacef, \\
					\ctmp{\Q}{^A}&=8\sqrt{2} \ctt{C}{^{AB}}\ct{C}{_B}=\pm 8\sqrt{2}  \ctt{D}{_{T}^{(B}}\ctc{\epsilon}{^{A)T}}\ct{C}{_B} \spacef,\\
					\ct{d}{_A}&=\ctc{\epsilon}{_{A}^E}\prn{\ct{\delta}{_E^B}D\pm C\ctc{\epsilon}{_E^B}}\ctc{C}{_B}\spacef.
					\end{align}
				\end{enumerate}
			\end{lemma}
			\begin{proof}
				All the points above follow directly from \cref{eq:Zp,eq:Zm,eq:Wp,eq:Wm,eq:Qp,eq:Qm,eq:z-z,eq:z+z,eq:w+w,eq:w-w,eq:QA+QA,eq:QA-QA}, using \cref{it:noQnoWnoZ} on page \pageref{it:noQnoWnoZ}.
			\end{proof}
		The next result follows by inspection of \cref{eq:poynting-radiant},
			\begin{lemma}\label{thm:radiant-poynting-vanishing}
			Consider the super-Poynting $ \cts{\P}{^a} $ associated to a timelike, unit vector $ \ct{u}{^\alpha} $ and a couple of independent lightlike vectors as in \cref{eq:u-kp-km}. Let $ \ctpm{\Q}{^\alpha} $ be their associated radiant supermomenta. Then, the necessary and sufficient conditions on the radiant quantities that make the super-Poynting vanish are
			\begin{align}\label{eq:noPoynting-noradiant}
			\begin{rcases}
			2\prn{\csm{\Z}-\csp{\Z}}-\csp{\W}+\csm{\W} & = 0\quad\\
			\sqrt{2}\prn{\ctps{\Q}{^A}+\ctms{\Q}{^A}}+12\ct{d}{^A}& =  0 \quad 
			\end{rcases}\iff \cts{\P}{^a} = 0\spacef.
			\end{align}
			\end{lemma}
			\begin{corollary}\label{thm:oneQnull-then}
				If one of the radiant supermomenta considered in \cref{thm:radiant-poynting-vanishing} vanishes, say $ \ctm{\Q}{^\alpha}=0 $, then
				\begin{equation}
					\ctp{\Q}{^\alpha}=0 \iff \cts{\P}{^a}=0\spacef.
				\end{equation}
				The same holds true by interchanging the $ + $ with the $ - $ sign.
			\end{corollary}
			\begin{proof}
				By \cref{it:noQnoWnoZ} on page \pageref{it:noQnoWnoZ}, $ \ctm{\Q}{^\alpha}=0 \iff \csm{\Z}=\csm{\W}=0$. Now,
				by that same property and \cref{thm:oneQvanishes}, $ \ctp{\Q}{^\alpha}=0 \iff \csp{\Z}=\csp{\W}=0\implies\ct{d}{_A}=0 $. Then, \cref{eq:noPoynting-noradiant} is trivially satisfied and $ \cts{\P}{^a}=0 $. For the converse, if $ \cts{\P}{^a}=0 $, by \cref{eq:noPoynting-noradiant} we get $ \csp{\Z}=-\csp{\W} $, but the only possibility is $ \csp{\Z}=\csp{\W}=0 $ because both quantities are non negative (\cref{eq:Wpdef,eq:Zpdef}). By \cref{it:noQnoWnoZ} on page \pageref{it:noQnoWnoZ}, $ \ctp{\Q}{^\alpha}=0 $.
			\end{proof}
		
			\begin{prop}
				Consider the super-Poynting $ \cts{\P}{^a} $ associated to a timelike, unit vector $ \ct{u}{^\alpha} $ and a couple of independent lightlike vectors as in \cref{eq:kp-def,eq:km-def}. Let $ \ctpm{\Q}{^\alpha} $ be the two associated radiant supermomenta. The following conditions are all equivalent:
					\begin{enumerate}
						\item $ \ctm{\Q}{^\alpha}=\ctp{\Q}{^\alpha}=0 $.\label{it:aux1}
						\item $ \ctm{\Q}{^\alpha}=0$ and $\cts{\P}{^a}=0 $.\label{it:aux2}
						\item $ \ctm{\Q}{^\alpha}=0$ and $\cts{\P}{^\alpha}\ct{r}{_\alpha}=0 $.\label{it:aux3}
						\item  $ \ctt{D}{_{AB}}=\ctt{C}{_{AB}}=0 $ and $ \ctc{D}{_A}=\ctc{C}{_A}=0 $.\label{it:aux4}
						\item In the basis $ \cbrkt{\ct{r}{^a},\ct{E}{^a_A}} $,\label{it:aux5}
							\begin{equation}\label{eq:propDCform}
							\prn{\ct{D}{_{ab}}}= \cs{D}
							\begin{pmatrix}
							1 & 0 & 0 \\ 
							0 & -1/2 & 0  \\ 
							0 & 0 & -1/2 
							\end{pmatrix}, \quad 
							\prn{\ct{C}{_{ab}}}= \cs{C}
							\begin{pmatrix}
							1 & 0 & 0\\ 
							0 & -1/2 & 0\\ 
							0 & 0 &  -1/2
							\end{pmatrix} \spacef.
							\end{equation}							
					\end{enumerate}
			\end{prop}
			\begin{remark}
				This case corresponds to the situation where the Weyl candidate tensor has Petrov type D at those points where the above conditions hold and $ \ctpm{k}{^\alpha} $ are the two double principal null directions.
			\end{remark}
			\begin{proof}{\ }
				\begin{itemize}
					\item \ref{it:aux1} $ \iff $ \ref{it:aux2}: it follows from \cref{thm:oneQnull-then}.
					\item \ref{it:aux2} $ \iff $ \ref{it:aux3}: point \ref{it:aux2} implies \ref{it:aux3} trivially; by \cref{eq:poynting-radiant}, point \ref{it:aux3} implies that the first line of \cref{eq:noPoynting-noradiant} in \cref{thm:radiant-poynting-vanishing} holds and, noting \cref{it:noQnoWnoZ} on page \pageref{it:noQnoWnoZ}, $ \csp{\W}=0=\csp{\Z} $. But, then, \cref{thm:oneQvanishes} tells us that $ \ct{d}{_A}=0=\ctp{\Q}{_A} $. Altogether, by \cref{thm:radiant-poynting-vanishing}, we have $ \cts{\P}{^a}=0 $.
					\item \ref{it:aux1} $ \iff $ \ref{it:aux4}: point \ref{it:aux1} implies \ref{it:aux4} by \cref{it:noQnoWnoZ} on page \pageref{it:noQnoWnoZ} and \cref{thm:oneQvanishes}. The converse is shown noting that \ref{it:aux4} implies $ \cspm{\W}=0=\cspm{\Z} $ by \cref{eq:z-z,eq:w-w,eq:z+z,eq:w-w} which, by  \cref{it:noQnoWnoZ} on page \pageref{it:noQnoWnoZ}, implies \ref{it:aux1}.
					\item \ref{it:aux4} $ \iff $ \ref{it:aux5}: point \ref{it:aux4} is saying explicitly that in the basis of point \ref{it:aux5}, the tensors $ \ct{C}{_{ab}} $, $ \ct{D}{_{ab}} $ have precisely the form displayed in \cref{eq:propDCform}.
				\end{itemize}
			\end{proof}

	\section{Conformal geometry and infinity}
			
		Studying isolated systems, among other aspects of gravity, requires investigating the asymptotic properties of the space-time. The geometrisation of infinity as the boundary of the conformal completion \cite{Penrose65} avoids the use of limits and facilitates the employment of covariant methods. Indeed, this boundary `attached' to the space-time is the suitable arena for describing the radiative degrees of freedom of the gravitational field. Many of the results that appear in this section are already known --see e.g. \cite{Szabados2015,Ashtekar2014,Kroon}-- and details are included for completeness.
			
			\subsection{Conformal completion}
				The physical space-times $ \prn{\ps{M},\pt{g}{_{\alpha\beta}}} $ that we consider admit a conformal completion (unphysical space-time) $ \prn{\cs{M},\ct{g}{_{\alpha\beta}}} $ {\em \`a la} Penrose with boundary $ \scri $ \cite{Kroon,Frauendiener2004}:
					\begin{properties}
						\item 	There exists an embedding $ \phi: \ps{M} \rightarrow \cs{M} $ such that $ \phi(\ps{M})=\cs{M}\setminus\scri $, and the physical metric is related to the conformal one as
						\begin{equation}
						\ct{g}{_{\alpha\beta}}= \Omega^2\pt{g}{_{\alpha\beta}},
						\end{equation} 
						where, abusing notation, we refer to the pullback of the conformal metric to the physical space-time, $ \prn{\phi^*g}_{\alpha\beta} $, by  $ \ct{g}{_{\alpha\beta}} $.
						\item $\Omega>0$ in $M\setminus\scri$ , $\Omega=0$ on $\scri$ and $\ct{N}{_\alpha}\defeq\cd{_\alpha}\Omega$ (the normal to $ \scri $) is non-vanishing there.\label{it:OmegaAtScri}
						\item $ \pt{g}{_{\alpha\beta}} $ is a solution of EFE \eqref{eq:efe} with $ \Lambda>0 $.\label{it:efesassumption}
						\item The energy-momentum tensor, $ \pt{T}{_{\alpha\beta}} $, vanishes at $ \scri $ and $ \ct{T}{_{\alpha\beta}}\defeq\Omega^{-1}\pt{T}{_{\alpha\beta}} $ is smooth there.\label{it:energytensorassumption}
					\end{properties}
				Depending on the specific matter content, \cref{it:energytensorassumption} can be strengthen. 
				Indeed for many relevant matter fields, $ \Omega^{-2}\pt{T}{_{\alpha\beta}} $ is smooth at $ \scri $. Notice that $ \cs{T}\defeq \ct{T}{^\mu_\mu}=\Omega^{-3}\pt{T}{^\mu_\mu}\defeq \Omega^{-3}\ps{T} $. Also, $\scri$ is not connected, in general; it is divided into `future' and `past' components, denoted by $\scri^\pm$ respectively. In this section, we use $ \scri $ generically to refer to any of them. In some of the subsequent sections we will work with $ \scri^+ $, though. 
				\\
				
				The connection of the unphysical space-time is written as
					\begin{equation}
						\ct{\Gamma}{^\alpha_{\beta\gamma}} = \pt{\Gamma}{^\alpha_{\beta\gamma}} +\ct{\gamma}{^\alpha_{\beta\gamma}}\spacef,
					\end{equation}
				where $ \pt{\Gamma}{^\alpha_{\beta\gamma}} $ is the connection of the physical space-time and
					\begin{equation}
						\ct{\gamma}{^\alpha_{\beta\gamma}}=\Omega^{-1}\prn{2\ct{\delta}{^\alpha_{(\beta}}\cd{_{\gamma)}}\Omega-\ct{g}{_{\gamma\beta}}\cd{^\alpha}\Omega}\spacef.
					\end{equation}
				Accordingly, the Ricci tensor and the scalar curvature are given by
					\begin{align}
						\ct{R}{_{\alpha\beta}}&=\pt{R}{_{\alpha\beta}}- 2\Omega^{-1}\cd{_\alpha}\ct{N}{_{\beta}}+3\Omega^{-2}\ct{g}{_{\alpha\beta}}\ct{N}{_{\mu}}\ct{N}{^\mu}-\Omega^{-1}\ct{g}{_{\alpha\beta}}\cd{_{\mu}}\ct{N}{^\mu}\spacef,\label{eq:conformalRicciTensor}\\
						\cs{R}&=\Omega^{-2}\ps{R}-6\Omega^{-1}\cd{_\mu}\ct{N}{^\mu}+12\Omega^{-2}\ct{N}{_\mu}\ct{N}{^\mu}\spacef.\label{eq:conformalRicciScalar}
					\end{align}
			\\
				The conformal completion is not unique: given $ \prn{\ps{M},\pt{g}{_{\alpha\beta}}} $ there is a conformal class of completions related by 
					\begin{equation}\label{eq:conformal-gauge}
						\Omega \rightarrow \ctg{\Omega}{}=\omega\Omega\quad \text{ with } \omega > 0 \text{ on } \cs{M}\spacef.
					\end{equation}
				This rescaling of the conformal factor is a gauge freedom and, as such, can be used to simplify matters. In the upcoming subsection we will fix it partially. A gauge transformation changes the geometry as					
					\begin{align}
						\ctg{g}{_{\alpha\beta}}&=\omega^2\ct{g}{_{\alpha\beta}}\spacef,\\
						\ctg{\Gamma}{^\alpha_{\beta\gamma}} &= \ct{\Gamma}{^\alpha_{\beta\gamma}} +\ct{C}{^\alpha_{\beta\gamma}}\spacef,\quad \ct{C}{^\alpha_{\beta\gamma}}=\omega^{-1}\ct{g}{^{\alpha\tau}}\prn{2\ct{g}{_{
									\tau(\beta}}\ct{\omega}{_{\gamma)}}-\ct{g}{_{\gamma\beta}}\ct{\omega}{_\tau}}\spacef,\label{eq:gauge-connection}\\
						\ctg{R}{_{\alpha\beta}}&=\ct{R}{_{\alpha\beta}}- 2\omega^{-1}\cd{_\alpha}\ct{\omega}{_{\beta}}-\omega^{-2}\ct{g}{_{\alpha\beta}}\ct{\omega}{_{\mu}}\ct{\omega}{^\mu}-\omega^{-1}\ct{g}{_{\alpha\beta}}\cd{_{\mu}}\ct{\omega}{^\mu}+4\omega^{-2}\ct{\omega}{_\alpha}\ct{\omega}{_\beta}\spacef,\label{eq:gaugeRicciTensor}\\
						\csg{R}&=\omega^{-2}\cs{R}-6\omega^{-3}\cd{_\mu}\ct{\omega}{^\mu} \spacef,\label{eq:gaugeRicciScalar}\\
						\ctg{N}{_{\alpha}}&=\omega\ct{N}{_{\alpha}}+\Omega\ct{\omega}{_\alpha}\label{eq:gauge-N}\quad 
					\end{align}
				where $ \ct{\omega}{_\alpha}\defeq\cd{_\alpha}\omega $. Here we have written the equations in terms of the original connection and metric. 
				Further gauge changes can be found in \cref{app:gauge-transformations}.
				
			\subsection{Fields at infinity}
				Einstein's field equations for a non-vanishing cosmological constant (here assumed to be positive) read
					\begin{equation}\label{eq:efe}
						\pt{R}{_{\alpha\beta}}-\frac{1}{2}\ps{R}\pt{g}{_{\alpha\beta}}+\Lambda\pt{g}{_{\alpha\beta}}=\varkappa \pt{T}{_{\alpha\beta}}\spacef,
					\end{equation} 
				with $ \varkappa\defeq 8\pi G c^{-4} $, where $ G $ is the gravitational constant and $ c $ the speed of light. Thus, the left-hand side extends smoothly to infinity. Using \cref{eq:conformalRicciTensor,eq:conformalRicciScalar} we get
					\begin{equation}\label{eq:efeMixed}
						\ct{R}{_{\alpha\beta}}-\frac{1}{2}\cs{R}\ct{g}{_{\alpha\beta}}+3\Omega^{-2}\ct{g}{_{\alpha\beta}}\ct{N}{_\mu}\ct{N}{^\mu}+2\Omega^{-1}\prn{\cd{_\alpha}\ct{N}{_{\beta}}-\ct{g}{_{\alpha\beta}}\cd{_\mu}\ct{N}{^\mu}}+\Omega^{-2}\Lambda\ct{g}{_{\alpha\beta}}=\varkappa\pt{T}{_{\alpha\beta}}\spacef.
					\end{equation}
				If we multiply by $ \Omega^2 $ and evaluate at $ \scri $ --i.e. set $ \Omega=0 $--, we obtain
					\begin{equation}\label{eq:normNscritensorial}
					3\ct{g}{_{\alpha\beta}}\ct{N}{_\mu}\ct{N}{^\mu}+\Lambda\ct{g}{_{\alpha\beta}}\eqs 0\spacef,
					\end{equation}
				from where 
					\begin{equation}\label{eq:normNscri}
					\ct{N}{_\mu}\ct{N}{^\mu}\eqs -\frac{\Lambda}{3}\spacef.
					\end{equation}
				This equation indicates the causal character of the conformal boundary: $ \scri $ can be a timelike ($ \Lambda<0 $), lightlike ($ \Lambda=0 $) or, in our case, spacelike ($ \Lambda>0 $) hypersurface. The normalised version of $ \ct{N}{_\alpha} $ reads
					\begin{equation}\label{eq:normalisedN}
					\ct{n}{_\alpha}\defeq \frac{1}{N}\ct{N}{_{\alpha}},
					\end{equation}	
				with $ N\defeq\sqrt{-\ct{N}{^\mu}\ct{N}{_\mu}} $. In general, this definition is valid only on a neighbourhood of $ \scri $, where $ \ct{N}{_\alpha} $ is timelike. In that neighbourhood we can introduce the projector to $ \scri $ (see \cref{app:spatial-hypersurfaces}):
					\begin{equation}\label{eq:projectorScri}
						\ct{P}{^\alpha_\beta}\defeq \ct{\delta}{^\alpha_\beta}-\ct{n}{^\alpha}\ct{n}{_\beta}\spacef.
					\end{equation}
				Now, multiply \cref{eq:efeMixed} by $ \Omega $ to get
					\begin{equation}
					\Omega\ct{R}{_{\alpha\beta}}-\frac{1}{2}\Omega\cs{R}\ct{g}{_{\alpha\beta}}+2\prn{\cd{_\alpha}\ct{N}{_{\beta}}-\ct{g}{_{\alpha\beta}}\cd{_\mu}\ct{N}{^\mu}}=\Omega\varkappa\pt{T}{_{\alpha\beta}}-\Omega^{-1}\prn{3\ct{N}{_\mu}\ct{N}{^\mu}+\Lambda}\ct{g}{_{\alpha\beta}}\quad 
					\end{equation}
				and observe that this equation is regular at $ \scri $, see \cref{eq:normNscritensorial}. Take its trace and evaluate at $ \Omega=0 $:
					\begin{equation}
					6\cd{_\mu}\ct{N}{^\mu} \eqs \Omega^{-1}4\prn{3\ct{N}{_\mu}\ct{N}{^\mu}+\Lambda}\spacef. 
					\end{equation}
				Then, insert this back into the previous equation. After evaluation at $ \scri $, we derive
					\begin{equation}\label{eq:derivativeN}
					\cd{_\alpha}\ct{N}{_\beta}\eqs  \frac{1}{4}\ct{g}{_{\alpha\beta}}\cd{_\mu}\ct{N}{^\mu}\spacef.
					\end{equation}
				It is well known \cite{Geroch1977,Wald1984,Penrose1986,Ashtekar2000} that the gauge can be chosen such that
					  \begin{equation}\label{eq:divNgauge}
					  \cd{_\mu}\ct{N}{^\mu}\eqs 0.
					  \end{equation}
				 To show this, compute the gauge change using \cref{eq:gauge-N,eq:gauge-connection}
					\begin{equation}
						\cdg{_\mu}\ctg{N}{^\mu}=2\omega^{-2}\ct{N}{^\mu}\cd{_\mu}\omega+ \omega^{-1}\cd{_\mu}\ct{N}{^\mu}-\omega^{-1}\ct{g}{^{\alpha\beta}}\ct{C}{^\mu_{\alpha\beta}}\prn{\ct{N}{_\mu}+\omega^{-1}\Omega\cd{_\mu}\omega}+\Omega\omega^{-2}\square\omega\spacef,
					\end{equation}
				evaluate at $ \scri $ and multiply by $ \omega^2 $
					\begin{align}
					\omega^2\cdg{_\mu}\ctg{N}{^\mu}&\eqs 2\ct{N}{^\mu}\cd{_\mu}\omega+\omega\cd{_\mu}\ct{N}{^\mu}-\omega\ct{g}{^{\alpha\beta}}\ct{N}{_\mu}\ct{C}{^\mu_{\alpha\beta}}\nonumber\\
					&\eqs \omega\cd{_\mu}\ct{N}{^\mu}+4\ct{N}{^\mu}\cd{_\mu}\omega\spacef.
					\end{align}
				Equating $ \cdg{_\mu}\ctg{N}{^\mu} $ to zero gives a differential equation for the possible gauge factors 
					\begin{equation}\label{eq:divergenc-freeEquation}
					4\ct{N}{^\mu}\partial_\mu \omega + \omega \square\Omega \eqs 0\quad
					\end{equation}
					which always has non-trivial solutions. From now on, we \emph{adopt this gauge fixing} that we call \emph{divergence-free} gauge. 
					 Nevertheless, note that the freedom is still large  and one can change from one conformal gauge to another by keeping \cref{eq:divNgauge}  with the additional restriction (given a solution $ \cst{\omega} $ of \cref{eq:divergenc-freeEquation}, $ \csg{\omega}\defeq \omega\cst{\omega} $ is a new solution)
					\begin{equation}
					\lied_{\vec{N}}\omega \eqs 0\spacef.
					\end{equation}		
				 From \cref{eq:derivativeN}, this gauge implies
					\begin{equation}
					\cd{_\alpha}\ct{N}{_\beta}\eqs 0\spacef.
					\end{equation}
				Consider the combination\footnote{Note that this is twice the Schouten tensor, whose standard definition in 4 dimensions is $ \frac{1}{2}\prn{\ct{R}{_{\alpha\beta}}-\frac{1}{6}\cs{R}\ct{g}{_{\alpha\beta}}} $.}
					\begin{equation}\label{eq:defSchouten}
					\ct{S}{_{\alpha\beta}} \defeq \ct{R}{_{\alpha\beta}}-\frac{1}{6}\cs{R}\ct{g}{_{\alpha\beta}}\spacef,
					\end{equation}
			 	and \cref{eq:efeMixed} as an equation for $ \cd{_\alpha}\ct{N}{_\beta} $. Substitute $ \kappa\pt{T}{}=-\ps{R}+4\Lambda$ together with $ \ps{T}=\Omega^{-3}\cs{T} $ and \cref{eq:conformalRicciScalar},
					\begin{equation}
					\cd{_\alpha}\ct{N}{_\beta}= -\frac{1}{2}\Omega\ct{R}{_{\alpha\beta}}+\frac{1}{8}\Omega\cs{R}\ct{g}{_{\alpha\beta}}+\frac{1}{4}\ct{g}{_{\alpha\beta}}\cd{_\mu}\ct{N}{^\mu}+\frac{1}{2}\varkappa\Omega^2\prn{\ct{T}{_{\alpha\beta}}-\frac{1}{4}\cs{T}\ct{g}{_{\alpha\beta}}}\spacef,
					\end{equation}
				and replace $ \ct{R}{_{\alpha\beta}} $ with $ \ct{S}{_{\alpha\beta}} $,
					\begin{equation}\label{eq:derivativeNaux}
					\cd{_\alpha}\ct{N}{_\beta}= -\frac{1}{2}\Omega\ct{S}{_{\alpha\beta}}+\frac{1}{24}\Omega\cs{R}\ct{g}{_{\alpha\beta}}+\frac{1}{4}\ct{g}{_{\alpha\beta}}\cd{_\mu}\ct{N}{^\mu}+\frac{1}{2}\varkappa\Omega^2\prn{\ct{T}{_{\alpha\beta}}-\frac{1}{4}\cs{T}\ct{g}{_{\alpha\beta}}}\spacef.
					\end{equation}
				It is convenient to define
					\begin{equation}
					\ct{\underline{T}}{_{\alpha\beta}}\defeq \ct{T}{_{\alpha\beta}}-\frac{1}{4}\cs{T}\ct{g}{_{\alpha\beta}}
					\end{equation}
				and introduce the scalar \cite{Friedrich2002}
					\begin{equation}\label{eq:friedrichscalar}
					\cs{f}\defeq \frac{1}{4}\cd{_\mu}\ct{N}{^\mu}+\frac{\Omega}{24}\cs{R}\spacef,
					\end{equation}
				which in our gauge vanishes at $ \scri $
					\begin{equation}\label{eq:fscalarScri}
						\cs{f}\eqs 0\spacef.
					\end{equation}
				In terms of $ \cs{f} $, \cref{eq:derivativeNaux} becomes
					\begin{equation}\label{eq:derivativeNcfe}
					\cd{_\alpha}\ct{N}{_\beta} = -\frac{1}{2}\Omega\ct{S}{_{\alpha\beta}}+ \cs{f}\ct{g}{_{\alpha\beta}}+ \frac{1}{2}\Omega^2\varkappa\ct{\underline{T}}{_{\alpha\beta}}\spacef.
					\end{equation}
			    Or course, from this equation one deduces directly \cref{eq:derivativeN}.
			     Once again, consider \cref{eq:efeMixed}, this time as an equation for $ \cd{_\mu}\ct{N}{^\mu} $. Take its trace and multiply by $ \Omega $:
					\begin{equation}\label{eq:normNcefeaux}
					\ct{N}{_\mu}\ct{N}{^\mu}=\frac{\Omega^3}{12}\varkappa \cs{T}-\frac{\Lambda}{3}+ \frac{\Omega^2}{12}\cs{R}+\frac{\Omega}{2}\cd{_\mu}\ct{N}{^\mu}\spacef.
					\end{equation}
				As a check, one recovers \cref{eq:normNscri} after evaluating at $ \scri $. 
				 Introducing $ f $ in \cref{eq:normNcefeaux}, we get
					\begin{equation}\label{eq:normNcefe}
					\ct{N}{_\mu}\ct{N}{^\mu}=\frac{\Omega^3}{12}\varkappa \cs{T}-\frac{\Lambda}{3}+ 2\Omega f\spacef.
					\end{equation}
				Note that the explicit form of $ \ct{n}{^\alpha} $ reads
					\begin{equation}
					\ct{n}{_\alpha}=\frac{\ct{N}{_\alpha}}{\sqrt{\frac{\Lambda}{3}-\frac{\Omega^3}{12}\varkappa \cs{T}- 2\Omega f}}\eqs\sqrt{\frac{3}{\Lambda}}\ct{N}{_\alpha}
					\end{equation}
				and, since
					\begin{equation}
					\cd{_\alpha}\cs{N}\eqs 0\spacef,
					\end{equation}
				the normalised $ \ct{n}{_\alpha} $ is covariantly constant at $ \scri $ as well
					\begin{equation}\label{eq:divergence-freeGauge}
					\cd{_\alpha}\ct{n}{_\beta}\eqs 0\spacef.
					\end{equation}
				If we contract \cref{eq:derivativeNcfe} with $ \ct{N}{^\beta} $,
					\begin{equation}
						\ct{N}{^\mu}\cd{_\alpha}\ct{N}{_\mu}=\frac{1}{2}\cd{_\alpha}\prn{\ct{N}{_\mu}\ct{N}{^\mu}}=-\frac{1}{2}\Omega\ct{S}{_{\alpha\mu}}\ct{N}{^\mu}+f\ct{N}{_\alpha}+\frac{1}{2}\varkappa\Omega^2 \ct{N}{^\mu}\ct{\underline{T}}{_{\alpha\mu}}\spacef,
					\end{equation}
				and take the covariant derivative of \cref{eq:normNcefe},
					\begin{equation}\label{eq:derivativeNormN}
					\frac{1}{2}\cd{_\alpha}\prn{\ct{N}{_\mu}\ct{N}{^\mu}}= \frac{1}{24}\Omega^3\varkappa\cd{_\alpha}\cs{T}+ \frac{1}{8}\Omega^2\varkappa \ct{N}{_\alpha}\cs{T}+\Omega\cd{_\alpha}f + f\ct{N}{_\alpha}
					\end{equation}	
				we arrive, equating both expressions, at
					\begin{equation}\label{eq:derivativefcefe}
					\cd{_\alpha}f= -\frac{1}{2}\ct{S}{_{\alpha\mu}}\ct{N}{^\mu}+\frac{1}{2}\Omega\varkappa\ct{N}{^\mu}\ct{\underline{T}}{_{\alpha\mu}}-\frac{1}{24}\Omega^2\varkappa\cd{_\alpha}\cs{T}- \frac{1}{8}\Omega\varkappa\ct{N}{_\alpha}\cs{T}\spacef.
					\end{equation}
				Using \cref{eq:fscalarScri}
					\begin{equation}
					\cd{_\alpha}f \eqs -\ct{N}{_\alpha}\ct{N}{^\rho}\cd{_\rho}f \eqs - \frac{3}{\Lambda}\ct{N}{_\alpha}\ct{N}{^\mu}\cd{_\mu}f\eqs \frac{3}{2\Lambda}\ct{N}{_\alpha}\ct{N}{^\rho}\ct{N}{^\mu}\ct{S}{_{\mu\rho}}\spacef,
					\end{equation}
				from where we deduce that
					\begin{equation}\label{eq:schoutenComponentNPScri}
					\ct{P}{^\alpha_\beta}\ct{N}{^\mu}\ct{S}{_{\alpha\mu}}\eqs 0\spacef.
					\end{equation}
				 If we want to extract information about the complete orthogonal component of $ \ct{S}{_{\alpha\beta}} $ at $ \scri $, we have to contract \cref{eq:derivativeNcfe} with $ \ct{N}{^\alpha} $  and substitute $ \ct{S}{_{\alpha\mu}}\ct{N}{^\mu} $ in terms of \cref{eq:derivativefcefe}
					\begin{align}
					\ct{N}{^\mu}\cd{_\mu}\ct{N}{_\alpha}&=-\frac{1}{2}\Omega\ct{N}{^\mu}\ct{S}{_{\mu\alpha}}+f\ct{N}{_\alpha}+\frac{1}{2}\Omega^2\varkappa\ct{N}{^\mu}\ct{\underline{T}}{_{\mu\alpha}}\nonumber\\
					&=\Omega\cd{_\alpha}f-\frac{1}{2}\Omega^2\varkappa\ct{N}{^\mu}\ct{\underline{T}}{_{\alpha\mu}}+\frac{1}{24}\Omega^3\varkappa\cd{_\alpha}\cs{T}+\frac{1}{8}\Omega^2\varkappa\cs{T}\ct{N}{_\alpha}\nonumber\\
					&+ f\ct{N}{_\alpha}+\frac{1}{2}\Omega^2\varkappa\ct{N}{^\mu}\ct{\underline{T}}{_{\alpha\mu}}\nonumber\\
					&= \Omega\cd{_\alpha}f+f\ct{N}{_\alpha}+\frac{1}{8}\Omega^2\varkappa\cs{T}\ct{N}{_\alpha}+\frac{1}{24}\Omega^3\varkappa\cd{_\alpha}\cs{T}\spacef.
					\end{align}
				After this, contract \cref{eq:derivativefcefe} with $ \ct{N}{^\mu} $ to find
					\begin{equation}
					\ct{N}{^\mu}\cd{_\mu} f = \ct{N}{^\mu}\ct{N}{^\rho}\prn{-\frac{1}{2}\ct{S}{_{\mu\rho}}+\frac{1}{2}\Omega\varkappa\ct{\underline{T}}{_{\mu\rho}}}	-\ct{N}{^\mu}\prn{\frac{1}{24}\varkappa\Omega^2\cd{_\mu}\cs{T}+\frac{1}{8}\Omega\varkappa\cs{T}\ct{N}{_{\mu}}}\quad 
					\end{equation}
				then take the covariant derivative of \cref{eq:derivativeNcfe} along $ \ct{N}{^\rho} $ taking into account the last two equations
					\begin{align}
					\ct{N}{^\rho}\cd{_\rho}\prn{\cd{_\alpha}\ct{N}{_\beta}}&= -\frac{1}{2}\Omega\ct{N}{^\rho}\cd{_\rho}\ct{S}{_{\alpha\beta}}+\frac{1}{2}\ct{N}{^2}\ct{S}{_{\alpha\beta}}+\ct{g}{_{\alpha\beta}}\ct{N}{^\rho}\cd{_\rho}f+\frac{1}{2}\varkappa\Omega^2\ct{N}{^\rho}\cd{_\rho}\ct{\underline{T}}{_{\alpha\beta}}+\Omega\varkappa\ct{N}{^2}\ct{\underline{T}}{_{\alpha\beta}}\spacef,\\
					&=-\frac{1}{2}\Omega\ct{N}{^\rho}\cd{_\rho}\ct{S}{_{\alpha\beta}}+\frac{1}{2}\ct{N}{^2}\ct{S}{_{\alpha\beta}}-\frac{1}{2}\ct{S}{_{\mu\nu}}\ct{N}{^\mu}\ct{N}{^\nu}\ct{g}{_{\alpha\beta}}+\frac{1}{2}\Omega\varkappa\ct{N}{^\mu}\ct{N}{^\nu}\ct{\underline{T}}{_{\mu\nu}}\ct{g}{_{\alpha\beta}}\nonumber\\
					&-\frac{1}{24}\Omega^2\varkappa\ct{N}{^\mu}\cd{_\mu}\cs{T}\ct{g}{_{\alpha\beta}}+\frac{1}{8}\Omega\varkappa\ct{N}{^2}\varkappa\cs{T}\ct{g}{_{\alpha\beta}}+\frac{1}{2}\Omega^2\varkappa\ct{N}{^\rho}\cd{_\rho}\ct{\underline{T}}{_{\alpha\beta}}+\Omega\varkappa\ct{N}{^2}\ct{\underline{T}}{_{\alpha\beta}}\spacef.
					\end{align} 
				Therefore, at $ \scri $
					\begin{equation}
					\ct{N}{^\rho}\cd{_\rho}\prn{\cd{_\alpha}\ct{N}{_\beta}}\eqs \frac{\Lambda}{6}\ct{S}{_{\alpha\beta}}-\frac{1}{2}\ct{N}{^\mu}\ct{N}{^\nu}\ct{S}{_{\mu\nu}}\ct{g}{_{\alpha\beta}}\spacef.					
					\end{equation}
				The contraction of this equation with $ \ct{N}{^\alpha}\ct{N}{^\beta} $ shows that
					\begin{equation}
					\ct{n}{^\mu}\ct{n}{^\nu}\ct{S}{_{\mu\nu}}\eqs	\ct{N}{^\mu}\ct{N}{^\nu}\ct{N}{^\rho}\cd{_\rho}\prn{\cd{_\mu}\ct{N}{_\nu}}\spacef.
					\end{equation}
				\\Now, we are going to prove that \emph{the Weyl tensor vanishes} at $ \scri $, which is crucial in the analysis of the asymptotic gravitational field. According to \cref{eq:defSchouten,eq:conformalRicciScalar,eq:conformalRicciTensor}, 
					\begin{equation}
					\pt{S}{_{\alpha\beta}}= \ct{S}{_{\alpha\beta}}+2\Omega^{-1}\cd{_{\alpha}}\ct{N}{_\beta}-\Omega^{-2}\ct{g}{_{\alpha\beta}}\ct{N}{_\mu}\ct{N}{^\mu}\spacef.
					\end{equation}
				Take the covariant derivative and antisymmetrise the first pair of indices,
					\begin{align}
					\cd{_{[\alpha}}\pt{S}{_{\beta]\gamma}}&=\cd{_{[\alpha}}\ct{S}{_{\beta]\gamma}}-2\Omega^{-2}\ct{N}{_{[\alpha}}\cd{_{\beta]}}\ct{N}{_\gamma}+ 2\Omega^{-1}\cd{_{[\alpha}}\cd{_{\beta]}}\ct{N}{_\gamma}+2\Omega^{-3}\ct{N}{_{[\alpha}}\ct{g}{_{\beta]\gamma}}\ct{N}{^\mu}\ct{N}{_\mu}\nonumber\\
					&-2\Omega^{-2}\ct{N}{^\mu}\ct{g}{_{\gamma[\beta}}\cd{_{\alpha]}}\ct{N}{_\mu}\spacef,
					\end{align}
				multiply by $ \Omega $,
					\begin{align}
					\cd{_{[\alpha}}\prn{\Omega\pt{S}{_{\beta]\gamma}}}-\ct{N}{_{[\alpha}}\pt{S}{_{\beta]\gamma}}&=\Omega\cd{_{[\alpha}}\ct{S}{_{\beta]\gamma}}-2\Omega^{-1}\ct{N}{_{[\alpha}}\cd{_{\beta]}}\ct{N}{_\gamma}+ 2\cd{_{[\alpha}}\cd{_{\beta]}}\ct{N}{_\gamma}+2\Omega^{-2}\ct{N}{_{[\alpha}}\ct{g}{_{\beta]\gamma}}\ct{N}{^\mu}\ct{N}{_\mu}\nonumber\\
					&-2\Omega^{-1}\ct{N}{^\mu}\ct{g}{_{\gamma[\beta}}\cd{_{\alpha]}}\ct{N}{_\mu}\spacef,
					\end{align}
				and replace the third term on the right-hand side using the Ricci identity and the Riemann tensor decomposition,
					\begin{equation}
					\ct{R}{_{\alpha\beta\gamma\mu}}=\ct{C}{_{\alpha\beta\gamma\mu}}+\ct{g}{_{\alpha[\gamma}}\ct{S}{_{\mu]\beta}}-\ct{g}{_{\beta[\gamma}}\ct{S}{_{\mu]\alpha}}\spacef,
					\end{equation}
				to obtain
					\begin{align}\label{eq:aux1}
					\cd{_{[\alpha}}\prn{\Omega\pt{S}{_{\beta]\gamma}}}-\ct{N}{_{[\alpha}}\pt{S}{_{\beta]\gamma}}&=\Omega\cd{_{[\alpha}}\ct{S}{_{\beta]\gamma}}-2\Omega^{-1}\ct{N}{_{[\alpha}}\cd{_{\beta]}}\ct{N}{_\gamma}+ \ct{C}{_{\alpha\beta\gamma}^\mu}\ct{N}{_\mu}+\ct{g}{_{\alpha[\gamma}}\ct{S}{_{\mu]\beta}}\ct{N}{^\mu}\nonumber\\
					&-\ct{g}{_{\beta[\gamma}}\ct{S}{_{\mu]\alpha}}\ct{N}{^\mu}+2\Omega^{-2}\ct{N}{_{[\alpha}}\ct{g}{_{\beta]\gamma}}\ct{N}{^\mu}\ct{N}{_\mu}-2\Omega^{-1}\ct{N}{^\mu}\ct{g}{_{\gamma[\beta}}\cd{_{\alpha]}}\ct{N}{_\mu}\spacef.
					\end{align}
				Notice that
					\begin{align}\label{eq:aux2}
					\ct{g}{_{\alpha[\gamma}}\ct{S}{_{\mu]\beta}}\ct{N}{^\mu}-\ct{g}{_{\beta[\gamma}}\ct{S}{_{\mu]\alpha}}\ct{N}{^\mu}&=\ct{g}{_{\gamma[\alpha}}\ct{S}{_{\beta]\mu}}\ct{N}{^\mu}-\ct{N}{_{[\alpha}}\ct{S}{_{\beta]\gamma}}\nonumber\\
					&=\ct{g}{_{\gamma[\alpha}}\pt{S}{_{\beta]\mu}}\ct{N}{^\mu}-2\Omega^{-1}\ct{N}{^\mu}\ct{g}{_{\gamma[\alpha}}\cd{_{\beta]}}\ct{N}{_\mu}+\Omega^{-2}\ct{g}{_{\gamma[\alpha}}\ct{N}{_{\beta]}}\ct{N}{_\mu}\ct{N}{^\mu}\nonumber\\
					&+2\Omega^{-1}\ct{N}{_{[\alpha}}\cd{_{\beta]}}\ct{N}{_\gamma}-\Omega^{-2}\ct{N}{_{[\alpha}}\ct{g}{_{\beta]\gamma}}\ct{N}{_\mu}\ct{N}{^\mu}\spacef,			
					\end{align}
				which  substituted in \cref{eq:aux1} produces
					\begin{equation}\label{eq:aux3}
					\cd{_{[\alpha}}\prn{\Omega\pt{S}{_{\beta]\gamma}}}= \Omega\cd{_{[\alpha}}\ct{S}{_{\beta]\gamma}}+\ct{C}{_{\alpha\beta\gamma}^\mu}\ct{N}{_\mu}+\ct{g}{_{\gamma[\alpha}}\pt{S}{_{\beta]\mu}}\ct{N}{^\mu}.
					\end{equation}
				 To see how the energy-momentum tensor enters into this equation, plug 
					\begin{equation}\label{eq:schoutenPhsycialMatter}
					\pt{S}{_{\alpha\beta}}= \pt{R}{_{\alpha\beta}}-\frac{1}{6}\ps{R}\pt{g}{_{\alpha\beta}}= \varkappa\pt{T}{_{\alpha\beta}}+\frac{1}{3}\prn{\Lambda-\varkappa\ps{T}}\pt{g}{_{\alpha\beta}}= \varkappa\Omega\ct{T}{_{\alpha\beta}}+\frac{1}{3}\prn{\Omega^{-2}\Lambda-\Omega\varkappa\ps{T}}\ct{g}{_{\alpha\beta}}
					\end{equation}
				into \cref{eq:aux3},
					\begin{align}
					\frac{1}{3}\Omega^{-2}\Lambda\ct{N}{_{[\beta}}\ct{g}{_{\alpha]\gamma}}+\varkappa\cd{_{[\alpha}}\prn{\Omega^2\ct{T}{_{\beta]\gamma}}}-\frac{1}{3}\varkappa\cd{_{[\alpha}}\prn{\Omega^2\cs{T}}\ct{g}{_{\beta]\gamma}}= \Omega\cd{_{[\alpha}}\ct{S}{_{\beta]\gamma}}+\ct{C}{_{\alpha\beta\gamma}^\mu}\ct{N}{_\mu}\nonumber\\
					+\Omega\varkappa\ct{g}{_{\gamma[\alpha}}\ct{T}{_{\beta]\mu}}\ct{N}{^\mu}-\frac{1}{3}\varkappa\ct{N}{^\mu}\ct{g}{_{\gamma[\alpha}}\ct{g}{_{\beta]\mu}}\Omega\cs{T}+\frac{1}{3}\Lambda\Omega^{-2}\ct{g}{_{\gamma[\alpha}}\ct{N}{_{\beta]}}\spacef,
					\end{align}
				and rearrange the terms to get
					\begin{align}\label{eq:aux5}
					2\Omega\varkappa\ct{N}{_{[\alpha}}\prn{\ct{T}{_{\beta]\gamma}}}+\Omega^2\varkappa\cd{_{[\alpha}}\ct{T}{_{\beta]\gamma}}-\Omega\varkappa\ct{N}{_{[\alpha}}\ct{g}{_{\beta]\gamma}}\cs{T}-\frac{1}{3}\Omega^2\varkappa\cd{_{[\alpha}}\prn{\cs{T}}\ct{g}{_{\beta]\gamma}} \nonumber\\
					=\Omega\cd{_{[\alpha}}\ct{S}{_{\beta]\gamma}}+\ct{C}{_{\alpha\beta\gamma}^\mu}\ct{N}{_\mu}+\Omega\varkappa\ct{g}{_{\gamma[\alpha}}\ct{T}{_{\beta]\mu}}\ct{N}{^\mu}\spacef,
					\end{align}
				where we have taken into account \cref{it:energytensorassumption} on page \pageref{it:energytensorassumption}. All the terms in the above equation are regular at $ \scri $, including $ \ct{C}{_{\alpha\beta\gamma}^\delta} $ by construction, as $(M,g_{\mu\nu})$ is smooth. 
				Therefore,
					\begin{equation}\label{aux4}
					\ct{C}{_{\alpha\beta\gamma}^\mu}\ct{N}{_{\mu}}\eqs 0\spacef.
					\end{equation}
				Recall that $ \scri $ is spacelike, which implies that the Weyl tensor is completely determined by the electric and magnetic parts in the standard 3+1 decomposition ---as generally described for a Weyl candidate tensor in \cref{sssec:orthogonal-decomposition},
					\begin{align}
					\ct{E}{_{\alpha\beta}}&\defeqs \ct{n}{^\mu}\ct{n}{^\nu}\ct{C}{_{\mu\alpha\nu\beta}},\\
					\ct{B}{_{\alpha\beta}}&\defeqs \ct{n}{^\mu}\ct{n}{^\nu}\ctru{*}{C}{_{\mu\alpha\nu\beta}}\spacef.
					\end{align}
				\Cref{aux4} clearly implies  $ \ct{E}{_{ab}}\eqs 0 \eqs \ct{B}{_{ab}} $, i. e.,
					\begin{equation}\label{eq:Weylvanishes}
					\ct{C}{_{\alpha\beta\gamma}^\delta}\eqs 0\spacef.
					\end{equation}	
				Therefore, one is led to define the \emph{rescaled Weyl tensor}
					\begin{equation}\label{eq:rescaledWeyl-def}
					\ct{d}{_{\alpha\beta\gamma}^\delta}\defeq\Omega^{-1} \ct{C}{_{\alpha\beta\gamma}^\delta}\spacef.
					\end{equation}
				This tensor plays a major role in the asymptotic study of the gravitational field and the properties of space-times from the point of view of their conformal extensions --- \cite{Geroch1977,Ashtekar81,Ashtekar-Dray81,Paetz2013,Mars2016,Mars2017} are just a few examples. It features the algebraic symmetries of the Weyl tensor and at $ \scri $ is completely determined by its electric and magnetic parts --see \cref{sssec:orthogonal-decomposition} and note that the notation used there for a Weyl candidate is used now for the rescaled Weyl tensor--,
					\begin{align}
					\ct{D}{_{\alpha\beta}}&\defeqs \ct{n}{^\mu}\ct{n}{^\nu}\ct{d}{_{\mu\alpha\nu\beta}}\spacef,\\
					\ct{C}{_{\alpha\beta}}&\defeqs \ct{n}{^\mu}\ct{n}{^\nu}\ctru{*}{d}{_{\mu\alpha\nu\beta}}\spacef.
					\end{align}

				The Cotton tensor, both of the physical and conformal space-time, is defined\footnote{Keep in mind the extra factor 2 with respect to the standard definition due to our definition of $ \ct{S}{_{\alpha\beta}} $.} as
					\begin{align}
					\pt{Y}{_{\alpha\beta\gamma}}&\defeq \pd{_{[\alpha}}\pt{S}{_{\beta]\gamma}}\spacef,\label{eq:cottonPhysical}\\
					\ct{Y}{_{\alpha\beta\gamma}}&\defeq \cd{_{[\alpha}}\ct{S}{_{\beta]\gamma}}\spacef.\label{eq:cottonUnphysical}
					\end{align}
				From the contracted Bianchi identity one can show that
					\begin{align}
					\cd{_\mu}\prn{\ct{C}{_{\alpha\beta\gamma}^\mu}}+\ct{Y}{_{\alpha\beta\gamma}}=0\spacef,\label{eq:bianchiIdConformal}\\
					\pd{_\mu}\prn{\pt{C}{_{\alpha\beta\gamma}^\mu}}+\pt{Y}{_{\alpha\beta\gamma}}=0\spacef. \label{eq:bianchiIdPhysical}
					\end{align}
				The first of these two equations can be rewritten in terms of the rescaled Weyl tensor,
					\begin{equation}\label{eq:aux8}
					\Omega\cd{_\mu}\ct{d}{_{\alpha\beta\gamma}^\mu}+\ct{d}{_{\alpha\beta\gamma}^\mu}\ct{N}{_\mu}+\ct{Y}{_{\alpha\beta\gamma}}=0,
					\end{equation}
				and evaluated at $ \scri $,
					\begin{equation}\label{eq:cottoncefeScri}
					\ct{d}{_{\alpha\beta\gamma}^\mu}\ct{N}{_\mu}+\ct{Y}{_{\alpha\beta\gamma}}\eqs 0\spacef.
					\end{equation}
				We can multiply now \cref{eq:aux5} by $ \Omega^{-1} $,
					\begin{align}
					2\varkappa\ct{N}{_{[\alpha}}\ct{T}{_{\beta]\gamma}}-\varkappa\ct{N}{_{[\alpha}}\ct{g}{_{\beta]\gamma}}\cs{T}+\Omega\varkappa\cd{_{[\alpha}}\ct{T}{_{\beta]\gamma}}-\frac{1}{3}\Omega\varkappa\ct{g}{_{\gamma[\beta}}\cd{_{\alpha]}}\cs{T}\nonumber\\ =\cd{_{[\alpha}}\ct{S}{_{\beta]\gamma}}+\ct{d}{_{\alpha\beta\gamma}^\mu}\ct{N}{_\mu}+\varkappa\ct{g}{_{\gamma[\alpha}}\ct{T}{_{\beta]\mu}}\ct{N}{^\mu}\quad				\label{eq:aux9}	
					\end{align}
				and evaluate it at $ \scri $ using \cref{eq:cottoncefeScri},
					\begin{equation}\label{eq:aux7}
					2\varkappa\ct{N}{_{[\alpha}}\ct{T}{_{\beta]\gamma}}-\varkappa\ct{N}{_{[\alpha}}\ct{g}{_{\beta]\gamma}}\cs{T}-\varkappa\ct{g}{_{\gamma[\alpha}}\ct{T}{_{\beta]\mu}}\ct{N}{^\mu}\eqs 0\spacef.
					\end{equation}
				Contracting with $ \ct{N}{^\alpha}\ct{N}{^\gamma}\ct{P}{^\beta_\delta} $ gives
					\begin{equation}\label{eq:aux10}
					\ct{n}{^\mu}\ct{P}{^\rho_\alpha}\ct{T}{_{\mu\rho}}\eqs 0 \spacef.
					\end{equation}
				Furthermore, if one contracts \cref{eq:aux7} with $ \ct{N}{^\alpha}\ct{P}{^\gamma_\chi}\ct{P}{^\beta_\delta} $,
					\begin{equation}
					\ct{N}{^\mu}\ct{N}{_\mu}\ct{P}{^\gamma_\chi}\ct{P}{^\beta_\delta}\ct{T}{_{\beta\gamma}}+\frac{1}{2}\ct{P}{_{\chi\delta}}\ct{N}{^\mu}\ct{N}{^\rho}\ct{T}{_{\mu\rho}}+\frac{1}{2}\ct{N}{^\mu}\ct{N}{_\mu}\cs{T}\ms{_{\chi\delta}}\eqs 0\spacef,
					\end{equation}
				and uses here \cref{eq:projectorScri},
					\begin{equation}\label{eq:aux6}
					\ct{N}{^\mu}\ct{N}{_\mu}\ct{P}{^\gamma_\chi}\ct{P}{^\beta_\delta}\ct{T}{_{\beta\gamma}}+\frac{1}{2}\ct{N}{^\tau}\ct{N}{_\tau}\ct{P}{_{\chi\delta}}\ct{P}{^{\rho\mu}}\ct{T}{_{\mu\rho}}\eqs 0\spacef.
					\end{equation}
				Then, contract one more time with $  \ct{P}{^{\chi\delta}} $ to get
					\begin{equation}
					\ct{P}{^{\mu\rho}}\ct{T}{_{\mu\rho}} \eqs 0\spacef.
					\end{equation}
				Inserting this last line into \cref{eq:aux6} the result is
					\begin{equation}\label{eq:energymomentumComponentPP}
					\ct{P}{^\mu_\alpha}\ct{P}{^\nu_\beta}\ct{T}{_{\mu\nu}}\eqs 0\spacef.
					\end{equation}
				Finally, apply $\ct{N}{^\alpha} \ct{P}{^{\beta\gamma}} $ to \cref{eq:aux7} to derive
					\begin{equation}
					\frac{3}{2}\ct{N}{^\alpha}\ct{N}{^\mu}\ct{T}{_{\alpha\mu}}\eqs \frac{3}{2}\ct{N}{_\nu}\ct{N}{^\nu}\cs{T}-\ct{N}{_\rho}\ct{N}{^\rho}\ct{P}{^{\beta\gamma}}\ct{T}{_{\beta\gamma}},
					\end{equation}
				or, equivalently,
					\begin{equation}\label{eq:energymomentumComponentNN}
					\ct{n}{^\rho}\ct{n}{^\mu}\ct{T}{_{\mu\rho}}\eqs - \cs{T}\spacef.
					\end{equation}
			 	We have just shown that, at $ \scri $, the tensor $ \ct{T}{_{\alpha\beta}} $ has only one non-vanishing component in general:
					\begin{equation}\label{eq:energymomentumConformalScri}
					\ct{T}{_{\alpha\beta}}\eqs -\cs{T}\ct{n}{_\alpha}\ct{n}{_\beta}\spacef.
					\end{equation}
				The energy-momentum tensor determines, through the field equations, the Cotton tensor. This appears explicitly from definition \eqfref{eq:cottonPhysical} and \cref{eq:schoutenPhsycialMatter},
					\begin{equation}
					\pt{Y}{_{\alpha\beta\gamma}}=\kappa\pd{_{[\alpha}}\pt{T}{_{\beta]\gamma}}-\frac{1}{3}\varkappa\pt{g}{_{\gamma[\beta}}\cd{_{\alpha]}}\ps{T}\spacef.
					\end{equation}
				In order to write this formula in terms of quantities in $ \cs{M} $, note that
					\begin{equation}
					\pd{_{[\alpha}}\pt{T}{_{\beta]\gamma}}=\Omega\cd{_{[\alpha}}\ct{T}{_{\beta]\gamma}}-\ct{N}{^\lambda}\ct{T}{_{\lambda[\beta}}\ct{g}{_{\alpha]\gamma}}+2\ct{N}{_{[\alpha}}\ct{T}{_{\beta]\gamma}}\spacef,
					\end{equation}
					\begin{equation}
					\pt{g}{_{\gamma[\beta}}\cd{_{\alpha]}}\cs{T}=3\ct{N}{_{[\alpha}}\ct{g}{_{\beta]\gamma}}\cs{T}+\Omega\ct{g}{_{\gamma[\beta}}\cd{_{\alpha]}}\cs{T}\spacef.
					\end{equation}
				Then,
					\begin{equation}\label{eq:cottonPhysicalMatter}
					\frac{1}{\varkappa}\pt{Y}{_{\alpha\beta\gamma}}=\Omega\cd{_{[\alpha}}\ct{T}{_{\beta]\gamma}}-\ct{N}{^\lambda}\ct{T}{_{\lambda[\beta}}\ct{g}{_{\alpha]\gamma}}+2\ct{N}{_{[\alpha}}\ct{T}{_{\beta]\gamma}}-\ct{N}{_{[\alpha}}\ct{g}{_{\beta]\gamma}}\cs{T}-\frac{1}{3}\Omega\ct{g}{_{\gamma[\beta}}\cd{_{\alpha]}}\cs{T}
					\end{equation}
				which at $ \scri $, after using \cref{eq:energymomentumConformalScri}, reduces to
					\begin{equation}
					\pt{Y}{_{\alpha\beta\gamma}}\eqs 0\spacef.
					\end{equation}
				Thus, it is natural to introduce the \emph{rescaled Cotton tensor}
					\begin{equation}
					\ct{y}{_{\alpha\beta\gamma}} \defeq \Omega^{-1}\pt{Y}{_{\alpha\beta\gamma}}\spacef,
					\end{equation}
				which is regular at $ \scri $. Notice that it is defined in terms of the physical Cotton tensor, which is not a conformal-invariant object. There exists a useful relation \cite{Kroon},
					\begin{equation}
					\cd{_\mu}\prn{\Omega^{-1}\ct{C}{_{\alpha\beta\gamma}^\mu}}= \Omega^{-1}\pd{_\mu}\ct{C}{_{\alpha\beta\gamma}^\mu}\quad
					\end{equation}
				which allows us to write equation \cref{eq:bianchiIdPhysical} as
					\begin{equation}\label{eq:divergencedcefe}
					\ct{y}{_{\alpha\beta\gamma}}+\cd{_{\mu}}\ct{d}{_{\alpha\beta\gamma}^\mu}= 0\spacef.
					\end{equation}
				To see which conditions on the matter content make the rescaled Weyl tensor divergence-free at $ \scri $, multiply \cref{eq:cottonPhysicalMatter} by $ \Omega^{-1} $
					\begin{equation}\label{eq:auxCotton}
						\frac{1}{\varkappa}\ct{y}{_{\alpha\beta\gamma}}=\cd{_{[\alpha}}\ct{T}{_{\beta]\gamma}}-\Omega^{-1}\ct{N}{^\lambda}\ct{T}{_{\lambda[\beta}}\ct{g}{_{\alpha]\gamma}}+2\Omega^{-1}\ct{N}{_{[\alpha}}\ct{T}{_{\beta]\gamma}}-\Omega^{-1}\ct{N}{_{[\alpha}}\ct{g}{_{\beta]\gamma}}\cs{T}-\frac{1}{3}\ct{g}{_{\gamma[\beta}}\cd{_{\alpha]}}\cs{T}\quad
					\end{equation}
				so that the following implication holds:
					\begin{equation}\label{eq:matter-decay-cotton}
					\pt{T}{_{\alpha\beta}}\lvert_{\scri}\sim \mathcal{O}\prn{\Omega^p} \text{ with } p>2  \implies \ct{y}{_{\alpha\beta\gamma}}\eqs 0\eqs\cd{_{\mu}}\ct{d}{_{\alpha\beta\gamma}^\mu}= 0\spacef.
					\end{equation}
				Substitution of \cref{eq:divergencedcefe} into \cref{eq:aux8} produces
					\begin{equation}\label{eq:cottoncefe}
					\ct{d}{_{\alpha\beta\gamma}^\mu}\ct{N}{_\mu}+\ct{Y}{_{\alpha\beta\gamma}}-\Omega\ct{y}{_{\alpha\beta\gamma}}=0\spacef.
					\end{equation}
				Continuing with our analysis, take equation \cref{eq:aux9}, apply $ \Omega^{-1}\ct{P}{^\beta_\delta}\ct{N}{^\alpha}\ct{N}{^\gamma} $ and evaluate at $ \scri $ using \cref{eq:cottoncefe,eq:auxCotton}:
					\begin{align}
					\frac{1}{6}\varkappa\ct{P}{^\beta_\delta}\ct{N}{^\alpha}\ct{N}{^\gamma}\cd{_{\beta}}\cs{T}\ct{g}{_{\alpha\gamma}}+\varkappa\ct{P}{^\beta_\delta}\ct{N}{^\alpha}\ct{N}{^\gamma}\cd{_{[\alpha}}\ct{T}{_{\beta]\gamma}} \nonumber\\
					\eqs -\ct{P}{^\beta_\delta}\ct{N}{^\alpha}\ct{N}{^\gamma}\cd{_\mu}\ct{d}{_{\alpha\beta\gamma}^\mu}-\frac{1}{2}\varkappa\ct{N}{_\mu}\ct{N}{^\mu}\ct{P}{^\beta_\delta}\ct{N}{^\rho}\Omega^{-1}\ct{T}{_{\beta\rho}}\spacef.
					\end{align}
				(We know, by \cref{eq:aux10}, that $\ct{P}{^\beta_\delta}\ct{N}{^\rho}\Omega^{-1}\ct{T}{_{\beta\rho}} $ is regular at $ \scri $). Dividing by $ \cs{N}^2 $,
					\begin{equation}\label{aux11}
					-\frac{1}{6}\varkappa\ct{P}{^\beta_\delta}\cd{_{\beta}}\cs{T}+\varkappa\ct{P}{^\beta_\delta}\ct{n}{^\alpha}\ct{n}{^\gamma}\cd{_{[\alpha}}\ct{T}{_{\beta]\gamma}}\eqs -\ct{P}{^\beta_\delta}\ct{n}{^\alpha}\ct{n}{^\gamma}\cd{_\mu}\ct{d}{_{\alpha\beta\gamma}^\mu}+\frac{1}{2}\varkappa\ct{P}{^\beta_\delta}\ct{N}{^\rho}\Omega^{-1}\ct{T}{_{\beta\rho}}.
					\end{equation}
				 Using our choice of gauge, the second term on the left-hand side can be rewritten as
					\begin{align}
					2\ct{P}{^\beta_\delta}\ct{n}{^\alpha}\ct{n}{^\gamma}\cd{_{[\alpha}}\ct{T}{_{\beta]\gamma}}&\eqs\ct{P}{^\beta_\delta}\ct{n}{^\alpha}\ct{n}{^\gamma}\cd{_\alpha}\ct{T}{_{\beta\gamma}}-\ct{P}{^\beta_\delta}\ct{n}{^\alpha}\ct{n}{^\gamma}\cd{_{\beta}}\ct{T}{_{\alpha\gamma}}\nonumber\\
					&\eqs\ct{n}{^\alpha}\cd{_\alpha}\prn{\ct{P}{^\beta_\delta}\ct{n}{^\gamma}\ct{T}{_{\beta\gamma}}}-\ct{n}{^\alpha}\underbrace{\cd{_{\alpha}}\prn{\ct{P}{^\beta_\delta}\ct{n}{^\gamma}}}_{\eqs 0}+\ct{P}{^\beta_\delta}\cd{_\beta}\cs{T}\nonumber\\
					&\eqs\ct{n}{^\alpha}\cd{_\alpha}\prn{\Omega\underbrace{\Omega^{-1}\ct{P}{^\beta_\delta}\ct{n}{^\gamma}\ct{T}{_{\beta\gamma}}}_{\text{Regular at }\scri}}+\ct{P}{^\beta_\delta}\cd{_\beta}\cs{T}\nonumber\\
					&\eqs\Omega^{-1}\ct{P}{^\beta_\delta}\ct{n}{^\gamma}\ct{T}{_{\beta\gamma}}\ct{n}{^\alpha}\ct{N}{_{\alpha}}+\ct{P}{^\beta_\delta}\cd{_\beta}\cs{T}.
					\end{align}
				Then, \cref{aux11} reads
					\begin{equation}\label{aux13}
					\frac{1}{3}\varkappa\ct{P}{^\beta_\delta}\cd{_\beta}\cs{T}-\varkappa\Omega^{-1}\ct{P}{^\beta_\delta}\ct{N}{^\gamma}\ct{T}{_{\beta\gamma}}\eqs -\ct{P}{^\beta_\delta}\ct{n}{^\gamma}\ct{n}{^\alpha}\cd{_\mu}\ct{d}{_{\alpha\beta\gamma}^\mu}\spacef.
					\end{equation}
				 Now, contract \cref{eq:aux9} with $ \Omega^{-1}\ct{\eta}{^{\alpha\beta}_{\rho\sigma}} \ct{N}{^\gamma}\ct{N}{^\rho}$ and evaluate at $ \scri $
					\begin{equation}
					\varkappa \ct{\eta}{^{\alpha\beta}_{\rho\sigma}}\ct{N}{^\rho}\ct{N}{^\gamma}\cd{_{[\alpha}}\ct{T}{_{\beta]\gamma}}\eqs -\ct{\eta}{^{\alpha\beta}_{\rho\sigma}}\ct{N}{^\rho}\ct{N}{^\gamma}\cd{_\mu}\ct{d}{_{\alpha\beta\gamma}^\mu}\spacef,
					\end{equation}
				where the rest of the terms vanish because $ \ct{\eta}{^{\alpha\beta}_{\rho\sigma}}\ct{N}{^\rho}\ct{N}{^\sigma}=0 $. Notice, also, that
					\begin{align}
					&\ct{\eta}{^{\alpha\beta}_{\rho\sigma}}\ct{N}{^\rho}\ct{N}{^\gamma}\cd{_{[\alpha}}\ct{T}{_{\beta]\gamma}}\eqs \ct{\eta}{_{\nu\mu\rho\sigma}}\ct{N}{^\rho}\ct{N}{^\gamma}\ct{P}{^{\mu\beta}}\ct{P}{^{\nu\alpha}}\cd{_{[\alpha}}\ct{T}{_{\beta]\gamma}}\nonumber\\
					&\eqs\frac{1}{2}\ct{\eta}{_{\nu\mu\rho\sigma}}\brkt{\ct{N}{^\rho}\ct{P}{^{\nu\alpha}}\underbrace{\cd{_{\alpha}}\prn{\ct{N}{^\gamma}\ct{P}{^{\mu\beta}}\ct{T}{_{\beta\gamma}}}}_{\propto \ct{N}{_\alpha}}-\ct{T}{_{\beta\gamma}}\ct{N}{^\rho}\ct{P}{^{\nu\alpha}}\underbrace{\cd{_{\alpha}}\prn{\ct{N}{^\gamma}\ct{P}{^{\mu\beta}}}}_{\eqs 0}-(\beta\leftrightarrow\alpha)}\eqs 0.
					\end{align}
				Then,
					\begin{equation}\label{aux12}
					\ct{n}{^\alpha}\ct{n}{^\gamma}\cd{_\mu}\ctru{*}{d}{_{\alpha\beta\gamma}^\mu}\eqs 0\spacef.
					\end{equation}
				\Cref{aux12,aux13} give us information about the divergence of $ \ct{D}{_{ab}} $ and $ \ct{C}{_{ab}} $ at $ \scri $. From the first it is easy to see that\footnote{The intrinsic covariant derivative on $  \scri $ is denoted by $ \cds{_a} $, while $ \prn{\ct{\omega}{_\alpha^a}} $ is a basis of linearly independent one-forms on $ \scri $ orthogonal to $ \ct{n}{^\alpha} $. For further details, see \cref{app:spatial-hypersurface}, where we introduce this notation for a general 3-dimensional hypersurface.}
					\begin{equation}\label{eq:divD}
					\ct{\omega}{_\delta^d}\cds{_m}\ct{D}{_d^m}\eqs \varkappa\Omega^{-1}\ct{P}{^\beta_\delta}\ct{N}{^\gamma}\ct{T}{_{\beta\gamma}}-\frac{1}{3}\varkappa\ct{P}{^\beta_\delta}\cd{_\beta}\cs{T}
					\end{equation}
				and from the latter,
					\begin{equation}\label{eq:divC}
					\cds{_m}\ct{C}{_a^m}\eqs 0\spacef.
					\end{equation}
				In particular, $ \ct{D}{_d^m} $ is divergence free too in vacuum and in presence of most of the relevant physical fields.\\
				
				To end this section, we summarise the relevant equations which are \cref{eq:cottoncefe,eq:derivativeNcfe,eq:derivativefcefe,eq:divergencedcefe,eq:normNcefe},
					\begin{align}
					&\cd{_\alpha}\ct{N}{_\beta} = -\frac{1}{2}\Omega\ct{S}{_{\alpha\beta}}+ \cs{f}\ct{g}{_{\alpha\beta}}+ \frac{1}{2}\Omega^2\varkappa\ct{\underline{T}}{_{\alpha\beta}}\spacef,\label{eq:cefesDerN}\\
					&\ct{N}{_\mu}\ct{N}{^\mu}=\frac{\Omega^3}{12}\varkappa \cs{T}-\frac{\Lambda}{3}+ 2\Omega f\spacef,\label{eq:cefesNormN}\\
					&\cd{_\alpha}f= -\frac{1}{2}\ct{S}{_{\alpha\mu}}\ct{N}{^\mu}+\frac{1}{2}\Omega\varkappa\ct{N}{^\mu}\ct{\underline{T}}{_{\alpha\mu}}-\frac{1}{24}\Omega^2\varkappa\cd{_\alpha}\cs{T}- \frac{1}{8}\Omega\varkappa\ct{N}{_\alpha}\cs{T}\spacef,\label{eq:cefesDerF}\\
					&\ct{d}{_{\alpha\beta\gamma}^\mu}\ct{N}{_\mu}+\cd{_{[\alpha}}\prn{\ct{S}{_{\beta]\gamma}}}-\Omega\ct{y}{_{\alpha\beta\gamma}}=0\spacef,\label{eq:cefesDerSchouten}\\
					&\ct{y}{_{\alpha\beta\gamma}}+\cd{_{\mu}}\ct{d}{_{\alpha\beta\gamma}^\mu}= 0\spacef,\label{eq:cefesDerWeyl}\\
					&\ct{R}{_{\alpha\beta\gamma\delta}} =\Omega\ct{d}{_{\alpha\beta\gamma\delta}}+ \ct{g}{_{\alpha[\gamma}}\ct{S}{_{\delta]\beta}}-\ct{g}{_{\beta[\gamma}}\ct{S}{_{\delta]\alpha}}\spacef.	\label{eq:cefesRiemann}						
					\end{align}
				whose evaluation at $ \scri $ is 
					\begin{align}
					&\cd{_\alpha}\ct{N}{_\beta} \eqs 0\spacef,\label{eq:cefesScriDerN}\\
					&\ct{N}{_\mu}\ct{N}{^\mu}\eqs -\frac{\Lambda}{3} \spacef,\label{eq:cefesScriNormN}\\
					&\cd{_\alpha}f\eqs -\frac{1}{2}\ct{S}{_{\alpha\mu}}\ct{N}{^\mu}\spacef,\label{eq:cefesScriDerF}\\
					&\ct{d}{_{\alpha\beta\gamma}^\mu}\ct{N}{_\mu}+\cd{_{[\alpha}}\prn{\ct{S}{_{\beta]\gamma}}}\eqs 0\spacef,\label{eq:cefesScriDerSchouten}\\
					&\ct{y}{_{\alpha\beta\gamma}}+\cd{_{\mu}}\ct{d}{_{\alpha\beta\gamma}^\mu}\eqs 0\spacef,\label{eq:cefesScriDerWeyl}\\
					&\ct{R}{_{\alpha\beta\gamma\delta}} \eqs\ct{g}{_{\alpha[\gamma}}\ct{S}{_{\delta]\beta}}-\ct{g}{_{\beta[\gamma}}\ct{S}{_{\delta]\alpha}}\spacef.	\label{eq:cefesScriRiemann}	
					\end{align}
					
				\Cref{eq:cefesDerF,eq:cefesDerN,eq:cefesDerSchouten,eq:cefesDerWeyl,eq:cefesNormN,eq:cefesRiemann} in vacuum -- $ \ct{T}{_{\alpha\beta}}=0=\ct{y}{_{\alpha\beta\gamma}} $ -- constitute the so called Metric Conformal Field Equations (MCFE) \cite{Paetz2013} which, when considered as a system of differential equations for the variables $ \prn{\Omega,\ct{d}{_{\alpha\beta\gamma}^\mu}, \cs{f},\ct{g}{_{\alpha\beta}},\ct{S}{_{\alpha\beta}}} $,  are equivalent to the physical vacuum EFE ---the Riemann components $ \ct{R}{_{\alpha\beta\gamma\delta}} $ are considered as functions of the metric components $ \ct{g}{_{\alpha\beta}} $.
				\begin{remark}\label{rem:dmasdegenerado}
				The Petrov type of $\ct{d}{_{\alpha\beta\gamma}^\mu}$ at $\scri$ is partly determined by the Petrov type of the Weyl tensor in the asymptotic region of the physical space-time in the sense that the type of the former can only be equally, or more, degenerate than that of the latter. To prove this, observe that from \eqref{eq:rescaledWeyl-def} the Petrov type of $\ct{d}{_{\alpha\beta\gamma}^\mu}$ outside $\scri$ in the physical supermomenta is precisely the same as that of $\pt{C}{_{\alpha\beta\gamma}^\mu}= \ct{C}{_{\alpha\beta\gamma}^\mu}$ there. However, at $\scri$ we have \eqref{eq:Weylvanishes} while $\ct{d}{_{\alpha\beta\gamma}^\mu}$ is regular and generally non-vanishing at $\scri$. Obviously, $\ct{d}{_{\alpha\beta\gamma}^\mu}|_\scri$ must have a Petrov type compatible with its Petrov type outside $\scri$. Using a reasoning based on the existence of PND, or one based on the invariants that are non-zero (or vanishing) for each Petrov type, the possible Petrov specializations of $\ct{d}{_{\alpha\beta\gamma}^\mu}|_\scri$ imply that its Petrov type is as degenerate, or more, than that of the physical Weyl tensor in a neighbourhood of $\scri$.
				\end{remark}
				
		

	\section{Asymptotic structure}
			
		This section is devoted to the study of the intrinsic asymptotic structure and its relation to the space-time fields. The approach that we follow is to treat $ \scri $ as a hypersurface and apply the formulae of \cref{app:spatial-hypersurfaces} --we also follow the notation for 3-dimensional hypersurfaces introduced there, here used for $ \scri $--. We will connect the intrinsic curvature with the kinematics of the congruence of timelike curves tangent to the vector field $ \ct{n}{^\alpha} $. Afterwards, the new gravitational radiation condition will be presented \cite{Fernandez-Alvarez_Senovilla20b} and compared with the $ \Lambda=0 $-limiting case. 
		\subsection{Infinity and its intrinsic geometry}\label{sec:intrinsicCurvature}
In the present scenario, $ \scri $ is a space-like three-dimensional hypersurface. Its topology is not fixed in general and typical examples include $ \mathbb{S}^3 $, $ \mathbb{S}^2\times \mathbb{R} $ or $ \mathbb{R}^3 $ --for some examples see \cite{Mars2017,Ashtekar2014}. Hence, one can always think of $ \scri $ as $ \mathbb{S}^3 $ or $ \mathbb{S}^3 $ after removing a set of points. Also, an important element in \cref{sec:news,sec:additional-structure} is the introduction of cuts; a cut $ \prn{\Sc, \mc{_{AB}}} $ on $ \scri $ is a two-dimensional Riemannian manifold $ \Sc \subset \scri$ equipped with the inherited metric $ \mc{_{AB}} $. 
			\begin{figure}[h!]
				\centering
				\includegraphics[width=0.8\textwidth]{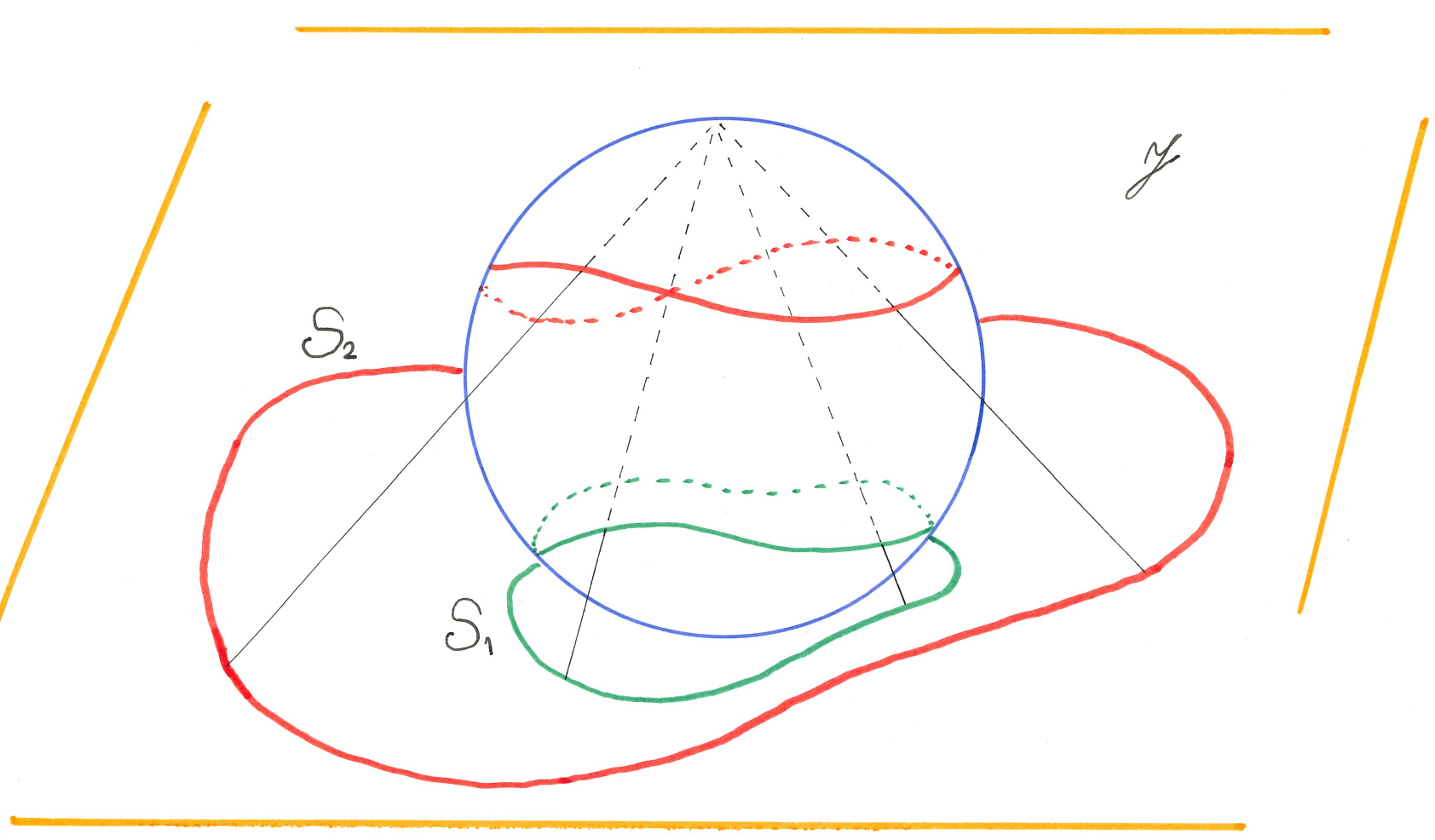}
					\caption[Stereographic projection of infinity]{In the presence of a positive cosmological constant, $ \scri $ usually has $ \mathbb{S}^3 $-topology or $ \mathbb{S}^3 $  without a set of points. Also, one can consider Riemannian surfaces, or cuts, denoted by $ \Sc $. The figure shows --wirh 1 dimension suppressed-- the stereographic projection of $ \scri $ to the plane, including a couple of cuts labelled by $ \Sc_{1} $ and $ \Sc_{2} $. Thus, one can picture $ \scri $ as $ \mathbb{R}^3 $, which is how it is represented in the rest of the figures.}
			\end{figure}
			
			Given \cref{eq:divergence-freeGauge}, the second fundamental form --\cref{eq:secondFundamentalFormI}-- vanishes,
				\begin{equation}\label{eq:extrinsicScri}
					\ct{\kappa}{_{ab}}= 0\spacef.
				\end{equation}
			The intrinsic Riemann and Ricci tensor and the scalar  curvature read, respectively -- \cref{eq:gaussrelIiii,eq:gaussrelIii,eq:gaussrelIi}--,
				\begin{align} 
				\cts{R}{_{abc}^d} &\eqs\ct{e}{^\alpha_a}\ct{e}{^\beta_b}\ct{e}{^\gamma_c}\ct{R}{_{\alpha\beta\gamma}^\delta}\ct{\omega}{_\delta^d}\spacef,\label{eq:gaussrelScriCo1Riemann}\\
				\cts{R}{_{ac}} &\eqs \ct{e}{^\alpha_a}\ct{e}{^\gamma_c}\ct{R}{_{\alpha\gamma}}+\ct{n}{^\beta}\ct{n}{_\delta}\ct{e}{^\alpha_a}\ct{e}{^\gamma_c}\ct{R}{_{\alpha\beta\gamma}^\delta}\spacef,\label{eq:gaussrelScriCo1RicciTensor}\\
				\cts{R}{}&\eqs \ct{R}{}+2\ct{n}{^\alpha}\ct{n}{^\gamma}\ct{R}{_{\alpha\gamma}}\spacef,\label{eq:gaussrelScriCo1RicciScalar}
				\end{align}		
			The intrinsic Schouten tensor in three dimensions is defined as
				\begin{equation}
				\cts{S}{_{ab}}\defeq \cts{R}{_{ab}}-\frac{1}{4}\cts{R}{}\ms{_{ab}}\spacef,
				\end{equation}
			and we can use the equations above in order to write it in terms of the space-time curvature
				\begin{equation}
				\cts{S}{_{ab}}\eqs \ct{e}{^\mu_a}\ct{e}{^\nu_b}\ct{S}{_{\mu\nu}}+\ct{n}{^\rho}\ct{n}{^\sigma}\ct{e}{^\mu_a}\ct{e}{^\nu_b}\ct{R}{_{\rho\mu\sigma\nu}}-\frac{1}{12}\cs{R}\ms{_{ab}}-\frac{1}{2}\ct{n}{^\mu}\ct{n}{^\nu}\ct{R}{_{\mu\nu}}\ms{_{ab}}\spacef.
				\end{equation}
			On $ \scri $ the space-time curvature is determined by $ \ct{S}{_{\alpha\beta}} $ completely, see \cref{eq:cefesScriRiemann}, and it is possible to write
				\begin{equation}
				\ct{n}{^\rho}\ct{n}{^\sigma}\ct{e}{^\mu_a}\ct{e}{^\nu_b}\ct{R}{_{\rho\mu\sigma\nu}}\eqs -\frac{1}{2}\ct{e}{^\mu_a}\ct{e}{^\nu_b}\ct{S}{_{\mu\nu}}+\frac{1}{2}\ms{_{ab}}\ct{n}{^\rho}\ct{n}{^\sigma}\ct{R}{_{\rho\sigma}}+\frac{1}{12}\cs{R}\ms{_{ab}},
				\end{equation}
			and use this to arrive at
				\begin{equation}\label{eq:intrinsicS}
				\cts{S}{_{ab}} \eqs \frac{1}{2}\ct{e}{^\mu_a}\ct{e}{^\nu_b}\ct{S}{_{\mu\nu}}\spacef.
				\end{equation}
			Indeed, by \cref{eq:intrinsicS,eq:gaussrelScriCo1Riemann,eq:cefesScriRiemann} one can write
			\begin{equation}\label{eq:intrinsicriemannScriSchoutenRelation}
			\cts{R}{_{abcd}}\eqs 2\ms{_{a[c}}\cts{S}{_{d]b}}-2\ms{_{b[c}}\cts{S}{_{d]a}}\spacef,
			\end{equation}	
		which is valid in general for dimension 3. Note that  on a neighbourhood of $ \scri $ where $ \ct{n}{_\alpha} $ is well defined, since $ \ct{P}{^\alpha_\beta} $ is defined there too, we can consider
				\begin{equation}
				\cts{S}{_{\alpha\beta}}=\frac{1}{2}\ct{P}{^\mu_\alpha}\ct{P}{^\mu_\beta}\ct{S}{_{\alpha\beta}}\spacef.
				\end{equation}
		 Also, we introduce the intrinsic Cotton tensor:
				\begin{equation}\label{eq:cottonIntrinsicSchoutenIntrinsic}
				\cts{Y}{_{abc}}\defeq 2\cds{_{[a}}\cts{S}{_{b]c}}\spacef,
				\end{equation}
			together with the Cotton-York tensor,
				\begin{equation}
				\cts{Y}{_{ab}}\defeq -\frac{1}{2}\ct{\epsilon}{_a^{pq}}\cts{Y}{_{pqb}}\spacef.
				\end{equation}
			The electric and magnetic parts of the rescaled Weyl tensor can be written explicitly in terms of $ \cts{S}{_{\alpha\beta}} $,
				\begin{align}
				\ct{C}{_{ab}}&\eqs \frac{1}{2}\ct{e}{^\alpha_a}\ct{e}{^\beta_b}\ct{\eta}{^{\rho\sigma}_{\alpha\nu}}\ct{n}{^\lambda}\ct{n}{^\nu}\ct{d}{_{\rho\sigma\beta\lambda}}\eqs -\frac{1}{2}\ct{e}{^\alpha_a}\ct{e}{^\beta_b}\ct{\epsilon}{^{\rho\sigma}_{\alpha}}\ct{n}{^\lambda}\ct{d}{_{\rho\sigma\beta\lambda}} \nonumber\\
				&\eqs -\frac{1}{2N}\ct{e}{^\alpha_a}\ct{e}{^\beta_b}\ct{e}{^\rho_p}\ct{e}{^\sigma_q}\ct{\epsilon}{^{pq}_{\alpha}}\ct{N}{^\lambda}\ct{d}{_{\rho\sigma\beta\lambda}}\eqs\frac{1}{2N}\ct{e}{^\alpha_a}\ct{e}{^\beta_b}\ct{e}{^\rho_p}\ct{e}{^\sigma_q}\ct{\epsilon}{^{pq}_{\alpha}} \cd{_{[\rho}}\ct{S}{_{\sigma]\beta}}\spacef.
				\end{align}
			In the second line we have used \cref{eq:cottoncefe}. A similar computation can be performed to write an equation for $ \ct{D}{_{ab}} $, and we end up with two important formulae:
				\begin{align}
				\ct{C}{_{ab}}&= \sqrt{\frac{3}{\Lambda}}\ct{\epsilon}{^{pq}_a}\cds{_{[p}}\cts{S}{_{q]b}}\spacef,\label{eq:magneticSchouten}\\
				\ct{D}{_{ab}}&\eqs -\sqrt{\frac{3}{\Lambda}}\ct{e}{^\alpha_a}\ct{e}{^\beta_b}\ct{n}{^\mu}\cd{_{[\alpha}}\ct{S}{_{\mu]\beta}}\eqs\sqrt{\frac{3}{\Lambda}}\ct{e}{^\alpha_a}\ct{e}{^\beta_b}\ct{n}{^\mu}\cd{_\mu}\cts{S}{_{\alpha\beta}}\spacef.\label{eq:electricSchouten}
				\end{align}
			where in the last line we have used \cref{eq:schoutenComponentNPScri}. Remarkably, \cref{eq:magneticSchouten} tell us that \textbf{the magnetic part of the rescaled Weyl tensor is completely determined by the geometry of $ \scri $}. In contrast, \cref{eq:electricSchouten} shows that \textbf{the electric part is unknown from the  intrinsic point of view}\footnote{Note that we are able to write $ \ct{D}{_{ab}} $ in terms of $ \cts{S}{_{\alpha\beta}} $  because it is defined on a neighbourhood of $ \scri $ and we can compute its derivative along $ \ct{n}{^\alpha} $.} --in agreement with known results  \cite{Friedrich1986a,Mars2016}. These two conclusions have direct implications in the search of the asymptotic radiative degrees of freedom with a positive cosmological constant and must be taken fully into account. \\

			To see what implication a vanishing $ \ct{C}{_{ab}} $ would have on the geometry of $ \scri $, use \cref{eq:cottonIntrinsicSchoutenIntrinsic} to write it as
				\begin{equation}
				\ct{C}{_{ab}} \eqs \frac{1}{2}\sqrt{\frac{3}{\Lambda}}\ct{\epsilon}{^{pq}_a}\cts{Y}{_{pqb}}\eqs -\sqrt{\frac{3}{\Lambda}}\cts{Y}{_{ab}}\spacef.
				\end{equation}
			It is well known (see \cite{Kroon}, for instance) that the Cotton-York tensor of the metric of a three dimensional manifold vanishes if and only if the metric is locally conformally flat. Thus, the vanishing of the magnetic part of the rescaled Weyl tensor strongly constraints the intrinsic geometry and the would-be degrees of  freedom of the gravitational field ---for a discussion on this matter, see \cite{Ashtekar2015}.

		\subsection{Kinematics of the normal to $ \scri $}	
			We are interested in relating the kinematic quantities --shear, acceleration and expansion\footnote{By definition, $ \ct{n}{_\alpha} $ is proportional to a gradient and, therefore, has vanishing rotation}-- of $ \ct{n}{_\alpha} $ with the Schouten tensor and with the electric and magnetic parts of the rescaled Weyl tensor through \cref{eq:magneticSchouten,eq:electricSchouten}. This gives an intuitive idea of the impact that $ \ct{C}{_{\alpha\beta}} $ and $ \ct{D}{_{\alpha\beta}} $ have on the congruence of curves that an asymptotic observer would follow. \\
			
			To start with, we have to compute the covariant derivative of $ \ct{n}{_\alpha} $ using \cref{eq:derivativeNcfe,eq:derivativeNormN,eq:derivativefcefe},
				\begin{align}
				\cd{_\alpha}\ct{n}{_\beta} &=\frac{1}{N}\cd{_\alpha}\ct{N}{_{\beta}}-\frac{1}{\cs{N}^2}\ct{N}{_{\beta}}\cd{_\alpha}\cs{N}= -\frac{\Omega}{2\cs{N}}\ct{S}{_{\alpha\beta}}+\frac{f}{N}\ct{g}{_{\alpha\beta}}+ \frac{1}{2}\Omega^2\frac{\varkappa}{\cs{N}}\ct{\underline{T}}{_{\alpha\beta}}\nonumber\\
				&-\frac{1}{2\cs{N}^3}\ct{N}{_{\beta}}\prn{-2\ct{N}{_\alpha}f-2\Omega\cd{_\alpha}f-\frac{1}{4}\Omega^2\varkappa\cs{T}\ct{N}{_{\alpha}}-\frac{1}{12}\Omega^3\varkappa\cd{_\alpha}\cs{T}}\nonumber\\
				&= -\frac{1}{2\cs{N}}\Omega\ct{S}{_{\alpha\mu}}\ct{P}{^\mu_\beta}+\frac{1}{\cs{N}}\ct{P}{_{\alpha\beta}}f+\frac{1}{2\cs{N}}\Omega^2\varkappa\ct{P}{^\mu_\beta}\ct{\underline{T}}{_{\alpha\mu}}\spacef.
				\end{align}
			It is easy to see that this vanishes at $ \scri $, as it must, given our choice of gauge. In other words, the kinematic quantities vanish at $ \scri $. Nevertheless, their `time derivatives' --along $ \ct{n}{^\alpha} $-- may be non-vanishing at $ \scri $. To begin with, consider the acceleration,
				\begin{equation}
				\ct{a}{_{\alpha}} = \ct{n}{^\mu}\cd{_\mu}\ct{n}{_{\alpha}}=-\frac{1}{2\cs{N}}\Omega\ct{P}{^\nu_\alpha}\ct{n}{^\mu}\ct{S}{_{\mu\nu}}+ 
				\frac{1}{2\cs{N}}\Omega^2\varkappa\ct{P}{^\nu_\alpha}\ct{n}{^\mu}\ct{\underline{T}}{_{\nu\mu}}\eqs 0\spacef,
				\end{equation}
				\begin{align}
				\ctd{a}{_\alpha}&\defeq\ct{n}{^\mu}\cd{_\mu}\ct{a}{_\alpha}=\frac{1}{2\cs{N}}\ct{P}{^\nu_\alpha}\ct{n}{^\mu}\ct{S}{_{\nu\mu}}-\Omega \ct{n}{^\rho}\cd{_\rho}\prn{\frac{1}{2\cs{N}}\ct{P}{^\nu_\alpha}\ct{n}{^\mu}\ct{S}{_{\nu\mu}}}\nonumber\\
				&-\Omega\varkappa\ct{P}{^\nu_\alpha}\ct{n}{^\mu}\ct{\underline{T}}{_{\nu\mu}}+\frac{1}{2}\Omega^2\varkappa\ct{n}{^\rho}\cd{_\rho}\prn{\frac{1}{N}\ct{P}{^\nu_\alpha}\ct{n}{^\mu}\ct{\underline{T}}{_{\nu\mu}}}\spacef,
				\end{align}
			and from \cref{eq:schoutenComponentNPScri} we deduce that $ \ctd{a}{_\alpha}\eqs 0 $. Next, consider the expansion
				\begin{equation}
				\cs{\theta}\defeq \cd{_\rho}\ct{n}{^\rho}=  -\frac{1}{2\cs{N}}\Omega\ms{^{\mu\rho}}\ct{S}{_{\mu\rho}}+\frac{3}{\cs{N}}f +\frac{1}{2\cs{N}}\Omega^2\varkappa\ms{^{\mu\rho}}\ct{\underline{T}}{_{\mu\rho}}\eqs 0\spacef.
				\end{equation}
				\begin{align}
				\cs{\dot{\theta}}&\defeq \ct{n}{^\rho}\cd{_\rho}\theta = \frac{1}{2}\ms{^{\mu\rho}}\ct{S}{_{\mu\rho}}-\Omega\ct{n}{^\rho}\cd{_\rho}\prn{\frac{1}{2\cs{N}}\ms{^{\mu\rho}}\ct{S}{_{\mu\rho}}}+\frac{3}{\cs{N}}\ct{n}{^\rho}\cd{_\rho}f-\frac{3}{\cs{N^2}}f\ct{n}{^\rho}\cd{_\rho}\cs{N}\nonumber\\
				&-\Omega\varkappa\ms{^{\mu\rho}}\ct{\underline{T}}{_{\mu\rho}}+\frac{1}{2}\varkappa\Omega^2\ct{n}{^\rho}\cd{_\rho}\prn{\frac{1}{\cs{N}}\ms{^{\mu\rho}}\ct{\underline{T}}{_{\mu\rho}}}=\frac{1}{2}\ct{P}{^{\mu\nu}}\ct{S}{_{\mu\rho}}+\\
				+&\Omega\brkt{-\ct{n}{^\rho}\cd{_\rho}\prn{\frac{1}{2\cs{N}}\ct{P}{^{\mu\nu}}\ct{S}{_{\mu\nu}}}
					+\varkappa\ct{P}{^{\mu\nu}}\ct{\underline{T}}{_{\mu\nu}}+\frac{3}{2}\ct{n}{^\mu}\ct{n}{^\nu}\ct{\underline{T}}{_{\mu\nu}}+\frac{3}{8}\varkappa\cs{T}}\nonumber\\
				&+3 f\ct{n}{^\rho}\cd{_\rho}\prn{\cs{N}^{-1}}-\frac{3}{2}\ct{S}{_{\mu\nu}}\ct{n}{^\mu}\ct{n}{^\nu}+\Omega^2\brkt{\frac{1}{2}\varkappa\ct{n}{^\rho}\cd{_{\rho}}\prn{\frac{1}{N}\ct{P}{^{\mu\nu}}\ct{\underline{T}}{_{\mu\nu}}}-\frac{1}{8\cs{N}}\varkappa\ct{n}{^\mu}\cd{_\mu}\cs{T}}\nonumber\\
				&\eqs \frac{1}{2}\ct{S}{^\mu_\mu}-\ct{S}{_{\rho\mu}}\ct{n}{^\rho}\ct{n}{^\mu}\spacef.
				\end{align}
			Finally, the shear,
				\begin{align}
				\ct{\sigma}{_{\alpha\beta}}&\defeq\prn{ \ct{P}{^\mu_\alpha}\ct{P}{^\nu_\beta}-\frac{1}{3}\ct{P}{_{\alpha\beta}}\ms{^{\nu\mu}}}\cd{_\mu}\ct{n}{_{\nu}}= -\frac{1}{2\cs{N}}\Omega\ct{P}{^\mu_\alpha}\ct{P}{^\nu_\beta}\ct{S}{_{\mu\nu}}+\frac{1}{N}\ct{P}{_{\alpha\beta}}f\nonumber\\
				& +\frac{1}{2N}\Omega^2 \ct{P}{^\mu_\alpha}\ct{P}{^\nu_\beta}\ct{\underline{T}}{_{\mu\nu}}+\prn{\frac{1}{6\cs{N}}\Omega
					\ct{P}{^{\mu\nu}}\ct{S}{_{\mu\nu}}-\frac{1}{\cs{N}}f-\frac{1}{6\cs{N}}\Omega^2\varkappa\ct{P}{^{\mu\nu}}\ct{\underline{T}}{_{\mu\nu}}}\ct{P}{_{\alpha\beta}}=\nonumber\\
				& = -\frac{1}{2\cs{N}}\Omega\ct{P}{^\mu_\alpha}\ct{P}{^\nu_\beta}\ct{S}{_{\mu\nu}}+\frac{1}{2N}\Omega^2 \ct{P}{^\mu_\alpha}\ct{P}{^\nu_\beta}\ct{\underline{T}}{_{\mu\nu}}+\prn{\frac{1}{6\cs{N}}\Omega
					\ct{P}{^{\mu\nu}}\ct{S}{_{\mu\nu}}-\frac{1}{6\cs{N}}\Omega^2\varkappa\ct{P}{^{\mu\nu}}\ct{\underline{T}}{_{\mu\nu}}}\ct{P}{_{\alpha\beta}}\nonumber\\
				&\eqs 0\spacef.
				\end{align}
				\begin{align}
				\ctd{\sigma}{_{\alpha\beta}}&\defeq\ct{n}{^\rho}\cd{_\rho}\ct{\sigma}{_{\alpha\beta}}=\frac{1}{2}\ct{P}{^\mu_\alpha}\ct{P}{^\nu_\beta}\ct{S}{_{\mu\nu}}-\frac{1}{6}\ct{P}{_{\alpha\beta}}\ct{P}{^{\mu\nu}}\ct{S}{_{\mu\nu}}\nonumber\\
				&+\Omega\brkt{-\ct{n}{^\rho}\cd{_\rho}\prn{\frac{1}{2\cs{N}}\ct{P}{^\mu_\alpha}\ct{P}{^\nu_\beta}\ct{S}{_{\mu\nu}}}+\frac{1}{6}\ct{n}{^\rho}\cd{_\rho}\prn{\frac{1}{N}\ct{P}{_{\alpha\beta}}\ct{P}{^{\mu\nu}}\ct{S}{_{\mu\nu}}}-\varkappa\ct{P}{^\mu_\beta}\ct{P}{^\nu_\alpha}\ct{\underline{T}}{_{\mu\nu}}+\frac{1}{3}\varkappa\ct{P}{^{\mu\nu}}\ct{\underline{T}}{_{\mu\nu}}}\nonumber\\
				&+\Omega^2\ct{n}{^\rho}\cd{_\rho}\prn{\frac{1}{2\cs{N}}\varkappa\ct{P}{^\mu_\beta}\ct{P}{^\nu_\alpha}\ct{\underline{T}}{_{\mu\nu}}-\frac{1}{6\cs{N}}\varkappa\ct{P}{^{\mu\nu}}\ct{\underline{T}}{_{\mu\nu}}\ct{P}{_{\alpha\beta}}}\nonumber\\
				&\eqs\frac{1}{2}\ct{P}{^\mu_\alpha}\ct{P}{^\nu_\beta}\ct{S}{_{\mu\nu}}-\frac{1}{6}\ct{P}{_{\alpha\beta}}\ct{P}{^{\mu\nu}}\ct{S}{_{\mu\nu}}  \spacef.
				\end{align}
			Note that this quantity is  different from zero (in general), completely tangent to $ \scri $ and coincides with the traceless part of the intrinsic Schouten tensor
				\begin{equation}\label{eq:sigmadscri}
				\ctd{\sigma}{_{ab}}\defeqs \cts{S}{_{ab}}-\frac{1}{3}\ms{_{ab}}\cts{S}{^c_c}\spacef.
				\end{equation}	
				It will be necessary, as we will see shortly, to have the second derivative too,
				\begin{align}
				\ct{\ddot{\sigma}}{_{\alpha\beta}}&\defeq \ct{n}{^\rho}\cd{_\rho}\ctd{\sigma}{_{\alpha\beta}}=\frac{1}{2}\ct{n}{^\rho}\cd{_\rho}\ct{s}{_{\alpha\beta}}-\frac{1}{6}\ct{P}{_{\alpha\beta}}\ct{n}{^\mu}\cd{_\mu}\ct{s}{^{\nu}_\nu}\nonumber\\
				&-\cs{N}\Big(-\frac{1}{2N}\ct{n}{^\rho}\cd{_\rho}\ct{s}{_{\alpha\beta}}+\frac{
					1}{6\cs{N}}\ct{P}{_{\alpha\beta}}\ct{n}{^\rho}\cd{_\rho}\ct{s}{^\mu_\mu}+\frac{1}{2\cs{N}^2}\ct{n}{^\rho}\cd{_\rho}\cs{N}\ct{s}{_{\alpha\beta}}-\frac{1}{6\cs{N}^2}\ct{n}{^\rho}\cd{_\rho}\prn{\cs{N}}\ct{s}{^\alpha_\beta}\ct{P}{_{\alpha\beta}}\nonumber\\
				&-\varkappa\ct{P}{^\mu_\beta}\ct{P}{^\nu_\alpha}\ct{\underline{T}}{_{\mu\nu}}+\frac{1}{3}\varkappa\ct{n}{^\mu}\ct{n}{^\nu}\ct{\underline{T}}{_{\mu\nu}}\ct{P}{_{\alpha\beta}}\Big)+\Omega\ct{A}{_{\alpha\beta}}+\Omega^2\ct{B}{_{\alpha\beta}}\spacef,
				\end{align}
			where $ A $ and $ B $ are regular (non-vanishing in general) symmetric tensors. Notice that contracting with $ \ct{n}{^\alpha}\ct{P}{^{\beta\gamma}}$ \cref{eq:cottoncefeScri} and using \cref{eq:schoutenComponentNPScri} 
				\begin{equation}
				\ct{n}{^\mu}\cd{_\mu}\cts{S}{^\nu_\nu}\eqs 0.
				\end{equation}
			Observe, also, that by \cref{eq:energymomentumComponentNN,eq:energymomentumComponentPP} we have
				\begin{align}
				\ct{P}{^\mu_\beta}\ct{P}{^\nu_\alpha}\ct{\underline{T}}{_{\mu\nu}}\eqs-\frac{1}{4}\cs{T}\ct{P}{_{\alpha\beta}}\spacef,\\
				\ct{n}{^\mu}\ct{n}{^\nu}\ct{\underline{T}}{_{\mu\nu}}\eqs-\frac{3}{4}\cs{T}\spacef.
				\end{align}
			Taking into account these last equations we arrive at
				\begin{equation}\label{eq:sigmaddscri}
				\ct{\ddot{\sigma}}{_{ab}}\eqs\ct{e}{^\alpha_a}\ct{e}{^\beta_b}\ct{\ddot{\sigma}}{_{\alpha\beta}}\eqs 2 \ct{e}{^\alpha_a}\ct{e}{^\beta_b}\ct{n}{^\rho}\cd{_\rho}\cts{S}{_{\alpha\beta}}\spacef.
				\end{equation}
			From \cref{eq:electricSchouten,eq:magneticSchouten,eq:sigmadscri,eq:sigmaddscri}, we get the desired relations:
				\begin{align}
				\ct{C}{_{ab}}&= \sqrt{\frac{3}{\Lambda}}\brkt{\ct{\epsilon}{^{pq}_a}\cds{_{[p}}\ctd{\sigma}{_{q]b}}+\frac{1}{2}\ct{\epsilon}{^p_{ba}}\cds{_c}\ctd{\sigma}{^c_p}}\spacef,\label{eq:magneticShear}\\
				\ct{D}{_{ab}}&=\frac{1}{2}\sqrt{\frac{3}{\Lambda}}\ct{\ddot{\sigma}}{_{ab}}\spacef.\label{eq:electricShear}
				\end{align} 
			For the first equation, we have used another interesting relation that can be obtained if one considers \cref{eq:sigmadscri} and takes the trace in \cref{eq:cottoncefeScri},
				\begin{equation}
				\cds{_c}\cts{S}{^{c}_a}\eqs\cds{_a}\cts{S}{^c_c}\eqs \frac{3}{2}\cds{_c}\ctd{\sigma}{_a^{c}}\spacef.
				\end{equation}

		\subsection{Characterisation of gravitational radiation at $\scri$}\label{ssec:radiation-condition}
			At this stage, we have presented the basic asymptotic structure with a positive cosmological constant (\cref{sec:intrinsicCurvature}) and the superenergy formalism (\cref{sec:supernergy}). Thus, we are ready to tackle the problem of gravitational radiation at infinity. In this subsection we formulate a radiation condition expanding the contents presented in \cite{Fernandez-Alvarez_Senovilla20b}. It is, to our knowledge, the first covariant, gauge-invariant criterion formulated in the presence of a positive cosmological constant.\\
			 
			As the Weyl tensor vanishes at $ \scri $, we need to define an appropriate superenergy tensor at infinity. The obvious choice is \emph{the rescaled Bel-Robinson tensor}:
				\begin{equation}
				\ct{\D}{_{\alpha\beta\gamma\delta}} \defeq \ct{d}{_{\alpha\mu\gamma}^\nu}\ct{d}{_{\delta\nu\beta}^\mu} + \ctru{*}{d}{_{\alpha\mu\gamma}^\nu}\ctru{*}{d}{_{\delta\nu\beta}^\mu}\spacef,
				\end{equation}		
			which is regular and, in general, non-vanishing at $ \scri $. Its divergence is easily computed using \cref{eq:div-superenergy-tensor,eq:cefesDerWeyl} and reads
				\begin{equation}\label{eq:div-rescaled-BR}
					\cd{_\mu}\ct{\D}{_{\alpha\beta\gamma}^\mu}= 2\ct{d}{_{\mu\gamma\nu\alpha}}\ct{y}{_{\beta}^{\nu\mu}}+2\ct{d}{_{\mu\gamma\nu\beta}}\ct{y}{_{\alpha}^{\nu\mu}}+ \ct{g}{_{\alpha\beta}}\ct{d}{^{\mu\nu\rho}_\gamma}\ct{y}{_{\mu\nu\rho}}\spacef. 
				\end{equation}
			In order to define a supermomentum, one needs to select an observer. Since we aim at an observer-independent characterisation of radiation, the optimal way would be to have a natural privileged `asymptotic observer'. But this is indeed given by the asymptotic geometry itself: the normal $ \ct{N}{^\alpha}|_\scri $ is the suitable vector field. Hence, a natural definition of \emph{asymptotic supermomentum}  is --see \cref{eq:supermomentum} for the general definition--
				\begin{equation}\label{eq:asymptotic-supermomentum} 
				\ct{p}{^\alpha}\defeq -\ct{N}{^\mu}\ct{N}{^\nu}\ct{N}{^\rho}\ct{\D}{^\alpha_{\mu\nu\rho}}\spacef,
				\end{equation}
			or its \emph{canonical} version
				\begin{equation}\label{eq:asymptotic-canonical-supermomentum}
				\ct{\P}{^\alpha}\defeq -\ct{n}{^\mu}\ct{n}{^\nu}\ct{n}{^\rho}\ct{\D}{^\alpha_{\mu\nu\rho}}\spacef.
				\end{equation}
			In a neighbourhood of $ \scri $, where $ \ct{n}{^\alpha} $ is well defined, these two vector fields are collinear, $ \ct{\P}{^\alpha}=\cs{N}^{-3}\ct{p}{^\alpha} $, and have the same causal orientation. The reason why we introduce them both is that \cref{eq:asymptotic-supermomentum} has a good behaviour in the limit $ \Lambda\rightarrow0 $ -- if the limit exists-- in contrast to \cref{eq:asymptotic-canonical-supermomentum}. This issue will be analysed in \cref{ssec:afs-limit}. Apart from this, the properties that will be listed next apply to both versions of the asymptotic supermomentum, unless explicitly said otherwise.\\
			
			The orthogonal splitting of $ \ct{\P}{^\alpha} $ at $\scri$ is given by 
			\begin{equation}\label{eq:canonical-se-decomposition}
					\ct{\P}{^\alpha}=-\cs{\W}\ct{n}{^\alpha}+\ct{e}{^\alpha_a}\cts{\P}{^a}\spacef.
				\end{equation}
			which defines
			\begin{itemize}
			\item the \emph{asymptotic canonical superenergy density}, $ \cs{\W}\defeq-\ct{n}{_\mu}\ct{\P}{^\mu} \geq 0$, 
			\item and the \emph{asymptotic canonical super-Poynting vector}, $ \cts{\P}{^\alpha}\defeq \ct{P}{^\alpha_\mu}\ct{\P}{^\mu}=\ct{e}{^\alpha_a}\cts{\P}{^a} $ ---see \cref{eq:supermomentum}---, which is a vector field tangent to $\scri$.
			\end{itemize}	
			From the general properties presented in \cref{ssec:basic-superenergy}, it follows that
				\begin{properties}
					\item $  \ct{\P}{^\alpha}$ is causal and future pointing at and around $\scri$, --see \cref{it:dominant-se-condition} on page \pageref{it:dominant-se-condition}.
				\item Using \cref{eq:auxCotton,eq:energymomentumConformalScri,eq:div-rescaled-BR}, the divergence of $ \ct{\P}{^\alpha} $ at $ \scri $ reads
						\begin{equation}
							\cd{_{\mu}}\ct{\P}{^\mu}\eqs N\varkappa\ctrd{1}{T}{_{ab}}\ct{D}{^{ab}}\spacef,
						\end{equation}
					where$ \ctrd{1}{T}{_{ab}}\defeqs\Omega^{-1}\ct{T}{_{\mu\nu}}\cts{e}{^\mu_{a}}\cts{e}{^\nu_{b}} $. In particular, if the energy-momentum tensor of the physical space-time $(\hat M,\hat g_{\mu\nu})$ behaves near $\scri$ as $ \pt{T}{_{\alpha\beta}}|_\scri \sim \mathcal{O}(\Omega^3)$ (which includes the vacuum case  $ \pt{T}{_{\alpha\beta}}=0 $), then 
					\begin{equation}
						\cd{_\mu}\ct{\P}{^\mu}\eqs 0 . 	
					\end{equation}
					This follows from \cref{eq:cefesScriDerWeyl}, recalling \cref{eq:divergence-freeGauge} and \cref{eq:matter-decay-cotton}.
					\item\label{it:gauge-supermomentum} Under gauge transformations, they change as
						\begin{align}
						\ct{\P}{^\alpha} \rightarrow& \frac{\omega^{-7}}{\prn{1-2\Omega N^{-2}\ct{N}{^\tau}\ct{\omega}{_\tau}-\Omega^2\omega^{-2}N^{-2}\ct{\omega}{_\tau}\ct{\omega}{^\tau}}^{3/2}}\bbrkt{\ct{\P}{^\alpha}\nonumber\\
							-&\prn{3\ct{\omega}{^{-1}}\Omega N^{-1}\ct{n}{^\rho}\ct{n}{^\nu}\ct{\omega}{^\mu}+3\omega^{-2}\Omega^2N^{-2}\ct{n}{^\rho}\ct{\omega}{^\nu}\ct{\omega}{^\mu}+\omega^{-3}N^{-3}\Omega^3\ct{\omega}{^\rho}\ct{\omega}{^\mu}\ct{\omega}{^\nu}}\ct{\D}{^\alpha_{\mu\nu\rho}}}.
						\end{align}
						\begin{align}
						\ct{p}{^\alpha} \rightarrow& \omega^{-7}\bbrkt{\ct{p}{^\alpha}-\prn{3\ct{\omega}{^{-1}}\Omega \ct{N}{^\rho}\ct{N}{^\nu}\ct{\omega}{^\mu}+3\omega^{-2}\Omega^2\ct{N}{^\rho}\ct{\omega}{^\nu}\ct{\omega}{^\mu}+\omega^{-3}\Omega^3\ct{\omega}{^\rho}\ct{\omega}{^\mu}\ct{\omega}{^\nu}}\ct{\D}{^\alpha_{\mu\nu\rho}}}.
						\end{align}
					This behaviour is deduced using \cref{eq:rescaledWeyl-def,eq:gauge-N,eq:gauge-NormN}, and the fact that the Weyl tensor is conformally invariant. At $ \scri $, the asymptotic supermomentum has good gauge-behaviour
						\begin{align}\label{eq:gauge-smomentum}
						\ct{\P}{^\alpha} &\stackrel{\scri}{\rightarrow}\omega^{-7}\ct{\P}{^\alpha}\spacef, \\
						\ct{p}{^\alpha} &\stackrel{\scri}{\rightarrow}\omega^{-7}\ct{p}{^\alpha}\spacef. 
						\end{align}
				\end{properties}
			 The divergence property of the canonical supermomentum can be expressed as
				\begin{equation}\label{eq:balance-law-diff}
				\cds{_e}\cts{\P}{^e}+\ct{n}{^\mu}\cd{_\mu}\left(\cs{\W}\right)\eqs N\varkappa\ctrd{1}{T}{_{ab}}\ct{D}{^{ab}}\spacef,\spacef.
				\end{equation}	
			Under appropriate conditions this expression leads to an integral balance-law ---see  \cref{ssec:balance-law}. Typically, kinematic terms associated to $ \ct{n}{^\alpha} $ enter this kind of equation \cite{Alfonso2008}, however, due to our partial gauge-fixing they vanish at $\scri$. Nevertheless, it is possible to write $ \cts{\P}{^\alpha} $ in terms of the derivatives of the shear by using  \cref{eq:electricShear,eq:magneticShear},
				\begin{equation}
				\cts{\P}{^a}\eqs -\frac{6}{\Lambda}\cds{^{[a}}\prn{\ctd{\sigma}{^{s]t}}}\ct{\ddot{\sigma}}{_{ts}} + \frac{3}{2\Lambda}\ct{\ddot{\sigma}}{^a_s}\cds{_c}\prn{\ctd{\sigma}{^{cs}}}\quad
				\end{equation}
			or, using \cref{eq:magneticSchouten,eq:electricSchouten}, in terms of the Schouten tensor,
				\begin{equation}
				\cts{\P}{^a}\eqs \frac{12}{\Lambda}\ct{e}{^\alpha_t}\ct{e}{^\beta_s}\cds{^{[s}}\left(\cts{S}{^{a]t}}\right)\ct{n}{^\mu}\cd{_\mu}\cts{S}{_{\alpha\beta}}\spacef.
				\end{equation}
			Our asymptotic gravitational-radiation condition is built upon this object. In order to characterise the presence of gravitational radiation at infinity, we aim at a criterion with the following features:
				\begin{properties}
					\item Gauge-invariant, as any physical statement should not depend on the choice of the representative within the conformal class of metrics.\label{it:crit-prop1}
					\item Observer-independent.\label{it:crit-prop2}
					\item \emph{Strictly} asymptotic, i. e., defined at $ \scri $.\label{it:crit-prop3}
					\item With the necessary and sufficient information encoded in $\prn{\scri,\ms{_{ab}},\ct{D}{_{ab}}}$. \label{it:crit-prop4}
				\end{properties}
				The last point \ref{it:crit-prop4} is central to our analysis, and marks a clear difference with other attempts in the literature. Its justification is based on (a) a fundamental result \cite{Friedrich1986a,Friedrich1986b} which states that a solution of the $\Lambda$-vacuum Einstein field equations is fully determined by initial/final data consisting of the conformal class of a 3-dimensional Riemmanian manifold  {\em plus} a traceless and divergence-free tensor $D_{ab}$; and (b) similar fundamental results that also hold for more general matter fields, and can be better understood in general from a classical simple study of the field equations in the presence of a positive cosmological constant $\Lambda$ \cite{Starobinsky1983}. Performing an expansion in powers of $e^{-\sqrt{\Lambda/3} t}$ as $t\rightarrow \infty$, one proves that the first three terms are a spatial metric, then its curvature (thus fully determined by the former), and a traceless tensor $D_{ab}$ whose divergence depends on the matter contents. And that these terms determine the whole expansion.
				
				{\em In summary, one cannot aspire to describe gravitational radiation at $\scri$ without taking $D_{ab}$ into account}.
				
			Our proposal was presented in \cite{Fernandez-Alvarez_Senovilla20b}, and reads
			
				\begin{crit}[Asymptotic gravitational-radiation condition with $ \Lambda>0 $]\label{def:criterionGlobal}
				Consider a 3-dimensional open connected subset $ \Delta \subset \scri $.
				There is gravitational radiation on $ \Delta $  if and only if the asymptotic super-Poynting does not vanish there. Equivalently,
					\begin{equation*}
					\cts{\P}{^\alpha}\eqsopen 0 \iff \text{No gravitational radiation on }\Delta .
					\end{equation*}
				\end{crit}		
			\begin{remark}\label{rmk:equivalent-statement-criterion}
				An equivalent statement is that, in absence of gravitational radiation and only in that case the supermomentum\footnote{The same applies to the canonical supermomentum. However, the characterisation in terms of $ \ct{p}{^\alpha} $ can be compared with the $ \Lambda=0 $ case, as we will see.} points along the normal $N^\alpha$ at $\scri$, or:
				\begin{itemize}
					\item  No gravitational radiation on $\Delta\subset \scri \iff \ct{p}{^\alpha}$ is orthogonal to all surfaces 
					within $\Delta$.
					\item No gravitational radiation on $\Delta\subset \scri \iff N^\alpha|_\Delta$ is a principal vector  of $\ct{d}{_{\alpha\beta\gamma}^\delta}|_\Delta$ .
				\end{itemize}
				Here we are using the concept of principal vector in the sense of Pirani and Bel, that is, those lying in the intersection of two principal planes, the latter defined by the eigen-2-forms of $\ct{d}{_{\alpha\beta\gamma\delta}}|_\Delta$. As a summary we can recall that, considering only the {\em causal} principal vectors and ignoring the spacelike ones, for Petrov type {\rm I} there is one principal timelike vector and no null one, for Petrov type D there are two null ones (the two multiple PND) and every timelike vector in the plane of these two PND is also a principal vector; and finally for Petrov types {\rm II, III, or N} there is just one null principal vector (the multiple PND) and no timelike one. See \cite{Pirani57,Bel1962,Ferrando1997} for further details.
			\end{remark}
			\begin{remark}\label{rmk:criterion}
				The criterion fulfils \cref{it:crit-prop1} as follows from \cref{eq:gauge-smomentum}; \cref{it:crit-prop2}, according to the discussion on the geometric nature of $ \ct{N}{^\alpha} $ at the beginning of this section; \cref{it:crit-prop3}, by  construction; \cref{it:crit-prop4}, since by \cref{eq:commutator-se} the presence of radiation is completely given by the interplay of $ \ct{D}{_{ab}} $ and $ \ct{C}{_{ab}} $, the latter being fully determined by the intrinsic geometry -- see \cref{eq:magneticSchouten}.
			\end{remark}
			\begin{remark}
				According to the previous remark, the presence/absence of radiation cannot be determined by the intrinsic geometry of $ \scri $ exclusively in general ---with the exception of the trivial cases with a conformally flat metric $ \ms{_{ab}} $. The cases with a vanishing $ \ct{D}{_{ab}} $ do not present radiation either, but this cannot be inferred from $(\scri,\ct{h}{_{ab}})$ alone.
			\end{remark}
			\begin{remark}\label{rmk:criterion-commutator}
				From \cref{eq:commutator-se}, the no-radiation condition is equivalent to the vanishing of the commutator of $ \ct{D}{_{ab}} $ and $ \ct{C}{_{ab}} $, and this is only possible if $ \ct{d}{^\alpha_{\beta\gamma\delta}}\rvert_{\scri} $ has Petrov-type I or D \cite{Bel1962,Alfonso2008}  in accordance with \cref{rmk:equivalent-statement-criterion}. The Petrov type-D situation arises when $ \ct{n}{^\alpha}|_\scri $ is coplanar with the two multiple PND.
				Observe, that according to remark \ref{rem:dmasdegenerado} all possible Petrov types (except 0) of the physical Weyl tensor are compatible with the existence of gravitational radiation at $\scri$, as well as with its absence.
			\end{remark}
			\begin{remark}\label{rmk:alfonso}
			Our criterion \ref{def:criterionGlobal} is different, but gained some inspiration, from definition \ref{Bel}. Had we chosen to inspire our criterion on definition \ref{Alfonso}, we would have had to use $\ct{Q}{_{abc}}|_\scri$ constructed with $n^\mu|_\scri$ instead of the asymptotic super-Poynting. The vanishing of  $\ct{Q}{_{abc}}|_\scri$ is equivalent to the electric and magnetic parts being proportional \cite{Alfonso2008}, that is
			\begin{equation}\label{CproptoD}
			A \ct{C}{_{ab}} + B \ct{D}{_{ab}}\eqs 0
			\end{equation}
			for some $A$ and $B$. This is always the case for Petrov type D. Thus, the small difference between both possibilities is that using $\ct{Q}{_{abc}}|_\scri$ there will be more radiative situations: those with the electric and magnetic parts commuting but not proportional to each other.
		
			\end{remark}
				\begin{figure}[h!]
				\centering
					\includegraphics[width=0.8\textwidth]{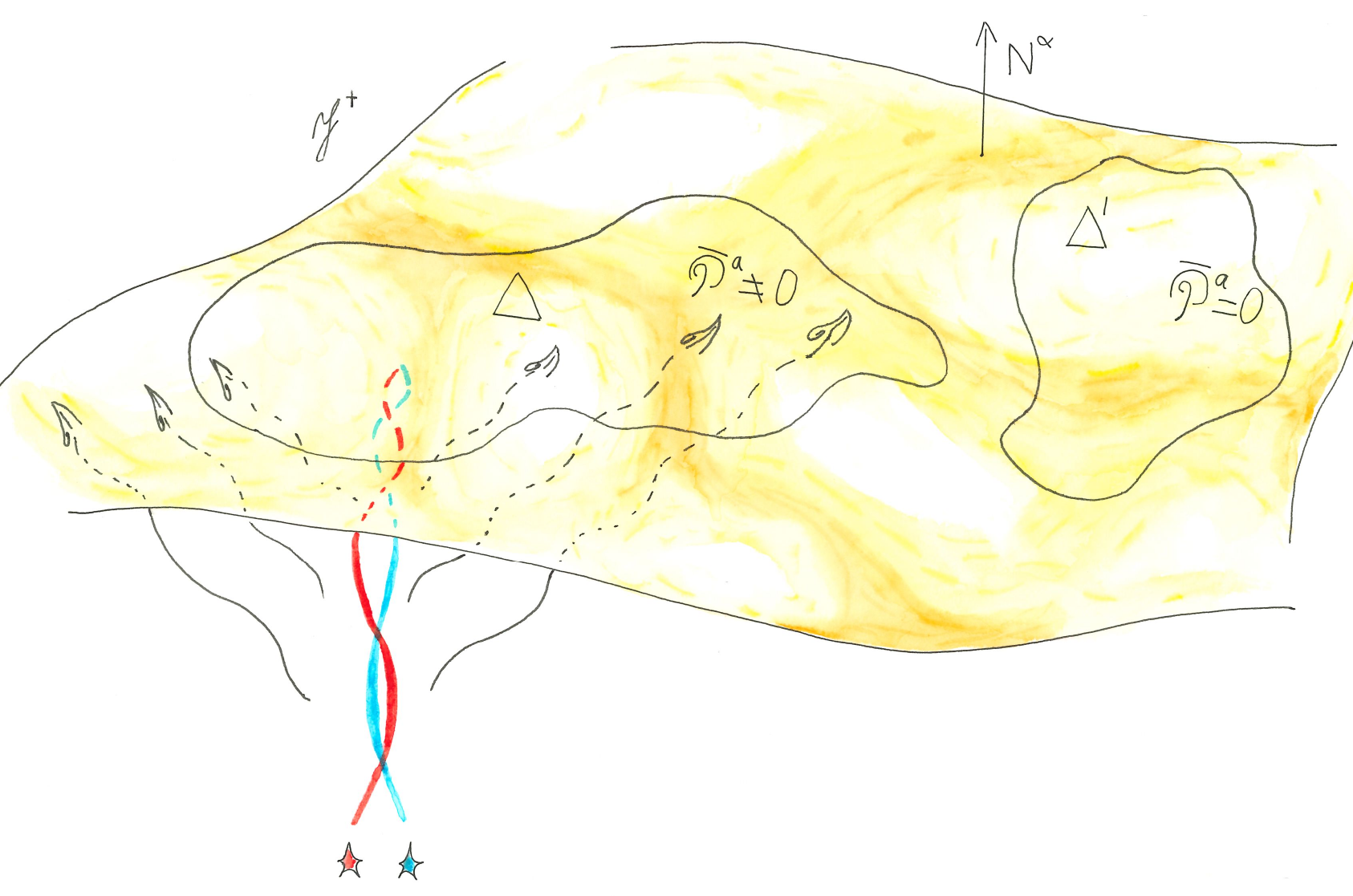}
					\caption[The gravitational radiation condition with a positive cosmological constant]{Gravitational radiation arrives at an open region $ \Delta $ on $ \scri^+ $ but does not at the open region $ \Delta' $. Our criterion states that the asymptotic super-Poynting is different from zero on $ \Delta $ and vanishes on $ \Delta' $.}
				\end{figure}
			Examples illustrating the soundness of this criterion were presented in \cite{Fernandez-Alvarez_Senovilla20b} and some of them will be expanded,  as well as new ones presented, in \cref{sec:Examples}. Furthermore, in \cite{Fernandez-Alvarez_Senovilla20b} we showed that criterion \ref{def:criterionGlobal} has an equivalent formulation in the asymptotically flat scenario. In that case, it has been proved to be successful and equivalent to the traditional one  based on the news tensor \cite{Fernandez-Alvarez_Senovilla20}, see also the companion paper \cite{Fernandez-Alvarez_Senovilla-afs} for further details. More on the limit to $ \Lambda=0 $ will be given in \cref{ssec:afs-limit} but, before that, we investigate the relation between the radiation condition and the radiant quantities.

			\subsection{Lightlike approach and the directional-dependence problem}\label{ssec:lightlike-condition-directional}
			We have presented a reliable condition that tells if gravitational waves arrive at infinity or not. Not only it is of special relevance by itself, but it constitutes a crucial step towards any deeper or stricter characterisation of gravitational radiation at $\scri$. One of the biggest challenges is the directional dependence that emerges when one approaches infinity in different lightlike directions \cite{Krtous2004}. Our \cref{def:criterionGlobal} already bypasses this difficulty. Actually, from \cref{def:criterionGlobal} follows that the presence of radiation cannot be determined by the rescaled Weyl scalar $ \ct{\phi}{_{4}}$ 
			only, as it is sometimes assumed in the literature --we are going to show this presently. We want to better understand this directional dependence in the presence/absence of radiation. 
			
			In our formalism, the logical way to proceed is to understand the role of the lightlike projections of the rescaled Bel-Robinson tensor, by defining --see \cref{eq:kp-def,eq:km-def}--
				\begin{align}\label{eq:n-kp-km}
				\ctp{k}{^\alpha}&\defeq\frac{1}{\sqrt{2}}\left(\ct{n}{^\alpha}+\ct{m}{^\alpha}\right)\\
				\ctm{k}{^\alpha}&\defeq\frac{1}{\sqrt{2}}\left(\ct{n}{^\alpha}-\ct{m}{^\alpha}\right)\quad\nonumber,
				\end{align}
				\begin{figure}[h]
					\centering
					\includegraphics[width=0.8\textwidth]{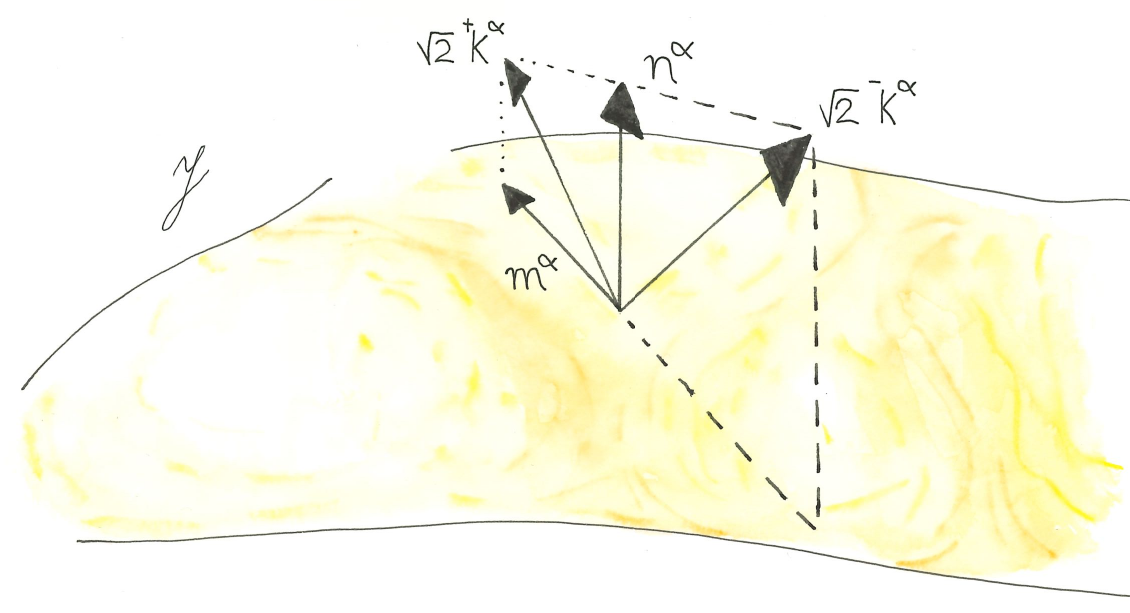}
					\caption[The lightlike decomposition]{The lightlike decomposition \eqref{eq:n-kp-km} on $ \scri $.}
				\end{figure}
			for some unit spacelike vector field tangent to $ \scri $, $ \ct{m}{^\alpha}\eqs\ct{e}{^\alpha_a}\ct{m}{^a} $. In these definitions $ \ct{n}{_\alpha} $ plays the role of $ \ct{u}{_\alpha} $ in \cref{ssec:lightlike-projections} and we denote by $ \cbrkt{\ctcn{E}{^\alpha_A}} $ 
			 the basis spanning the two-dimensional space of vectors orthogonal to $ \ct{m}{^\alpha} $ and $ \ct{n}{^\alpha} $ --see \cref{app:congruences} for more details on that. The algebraic, lightlike decomposition of \cref{ssec:lightlike-projections} applies the same now, though we substitute the over ring by an underbar in quantities projected with $ \ctcn{E}{^\alpha_A} $ in order to distinguish them from objects projected with $ \ct{E}{^\alpha_A} $\footnote{The point of making this change of notation is to distinguish the quantities associated to $ \ct{r}{^a} $ of \cref{ssec:lightlike-projections}, which is not in general a field on $ \scri $, from those associated with $ \ct{m}{^\alpha} $, which is a vector field on $ \scri $.}
			--e.g., given any one-form $ \ct{v}{_\alpha} $ on $ \scri $, $ \ctcn{v}{_A}\defeq \ctcn{E}{^\alpha_A}\ct{v}{_\alpha} $ whereas $  \ctc{v}{_A}\defeq \ct{E}{^\alpha_A}\ct{v}{_\alpha} $. Also, we define the radiant supermomenta associated to the vector fields of \cref{eq:n-kp-km} using the rescaled Bel-Robinson tensor:
				\begin{align}
				\ctp{\Q}{^\alpha}&\defeq -\ctp{k}{^\mu}\ctp{k}{^\nu}\ctp{k}{^\rho}\ct{\D}{^\alpha_{\mu\nu\rho}}=\csp{\W}\ctm{k}{^\alpha}+\ctps{\Q}{^\alpha}=\csp{\W}\ctm{k}{^\alpha}+\ctps{\Q}{^a}\ctp{e}{^\alpha_a}\spacef,\label{eq:Qpdef-m}\\
				\ctm{\Q}{^\alpha}&\defeq -\ctm{k}{^\mu}\ctm{k}{^\nu}\ctm{k}{^\rho}\ct{\D}{^\alpha_{\mu\nu\rho}}=\csm{\W}\ctp{k}{^\alpha}+\ctms{\Q}{^\alpha}=\csm{\W}\ctp{k}{^\alpha}+\ctms{\Q}{^k}\ctm{e}{^\alpha_k}\spacef.\label{eq:Qmdef-m}
				\end{align}
			\\ 
			
			Thus, the first step is to write \cref{def:criterionGlobal} in terms of the radiant quantities.
				\begin{lemma}[Radiant formulation of the no-radiation condition]\label{thm:radiantFormulCondition}
					Consider a three-dimensional open connected subset $ \Delta \subset \scri $, then 
					\begin{align}\label{eq:norad-radiant}
					\begin{rcases}
						2\prn{\csm{\Z}-\csp{\Z}}-\csp{\W}+\csm{\W} &\eqsopen 0\quad\\
						\sqrt{2}\prn{\ctps{\Q}{^A}+\ctms{\Q}{^A}}+12\ct{d}{^A}& \eqsopen 0 \quad 
					\end{rcases}\iff \text{No gravitational radiation on }\Delta\spacef.
					\end{align}
					\begin{proof}
						It follows directly by application of \cref{thm:radiant-poynting-vanishing}.
					\end{proof}
				\end{lemma}
			
					\begin{remark}In terms of Weyl scalars, the no-radiation condition in \cref{eq:norad-radiant} reads:
					\begin{align}
						8\ct{\phi}{_1}\ct{\bar{\phi}}{_1}-8\ct{\phi}{_3}\ct{\bar{\phi}}{_3}-4\ct{\phi}{_4}\ct{\bar{\phi}}{_4}+4\ct{\phi}{_0}\ct{\bar{\phi}}{_0} &= 0\spacef,\\
						\ct{\phi}{_3}\ct{\bar{\phi}}{_4}+\ct{\phi}{_0}\ct{\bar{\phi}}{_1}-3\ct{\phi}{_1}\ct{\overline{\phi}}{_2}-3\ct{\phi}{_2}\ct{\overline{\phi}}{_3}& = 0 \spacef.
					\end{align} 
				This is easily deduced using the formulae of \cref{app:NPformulation}.
				\end{remark}
			 The directional freedom translates into the choice of $ \ct{m}{^a} $, which then automatically gives $ \ctpm{k}{^\alpha} $ by \cref{eq:n-kp-km}. Indeed, this vector field may serve to define an intrinsic `evolution' direction on $ \scri $, if selected properly. Thus, one needs some physical criteria underlying one choice or another. We put forward two possible choices of increasing specialization that we call \emph{orientations},
				\begin{deff}[Weak orientation]\label{def:orientationWeak}
					We say that $ \ct{m}{^a} $ defines a weak orientation when $ \ctm{k}{^\alpha} $ is aligned with a principal null direction (PND) of the rescaled Weyl tensor at $\scri$.
				\end{deff}
				\begin{remark}
					For Petrov-type-I $ \ct{d}{_{\alpha\beta\gamma}^\delta}|_\scri $ there are four possible, non-equivalent, weak orientations; one for each PND. For type II, there are three; for type D and III, two; for type N, just one.  
				\end{remark}
				\begin{remark}
					 The vector $ \ct{m}{^a} $ defines a weak orientation if and only if $ \csm{\W}=0 $. See \cref{eq:Wmdef} and the Petrov characterisation on page \pageref{it:petrov-class-BR}.
				\end{remark}
				\begin{deff}[Strong orientation]\label{def:orientationStrong}
					We say that $ \ct{m}{^a} $ defines a strong orientation when $ \ctm{k}{^\alpha} $ is aligned with a PND  of highest multiplicity of the rescaled Weyl tensor at $\scri$.
				\end{deff}			
				\begin{remark}\label{rmk:number-strong-orientations}
					The strong orientation is a particular case of weak orientation. If $ \ct{d}{_{\alpha\beta\gamma}^\delta}|_\scri $ has Petrov type I, any strong orientation is a weak orientation too, hence there are four non-equivalent possibilities; for type II, III and N, there is one single strong orientation; for type D, there are two. 
				\end{remark}
				\begin{remark}\label{rmk:strong-orientation}
					For algebraically special $ \ct{d}{_{\alpha\beta\gamma}^\delta} $, the vector $ \ct{m}{^a} $ defines a strong orientation if and only if $ \csm{\W}=0=\csm{\Z} $. To show this, note the definitions in \cref{eq:Wmdef,eq:Zmdef} and see the Petrov characterisation on page \pageref{it:petrov-class-BR}. Additionally, by \cref{it:noQnoWnoZ} on page \pageref{it:noQnoWnoZ} and \cref{eq:Qmdef}, this is equivalent to $ \ctm{\Q}{^\alpha}\eqs 0 $.
				\end{remark}
			 An immediate result that follows by applying these definitions is the characterisation of the Petrov type of $ \ct{d}{^\alpha_{\beta\gamma\delta}} $ in the absence of radiation at infinity by means of the radiant superenergy quantities: 
				\begin{lemma}[Radiation condition and Petrov types]
					Consider a three-dimensional open connected subset $ \Delta \subset \scri $. Choose $ \ct{m}{^a} $ defining a weak orientation according to \cref{def:orientationWeak} and define $ \ctpm{k}{^\alpha} $ as in \cref{eq:n-kp-km}. Let $ \cts{\P}{^a} $ and $ \ctpm{\Q}{^\alpha} $ be the canonical asymptotic super-Poynting vector and the radiant supermomenta associated to $ \ctpm{k}{^\alpha} $, respectively. Then,
						\begin{align}\label{eq:noRad-radiant-Petrov}
						\begin{cases}
						\begin{rcases}
						2\prn{\csm{\Z}-\csp{\Z}}-\csp{\W} \eqsopen 0&\quad\\
							\sqrt{2}\ctps{\Q}{^A}+12\ct{d}{^A} \eqsopen 0&\quad \\
						\ctp{\Q}{^\alpha}\neqsopen 0 \neqsopen\ctm{\Q}{^\alpha}&
						\end{rcases}
						\end{cases}
						&\iff \cbrkt{\cts{\P}{^a}\eqsopen 0 \quad \text{and }\ct{d}{^\alpha_{\beta\gamma\delta}} \text{ Petrov type I on } \Delta},\\
						\quad\nonumber\\
							\cbrkt{	\ctp{\Q}{^\alpha}\eqsopen 0 \eqsopen\ctm{\Q}{^\alpha}}\quad&\iff \cbrkt{\cts{\P}{^a}\eqsopen 0 \quad \text{and } \ct{d}{^\alpha_{\beta\gamma\delta}} \text{ Petrov type D on } \Delta}.
						\end{align}
				\end{lemma}	
				 \begin{proof}
				 	For $ \ct{d}{_{\alpha\beta\gamma}^\delta} $ of Petrov type I, set $ \csm{\W}=0 $ in \cref{eq:norad-radiant} which, by \cref{it:normQpmA} on page \pageref{it:normQpmA}, gives the first two lines in \cref{eq:noRad-radiant-Petrov}. If $ \csm{Z}=0 $, then $ \ctm{k}{^\alpha} $ is a repeated principal null direction of $  \ct{d}{_{\alpha\beta\gamma}^\delta}  $, which is incompatible with Petrov-type I. The same occurs if $ \ctp{\Q}{^\alpha} =0$. Thus, the third line in \cref{eq:noRad-radiant-Petrov} follows. The case of Petrov type-D $  \ct{d}{_{\alpha\beta\gamma}^\delta}  $ is a consequence of weak orientation, together with what it is said at the end of \cref{rmk:criterion-commutator}.
				 \end{proof}
			 	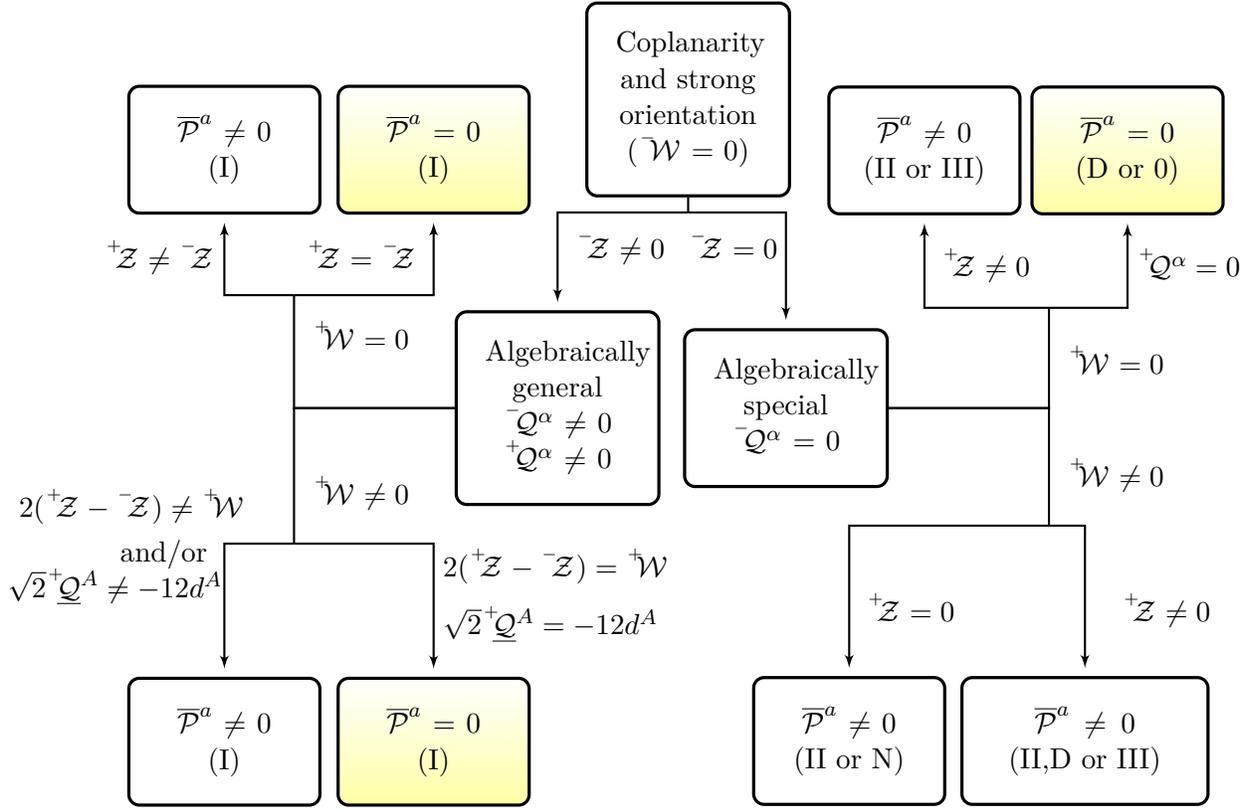
\begin{figure}[ht!]
			 	\centering
			 	\begin{tikzpicture}[node distance=0.3cm, auto]  
			 	\tikzset{
			 		mynode/.style={rectangle,rounded corners,draw=black, top color=white, bottom color=white,very thick, inner sep=1em, minimum 				size=1em, text centered,text width=5em},
			 		mynode2/.style={rectangle,rounded corners,draw=black, top color=white, bottom color=yellow!35,very thick, inner sep=1em, minimum 				size=1em, text centered,text width=5em},
			 		myarrow/.style={->, >=latex', shorten >=3pt, thick},
			 		myarrow2/.style={-, >=latex', shorten >=0pt, thick},
			 		mylabel/.style={text width=12em, text centered},
			 		mynodet/.style={rectangle,rounded corners,draw=black, top color=white, bottom color=white,very thick, inner sep=1em, minimum 				size=1em, text centered,text width=4.6em},
			 		mynode2t/.style={rectangle,rounded corners,draw=black, top color=white, bottom color=yellow!35,very thick, inner sep=1em, minimum 				size=1em, text centered,text width=4.6em},
			 		mynodew/.style={rectangle,rounded corners,draw=black, top color=white, bottom color=white,very thick, inner sep=1em, minimum 				size=1em, text centered,text width=6.5em} 
			 	}  
			 	\node[mynode] (conditions) {Coplanarity and strong orientation \\($ \csm{\W}=0 $)};
			 	\node[mynode,{below left=1.5cm and -1cm of conditions}] (caseI) {Algebraically general\\$ \ctm{\Q}{^\alpha}\neq 0$\\$\ctp{\Q}{^\alpha}\neq 0 $};  
			 	\node[mynode, right=of caseI] (special) {Algebraically special\\ $ \ctm{\Q}{^\alpha}=0 $}; 			
			 	\node[label,{below right=0.3cm and -1.5cm of conditions}] (label5) {$ \csm{\Z}=0 $};
			 	\node[label,{below right=0.3cm and -3.0cm of conditions}] (label6) {$ \csm{\Z}\neq0 $};
			 	\node[below=0.2 cm of conditions] (dummy){};			
			 	\draw[myarrow2] (conditions.south) -- ++(0,0) -- ++(0,0) -| (dummy.north);
			 	\draw[myarrow] (dummy.north)-- ++(0,0) -- ++(0,0) -| (caseI.north);
			 	\draw[myarrow] (dummy.north)-- ++(0,0) -- ++(0,0) -| (special.north);	
			 	
			 	\node[above left=0.2cm and 2cm of caseI] (dummy3) {};  
			 	\node[below left=0.5cm and 2cm of caseI] (dummy4) {};  
			 	\node[mynodet,{above left=0.8cm and -0.5cm of dummy3}] (caseIrad1) {$ \cts{\Pc}{^a}\neq 0 $ \\(I) }; 
			 	\node[mynode2t,{right = 0.2cm of caseIrad1}] (caseInorad1) {$ \cts{\Pc}{^a} = 0 $ \\(I) }; 
			 	\node[label,{below left=0.2cm and -1.3cm of caseIrad1}] (label7) {$ \csp{\Z}\neq\csm{\Z} $};
			 	\node[label,{below left=0.2cm and -1.2cm of caseInorad1}] (label8) {$ \csp{\Z}=\csm{\Z} $};
			 	\node[mynodet,{below left=1.5cm and -0.5cm of dummy4}] (caseIrad2) {$ \cts{\Pc}{^a}\neq 0 $ \\(I) }; 
			 	\node[mynode2t,{right = 0.2cm of caseIrad2}] (caseInorad2) {$ \cts{\Pc}{^a} = 0 $\\ (I) };
			 	\node[label,{above left=1.9cm and -1.7cm of caseIrad2}] (label9) {$ 2(\csp{\Z}-\csm{\Z})\neq\csp{\W} $};
			 	\node[label,{above left=1.3cm and -1.2cm of caseIrad2}] (label13) {and/or};
			 	\node[label,{above left=0.8cm and -1.4cm of caseIrad2}] (label11) {$\sqrt{2}\ctps{\Q}{^A}\neq- 12\ct{d}{^A}$};
			 	\node[label,{above right=1.1cm and -1.3cm of caseInorad2}] (label10) {$ 2(\csp{\Z}-\csm{\Z})=\csp{\W} $};
			 	\node[label,{above right=0.3cm and -1.3cm of caseInorad2}] (label12) {$\sqrt{2}\ctps{\Q}{^A}=-12\ct{d}{^A}$};
			 	
			 	\draw[myarrow] (dummy3.south)-- ++(0,0) -- ++(0,0) -| (caseIrad1.south);
			 	\draw[myarrow] (dummy3.south)-- ++(0,0) -- ++(0,0) -| (caseInorad1.south);
			 	\draw[myarrow] (dummy4.north)-- ++(0,0) -- ++(0,0) -| (caseIrad2.north);
			 	\draw[myarrow] (dummy4.north)-- ++(0,0) -- ++(0,0) -| (caseInorad2.north);
			 	\draw[myarrow2] (caseI.west)-- ++(0,0) -- ++(0,0) -| (dummy3.south);
			 	\draw[myarrow2] (caseI.west)-- ++(0,0) -- ++(0,0) -| (dummy4.north);	
			 	\node[label,{below left=0.2cm and -1.8cm of dummy3}] (label5) {$ \csp{\W}=0 $};
			 	\node[label,{above left=0.3cm and -1.8cm of dummy4}] (label5) {$ \csp{\W}\neq0 $};
			 	
			 	\node[mynode2t,{above left=1.5cm and -7.2cm of special}] (norad) {$ \cts{\Pc}{^a}= 0 $ (D or 0)};
			 	\node[mynodet,{below left=2.5cm and -3.5cm of special}] (rad1) {$ \cts{\Pc}{^a}\neq 0 $ (II or N)};
			 	\node[mynodew,{below left=2.5cm and -7cm of special}] (rad2) {$ \cts{\Pc}{^a}\neq 0 $ (II,D or III)};  
			 	\node[mynodet,{above left=1.5cm and -4.5cm of special}] (rad3) {$ \cts{\Pc}{^a}\neq 0 $ (II or III)};  
			 	
			 	\node[label,{below right=0.3cm and -1.3cm of norad}] (label5) {$ \ctp{\Q}{^\alpha}=0 $};
			 	\node[label,{above right=0.5cm and -1.2cm of rad1}] (label6) {$ \csp{\Z}=0$ };
			 	\node[label,{above right=0.5cm and -1.3cm of rad2}] (label6) {$ \csp{\Z}\neq 0$ };
			 	\node[label,{below right=0.3cm and -1.2cm of rad3}] (label8) {$ \csp{\Z}\neq 0$ };
			 	\node[above left=0cm and -5cm of special] (dummy2){};
			 	\node[below left=0.5cm and -5cm of special] (dummy5) {};
			 	\node[label,{below left=0.1cm and -1.8cm of dummy2}] (label5) {$ \csp{\W}=0 $};
			 	\node[label,{above left=0.3cm and -1.8cm of dummy5}] (label5) {$ \csp{\W}\neq0 $};
			 	%
			 	\draw[myarrow2] (special.east) -- ++(0,0) -- ++(0,0) -| (dummy2.north);
			 	\draw[myarrow2] (special.east)-- ++(0,0) -- ++(0,0) -| (dummy5.north);
			 	\draw[myarrow] (dummy2.north)-- ++(0,0) -- ++(0,0) -| (norad.south);
			 	\draw[myarrow] (dummy5.north)-- ++(0,0) -- ++(0,0) -| (rad1.north);	
			 	\draw[myarrow] (dummy5.north)-- ++(0,0) -- ++(0,0) -| (rad2.north);	 
			 	\draw[myarrow] (dummy2.north)-- ++(0,0) -- ++(0,0) -| (rad3.south);	
			 	\end{tikzpicture} 
			 	\medskip
			 	\caption[Flow of the asymptotic superenergy]{\textbf{Flow of the asymptotic superenergy  quantities}. One starts from the above middle node: strong orientation is chosen ($ -\ct{m}{^a} $ points along the spatial projection to $ \scri $ of a  PND of the rescaled Weyl tensor with highest multiplicity). Then, either the rescaled Weyl tensor is algebraically general (left-hand side of the diagram) or it is special (right-hand side of the diagram). Moving to the left, either the radiant superenergy $ \csp{\W} $ vanishes (above left-hand side) or not (below left-hand side). Thus, for an algebraically general rescaled Weyl tensor on $ \scri $, there are four configurations of asymptotic radiant superenergy: in two of them, there is gravitational radiation (one with $ \csp{\W}\neq 0 $, the other one with $ \csp{\W}=0 $); in the other two  there is no gravitational radiation (the shaded nodes). Moving to the right, one finds the algebraically special cases. There are four possibilities, from which just one corresponds to no radiation (the shaded node, for Petrov type D or 0, the only case in which both radiant supermomenta vanish).}\label{fig:superenergy-flow}
			 \end{figure}
			More can be said on the direction of propagation of the superenergy, in this case applying strong orientation,
				\begin{lemma}
					 Choose $ \ct{m}{^a} $ defining a strong orientation according to \cref{def:orientationStrong}, and define $ \ctpm{k}{^\alpha} $ as in \cref{eq:n-kp-km}. Let $ \cts{\P}{^a} $ and $ \ctpm{\Q}{^\alpha} $ be the canonical asymptotic super-Poynting vector and the radiant supermomenta associated to $ \ctpm{k}{^\alpha} $, respectively. Then, the canonical asymptotic super-Poynting vector takes the form
					 	\begin{equation}\label{eq:sPspecial}
					 	\cts{\P}{^a}\eqs -\prn{\frac{1}{2}\csp{\Z}+\frac{1}{4}\csp{\W}}\ct{m}{^a}+\prn{\frac{1}{2\sqrt{2}}\ctps{\Q}{^A}+3\ct{d}{^A}}\ctcn{E}{^a_A}\spacef,
					 	\end{equation}		
					hence, the superenergy flux cannot propagate in directions orthogonal to $ \ct{m}{^a} $ on $ \scri $. Furthermore, 
					 \begin{equation}\label{eq:flux-direction}
					 	\ct{m}{_s}\cts{\P}{^s}\le 0
					 \end{equation}
		equality holding only if $\cts{\P}{^s}=0$.
				\end{lemma}
				\begin{proof}
					For the first part, one only has to plug $ \csm{\W}=0=\csm{\Z} $ in \cref{eq:poynting-radiant}. For the second part, on the one hand, if $ \ct{m}{_s}\cts{\P}{^s}= 0 $, then $ \csp{\W}=0=\csp{\Z} $ and, by \cref{it:normQpmA} on page \pageref{it:normQpmA} and \cref{eq:Qpdef}, $ \ctp{\Q}{^\alpha}=0 $. But then, since strong orientation requires $ \ctm{\Q}{^\alpha}=0 $ (\cref{rmk:strong-orientation}), using \cref{thm:radiant-poynting-vanishing} $ \cts{\P}{^a}=0 $ follows. In that case there is no radiation according to \cref{def:criterionGlobal}. On the other hand, if $ \ct{m}{_s}\cts{\P}{^s}\neq 0 $, by the positivity of $ \csp{\Z} $ and $ \csp{\W} $, $ \ct{m}{_s}\cts{\P}{^s}<0 $ necessarily.
				\end{proof}
				\begin{remark}
					\Cref{eq:flux-direction} supports the idea of considering any $ \ct{m}{^a} $ that defines a strong orientation as a good candidate for intrinsic `evolution' direction. The reason is that directions orthogonal to  $ \ct{m}{^a} $ are \emph{transversal} to the flux of superenergy, which can be thought to be associated with `changes in the gravitational system'.
				\end{remark}
				\begin{remark}
					For $ \ct{d}{_{\alpha\beta\gamma}^\delta}|_\scri $ of Petrov type I, the sign of $ \ct{m}{_s}\cts{\P}{^s} $ is not defined because there are four non-equivalent strong  orientations and, by positivity of the radiant superenergy quantities, it is not determined in the general expression of \cref{eq:poynting-radiant}.
				\end{remark}
				The Petrov characterisation of the rescaled Weyl tensor at $ \scri $ in terms of the asymptotic superenergy quantities is summarised in \cref{fig:superenergy-flow}.\\

			The idea of having a preferred, intrinsic, `evolution' direction, $ \ct{m}{^a} $, at $ \scri $ is conceptually important. Indeed, the existence of a congruence of curves intrinsic to $ \scri $ will serve to define further structure related to \emph{absence of incoming radiation} and to  extended symmetries in \cref{sec:additional-structure}. 
			
			 Write the decomposition of the covariant derivative of this vector field as --see \cref{app:congruences} for details--
				\begin{equation}
					\cds{_a}\ct{m}{_b}= \ct{m}{_a}\ctcn{a}{_b}+\ctcn{\kappa}{_{ab}}+\ctcn{\omega}{_{ab}}\spacef,	
				\end{equation}
			where the shear of $ \ct{m}{_a} $ is defined as the traceless part of $ \ctcn{\kappa}{_{ab}} $,
				\begin{equation}
					\ctcn{\Sigma}{_{ab}}\defeq \ctcn{\kappa}{_{ab}}-\frac{1}{2}\ctcn{P}{_{ab}}\ctcn{P}{^{cd}}\ctcn{\kappa}{_{cd}}\spacef.
				\end{equation}
			Also, define the symmetric, traceless part of the symmetrised, projected derivative of $ \ctm{k}{_{\alpha}} $,
			\begin{equation}\label{eq:sigmam}
			\ctcnm{\sigma}{_{\alpha\beta}}\defeqs \ctcn{P}{^\mu_\alpha}\ctcn{P}{^\mu_\beta}\cd{_{(\mu}}\ctm{k}{_{\nu)}}-\frac{1}{2}\ctcn{P}{_{\alpha\beta}}\ctcn{P}{^{\mu\nu}}\cd{_\mu}\ctm{k}{_\nu}\spacef.
			\end{equation}
			This is, of course, the shear associated to $ \ctm{k}{^\alpha} $. It coincides up to a factor with the shear of $ \ct{m}{_a} $,
				\begin{equation}\label{eq:sigmam-Sigma}
					\sqrt{2}\ctcnm{\sigma}{_{\alpha\beta}}\eqs -\ctcn{P}{^\mu_\alpha}\ctcn{P}{^\mu_\beta}\cd{_{(\mu}}\ct{m}{_{\nu)}}+\frac{1}{2}\ctcn{P}{_{\alpha\beta}}\ctcn{P}{^{\mu\nu}}\cd{_\mu}\ct{m}{_\nu}\eqs -\ctcn{\Sigma}{_{\alpha\beta}},
				\end{equation}
			where we have used \cref{eq:derivativeNcfe,eq:extrinsicScri} and $ \ctcn{\Sigma}{_{\alpha\beta}}\defeq \ct{\omega}{_\alpha^a}\ct{\omega}{_\beta^b} \ctcn{\Sigma}{_{ab}}$. In addition, let us introduce the expansion of $ \ctm{k}{^\alpha} $,
				\begin{equation}
					\csm{\theta}\defeq \ctcn{P}{^{\mu\nu}}\cd{_\mu}\ctm{k}{_\nu}\spacef.
				\end{equation}
			It is possible to formulate an asymptotic `Goldberg-Sachs' theorem:
			\begin{lemma}\label{thm:gslikethm}
				On the neighbourhood of $ \scri $ where $ \ct{n}{_\alpha} $ is well defined choose an extension of $ \ct{m}{_\alpha} $ such that $ \ct{n}{_\alpha}\ct{m}{^\alpha}=0 $ and $ \ct{m}{_\alpha}\ct{m}{^\alpha}=1 $ there. Assume that $ \ct{d}{_{\alpha\beta\gamma}^\delta}\neqs 0 $ and $ \ctp{k}{^\beta}\ct{y}{_{A\beta C}}\eqs\ctm{k}{^\beta}\ct{y}{_{A\beta C}}\eqs0\eqs\ctm{k}{^\beta}\ctp{k}{^\gamma}\ct{y}{_{A\beta\gamma}} $. Then,
				\begin{equation*}
				\lied_{\vec{n}}\ctcnm{D}{_{\alpha\beta}}\eqs\ctcnm{D}{_{\alpha\beta}}\eqs 0,\quad\lied_{\vec{n}}\ctcnm{D}{_\alpha}\eqs \ctcnm{D}{_\alpha}\eqs 0 \implies \ctcnm{\sigma}{_{\alpha\beta}}\eqs 0.
				\end{equation*}
			\end{lemma}
			\begin{remark}\label{rmk:asymp-gs-remarkPND}
				The condition $ \ctcnm{D}{_{\alpha\beta}}\eqs \ctcnm{D}{_\alpha} \eqs 0 $ is equivalent to $ \ctm{\Q}{_\alpha}=0 $ and, therefore, to saying that $ \ct{m}{_\alpha} $ defines a strong orientation on $ \scri $ and $ \ctm{k}{^\alpha} $ is a repeated principal null direction of $ \ct{d}{_{\alpha\beta\gamma}^\delta} $. 
			\end{remark} 
			\begin{remark}\label{rmk:asymp-gs-remarkY}
				The assumption on the components of the Cotton-York tensor,  $ \ctp{k}{^\beta}\ct{y}{_{A\beta C}}\eqs\ctm{k}{^\beta}\ct{y}{_{A\beta C}}\eqs0\eqs\ctm{k}{^\beta}\ctp{k}{^\gamma}\ct{y}{_{A\beta\gamma}} $, is satisfied if the rescaled energy momentum tensor $ \ct{T}{_{\alpha\beta}} $ fulfils the corresponding equations coming form \cref{eq:auxCotton}. In particular, given \cref{eq:matter-decay-cotton}, the assumption is satisfied in vacuum or if the physical energy-momentum tensor $ \pt{T}{_{\alpha\beta}} $ decays towards infinity as $ \pt{T}{_{\alpha\beta}}\lvert_{\scri}\sim \mathcal{O}\prn{\Omega^p} \text{ with } p>2   $.
			\end{remark}
			\begin{proof}
				We will need the Bianchi identities written in terms of the lightlike components of the rescaled Weyl tensor, which can be found in \cref{app:bianchi-id} --recall that one has to substitute the over ring by an underbar in quantities carrying uppercase Latin indices $ A, B, C, $ etc. Under the assumptions above $ \ctcnm{t}{_{ABC}}\eqs 0  $ and, using \cref{eq:tABC}, \cref{eq:idbianchi7} reads
					\begin{align}
					0&\eqs -\ctcn{E}{^\omega_A}\ctcn{E}{^\sigma_C}\ctm{k}{^\mu}\cd{_\mu}\ctcnmp{D}{_{\omega\sigma}}+\sqrt{2}\ctcnp{D}{_C}\ctcn{E}{^\omega_A}\ctm{k}{^\mu}\cd{_\mu}\ctm{k}{_\omega}\nonumber\\
					&+2\sqrt{2}\mcn{_{A[C}}\ctcn{E}{^\sigma_{M]}}\ctcnp{D}{^M}\ctm{k}{^\mu}\cd{_\mu}\ctm{k}{_{\sigma}}-\ctcnmp{D}{_{AC}}\ctcn{P}{^{\mu\tau}}\cd{_\mu}\ctm{k}{_\tau}\nonumber\\
					&+\ctcnmp{D}{_{A\tau}}\ctcn{P}{^{\mu\tau}}\ctcn{E}{^\sigma_C}\cd{_\mu}\ctm{k}{_\sigma}-2\ctcnmp{D}{_{[\tau C]}}\ctcn{P}{^{\mu\tau}}\ctcn{E}{^\omega_A}\cd{_\mu}\ctm{k}{_\omega}\nonumber\\
					&-\cs{D}\prn{ \ctcn{E}{^\lambda_C}\ctcn{E}{^\mu_A}-\mcn{_{AC}}\ctcn{P}{^{\lambda\mu}}}\cd{_\mu}\ctm{k}{_\lambda}\spacef.\label{eq:aux24}
					\end{align}
				Taking the symmetric traceless part of this equation and noting \cref{it:nullDecompProp17} on page \pageref{it:nullDecompProp17}, after some manipulation, \cref{eq:aux24} is expressed as
					\begin{equation}\label{eq:aux29}
					 0 \eqs 	-\frac{3}{2}D\ctcnm{\sigma}{_{AC}} +C\frac{3}{2}\ctcn{E}{^\mu_D}\ctcn{\epsilon}{_{E(A}}\ctcn{E}{^\sigma_{C)}}\mcn{^{ED}}\cd{_\mu}\ctm{k}{_\sigma}-\frac{3}{4}\mc{_{AC}}\ctcn{\epsilon}{^{MN}}\ctcnm{\omega}{_{MN}}\spacef,
					\end{equation}
				where we have split $ \ctcn{E}{^\mu_A}\ctcn{E}{^\nu_B}\cd{_\mu}\ctm{k}{_\nu} $ into its symmetric and antisymmetric parts, introduced \eqref{eq:sigmam} and defined $ \ctcnm{\omega}{_{AB}}\defeq \ctcn{E}{^\mu_{[A}}\ctcn{E}{^\nu_{B]}}\cd{_\mu}\ctm{k}{_\nu}$. Note that in two dimensions we have
				\begin{equation}\label{eq:auxomegaId}
				\ctcnm{\omega}{_{AB}}=\frac{1}{2}\ctcn{\epsilon}{_{AB}}\ctcn{\epsilon}{^{CD}}\ctcnm{\omega}{_{CD}}\spacef,
				\end{equation}
				which after substitution into \cref{eq:aux29} leads to
					\begin{equation}\label{eq:aux25}
					0\eqs -\frac{3}{2}\prn{D\ctcnm{\sigma}{_{AC}}+C\ctcnm{\sigma}{_{E(C}}\ctc{\epsilon}{_{A)}^E}}\spacef.
					\end{equation}	
			 \Cref{eq:aux25} requires either $ \ctcnm{\sigma}{_{AB}}\eqs 0 $ or
					\begin{equation}\label{eq:aux26}
						\ctcnm{\sigma}{_{AB}}\neqs 0,\quad C\eqs D\eqs 0.
					\end{equation}
				 If condition \eqref{eq:aux26} holds, then $ \ctcnmp{D}{_{\alpha\beta}}=0 $ by \cref{it:nullDecompProp17} on page \pageref{it:nullDecompProp17}, and we have to consider \cref{eq:idbianchi3} --taking into account \cref{eq:aux26} and using \cref{eq:tABC}--,
					\begin{align}
					0\eqs 2\sqrt{2}\ctcnp{D}{^M}\ctcn{E}{^\mu_{(M}}\ctcn{E}{^\omega_{A)}}\cd{_\mu}\ctm{k}{_\omega}-\sqrt{2}\ctcnp{D}{_A}\ctcn{P}{^{\mu\sigma}}\cd{_\mu}\ctm{k}{_{\sigma}}\spacef,
					\end{align}
				and in terms of $ \ctcnm{\sigma}{_{AB}} $,
					\begin{equation}\label{eq:aux27}
						0\eqs \ctcnm{\sigma}{_{AM}}\ctcnp{D}{^M}\spacef.
					\end{equation}
				Because we are working in 2 dimensions and $ \ctcnm{\sigma}{_{AB}} $ is traceless, it cannot have eigenvectors with zero eigenvalue and, thus, \cref{eq:aux27} implies, if $ \ctcnm{\sigma}{_{AB}}\neqs 0  $, that 
					\begin{equation}\label{eq:aux28}
					 \ctcnm{\sigma}{_{AB}}\neqs 0 ,\quad	\ctcnp{D}{_A}\eqs 0\spacef.
					\end{equation}
				If condition \eqref{eq:aux28} holds, \cref{eq:idbianchi1} reads ($ \ctm{k}{^\mu}\ct{g}{^{\alpha\beta}}\ct{y}{_{\mu\alpha\beta}}\eqs -\ctm{k}{^\mu}\ctp{k}{^\alpha}\ctm{k}{^\beta}\ct{y}{_{\mu\alpha\beta}}+\ctm{k}{^\mu}\ctcn{P}{^{\lambda\nu}}\ct{y}{_{\mu\lambda\nu}}\eqs\\
				-\ctm{k}{^\mu}\ctp{k}{^\alpha}\ctm{k}{^\beta}\ct{y}{_{\mu\alpha\beta}}\eqs 0 $)
					\begin{equation}\label{eq:aux31}
						0\eqs \ctm{k}{^\mu}\cd{_\mu}\cs{D}\spacef,
					\end{equation}
				which, recalling \cref{eq:aux26}, tells us that $ \lied_{\vec{n}} \cs{D}\eqs 0\eqs \ctp{k}{^\mu}\cd{_\mu}\cs{D}$. Considering \cref{eq:aux26,eq:aux27,eq:aux28}, \cref{eq:idbianchi2} gives 
					\begin{equation}
					0\eqs \ctcnp{D}{_{\alpha\tau}}\ctcn{P}{^{\tau\mu}}\cd{_\mu}\ctm{k}{^\alpha}\spacef,
					\end{equation}
				or by means of $ \ctcnm{\sigma}{_{AB}} $, recalling \cref{it:nullDecompProp3} on page \pageref{it:nullDecompProp3} and $ \ctcnp{D}{_{AB}}=\ctcnp{D}{_{BA}}$,
					\begin{equation}\label{eq:aux29a}
					0\eqs \ctcnp{D}{^{CM}}\ctcnm{\sigma}{_{CM}}\spacef.
					\end{equation}	
				Next, taking into account all the quantities that vanish so far, it can be shown that the trace of \cref{eq:idbianchi8} gives \cref{eq:aux29a} again, while contracting it with $ \ctcn{\epsilon}{^{AC}} $ gives ($ \ctp{k}{^\mu}\ct{g}{^{\alpha\beta}}\ct{y}{_{\alpha\mu\beta}}\eqs -\ctp{k}{^\mu}\ctm{k}{^\alpha}\ctm{k}{^\beta}\ct{y}{_{\alpha\mu\beta}}+\ctp{k}{^\mu}\ctcn{P}{^{\lambda\nu}}\ct{y}{_{\lambda\mu\nu}}\eqs-\ctp{k}{^\mu}\ctp{k}{^\alpha}\ctm{k}{^\beta}\ct{y}{_{\alpha\mu\beta}}\eqs 0 $)
					\begin{equation}\label{eq:aux30}
						0\eqs-\ctp{k}{^\mu}\cd{_\mu}C+\ctcn{\epsilon}{^{AC}}\ctcp{D}{_{A}^E}\prn{\ctcnm{\omega}{_{EC}}+\ctcnm{\sigma}{_{EC}}},
					\end{equation}
				where we have taken into account $ \ctcnp{D}{_{AB}}=\ctcnp{D}{_{BA}} $. \Cref{eq:auxomegaId}
				and \cref{it:nullDecompProp3} on page \pageref{it:nullDecompProp3} simplify \cref{eq:aux30} to
					\begin{equation}\label{eq:aux32}
						0\eqs-\ctp{k}{^\mu}\cd{_\mu}C+\ctcn{\epsilon}{^{AC}}\ctcp{D}{_{A}^E}\ctcnm{\sigma}{_{EC}}\spacef.
					\end{equation}
				Back to \cref{eq:aux24}, using \cref{eq:aux26,eq:aux31,eq:aux28}, we arrive at
					\begin{equation}
						\ct{n}{^\mu}\cd{_\mu}C\eqs 0\spacef.
					\end{equation}
				Then, \cref{eq:aux30} reads simply		
					\begin{equation}\label{eq:aux29b}
						0 \eqs \ctcn{\epsilon}{^{AC}}\ctcp{D}{_{A}^E}\ctcnm{\sigma}{_{EC}}.
					\end{equation}
				It is easily shown, given that $ \ctcnm{\sigma}{_{AB}} $ and $ \ctcnp{D}{_{AB}}  $ are both symmetric and traceless, that \cref{eq:aux29a,eq:aux29b} imply --e.g., by writing these equations in components $ A=2,3 $--
					\begin{equation}
						\ctcnm{\sigma}{_{AB}}\neqs 0 \implies \ctcnp{D}{_{AB}}\eqs 0.
					\end{equation}
				But $ \ctcnp{D}{_{AB}}\eqs 0 $ together with \cref{eq:aux26,eq:aux28} and the assumptions $ \ctcnm{D}{_{AB}}\eqs 0 \eqs \ctcnm{D}{_A} $ leads to $ \ct{d}{_{\alpha\beta\gamma}^\delta}\eqs 0 $. This follows from the fact that in this case $ \csmp{\W}\eqs\csmp{\Z}\eqs \cs{\V}\eqs 0$ --see \cref{eq:Wmdef,eq:Wpdef,eq:Zmdef,eq:Zpdef,eq:Vdef}-- which, by  \cref{thm:noRadCulSE-noSE,eq:noWnoT} implies $ \ct{d}{_{\alpha\beta\gamma}^\delta}\eqs 0 $. Alternatively, use \cref{eq:Cab-decomposition,eq:Dab-decomposition} to show that $ \ct{D}{_{ab}}\eqs\ct{C}{_{ab}}\eqs 0 $($ \iff \ct{d}{_{\alpha\beta\gamma}^\delta}\eqs 0 $). However, this contradicts one of the assumptions of the theorem. Therefore, the only possibility is
					\begin{equation}
						\ctcnm{\sigma}{_{AB}}\eqs 0\spacef.
					\end{equation}
			\end{proof}
			\Cref{thm:gslikethm} is in fact a result on $ \ct{m}{^a} $ as well, noting \cref{eq:sigmam-Sigma}:
			\begin{corollary}\label{thm:noSigma}
			Under the same assumptions of \Cref{thm:gslikethm}, its conclusion can be equivalently stated as 
				\begin{equation*}
					\ctcn{\Sigma}{_{ab}}\eqs 0.
				\end{equation*}
			\end{corollary}

		\subsection{The $ \Lambda=0 $ limit}\label{ssec:afs-limit}
		In order to gain a deeper insight into the differences and analogies between the $ \Lambda>0 $ and $ \Lambda=0 $ scenarios, and in particular concerning the characterisation of gravitational radiation at $\scri$, one can study the limit to $ \Lambda=0 $ of \cref{def:criterionGlobal}. To that end, in this subsection we assume that $ \limitl\ct{g}{_{\alpha\beta}} $ exists and defines a good Lorentzian metric.\\
		
		The limit of the normal to $ \scri $,  $ \ct{N}{_\alpha}\evalat{_{\Lambda=0}} $,  coincides with the normal to $ \scri_0 $, the conformal boundary for $ \Lambda=0 $. Also, we have already mentioned in \cref{ssec:radiation-condition} that the asymptotic supermomenta $ \ct{p}{^\alpha} $ \eqref{eq:asymptotic-supermomentum} has  a good limit to $ \Lambda=0 $,
			\begin{equation}\label{eq:limit-asymptotic-supermomentum}
		 		\limitl\ct{p}{^\alpha}\eqsflat\ct{\Q}{^\alpha}\spacef,
			\end{equation}
		where $ \ct{\Q}{^\alpha} $ is the asymptotic \emph{radiant} supermomentum at $ \scri_0 $ \cite{Fernandez-Alvarez_Senovilla20,Fernandez-Alvarez_Senovilla-afs}
			\begin{equation}\label{eq:asymptotic-radiant-sm-flat}
				\ct{\Q}{^\alpha}\eqsflat-\prn{\ct{N}{^\mu}\ct{N}{^\nu}\ct{N}{^\rho}\ct{\D}{^\alpha_{\mu\nu\rho}}}\evalat{_{\Lambda=0}}\spacef.
			\end{equation}
		Therefore, the absence of gravitational radiation in the $ \Lambda>0 $ case according to \cref{def:criterionGlobal} implies that the asymptotic radiant supermomentum $ \ct{\Q}{^\alpha} $ vanishes in the $ \Lambda=0 $ counterpart and, as a consequence, that the news tensor vanishes there  \cite{Fernandez-Alvarez_Senovilla20,Fernandez-Alvarez_Senovilla-afs} so that there is no radiation either. This limit reinforces the validity of \cref{def:criterionGlobal}.\\
			
		Apart from $ \ct{p}{^\alpha} $, it is possible to study the limit of the radiant supermomenta of \cref{eq:Qmdef-m,eq:Qpdef-m}. The first thing to do is to define a couple of lightlike vector fields on $ \scri $ in a way that their limit to $ \Lambda=0 $ is well-behaved. This can be achieved by multiplying the expressions on the right-hand side of \cref{eq:kp-def-m,eq:km-def-m} by $ N $,
				\begin{align}
				\ctp{K}{^\alpha}&\defeq\frac{1}{\sqrt{2}}\left(\ct{N}{^\alpha}+\ct{M}{^\alpha}\right)\spacef,\label{eq:Kp-def-m}\\
				\ctm{K}{^\alpha}&\defeq\frac{1}{\sqrt{2}}\left(\ct{N}{^\alpha}-\ct{M}{^\alpha}\right)\spacef,\label{eq:Km-def-m}
				\end{align}	
		where $ \ct{M}{^\alpha}\defeq N \ct{m}{^\alpha} $ with the following normalisations:
			\begin{equation}
				\ct{g}{_{\mu\nu}}\ctm{K}{^{\mu}}\ctp{K}{^\nu}=N^2,\quad\ct{g}{_{\mu\nu}}\ct{M}{^\mu}\ct{M}{^\nu}=N^2,\quad\ct{g}{_{\mu\nu}}\ct{N}{^\mu}\ct{M}{^{\nu}}=0.
			\end{equation}			
		Vector fields on $ \scri $ of the kind of $ \ct{M}{^\alpha} $ obey:
		\begin{lemma}\label{thm:limitM}
				Assume that $ \limitl\ct{g}{_{\alpha\beta}} $ exists and let $ \ct{M}{^\alpha} $ be any vector field on $ \scri $ whose norm is proportional to a positive power of the cosmological constant $ \Lambda $. Then, 
				\begin{equation}\label{eq:limit-M}
					\limitl\ct{M}{^\alpha}\eqsflat B\ct{N}{^\alpha}\evalat{_{\Lambda=0}},
				\end{equation}
		for some function $ B $ which may have zeros.
			\end{lemma}
			\begin{proof}
				We know that the limit $ \ct{N}{^\alpha}\evalat{_{\Lambda=0}} $ does not vanish and is lightlike at $ \scri_0 $. Then, we have
					\begin{align}
						\limitl\prn{\ct{g}{_{\mu\nu}}\ct{M}{^{\mu}}\ct{M}{^\nu}}&=\limitl f\Lambda^p=0\spacef,\\
						\limitl\prn{\ct{g}{_{\mu\nu}}\ct{M}{^\mu}\ct{N}{^\nu}}&=\limitl 0= 0\eqsflat\limitl\prn{\ct{g}{_{\mu\nu}}\ct{M}{^\mu}}\ct{N}{^\nu}\evalat{_{\Lambda=0}}\spacef,
					\end{align}
				where $ f $ is a function and $ p\in \mathbb{R} $, $ p>0 $. The first of this formulae implies that the limit of $ \ct{M}{^\alpha} $ is either lightlike or zero at $ \scri_0 $. Taking this into account, the second formula indicates that, if different from zero, the limit of $ \ct{M}{^\alpha} $ has to be proportional to $ \ct{N}{^\alpha} $ --as the scalar product of two non-vanishing lightlike vector fields is zero if and only if they are collinear.
			\end{proof}
			\begin{figure}[h!]
				\centering
				\includegraphics[width=0.9\textwidth]{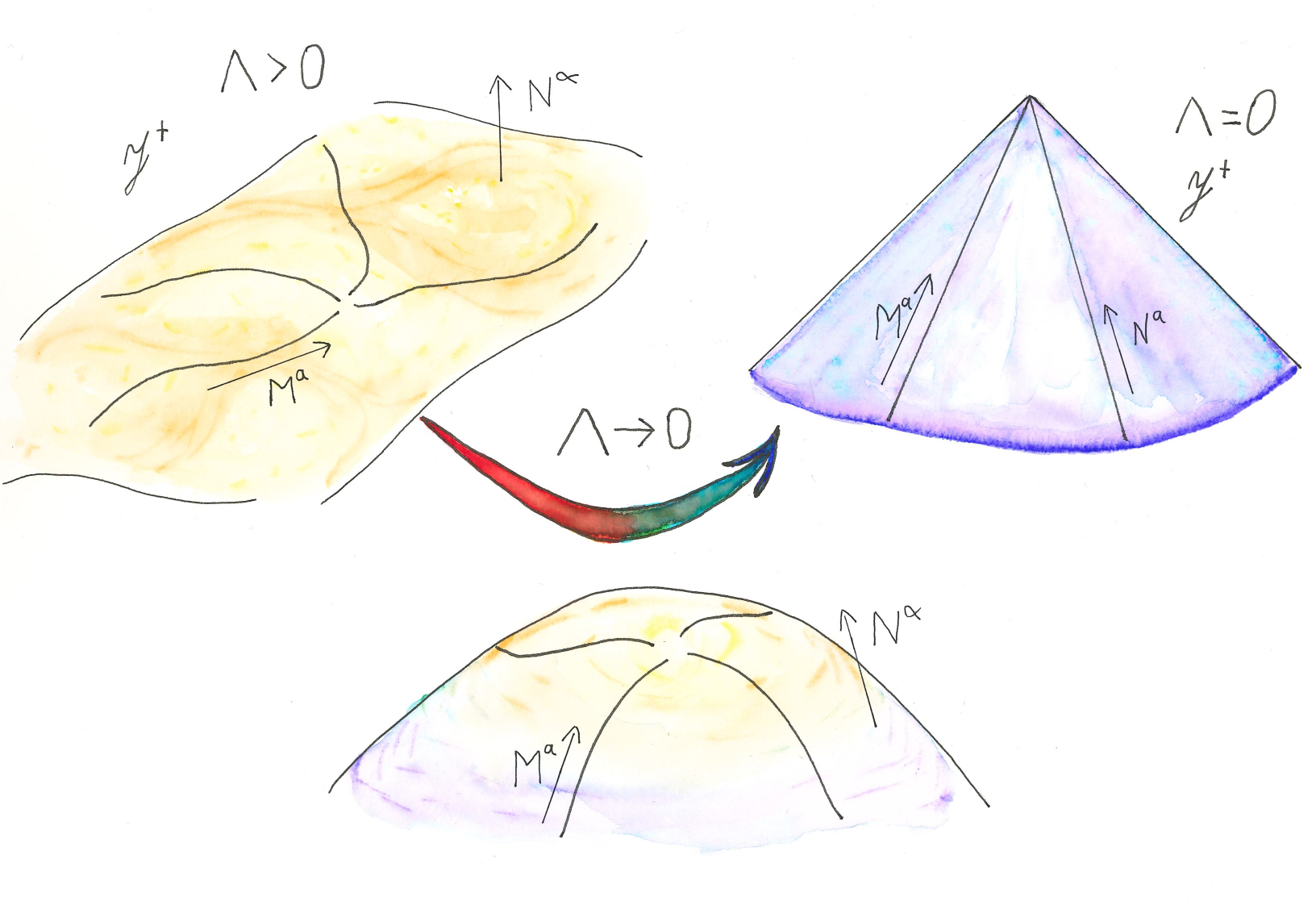}
				\caption[The limit to $ \Lambda=0 $ on $ \scri $]{In the $ \Lambda=0 $ limit vector fields of the class described in \cref{thm:limitM} become collinear with $ \ct{N}{^\alpha}\evalat{_{\Lambda=0}} $, the vector field tangent to the null generators of $ \scri_0 $.}
			\end{figure}
			Then, by \cref{thm:limitM}, the limit of $ \ctpm{K}{^\alpha} $ reads
				\begin{equation}\label{eq:limitK}
					\limitl\ctpm{K}{^\alpha}\eqsflat \frac{1}{\sqrt{2}}\prn{1\pm B}\ct{N}{^\alpha}\evalat{_{\Lambda=0}}\spacef.
				\end{equation}
			After this, define the radiant supermomenta associated to $ \ctpm{K}{^\alpha} $,
				\begin{align}
				\ctp{q}{^\alpha}&\defeq -\ctp{K}{^\mu}\ctp{K}{^\nu}\ctp{K}{^\rho}\ct{\D}{^\alpha_{\mu\nu\rho}}\spacef,\label{eq:qpdef}\\
				\ctm{q}{^\alpha}&\defeq -\ctm{K}{^\mu}\ctm{K}{^\nu}\ctm{K}{^\rho}\ct{\D}{^\alpha_{\mu\nu\rho}}\spacef.\label{eq:qmdef}
				\end{align} 
			These are nothing else than the radiant supermomenta given in \cref{eq:Qpdef-m,eq:Qmdef-m} appropriately rescaled by a factor $ N^3 $.
			\begin{lemma}\label{thm:limit-q}
				Assume that $ \limitl\ct{g}{_{\alpha\beta}} $ exists. The radiant supermomenta \lmultieqref{eq:qpdef}\rmultieqref{eq:qmdef}  have, respectively, the regular limit
					\begin{equation}
						\limitl  \ctpm{q}{^\alpha}\eqsflat\frac{1}{2\sqrt{2}}\prn{1\pm B}^3\ct{\Q}{^\alpha}\quad
					\end{equation}
				where $ \ct{\Q}{^\alpha} $ is the asymptotic radiant supermomentum \eqref{eq:asymptotic-radiant-sm-flat} on $ \scri_0 $ for a vanishing cosmological constant and $ B $ is a function which may have zeros. Moreover,
					\begin{equation*}
					\limitl\ctp{q}{^\alpha}= 0 =\limitl\ctm{q}{^\alpha} \iff \ct{\Q}{^\alpha}\eqsflat 0 \iff \text{No gravitational radiation at $ \scri_0 $.}
					\end{equation*}
			\end{lemma}
			\begin{proof}
				The limit of $ \ctpm{q}{^\alpha} $ is computed using \cref{eq:limitK}. Then, one notices that for non-vanishing $ \ctpm{q}{^\alpha} $, it is not possible that both radiant supermomenta vanish simultaneously in the limit unless $ \ct{\Q}{^\alpha}\eqsflat 0 $. The reason is that
					\begin{align}
					B=&1 \implies \limitl \ctm{q}{^\alpha}= 0,\quad\limitl\ctp{q}{^\alpha}\eqsflat \frac{1}{2\sqrt{2}}\prn{1+ B}^3\ct{\Q}{^\alpha}\spacef,\\
					B=&-1 \implies \limitl \ctm{q}{^\alpha}\eqsflat \frac{1}{2\sqrt{2}}\prn{1- B}^3\ct{\Q}{^\alpha},\quad\limitl\ctp{q}{^\alpha}= 0\spacef,\\
					B\neq&\pm 1 \implies \limitl \ctm{q}{^\alpha}\eqsflat \frac{1}{2\sqrt{2}}\prn{1- B}^3\ct{\Q}{^\alpha},\quad\limitl\ctp{q}{^\alpha}\eqsflat \frac{1}{2\sqrt{2}}\prn{1+ B}^3\ct{\Q}{^\alpha}\spacef.
					\end{align}
				Hence, if we assume $ B=1 $, $ \limitl\ctp{q}{^\alpha}= 0\iff \ct{\Q}{^\alpha}\eqsflat 0 $. But if $ B=-1 $, then  $ \limitl\ctm{q}{^\alpha}= 0\iff \ct{\Q}{^\alpha}\eqsflat 0 $. Finally, if $ B\neq \pm1 $ the only possibility is $ \limitl \ctm{q}{^\alpha}=0=  \limitl\ctp{q}{^\alpha} \iff \ct{\Q}{^\alpha}\eqsflat 0$.
			\end{proof}
		These results have a particularly interesting interpretation regarding  incoming versus outgoing radiation and intrinsic evolution directions that will be presented in \cref{ssec:noincomingrad}.
		    	\begin{corollary}
			    	If one (and only one) of the asymptotic radiant supermomenta $ \ctpm{q}{^\alpha} $ of \cref{eq:qpdef,eq:qmdef} vanishes, then $ B=\mp 1 $ and
			    		\begin{align}
		    				\limitl\ctpm{M}{^\alpha}&\eqsflat\pm\ct{N}{^\alpha}\evalat{_{\Lambda=0}}\spacef,\label{eq:limit-aux-1}\\
			    			\limitl\ctpm{K}{^\alpha}&\eqsflat 0\spacef,\label{eq:limit-aux-2}\\
			    			\limitl\ctmp{K}{^\alpha}&\eqsflat \sqrt{2}\ct{N}{^\alpha}\evalat{_{\Lambda=0}}\spacef.\label{eq:limit-aux-3}
			    		\end{align}
		    	\end{corollary}
		    	\begin{proof}
			    	From the proof in \cref{thm:limit-q}, if $ \ctpm{q}{^\alpha}=0 $,  one has $ B=\mp1 $. Setting the corresponding value of $ B $ in \cref{eq:limitK,eq:limit-M} gives \cref{eq:limit-aux-1,eq:limit-aux-2,eq:limit-aux-3}.
		    	\end{proof}


	\section{In search for `News'}\label{sec:news}
			
		 Once we have presented our gravitational-radiation condition at infinity and its interplay with the asymptotic Petrov type of the rescaled Weyl tensor,
		a next logical step forward 
		would be to find the analogous of the \emph{news tensor} available in the case with $\Lambda=0$, i.e., an object describing in the \emph{full, covariant} theory the two radiative degrees of freedom of the gravitational field arriving at $\scri$. In the $ \Lambda=0 $ scenario the news tensor (or its traditional equivalent the `news complex function')  is a covariant, rank-2 trace-free tensor field orthogonal to the null generators of $\scri$ such that some tidal field (of type $\ctcnp{C}{_{ab}}$) acts as its source \cite{Geroch1977}, see also \cite{Fernandez-Alvarez_Senovilla20} and the companion paper \cite{Fernandez-Alvarez_Senovilla-afs}. 
		We wonder if a similar tensor may exist in the presence of a positive cosmological constant and, if so, under which conditions. We are going to prove that a complete equivalent cannot exist, for various reasons: mainly due to the mingling of in- and out-going radiation, and to the different structure of $\scri$ when $\Lambda=0$. However, we will also show that there certainly exist (at least partial) analogous tensors {\em for each cut} on $\scri$ that retain the essentials of the news tensor.\\
		
		To that end,we work with an arbitrary cut $ \prn{\Sc, \mc{_{AB}}} $ on $ \scri $, that is, a two-dimensional Riemannian manifold $ \Sc \in \scri$ equipped with the metric $ \mc{_{AB}} $ inherited from the ambient metric $ \ms{_{ab}} $. We denote by  $ \cbrkt{\ct{E}{^\alpha_A}} $ a basis of vector fields on $ \Sc $ 
		and by $ \ct{r}{^a} $ the unit normal to the cut within $ \scri $. Let us emphasise that $ \ct{r}{^a} $ is defined at least on $ \Sc $ but not necessarily outside.
		We denote the second fundamental form, its trace and trace-free parts (the shear) by $ \ctc{\kappa}{_{AB}} $, $ \csC{\kappa} $ and $ \ctc{\Sigma}{_{AB}} $, respectively. Also, $ \ctc{\epsilon}{_{AB}} $ is the intrinsic, canonical, volume two-form relative to the metric $ \mc{_{AB}} $. For more details, see \cref{ssec:foliationAndCuts}. 
		
		In the same fashion as \cref{eq:n-kp-km}, we introduce a pair of vector fields $ \ctpm{k}{^\alpha} $, defined at least on $ \Sc $. Notice that $ \ctpm{k}{_\alpha}\ct{E}{^\alpha_A} =0$. Let us present some useful relations involving the intrinsic Schouten tensor $ \cts{S}{_{ab}} $ and the extrinsic curvature of $ \Sc $. First, define the tangent and orthogonal components to $ \Sc $ --according to the general notation \lmultieqref{eq:notation-symmetric-tensor}\rmultieqref{eq:notation-symmetric-tensorLatin}-- as 
			\begin{equation}\label{eq:decomp-Schouten-r}
			\cts{S}{_{ab}}\eqc \csS{S}\ct{r}{_a}\ct{r}{_b}+2\ctc{S}{_B}\ct{r}{_{(a}}\ct{W}{_{b)}^B}+\ctc{S}{_{AB}}\ct{W}{_a^A}\ct{W}{_b^B}\spacef.
			\end{equation}
		After this, project \cref{eq:intrinsicriemannScriSchoutenRelation} thrice with $ \ct{E}{^a_A} $ and once with $ \ct{r}{^a} $ and use \fblue{\cref{eq:cod-cuts}} to obtain
			\begin{equation}
			\mc{_{A[C}}\ctc{S}{_{B]}}= \cdc{_{[C}}\ctc{\kappa}{_{B]A}}\spacef,
			\end{equation} 
		whose trace reads
			\begin{equation}\label{eq:SAkappa}
			\ctc{S}{_{B}}= \cdc{_C}\ctc{\kappa}{_B^C}-\cdc{_B}\csC{\kappa}\spacef.
			\end{equation}
		Also, by  \cref{eq:intrinsicS} and \cref{eq:gaussrelCiii,eq:gaussrelCii}, it can be seen that
			\begin{align}
			\csS{S}&\eqc-\frac{1}{2}\cs{K}+\cts{R}{_{ab}}\ct{r}{^a}\ct{r}{^b}\eqc\frac{1}{2}\cs{K}-\ctc{S}{^E_E}\spacef,\\
			\ctc{S}{^E_E}&= \cs{K}+\frac{1}{2}\csC{\Sigma}^2-\frac{1}{4}\csC{\kappa}^2\quad\label{eq:traceSchoutenCut}.
			\end{align}
		 Here, $ \cs{K} $ is the Gaussian curvature of the  cut, which is related to its scalar curvature by $ \cs{K}=\csC{R}/2 $ ---see \cref{ssec:foliationAndCuts} for more details. \\
		 
		\subsection{General considerations}
			In the asymptotically flat case,  the news tensor vanishes if and only if the asymptotic (radiant) supermomenta vanishes; indeed, the asymptotic superenergy acts as source for the news tensor \cite{Fernandez-Alvarez_Senovilla20,Fernandez-Alvarez_Senovilla-afs}. In the presence of a positive cosmological constant, however, the asymptotic supermomentum is not radiant. Thus, a question arises: do we look for a news tensor which can be associated to a radiant supermomentum in a similar fashion as in the $ \Lambda=0 $ case or, alternatively, one that vanishes if and only if the asymptotic super-Poynting vanishes? In \cref{ssec:general-approach} we will present a general programme valid for both possibilities, while in \cref{ssec:conditions-news} we will explore thoroughly the first one.\\
			
			Generically, we expect any news-like object to have some basic properties. First of all, the would-be news tensor must appear at the energy-density level. From this point of view, it is reasonable to think that the gravitational radiative degrees of freedom cannot be extracted by local methods alone --- for a discussion in the asymptotically flat case, see \cite{Penrose1986}. For $ \Lambda=0 $, $ \scri $ is naturally foliated by two dimensional cuts; this is not the case for $ \Lambda> 0$ in general, and for that reason we are just considering a single cut $ \Sc $. Another important difference is that in the $ \Lambda=0 $ case any cut has a unique, lightlike, orthogonal (outgoing  for $\scri^+$) direction that escapes from the space-time and is linearly independent of the (incoming  for $\scri^+$) lightlike direction given by the generators of $ \scri $. For $ \Lambda>0 $, there are always two independent, future lightlike directions orthogonal to any cut $ \Sc $ on $ \scri $ pointing out of (or into) the space-time, $ \ctpm{k}{^\alpha} $. Therefore, a priori there is no reason why there should be only one  news-like tensor for each cut instead of two, one for each radiant supermomentum associated to $  \ctpm{k}{^\alpha} $. Secondly, to describe the radiative sector
			the would-be news tensor(s) should have two degrees of freedom. The most plausible object is a symmetric, traceless, rank-2 tensor. Thirdly, it has to be gauge invariant to have physical significance. Finally, a key feature that will guide us is that we want it to vanish if and only if some meaningful superenergy quantity vanishes, such as the radiant super-Poynting $ \ctpms{\Q}{^\alpha} $ or the canonical asymptotic super-Poynting $ \cts{\Pc}{^a} $. Thus, according to \cref{eq:Zmdef,eq:Zpdef,eq:commutator-se}, the news tensor has to carry information from \emph{both} the magnetic $ \ct{C}{_{ab}} $ and the electric $ \ct{D}{_{ab}} $ parts of the rescaled Weyl tensor.\\
			
			More concisely, the properties that the would-be news tensor is expected to have are:
				\begin{properties}
					\item Rank-2 tensor field on $ \Sc $.\label{it:propertyNewsTwoDim}
					\item Symmetric.\label{it:propertyNewsSymmetric}
					\item Traceless.\label{it:propertyNewsTraceless}
					\item Gauge invariant.\label{it:propertyNewsGaugeInvariant}
					\item Contain information related to  $ \ct{C}{_{ab}} $ and $ \ct{D}{_{ab}} $. \label{it:propertyNewsCandF}
					\item Vanish if and only if some meaningful superenergy quantity does (e.g., $ \csp{\Z}=0 $ or/and $ \csm{\Z} =0$, or $ \cts{\Pc}{^a}=0 $).\label{it:propertyNewsZ}
				\end{properties}
				
				\subsection{A geometric result: the counterpart of Geroch's \emph{tensor} $ \rho $}
			Here we present an intermediate and crucial step in our search. It begins with the following lemma, where $ \ctc{\omega}{_A} \defeqc \cdc{_A}\cs{\omega}$:
				\begin{lemma}\label{thm:general-gauge-trace}
					Let $ \ct{t}{_{AB}} $ be any symmetric tensor field on $ \Sc $ whose behaviour under conformal rescalings \eqref{eq:gauge-mc} is
						\begin{equation}\label{eq:gauge-behaviour-appropriate}
								\ctg{t}{_{AB}}=\ct{t}{_{AB}}-a\frac{1}{\omega}\cdc{_A}\ctc{\omega}{_B}+\frac{2a}{\omega^2}\ctc{\omega}{_A}\ctc{\omega}{_B}-\frac{a}{2\omega^2}\ctc{\omega}{_C}\ctc{\omega}{^C}\mc{_{AB}}
						\end{equation}
					for some fixed constant $ a\in\mathbb{R} $. Then,
						\begin{equation}\label{eq:gauge-derivative-appropiate-behaviour}
							\cdcg{_{[C}}\ctg{t}{_{A]B}}=\cdc{_{[C}}\ct{t}{_{A]B}}+\frac{1}{\omega}\prn{a\cs{K}-\ct{t}{^E_E}}\ctc{\omega}{_{[C}}\mc{_{A]B}}\spacef,
						\end{equation}						
					where $ K $ is the Gaussian curvature of $ \prn{\Sc,\mc{_{AB}}} $. 	In particular, for any symmetric gauge-invariant tensor field $ \ct{B}{_{AB}} $ on $ \Sc $,
						\begin{equation}\label{eq:aux23}
							\cdcg{_{[C}}\ctg{B}{_{A]B}}=\cdcg{_{[C}}\ct{B}{_{A]B}}= \cdc{_{[C}}\ct{B}{_{A]B}}-\frac{1}{\omega}\ct{B}{^E_E}\ctc{\omega}{_{[C}}\mc{_{A]B}}\quad
						\end{equation}
				\end{lemma}
				\begin{remark}
					This result applies locally and it is valid for any Riemannian surface, independently of the topology.
				\end{remark}
				\begin{proof}
					Using the formulae in \cref{app:gauge-transformations} for cuts, a direct calculation yields
						\begin{equation}
							\cdcg{_{[C}}\ctg{t}{_{A]B}}=\cdc{_{[C}}\ct{t}{_{A]B}}+\frac{1}{\omega}\ct{t}{_{B[C}}\ctc{\omega}{_{A]}}+\frac{1}{\omega}\mc{_{B[C}}\ct{t}{_{A]}^D}\ctc{\omega}{_D}+\frac{1}{\omega} a\cs{K}\ctc{\omega}{_{[C}}\mc{_{A]B}}\spacef.
						\end{equation}
					Then, one uses the two-dimensional identity \cite{Edgar2002}
						\begin{equation}\label{eq:dimensional-identity}
							\ct{A}{_{CA}^E}=2\delta^E_{[A}\ct{A}{_{C]D}^D},\text{ for any tensor  such that }\ct{A}{_{CA}^E}=-\ct{A}{_{AC}^E}\quad
						\end{equation}
					in order to write
						\begin{equation}
						 	\frac{1}{\omega}\ct{t}{_{B[C}}\ctc{\omega}{_{A]}}+\frac{1}{\omega}\mc{_{B[C}}\ct{t}{_{A]}^D}\ctc{\omega}{_D}=-\frac{1}{\omega}\ct{t}{^E_E}\ctc{\omega}{_{[C}}\mc{_{A]B}}\spacef,
						\end{equation}
					arriving at the final result. For a gauge invariant tensor $ a=0 $ in \cref{eq:gauge-behaviour-appropriate}, therefore one only has to set this value in \cref{eq:gauge-derivative-appropiate-behaviour} to obtain \cref{eq:aux23}.
				\end{proof}
				\begin{corollary}\label{thm:gauge-invariant-diff}
					A symmetric gauge-invariant tensor field $ \ct{m}{_{AB}} $ on $ \Sc $ satisfies
						\begin{equation}
							\cdcg{_{[C}}\ctg{m}{_{B]A}}=\cdc{_{[C}}\ct{m}{_{B]A}}
						\end{equation}
					if and only if $ \ct{m}{^E_E}=0 $.
				\end{corollary}
				\begin{corollary}[The tensor $ \rho $]\label{thm:rho-tensor}
					If $ \Sc $ has $ \mathbb{S}^2 $-topology, there is a unique symmetric tensor field $ \ct{\rho}{_{AB}} $  whose behaviour under conformal rescalings \eqref{eq:gauge-mc} is as in \eqref{eq:gauge-behaviour-appropriate} and satisfies the equation
						\begin{equation}\label{eq:rho-diff-eq}
							\cdc{_{[C}}\ct{\rho}{_{A]B}}=0
						\end{equation}
					in any conformal frame. This tensor field must have a trace $ \ct{\rho}{^E_E}=aK $ and obeys  
						\begin{equation}\label{eq:rho-lie-CKVF}
							\lied_{\vec{\chi}}\ct{\rho}{_{AB}}=-a\cdc{_{A}}\cdc{_{B}}\phi
						\end{equation}
					independently of the conformal frame, where $ \ct{\chi}{^A} $ is any CKVF of $ \prn{\Sc,\mc{_{AB}}} $ and $ \phi\defeq\cdc{_M}\ct{\chi}{^M}/2 $. Specifically, it is invariant under transformations generated by KVF (and homothetic Killing vectors) of $ \prn{\Sc,\mc{_{AB}}} $. Furthermore, it is given for round spheres by $ \ct{\rho}{_{AB}}=\mc{_{AB}}aK/2 $.
				\end{corollary}	
				\begin{proof}
						Existence is proved by using the (trivial) $ L^2 $-orthogonality of the right-hand side of \cref{eq:rho-diff-eq} with all conformal Killing vectors on $ \Sc $ (see for instance \cite{Besse1987}, appendix H) or, more directly, by noticing that $ \ct{\rho}{_{AB}}=\mc{_{AB}}aK/2 $ fulfils $ \cdc{_A}\ct{\rho}{_{BC}} =0$ in the round metric sphere. Concerning uniqueness, notice that \cref{thm:general-gauge-trace} fixes the trace of $ \ct{\rho}{_{AB}} $ to $ \ct{\rho}{^E_E}=aK $, and recall the assumption that \cref{eq:rho-diff-eq} holds in any gauge. Then, if two different solutions $ \ctrd{1}{\rho}{_{AB}} $ and $ \ctrd{2}{\rho}{_{AB}} $ exist, $ \cdc{_{[C}}\prn{\ctrd{1}{\rho}{_{A]B}}-\ctrd{2}{\rho}{_{A]B}}}=0 $. However, in that case, the difference $ \ctrd{1}{\rho}{_{AB}}-\ctrd{2}{\rho}{_{AB}} $ is a traceless, Codazzi tensor on $ \mathbb{S}^2 $ and, as a consequence of the uniqueness of this kind of tensors \cite{Liu1998}, $ \ctrd{1}{\rho}{_{AB}}-\ctrd{2}{\rho}{_{AB}}=0 $. To show \cref{eq:rho-lie-CKVF}, first define $ \ct{M}{_{AB}}\defeq \lied_{\vec{\chi}}\ct{\rho}{_{AB}} +a\cdc{_{A}}\cdc{_{B}}\phi $. This tensor field is gauge invariant and using $  \cdc{_{M}}\cdc{^M}\phi=-\lied_{\vec{\chi}}K-2\phi K $ (see \cref{eq:LieK}) and $ \ct{\rho}{^C_C} =aK$ one derives $ \ct{M}{^C_C}=0 $. Also, write the formula for the commutator $ \commute{\lied_{\vec{\chi}}}{\cdc{_A}} $ (see e.g. \cite{Yano1957}) acting on $ \ct{\rho}{_{AB}} $ 
							\begin{equation}
								\prn{\lied_{\vec{\chi}}\cdc{_C}-\cdc{_C}\lied_{\vec{\chi}}}\ct{\rho}{_{AB}}=-\ct{\rho}{_{EA}}\lied_{\vec{\chi}}\ctc{\Gamma}{^E_{CB}}-\ct{\rho}{_{EB}}\lied_{\vec{\chi}}\ctc{\Gamma}{^E_{CA}},
							\end{equation}
						which, noting that
							\begin{equation}
								\lied_{\vec{\chi}}\ctc{\Gamma}{^E_{CB}}=\delta^E_C \cdc{_{B}}\phi+\delta^E_B\cdc{_C}\phi-\mc{_{BC}}\mc{^{FE}}\cdc{_E}\phi\spacef,
							\end{equation}
						can be antisymetrised to get
							\begin{equation}
								\prn{\lied_{\vec{\chi}}\cdc{_{[C}}-\cdc{_{[C}}\lied_{\vec{\chi}}}\ct{\rho}{_{A]B}}=-\ct{\rho}{_{B[A}}\cdc{_{C]}}\phi +\mc{_{B[C}}\ct{\rho}{_{A]}^E}\cdc{_E}\phi\spacef.
							\end{equation}
					 Making use of \cref{eq:rho-diff-eq} one arrives at
					 		\begin{equation}
					 			\cdc{_{[C}}\ct{M}{_{A]B}}=\ct{\rho}{_{B[A}}\cdc{_{C]}}\phi -\mc{_{B[C}}\ct{\rho}{_{A]}^E}\cdc{_E}\phi+a\frac{1}{2}K\prn{\delta_{[A}^E\mc{_{C]B}}-\mc{_{B[A}}\delta_{C]}^E}=0
					 		\end{equation}
					 where the first equality follows by $ 2\cdc{_{[C}}\cdc{_{A]}}\cdc{_B}\phi=\ctc{R}{_{CAB}^E}\cdc{_{E}}\phi $, and the second using the identity \eqref{eq:dimensional-identity} together with $ \ct{\rho}{^C_C}=aK $. Because $ \ct{M}{_{AB}} $ is symmetric, traceless and divergence free  (a `TT-tensor') 
					 on the compact  two-dimensional $\mathbb{S}^2$ necessarily
					 $ \ct{M}{_{AB}}=0 $. For (homothetic) KVF, $ \phi= $ constant and  \cref{eq:rho-lie-CKVF} reads $ \lied_{\vec{\chi}}\ct{\rho}{_{AB}}=0 $, i.e., $ \ct{\rho}{_{AB}} $ is left invariant by  (homothetic) KVF. 
				\end{proof}
				\begin{remark}
					Let $ \vec{\chi} $ be a CKVF on $ \prn{\Sc,\mc{_{AB}}} $,
						\begin{equation}
							\lied_{\vec{\chi}}\mc{_{AB}}= 2\phi\mc{_{AB}},\label{eq:conformal-chi}
						\end{equation}
					generating a one-parameter group of local conformal transformations $ \cbrkt{\csrd{\epsilon}{\Psi}} $ on $ \Sc $ ($ \prn{\csrd{\epsilon}{\Psi}^* \mc{}}_{AB}=\csrd{\epsilon}{\Phi}^2\mc{_{AB}} $) with $ \phi\defeq\df{\csrd{\epsilon}{\Phi}}/\df\epsilon\evalat{_{\epsilon=0}} $ and $ {\csrd{\epsilon}{\Phi}}\evalat{_{\epsilon=0}}=1 $. 	
					Then, the finite change of $ \ct{\rho}{_{AB}} $ under these conformal transformations is
				 		\begin{equation}\label{eq:conformal-behaviour-rho}
						\ctg{\rho}{_{AB}}=\ct{\rho}{_{AB}}-a\frac{1}{{\csrd{\epsilon}{\Phi}}}\cdc{_A}\ct{{\csrd{\epsilon}{\Phi}}}{_B}+\frac{2a}{{\csrd{\epsilon}{\Phi}}^2}\ct{{\csrd{\epsilon}{\Phi}}}{_A}\ct{{\csrd{\epsilon}{\Phi}}}{_B}-\frac{a}{2{\csrd{\epsilon}{\Phi}}^2}\ct{{\csrd{\epsilon}{\Phi}}}{_C}\ct{{\csrd{\epsilon}{\Phi}}}{^C}\mc{_{AB}}
						\end{equation}
					with $ \ct{{\csrd{\epsilon}{\Phi}}}{_A}\defeq \cdc{_A}\csrd{\epsilon}{\Phi} $. Expression \eqref{eq:conformal-behaviour-rho} follows from \cref{eq:rho-lie-CKVF} and the exponential map from the Lie algebra to the finite group of conformal transformations.
				\end{remark}
				\begin{corollary}[The tensor $ \rho $ for non-$ \mathbb{S}^2 $ manifolds]\label{thm:rho-tensor-noncompact}
					Let $\prn{\Sc,\mc{_{AB}}}$ be a 2-dimensional Riemannian manifold, no necessarily with $\mathbb{S}^2$ topology, and  such that there exists a CKVF $ \ct{\chi}{^A} $ with a fixed point. Then, there is a unique symmetric tensor field $ \ct{\rho}{_{AB}} $ on $ \Sc $  whose behaviour under conformal rescalings \eqref{eq:gauge-mc} is as in \eqref{eq:gauge-behaviour-appropriate} and satisfies the equations
					\begin{align}\label{eq:rho-diff-eq-axi}
					\cdc{_{[C}}\ct{\rho}{_{A]B}}=0\spacef,\\
					\lied_{\vec{\chi}}\ct{\rho}{_{AB}}=-a\cdc{_{A}}\cdc{_{B}}\phi\spacef,\label{eq:rho-lie-CKVF-nonS}
					\end{align}
					in any conformal frame, where $ \phi\defeq\cdc{_M}\ct{\chi}{^M}/2 $. Furthermore, this tensor field must have a trace $ \ct{\rho}{^E_E}=aK $, is given for the metric with constant positive Gaussian curvature by $ \ct{\rho}{_{AB}}=\mc{_{AB}}aK/2 $, vanishes for the flat Euclidean metric and is invariant under transformations generated by $ \ct{\chi}{^A} $ when this is a KVF (that is, when $\phi=0$).
				\end{corollary}
				\begin{remark}
					In two dimensions the CKVF $ \ct{\chi}{^A} $ with a fixed point generates an axial symmetry locally around the fixed point (see \cite{Mars1993} or Appendix \ref{app:conformal-miscellaneous}). The existence of such vector field is ensured for $ \Sc=\mathbb{S}^2$, $ \Sc=\mathbb{S}^2\setminus\cbrkt{p_{1}}=\mathbb{R}^2$ and $ \Sc= \mathbb{S}^2\setminus\cbrkt{p_{1},p_{2}}=\mathbb{S}^1\times\mathbb{R} $ --see \cref{app:conformal-miscellaneous}.			
				\end{remark}
				\begin{remark}
						The further requirement of \cref{eq:rho-lie-CKVF-nonS} with respect to \cref{thm:rho-tensor} provides the uniqueness of $ \ct{\rho}{_{AB}} $. Note that this is a natural condition to be imposed. Actually, the validity of \eqref{eq:rho-lie-CKVF-nonS} for any CKVF would be motivated on physical arguments as well, for it makes the tensor $ \ct{\rho}{_{AB}} $ respect the symmetries of the cut. This also would fix the behaviour under finite conformal transformations to be of type  \eqref{eq:conformal-behaviour-rho}.
				\end{remark}
				\begin{proof}
							Existence is proved by noticing that $ \ct{\rho}{_{AB}}=\mc{_{AB}}aK/2 $ fulfils $ \cdc{_A}\ct{\rho}{_{BC}} =0$ in the metric with constant positive Gaussian curvature, and one can check using the gauge change \eqref{eq:gauge-behaviour-appropriate} and \cref{eq:rho-lie-CKVF-nonS} that this gives the vanishing tensor for the flat metric. Concerning uniqueness, the proof follows along the same lines of \cref{thm:colRhothm} and we also arrive at $ \cdc{_{[C}}\prn{\ctrd{1}{\rho}{_{A]B}}-\ctrd{2}{\rho}{_{A]B}}}=0 $, if two different solutions $ \ctrd{1}{\rho}{_{AB}} $ and $ \ctrd{2}{\rho}{_{AB}} $ exist. Then, choose the conformal frame such that $ \ct{\chi}{^A} $ becomes a KVF (which necessarily keeps the fixed point). To see that $ \ct{\rho}{_{AB}} $ is left invariant by $ \ct{\chi}{^A} $ in this conformal frame, one only has to set $ \phi=0 $ in \cref{eq:rho-lie-CKVF-nonS}. Now, the difference $ \ctrd{1}{\rho}{_{AB}}-\ctrd{2}{\rho}{_{AB}} $ is trace- and divergence-free, i.e., a TT-tensor which also fulfils the so called KID equations \cite{Paetz2016} for $ \ct{\chi}{^A} $ because of its invariance by this KVF and we are working in 2 dimensions. Now, a result in \cite{MarsPeon2020} states that the only solution to this problem if the KVF has fixed points --as it is the case of $ \ct{\chi}{^A} $-- is the trivial one. Hence, $ \ctrd{1}{\rho}{_{AB}}-\ctrd{2}{\rho}{_{AB}}=0 $.
							To see that uniqueness holds in any conformal frame, recall that if two solutions exist, they have to coincide in the particular frame(-family) in which $ \ct{\chi}{^A} $ is a KVF. Since the change of any two solutions to that frame is the same (given by \cref{eq:gauge-behaviour-appropriate}), the only possibility is $ \ctrd{1}{\rho}{_{AB}}=\ctrd{2}{\rho}{_{AB}}$ in any conformal frame. The proof that $ \ct{\rho}{_{AB}} $ vanishes for the flat metric will be completed below.
				\end{proof}
			By these means we can recover, in a direct manner, a non-trivial result on the sphere $ \mathbb{S}^2 $ (with any metric) ---first proven in \cite{Bourguignon1987},
				\begin{corollary}\label{thm:colRhothm}
					Let, as before, $ \prn{\Sc,\mc{_{AB}}} $ be any Riemannian manifold, topologically $ \mathbb{S}^2 $, with metric $ \mc{_{AB}} $. Then, for every conformal Killing vector field $ \ct{\chi}{^{A}} $ 
						\begin{equation}
							\int_\Sc \lied_{\vec{\chi}}K \csC{\epsilon}= 0\spacef.
						\end{equation}
				\end{corollary}
				\begin{proof}
					\Cref{eq:rho-diff-eq} is equivalent to its trace,
						\begin{equation}
						 \cdc{_C}\prn{\ct{\rho}{_A^C}-a\ct{\delta}{^C_A}K}= 0,
						\end{equation}
					which is a gauge invariant equation too. Contracting with $ \ct{\chi}{^A} $ and integrating over $ \Sc $ one obtains the desired result, noting that $ 2\cdc{_{(A}}\ct{\chi}{_{B)}}=\ctc{q}{_{AB}}\cdc{_C}\ct{\chi}{^C} $.
				\end{proof}
			From now on, we will use $ \ct{\rho}{_{AB}} $ to denote this tensor field in the case $ a=1 $.
			
			 We will later need the gauge change of the tensor $\rho_{AB}$ but using the covariant derivative $\cdcg{_{A}}$ instead of $\cdc{_{A}}$. To that end, we can use (\ref{eq:gauge-behaviour-appropriate}) with $a=1$ but applied to the conformal change $\mc{_{AB}}= \omega^{-2} \mcg{_{AB}}$, so that
			\begin{equation}
								\ct{\rho}{_{AB}}=\ctg{\rho}{_{AB}}-\omega\cdcg{_A}{\cdcg{_B}{\omega^{-1}}}+2 \omega^2 \cdcg{_A}{\omega^{-1}}\cdcg{_B}{\omega^{-1}}-\frac{\omega^2}{2}\mcg{^{CD}}\cdcg{_C}{\omega^{-1}}\cdcg{_D}{\omega^{-1}}\mcg{_{AB}}
						\end{equation}
			and expand the righthand side to get 
			\begin{equation}\label{eq:gauge-behaviour-inverse}
			\ct{\rho}{_{AB}}=\ctg{\rho}{_{AB}}+\frac{1}{\omega} \cdcg{_A} \ctc{\omega}{_B}-\frac{1}{2\omega^2} \mcg{^{CD}}\ctc{\omega}{_C}\ctc{\omega}{_D}\mcg{_{AB}}
			\end{equation}
			which is the sought relation.
			
			\subsubsection{An interpretation of $\rho_{AB}$}\label{sssec:rho-interpretation}
			$\ct{\rho}{_{AB}}$ is reminiscent of a tensor field defined by Geroch in the asymptotically flat case \cite{Geroch1977} that was key for the definition of the news tensor field. In our case, its role in the existence of news-like objects will be made clear in a later subsection. Geroch presented the tensor with little explanations, just providing its properties. $\ct{\rho}{_{AB}}$ has later been used in several papers but just referring to \cite{Geroch1977} for its significance. In what follows, we intend to give a geometrical explanation of how it comes about and the reasons behind its properties.
			
			Consider the round sphere $(\mathbb{S}^2,\mc{_{\rm round}})$ with constant Gaussian curvature $K$ as given in \cref{eq:round} of the Appendix \ref{app:conformal-miscellaneous}. In that Appendix, relation \cref{eq:gradientCKV} provides three solutions, $n^{(i)}$, of the equation
			\begin{equation}\label{eq:Hess}
			\cdc{_{A}}\cdc{_{B}}\Phi -\frac{1}{2}  \cdc{_{C}}\cdc{^{C}}\Phi \prn{\mc{_{\rm round}}}_{AB} =0
			\end{equation}
			for the unknown $\Phi$. Observe that $\cdc{^A}\Phi $ are then gradient CKVs of the round metric.  A fourth solution of the equation is obviously given by $\Phi=$ constant and these are actually the four independent solutions of \cref{eq:Hess}. The questions are: what are the corresponding solutions in any other metric on the sphere? Is there a conformally invariant version of \cref{eq:Hess}?
			
			To answer them, start with any two conformally related metrics as in \eqref{eq:gauge-mc} and compute the lefthand side of \eqref{eq:Hess} for both of them assuming that any solution $\Phi$ has a gauge behaviour of type $\tilde{\Phi} = h(\omega) \Phi $ for some function $h$ of the conformal factor $\omega$. The relation between them involves several terms of type $\cdc{_A} h - h \ctc{\omega}{_A} /\omega$ which prevent the existence of a well-behaved relationship between the two. Thus, one is led to fix $h= \omega$, which remove those terms and where a multiplicative constant has been set to 1, without loss of generality, by a simple redefinition of $\Phi$. The remaining relation reads then
			\begin{eqnarray*}
			\cdcg{_A} \cdcg{_B} \tilde{\Phi}-\frac{1}{2} \mcg{_{AB}} \tilde\Delta \tilde\Phi -\omega\left(
			\cdc{_A} \cdc{_B} \Phi -\frac{1}{2} \mc{_{AB}} \Delta\Phi \right)\\
			= \Phi \left( \cdc{_A} \cdc{_B} \omega -\frac{2}{\omega} \ctc{\omega}{_A}\ctc{\omega}{_B} +\frac{1}{\omega} \mc{^{CD}} \ctc{\omega}{_C} \ctc{\omega}{_D} \mc{_{AB}} -\frac{1}{2} \Delta \omega \mc{_{AB}} \right)
			\end{eqnarray*}
			and introducing here the conformal behaviour of $K$ given in \eqref{eq:Kprimed} of the Appendix, one gets
			\begin{eqnarray}
			\cdcg{_A} \cdcg{_B} \tilde{\Phi}-\frac{1}{2} \mcg{_{AB}} \tilde\Delta \tilde\Phi -\frac{\tilde{K}}{2} \mcg{_{AB}} \tilde\Phi-\omega\left(\cdc{_A} \cdc{_B} \Phi -\frac{1}{2} \mc{_{AB}} \Delta\Phi -\frac{K}{2} \mc{_{AB}} \Phi \right)\nonumber \\
			= \Phi \left( \cdc{_A} \cdc{_B} \omega -\frac{2}{\omega} \ctc{\omega}{_A}\ctc{\omega}{_B} +\frac{1}{2\omega} \mc{^{CD}} \ctc{\omega}{_C} \ctc{\omega}{_D} \mc{_{AB}} \right)\label{eq:step}
			\end{eqnarray}
			If we want this expression to lead to a well-behaved one under conformal rescalings \eqref{eq:gauge-mc} the second line must be the difference between a symmetric tensor and its tilded version except for a factor $\omega$, that is to say
			\begin{equation}\label{eq:rho-rho'}
			\ct{\rho}{_{AB}} -\ctg{\rho}{_{AB}} = \frac{1}{\omega} \cdc{_A} \cdc{_B} \omega -\frac{2}{\omega^2} \ctc{\omega}{_A}\ctc{\omega}{_B} +\frac{1}{2\omega^2} \mc{^{CD}} \ctc{\omega}{_C} \ctc{\omega}{_D} \mc{_{AB}}
			\end{equation} 
			so that \eqref{eq:step} becomes finally
			\begin{equation*}
			\cdcg{_A} \cdcg{_B} \tilde{\Phi}-\frac{1}{2} \mcg{_{AB}} \tilde\Delta \tilde\Phi +\left( \ctg{\rho}{_{AB}} -\frac{\tilde{K}}{2} \mcg{_{AB}}\right) \tilde\Phi=\omega\left[\cdc{_A} \cdc{_B} \Phi -\frac{1}{2} \mc{_{AB}} \Delta\Phi +\left( \ct{\rho}{_{AB}} -\frac{K}{2} \mc{_{AB}}\right) \Phi \right] .
			\end{equation*}
			This is what we were looking for: in simpler words, this states that if $\Phi$ is any solution of the equation
			\begin{equation}\label{eq:Hess-gen}
			\cdc{_A} \cdc{_B} \Phi -\frac{1}{2} \mc{_{AB}} \Delta\Phi +\left( \ct{\rho}{_{AB}} -\frac{K}{2} \mc{_{AB}}\right)\Phi =0
			\end{equation} 
			then $\tilde{\Phi} =\omega \Phi$ is the corresponding solution in any other conformal gauge defined by $\omega$. The trace of \eqref{eq:Hess-gen} leads to 
			\begin{equation}\label{eq:traceOFrho}
			\mc{^{AB}} \ct{\rho}{_{AB}} = K
			\end{equation}
			which, taking \cref{eq:rho-rho'} into account also holds in any gauge. Finally, if we wish to recover \cref{eq:Hess} in the round gauge, \cref{eq:Hess-gen} requires that in that gauge $\ct{\rho}{_{AB}} =(K/2) \mc{_{AB}}$ so that $\cdc{_C}\ct{\rho}{_{AB}}=0$ holds in the round gauge, which obviously implies
			\begin{equation}\label{eq:Derivadarho} 
			\cdc{_{[C}} \ct{\rho}{_{A]B}} =0.
			\end{equation}
			But this formula holds in any gauge due to \eqref{eq:rho-rho'} and \eqref{eq:traceOFrho}. Properties \eqref{eq:rho-rho'} and \eqref{eq:Derivadarho} uniquely determine the tensor $\ct{\rho}{_{AB}}$ according to corollary \ref{thm:rho-tensor}.

			\subsubsection{The tensor $ \rho $ for axially symmetric 2-dimensional cuts}\label{sssec:rho-axisymm}
				One can give the explicit form of $ \ct{\rho}{_{AB}} $ for any 2-dimensional metric with axial symmetry $ \mc{_{AB}} $. Lets choose coordinates $ x^A=\cbrkt{p,\varphi} $ such that
					\begin{equation}\label{eq:axi-metric}
						\mc{}=F\prn{p}\df{p}^2+G\prn{p}\df{\varphi}^2
					\end{equation}
				and $ \dpart{\varphi} $ is the axial KVF. This metric is locally conformal to the round metric with constant positive Gaussian curvature $ \cs{K} $
					\begin{equation}\label{eq:qround}
						\ctc{q}{}=\frac{1}{\cs{K}}\prn{\df{\theta}^2+\sin^2\theta\df{\varphi}^2}\spacef.
					\end{equation}
				Assume that the conformal factor $ \omega $ relating both metrics ($ \mc{_{AB}}=\omega^2\ctc{q}{_{AB}} $) respects the axial symmetry. Then,
					\begin{equation}
						G\prn{p}=\frac{\omega^2}{K}\sin^2\theta,\quad F\prn{p}=\frac{\omega^2}{K}{\theta'}^2,\quad\theta'=\frac{\df{\theta}}{\df{p}},
					\end{equation}
				yields
					\begin{equation}\label{eq:p-thetha}
						\tan\frac{\theta}{2}=Ce^{\epsilon\int\sqrt{F/G}\df{p}},\quad\sin\theta=\frac{2Ce^{\epsilon\int\sqrt{F/G}\df{p}}}{1+C^2e^{2\epsilon\int\sqrt{F/G}\df{p}}},\quad\cos \theta=\frac{1-\tan^2\prn{\theta/2}}{1+\tan^2\prn{\theta/2}},
					\end{equation}
				where $ C $ has to be fixed (making the value of $ p $ at the fixed point of $ \dpart{\phi} $ correspond to $ \theta=0 $ or $ \theta=\pi $) and $ \epsilon^2=1 $. With this, the conformal factor reads
					\begin{equation}\label{eq:omega-axial}
						\omega=\sqrt{K}\frac{\sqrt{G}}{\sin\theta}.
					\end{equation}
				Its first and second derivatives (using the connection of $ \mc{_{AB}} $) read
					\begin{align}
						\frac{1}{\omega}\omega_{A}&=\delta_A^p\psi,\quad\text{with}\quad\psi\defeq\prn{\frac{G'}{2G}-\epsilon\sqrt{\frac{F}{G}}\cos\theta(p)},\quad G'\defeq \frac{\df{G}}{\df{p}},\\
						\frac{1}{\omega}\cdc{_{A}}\omega_{B}\df{x}^A\df{x}^B&=\prn{\psi'-\frac{F'}{2F}\psi+\psi^2}\df{p}^2+\frac{G'}{2F}\psi\df{\varphi}^2,\quad\text{with}\quad F'\defeq\frac{\df{F}}{\df{p}},\quad\psi'\defeq\frac{\df{\psi}}{\df{p}},
					\end{align}	
				Setting  $ \ctc{\rho}{_{AB}}=\frac{1}{2}\cs{K}\ctc{q}{_{AB}} $ and using the inverse conformal behaviour 
				\cref{eq:gauge-behaviour-inverse} one gets 
					\begin{equation}
						\ct{\rho}{_{AB}}=\frac{1}{2\omega^2}\prn{K+\frac{\omega^2}{F}\psi^2}\mc{_{AB}}-\frac{1}{\omega}\cdc{_{A}}\omega_{B}\spacef.
					\end{equation}
				Now, inserting \cref{eq:omega-axial} in this last expressions gives the explicit form of $ \ct{\rho}{_{AB}} $ for any axially symmetric $ \mc{_{AB}} $ \eqref{eq:axi-metric}:
					\begin{equation}\label{eq:rho-axial}
						\ct{\rho}{_{AB}}\df{x}^A\df{x}^B=\prn{\frac{F}{2G}\sin^ 2\theta-\psi'+\frac{F'}{2F}\psi-\frac{1}{2}\psi^2}\df{p}^2+\prn{\frac{1}{2}\sin^2\theta+\frac{\psi}{2F}\prn{G\psi-G'}}\df{\varphi}^2,
					\end{equation}
				where $ \sin^2\theta $ must be understood as the function of $ p $ given in \cref{eq:p-thetha}.
				
			We can apply the formula above to compute $\ct{\rho}{_{AB}}$ for the flat Euclidean metric written in polar coordinates. This is simply \cref{eq:axi-metric} with $F(p)=1$ and $G(p)=p^2$, while the conformal factor can be extracted from formulas \cref{eq:q-round} and \cref{eq:round} in the Appendix: $\omega = 1+(K/4) p^2$. Application of \cref{eq:rho-axial}, where one must fix $\epsilon=-1$ and $C=2/\sqrt{K}$, then leads readily to
			$$
			\ct{\rho}{_{AB}}|_{flat} =0.
			$$
			This finishes the proof of \cref{thm:rho-tensor-noncompact}. 
			
		\subsection{General approach to equations for gauge-invariant traceless symmetric tensor fields on $ \scri $}\label{ssec:general-approach}
			We present a way of constructing equations for tensor fields fulfilling \cref{it:propertyNewsGaugeInvariant,it:propertyNewsSymmetric,it:propertyNewsTraceless,it:propertyNewsTwoDim} on page \pageref{it:propertyNewsTwoDim}. To that end, we take as starting point  \cref{eq:magneticSchouten} and contract it with $ \ct{r}{^a}\ct{E}{^b_B} $ to obtain
			\begin{equation}\label{eq:CA-schouten}
			N\ctc{C}{^A}= \ctc{\epsilon}{^{CD}}\prn{\cdc{_{[C}}\ctc{S}{_{D]}^A}+\ctc{\kappa}{_{[C}^A}\ctc{S}{_{D]}}}\spacef.
			\end{equation}	
			By \cref{eq:SAkappa}, the right-hand side of \cref{eq:CA-schouten} can be rearranged as
				\begin{equation}\label{eq:the-rhs}
				\ctc{\epsilon}{^{CD}}\prn{\cdc{_{[C}}\ctc{U}{_{D]}^A}-\ct{T}{_{CD}^A}}\spacef,					
				\end{equation}
			with 
				\begin{align}
				\ct{U}{_{AB}}&\defeq \ctc{S}{_{AB}}+\frac{1}{2}\csC{\kappa}\ctc{\Sigma}{_{AB}}+\prn{\frac{1}{8}\csC{\kappa}^2-\frac{1}{4}\csC{\Sigma}^2}\mc{_{AB}}=\ct{U}{_{(AB)}}\spacef,\label{eq:UABdef}\\
				\ct{T}{_{AB}^C}&\defeq  \frac{1}{2}\brkt{\ct{\delta}{^C_{[A}}\cdc{_{B]}}\csC{\Sigma}^2-\cdc{^C}\prn{\ctc{\Sigma}{^D_{[B}}}\ctc{\Sigma}{_{A]D}}}=\ct{T}{_{[AB]}^C}\spacef,\label{defTB}
				\end{align}	
			and we can write \cref{eq:CA-schouten} in the equivalent form
				\begin{equation}\label{eq:CA-U}
					\frac{1}{2}N\ctc{\epsilon}{_{CA}}\ctc{C}{_B}=\cdc{_{[C}}\ct{U}{_{A]B}}-\ct{T}{_{CAB}}\spacef.
				\end{equation}
			Observe that since $ \ct{T}{_{AB}^C} $ is antisymmetric on its two covariant indices, by the identity  \eqref{eq:dimensional-identity} it is completely determined by its trace
				\begin{equation}\label{eq:TA}
					\ct{T}{_{B}}\defeq \ct{T}{_{CB}^C}=\frac{1}{4}\brkt{\cdc{_B}\csC{\Sigma}^2-\ctc{\Sigma}{^{CD}}\cdc{_C}\prn{\ctc{\Sigma}{_{DB}}}+\ctc{\Sigma}{_B^D}\cdc{_C}\prn{\ctc{\Sigma}{_D^C}}}\spacef.
				\end{equation}
			The point of this decomposition is that $ \ct{T}{_{ABC}} $	is gauge invariant
				\begin{equation}
				\ctg{T}{_{ABC}}= \ct{T}{_{ABC}}\quad
				\end{equation}
			and that $ \ct{U}{_{AB}} $ transforms as 
				\begin{equation}\label{eq:Ugauge-transformation}
				\ctg{U}{_{AB}}= \ct{U}{_{AB}} +\frac{2}{\omega^2}\ctc{\omega}{_A}\ctc{\omega}{_B}-\frac{1}{\omega}\cdc{_A}\ctc{\omega}{_B}-\frac{1}{2\omega^2}\ctc{\omega}{_P}\ctc{\omega}{^P}\mc{_{AB}}\spacef,
				\end{equation}
			where we have used the formulae of \cref{app:gauge-transformations}. In addition to that, taking \cref{eq:traceSchoutenCut} into account, the trace of $ \ct{U}{_{AB}} $ reads
				\begin{equation}\label{eq:Utrace}
					\ct{U}{^E_E}=\frac{\csC{R}}{2}=K\spacef.
				\end{equation}
			Remarkably, these are the gauge behaviour and the trace of the tensor required to prove the following:
				\begin{lemma}\label{thm:dU-invariant}
					The tensor field $ \cdc{_{[C}}\ct{U}{_{D]A}} $ is gauge invariant,
						\begin{equation}
							\cdcg{_{[C}}\ctg{U}{_{D]A}}= \cdc{_{[C}}\ct{U}{_{D]A}}\spacef.
						\end{equation}
				\end{lemma}
				\begin{proof}
					It follows from \cref{thm:general-gauge-trace}, noting the gauge behaviour of $ \ct{U}{_{AB}} $ --\cref{eq:Ugauge-transformation}-- and its trace --\cref{eq:Utrace}. 
				\end{proof}
				\begin{remark}
					In particular, the combination $ \cdc{_{[C}}\ct{U}{_{D]A}}+\ct{T}{_{DCA}} $ is gauge invariant too, as follows from the gauge invariance of $ \ct{T}{_{ABC}} $.    This can also be proven looking at \cref{eq:CA-schouten} and noting the gauge behaviour of $ \ctc{C}{^A} $ (see \cref{app:gauge-transformations}).
				\end{remark}
			We write now an important result:

\begin{prop}[A first component of `news']\label{thm:onepiece-news}
				 Let $\prn{\Sc,\mc{_{AB}}}$ be a 2-dimensional Riemannian manifold endowed with the metric $ \mc{_{AB}} $. If $ \mc{_{AB}} $ has a CKVF with a fixed point, the tensor field
					\begin{equation}
					\ct{V}{_{AB}}\defeq \ct{U}{_{AB}}-\ct{\rho}{_{AB}}\spacef,
					\end{equation}
				is symmetric, traceless, gauge invariant and satisfies the gauge-invariant equation
					\begin{equation}\label{eq:diffUAB}
					\cdc{_{[A}}\ct{U}{_{B]C}}=\cdc{_{[A}}\ct{V}{_{B]C}}\spacef,
					\end{equation}
					where $ \ct{\rho}{_{AB}} $ is the tensor field of  \cref{thm:rho-tensor-noncompact} (for $ a=1 $). Besides, $ \ct{V}{_{AB}} $ is unique with these properties.
				\end{prop}
				\begin{proof}
					The tensor field $ \ct{V}{_{AB}} $ is symmetric, traceless and gauge invariant as a consequence of \Cref{eq:UABdef,eq:Ugauge-transformation,eq:Utrace} and \cref{thm:rho-tensor-noncompact}. The uniqueness of $ \ct{V}{_{AB}} $ follows from \cref{thm:rho-tensor-noncompact} too and \Cref{eq:diffUAB}.
				\end{proof}
				\begin{remark}\label{rmk:fixed-points}
					The existence of a CKVF with a fixed point is warranted for the topologies $ \Sc=\mathbb{S}^2 $, $ \Sc=\mathbb{R}\times\mathbb{S}^1 $ and $ \Sc=\mathbb{R}^2 $ --see \cref{app:conformal-miscellaneous}.
				\end{remark}
			In passing, notice the identity that follows taking the trace of \cref{eq:CA-U} and applying \cref{thm:onepiece-news}:
			\begin{equation}
			\int_\Sc \ct{\chi}{^B}\prn{N\ctc{\epsilon}{_{BE}}\ctc{C}{^E}-2\ct{T}{_{B}} }=0\quad\forall \text{ CKVF } \ct{\chi}{^B}\text{ on a topological-}\mathbb{S}^2\quad\Sc.
			\end{equation}
			 We consider $ \ct{V}{_{AB}} $ an essential component of any news-like tensor
			as will be justified in \cref{ssec:conditions-news}.
			In general, one has
				\begin{prop}
					If the equation
						\begin{equation}\label{eq:unknown-equation}
							\cdc{_{[C}}\ct{Z}{_{A]B}}=\ct{Y}{_{CAB}}
						\end{equation}
					for a given gauge-invariant tensor field $ \ct{Y}{_{CAB}}=\ct{Y}{_{[CA]B}} $ has a solution for $ \ct{Z}{_{AB}}=\ct{Z}{_{(AB)}} $ whose gauge behaviour is given by \eqref{eq:gauge-behaviour-appropriate} with $ a=1 $, then this solution is unique and given by
						\begin{equation}\label{eq:Z-U-M}
							\ct{Z}{_{AB}}=\ct{U}{_{AB}}+\ct{X}{_{AB}}
						\end{equation}
					where $ \ct{X}{_{AB}} $	is the unique traceless gauge invariant symmetric tensor field solution of
						\begin{equation}\label{eq:aux34}
							\cdc{_{[C}}\ct{X}{_{A]B}}=\ct{Y}{_{CAB}}-\frac{1}{2}N\ctc{\epsilon}{_{CA}}\ctc{C}{_B}-\ct{T}{_{CAB}}\spacef.
						\end{equation}
				\end{prop}
				\begin{proof}
					$ \ct{X}{_{AB}} $ must be symmetric and gauge invariant by \cref{eq:Z-U-M}, noting that $ \ct{U}{_{AB}} $ and $ \ct{Z}{_{AB}} $ are symmetric tensor fields with the same gauge behaviour \eqref{eq:gauge-behaviour-appropriate}. To see that $ \ct{X}{_{AB}} $ is traceless, use \cref{thm:general-gauge-trace} which, taking into account that $ \ct{Y}{_{ABC}} $ is assumed to be gauge invariant and that  $ \ct{Z}{_{AB}} $ is symmetric with gauge behaviour \eqref{eq:gauge-behaviour-appropriate}, implies that $ \ct{Z}{^C_{C}}=K=\ct{U}{^C_C} $, hence $ \ct{X}{^C_{C}}=0 $. For the second part, note that by \cref{eq:CA-U}
						\begin{equation}
							\cdc{_{[C}}\ct{Z}{_{A]B}}=\cdc{_{[C}}\ct{X}{_{A]B}}+\frac{1}{2}N\ctc{\epsilon}{_{CA}}\ctc{C}{_B}+\ct{T}{_{CAB}}\spacef,
						\end{equation}
					from where \cref{eq:aux34} follows immediately. If two different solutions $ \ctrd{1}{Z}{_{AB}} $ and $ \ctrd{2}{Z}{_{AB}} $ exist, one has $ \cdc{_{[C}}\prn{\ctrd{1}{Z}{_{A]B}}-\ctrd{2}{Z}{_{A]B}}}=0 $. Then, because their difference is a traceless Codazzi tensor on $ \mathbb{S}^2 $, the only possibility \cite{Liu1998} is $\ctrd{1}{Z}{_{AB}}-\ctrd{2}{Z}{_{AB}}=0 $.
				\end{proof}
				
			\begin{remark}
					 The $\mathbb{S}^2$ topology can be dropped from the assumptions if $\prn{\Sc,\mc{_{AB}}}$ is a 2-dimensional Riemannian manifold such that there exists a CKVF $ \ct{\chi}{^A} $ with a fixed point and $ \ct{Z}{_{AB}} $ fulfils the KID equations \cite{Paetz2016}. To prove this, one applies the result of \cite{MarsPeon2020} which was used in the proof of \cref{thm:rho-tensor-noncompact} to show that $\cdc{_{[C}}\prn{\ctrd{1}{Z}{_{A]B}}-\ctrd{2}{Z}{_{A]B}}}=0 $ implies $ \ctrd{1}{Z}{_{A]B}}-\ctrd{2}{Z}{_{A]B}}=0 $ .
				\end{remark}
				
				\begin{remark}\label{rmk:generalNews}
					By \cref{thm:onepiece-news} (or with different appropriate assumptions, \cref{thm:rho-tensor-noncompact}) and \cref{eq:Z-U-M}, the general  \cref{eq:unknown-equation} is written as
						\begin{equation}\label{eq:unknownNewsdiff}
							\cdc{_{[C}}\ct{N}{_{A]B}}=\ct{Y}{_{CAB}}\spacef,
						\end{equation}
					where we have defined the gauge-invariant traceless symmetric tensor field
						\begin{equation}\label{eq:general-news}
							\ct{N}{_{AB}}\defeq \ct{V}{_{AB}}+ \ct{X}{_{AB}}\spacef.
						\end{equation} 
					\Cref{eq:unknownNewsdiff} is equivalent to its trace,
						\begin{equation}\label{eq:unknownNewsDiv}
							\cdc{_C}\ct{N}{_A^C}=2\ct{Y}{_{A}},
						\end{equation}
					with
						\begin{equation}
							\ct{Y}{_{C}}\defeq \ct{Y}{_{CA}^A}\spacef.
						\end{equation}
				\end{remark}
				\begin{remark}\label{rmk:existence-solutions-NAB}
					For $ \Sc $ topologically $ \mathbb{S}^2 $, solutions to \cref{eq:unknownNewsdiff} exist if and only if 
						\begin{equation}
						 	\int_\Sc \ct{\chi}{^B} \ct{Y}{_B} \csC{\epsilon}= 0\spacef,\label{eq:conditionY}
						\end{equation}
					for any CKVF on  $( \Sc ,\mc{_{AB}})$ (see \cite{Besse1987}, appendix H). In general, one can always prescribe $ \ct{Y}{_{CAB}} $ (equivalently, $ \ct{Y}{_A} $) such that solutions exist, one plausible option is
						\begin{equation}\label{eq:plausible-Y}
						\ct{Y}{_B}\defeq\Delta\cs{y}\,\cdc{_B}\cs{y},\quad\forall y\in C^2\prn{\Sc}
						\end{equation}
					which follows from a result proven in \cite{Bourguignon1987}: if assumptions in \cref{thm:colRhothm} hold, then
						\begin{equation}
						\int_{\Sc}\Delta y \lied_\chi y \csC{\epsilon}=0,\quad\forall y\in C^2\prn{\Sc}
						\end{equation}
					and this statement is conformally invariant.
				\end{remark}
				 $\ct{N}{_{AB}}$ as defined in \eqref{eq:general-news} is our candidate for the news-like object we are seeking. It has two `components', one given by $\ct{V}{_{AB}}$ which is fully determined on each cut (see next parapgraph), and another component, yet to be uncovered, which depends on the choice of $\ct{Y}{_A}$.
				 Note that $ \ct{N}{_{AB}} $ fulfils \cref{it:propertyNewsGaugeInvariant,it:propertyNewsSymmetric,it:propertyNewsTraceless,it:propertyNewsTwoDim} on page \pageref{it:propertyNewsSymmetric}. According to \cref{rmk:existence-solutions-NAB},  prescriptions of $ \ct{Y}{_{ABC}} $ are always possible such that these kind of tensor fields exist as solutions of \cref{eq:unknownNewsdiff}. Nevertheless, the great difficulty  one has to face stems from the need to \emph{fix $ \ct{Y}{_{ABC}} $ such that $ \ct{N}{_{AB}} $ makes a reasonable news tensor} that satisfies all the requirements on page \pageref{it:propertyNewsSymmetric}, including \cref{it:propertyNewsCandF,it:propertyNewsZ} too.
				 
				 At this stage, there is no reason to ensure that there exists some function $ y $ such that the choice \eqref{eq:plausible-Y} meets all these points in general. Observe,  in this sense, that \cref{eq:CA-U} in terms of $ \ct{V}{_{AB}} $ reads
	 	 		\begin{equation}\label{eq:CA-Vi}
			 	 	\frac{1}{2}N\ctc{\epsilon}{_{CA}}\ctc{C}{_B}=\cdc{_{[C}}\ct{V}{_{A]B}}-\ct{T}{_{CAB}}\quad
		 	 	\end{equation}
		 	 and, therefore, $ \ct{V}{_{AB}} $ is completely determined by 
			  the geometry of the cut $\Sc\subset \scri $, everything intrinsic to $(\scri,h_{ab})$. Hence, in order to achieve a $ \ct{N}{_{AB}} $ satisfying \cref{it:propertyNewsCandF}  too 
			 the choice of $ \ct{Y}{_{ABC}} $ {\em has to incorporate the dependence on $ \ct{D}{_{ab}} $}.  And, in addition, it has to vanish in accordance with some meaningful superenergy quantity. As mentioned earlier, there are several options for this quantity, such as the asymptotic super-Poynting, or radiant supermomenta. The problem of in- and out-going radiative sectors seems to make it difficult to find a second component of $\ct{N}{_{AB}}$ associated to the former, as it contains information from both sectors. On the other hand, the difficulty with supermomenta is that there are many choices for them. Next subsection deals with these issues by proposing a particular fixing of $ \ct{Y}{_{ABC}} $.
			 
		\subsection{On the second component of $\ct{N}{_{AB}}$}\label{ssec:conditions-news}
			Now we take as a guide \cref{it:propertyNewsZ} on page \pageref{it:propertyNewsZ}; throughout this subsection, in particular we choose to focus on the vanishing of $ \cspm{\Z} $. In the light of \cref{eq:Zmdef,eq:Zpdef}, it is clear that it is convenient to work with the quantities $ \ctcupm{C}{_A} $. Observe that \cref{eq:CA-Vi} can be rewritten using  \cref{it:nullDecompProp14,it:nullDecompProp15}  on page \pageref{it:nullDecompProp14} as
				\begin{align}
				N\ctc{C}{^A} &= \ctc{\epsilon}{^{CD}}\prn{\cdc{_{[C}}\ct{V}{_{D]}^A}+\ct{T}{_{DC}^A}}\spacef,\label{eq:CA-V}\\
				2N\ctcp{C}{^A}+N\ctc{\epsilon}{^{AC}}\ctc{D}{_C}&= \ctc{\epsilon}{^{CD}}\prn{\cdc{_{[C}}\ct{V}{_{D]}^A}+\ct{T}{_{DC}^A}}\spacef, \label{eq:CApV}\\
				2N\ctcm{C}{^A}-N\ctc{\epsilon}{^{AC}}\ctc{D}{_C}&=\ctc{\epsilon}{^{CD}}\prn{\cdc{_{[C}}\ct{V}{_{D]}^A}+\ct{T}{_{DC}^A}}\spacef, \label{eq:CAmV}
				\end{align}
			or equivalently
				\begin{align}
				N\ctc{\epsilon}{_{BE}}\ctc{C}{^E}&= -\cdc{_{E}}\ct{V}{_{B}^E}+2\ct{T}{_{B}} \label{eq:CASchoutenDualV}\spacef,\\
				2N\ctc{\epsilon}{_{BE}}\ctcp{C}{^E}-N\ctc{D}{_B}&= -\cdc{_{E}}\ct{V}{_{B}^E}+2\ct{T}{_{B}} \label{eq:CApSchoutenDualV}\spacef,\\
				2N\ctc{\epsilon}{_{BE}}\ctcm{C}{^E}+N\ctc{D}{_B}&= -\cdc{_{E}}\ct{V}{_{B}^E}+2\ct{T}{_{B}} \label{eq:CAmSchoutenDualV}\spacef.
				\end{align}	
			Note that \cref{eq:CApSchoutenDualV,eq:CAmSchoutenDualV,eq:CApV,eq:CAmV} are nothing more than \cref{eq:CASchoutenDualV,eq:CA-V} expressed in terms of $ \ctcupm{C}{_A} $. It is useful to have them at hand, though. 	\\
		
			One approach is to look for the necessary and sufficient conditions such that 
				\begin{align}
					-2N\ctc{\epsilon}{_{B}^E}\ctp{C}{_E}&=\cdc{_C}\ctp{n}{_B^C} \label{eq:CApNewsp}\spacef,\\
					-2N\ctc{\epsilon}{_{B}^E}\ctm{C}{_E}&=\cdc{_C}\ctm{n}{_B^C} \label{eq:CAmNewsm}\spacef,
				\end{align}
			for $ \ctpm{n}{_{AB}} $ symmetric traceless gauge invariant tensor fields on $ \Sc $. These are the particular versions of the general $\ct{N}{_{AB}}$ for the choices \eqref{eq:CApNewsp} and \eqref{eq:CAmNewsm}, as we prefer to keep the generic name $\ct{N}{_{AB}}$ for the general method. The left-hand side of these  equations correspond to two different --compatible-- choices of $ \ct{Y}{_{A}} $ in \cref{eq:unknownNewsDiv}, respectively. Hence, we define 
					\begin{equation}\label{eq:Y+-}
					\ctp{Y}{_B}\defeq N\ctc{\epsilon}{_{B}^E}\ctp{C}{_E},\quad
					\ctm{Y}{_B}\defeq  N\ctc{\epsilon}{_{B}^E}\ctm{C}{_E}.
					\end{equation}
			 Let us emphasise once more  that $ \ct{V}{_{AB}} $ fulfils \cref{it:propertyNewsGaugeInvariant,it:propertyNewsSymmetric,it:propertyNewsTraceless,it:propertyNewsTwoDim} on page \pageref{it:propertyNewsGaugeInvariant}. As explained in the previous section, it does not satisfy \cref{it:propertyNewsCandF}  
			 because it  carries no information about $ \ct{D}{_{ab}} $. Thus, intuitively one would expect \emph{the second component of $ \ctpm{n}{_{AB}} $}, $ \ctpm{X}{_{AB}} $, coming from an equation for $ \ct{D}{_{ab}} $, such that the generic expression \eqref{eq:general-news} becomes now
				\begin{align}
				\ctp{n}{_{AB}}\defeq \ct{V}{_{AB}}+\ctp{X}{_{AB}}\spacef,\label{eq:nABplus}\\
				\ctm{n}{_{AB}}\defeq \ct{V}{_{AB}}+\ctm{X}{_{AB}}\spacef,\label{eq:nABminus}
				\end{align}
			where $ \ctpm{X}{_{AB}} $ are unknown symmetric traceless gauge invariant tensor fields on $ \Sc $. 
			It can be checked by direct computation that the necessary and sufficient conditions for \cref{eq:CApNewsp,eq:CAmNewsm} to hold are
				\begin{align}
					-\frac{1}{2}N\ctc{D}{_B}&= \ct{T}{_B}+\frac{1}{2}\cdc{_C}\ctp{X}{_B^C}\spacef,\label{eq:conditionPlus}\\
					\frac{1}{2}N\ctc{D}{_B}&= \ct{T}{_B}+\frac{1}{2}\cdc{_C}\ctm{X}{_B^C}\spacef,\label{eq:conditionMinus}
				\end{align}
			which are satisfied if and only if for any CKVF $ \ct{\chi}{^B} $ on $ \Sc $ 
				\begin{align}
				\int_\Sc \ct{\chi}{^B} \ctc{\epsilon}{_{B}^E}\ctp{C}{_E} \csC{\epsilon}= 0\spacef,\label{eq:conditionPlusChi}\\
				\int_\Sc \ct{\chi}{^B}\ctc{\epsilon}{_{B}^E}\ctm{C}{_E} \csC{\epsilon}= 0\spacef,\label{eq:conditionMinusChi}
				\end{align}
			which of course fits the general construction --see \cref{rmk:existence-solutions-NAB}.
			Summarising, if  \cref{eq:conditionMinus,eq:conditionPlus}  hold,  \cref{eq:CApNewsp,eq:CAmNewsm} are satisfied for $ \ctpm{n}{_{AB}} $ as in \cref{eq:nABminus,eq:nABplus} and their square produces the following formulae:
				\begin{align}
				N^2\csp{\Z}&= \cdc{_C}\prn{\ctp{n}{_B^C}}\cdc{_D}\prn{\ctp{n}{^{BD}}}\spacef,\\
				N^2\csm{\Z}&= \cdc{_C}\prn{\ctm{n}{_B^C}}\cdc{_D}\prn{\ctm{n}{^{BD}}}\spacef.
				\end{align}				
			Let us remark that the tensor fields $ \ctpm{X}{_{AB}} $ satisfying \cref{eq:conditionMinus,eq:conditionPlus} do not necessarily exist in general. They are the two \emph{second components} of $ \ctpm{n}{_{AB}} $ according to the next result:
				\begin{prop}[Radiant news]\label{thm:theTwoNewsProp}
					If the condition of \cref{eq:conditionPlus} (\cref{eq:conditionMinus}) holds on a cut $ \Sc $ with $ \mathbb{S}^2 $-topology, then
					\begin{align}
					\ctp{n}{_{AB}}= 0 \iff \csp{\Z}= 0\spacef.\quad\\
					\prn{\ctm{n}{_{AB}}= 0 \iff \csm{\Z}= 0}\spacef. 	
					\end{align}
					Hence, $ \ctp{n}{_{AB}} $ ($ \ctm{n}{_{AB}} $) 
					fulfills \cref{it:propertyNewsCandF,it:propertyNewsGaugeInvariant,it:propertyNewsSymmetric,it:propertyNewsTraceless,it:propertyNewsTwoDim,it:propertyNewsZ} on page \pageref{it:propertyNewsCandF}. Therefore, $ \ctp{n}{_{AB}} $ ($ \ctm{n}{_{AB}} $) can be seen as a news-like tensor for $ \ctps{\Q}{^a} $ ($ \ctms{\Q}{^k}  $) and we call them {\em radiant news}.
					\begin{proof}
						 \Cref{it:propertyNewsTwoDim,it:propertyNewsSymmetric,it:propertyNewsTraceless,it:propertyNewsGaugeInvariant} are fulfilled by the definition of $ \ctp{n}{_{AB}} $ ($ \ctm{n}{_{AB}} $), see \cref{eq:nABplus} (\cref{eq:nABminus}) where $ \ct{V}{_{AB}} $ is the piece of news of \cref{thm:onepiece-news}. From \cref{eq:conditionPlus,eq:CApV} (\cref{eq:conditionMinus,eq:CAmV}), \cref{it:propertyNewsCandF} is fulfilled, as well. Since $ \ctp{n}{_{AB}}$ ($ \ctm{n}{_{AB}} $) is a symmetric traceless tensor field on the sphere, $ \ctp{n}{_{AB}}= 0\iff \cdc{_C}\ctp{n}{_B^C}= 0 $ ($ \ctm{n}{_{AB}}= 0\iff \cdc{_C}\ctm{n}{_B^C}= 0 $), and that, by \cref{eq:CApNewsp} (\cref{eq:CAmNewsm}), this holds if and only if $ \ctcp{C}{_A}= 0 $ ($ \ctcm{C}{_A}= 0$) ---which by \cref{eq:Zpdef} (\cref{eq:Zmdef}) and \cref{it:noQnoWnoZ} on page \pageref{it:noQnoWnoZ}, holds if and only if $ \csp{\Z}=0 $ ($ \csm{\Z}=0 $).
					\end{proof}
				\end{prop}
						\begin{prop}[Radiant pseudo-news for non-$ \mathbb{S}^2 $ cuts]\label{thm:theTwoNewsProp-non-S2}
						Assume that the condition of \cref{eq:conditionPlus} (\cref{eq:conditionMinus}) holds on a 2-dimensional Riemannian manifold $\prn{\Sc,\mc{_{AB}}}$ whose topology is non-necessarily $\mathbb{S}^2$ and that the metric $ \mc{_{AB}} $ possesses a CKVF $ \ct{\chi}{^A}$  with a fixed point. Then, 
						\begin{align}
						\ctp{n}{_{AB}}= 0 \implies \csp{\Z}= 0\spacef.\quad\\
						\prn{\ctm{n}{_{AB}}= 0 \implies\csm{\Z}= 0}\spacef, 	
						\end{align}
						and $ \ctp{n}{_{AB}} $ ($ \ctm{n}{_{AB}} $) has all the  \cref{it:propertyNewsCandF,it:propertyNewsGaugeInvariant,it:propertyNewsSymmetric,it:propertyNewsTraceless,it:propertyNewsTwoDim} on page \pageref{it:propertyNewsCandF} but not \cref{it:propertyNewsZ}. 
						\begin{proof}
							The first part of the proof follows the same lines as in \cref{thm:theTwoNewsProp}, where now the tensor $ \ct{V}{_{AB}} $ is the one of \cref{thm:rho-tensor-noncompact}. Then, by \cref{eq:CApNewsp} (\cref{eq:CAmNewsm}), $ \ctcp{n}{_{AB}}=0\implies \ctcp{C}{_A}= 0 $ ($ \ctm{n}{_{AB}}= 0\implies \ctcm{C}{_A}= 0 $) ---which by \cref{eq:Zpdef} (\cref{eq:Zmdef}) and \cref{it:noQnoWnoZ} on page \pageref{it:noQnoWnoZ}, holds if and only if  $ \csp{\Z}=0 $ ($ \csm{\Z}=0 $). The converse is not true in general, as the topology of $ \Sc $ can be other than $ \mathbb{S}^2 $.
						\end{proof}
					\end{prop}
				
				\begin{remark}
					Given  any cut $ \Sc $ 	with $ \mathbb{S}^2 $-topology on $ \scri $, there always exists a unique (intrinsic) first component of news, $ \ct{V}{_{AB}} $, which is determined by the intrinsic geometry of $ \prn{\scri,\ms{_{ab}}} $ and the cut. The existence of the  entire news-like tensor depends on information extrinsic to $ \scri $ and, especifically in this section's approach, on $ \ctc{D}{_A} $, from where two different second components, $ \ctpm{X}{_{AB}} $, can emerge. Note that these tensor fields are extrinsic to $ \scri $, in the sense of not being determined by $\prn{\scri,\ms{_{ab}}}$. Eventually, one ends up with none, one or two different radiant news, $ \ctpm{n}{_{AB}} $, which are each one the sum of the corresponding  first and the second component --- see \cref{eq:nABplus,eq:nABminus}.
				\end{remark}
			There are simple cases in which the conditions for the existence of news, \cref{eq:conditionMinus,eq:conditionPlus} (equivalently, \cref{eq:conditionMinusChi,eq:conditionPlusChi}) are fulfilled. The following result shows this:
				\begin{lemma}\label{thm:oneNewsUmbilical}
					Consider any \emph{umbilical} cut $ \Sc $ with $ \mathbb{S}^2 $-topology on $ \scri $ such that $ \ct{r}{_a} $ defines a strong orientation on $ \Sc $, i. e., $ \csm{\Z}\eqc 0 $. Then, there always exists the radiant news given by
						\begin{equation}
							\ctp{n}{_{AB}}=2\ct{V}{_{AB}}\spacef.
						\end{equation}
				\end{lemma}
				\begin{proof}
					On the one hand, the umbilical property ($ \ctc{\Sigma}{_{AB}}=0 $) implies that $ \ct{T}{_{A}}=0 $ (see \cref{eq:TA}). On the other hand, $ \csm{\Z}=0 \iff \ctc{D}{_A}=0\iff \ctc{C}{_A}=\ctcp{C}{_A}$. These two conditions together make \cref{eq:CA-V,eq:CASchoutenDualV} read
						\begin{align}
								N\ctcp{C}{^A} &= \ctc{\epsilon}{^{CD}}\cdc{_{[C}}\ct{V}{_{D]}^A}\spacef,\\
								N\ctc{\epsilon}{_{BE}}\ctcp{C}{^E}&= -\cdc{_{E}}\ct{V}{_{B}^E}\spacef.
						\end{align}
					Thus, \cref{eq:CApNewsp} reads
						\begin{equation}\label{eq:exampleBetaLambda}
							2\cdc{_E}\ct{V}{_B^E}=\cdc{_C}\ctp{n}{_B^C}\spacef.
						\end{equation}	
				 	Then, one finds $ \cdc{_{E}}\prn{\ctcp{n}{_{B}^E}-2\ct{V}{_{B}^E}} =0$, which holds if and only if $ \ctp{n}{_{AB}}=2\ct{V}{_{AB}} $ ---the divergence of a traceless, symmetric, two-dimensional tensor field on the sphere vanishes if and only if the tensor itself vanishes.	
				\end{proof}
			In \cref{sec:additional-structure}, we will study how to extend these objects to tensor fields on $ \scri $ by introducing additional structure on $ \scri $. The relation between the radiant news and the no-radiation condition will be analysed in \cref{ssec:news-congruences}.
				\subsubsection{Possible generalisation}\label{sssec:possible-generalisation}
						There is a generalisation of the approach we have presented. Since each cut $ \Sc $ is two-dimensional
						the vanishing of $ \ctcupm{C}{_A} $  is trivially equivalent to the vanishing of any linear combination
					\begin{equation}\label{eq:lb-possCombi}
					\cspm{\beta} \ctcupm{C}{_B}+\cspm{\lambda}\ctc{\epsilon}{_{B}^E}\ctcupm{C}{_E}\spacef,
					\end{equation}
					where the coefficients  $ \cspm{\lambda} $ and $ \cspm{\beta} $ are such that they do not vanish simultaneously, i. e.,
					\begin{align}
					\cspm{\lambda}= 0 \implies \cspm{\beta}\neq 0\spacef,\\
					\cspm{\beta}= 0 \implies \cspm{\lambda}\neq 0\spacef.
					\end{align}
					In other respects, $ \cspm{\lambda} $ and $ \cspm{\beta} $ are arbitrary real functions. One can ask these coefficients to fulfil 
					\begin{align}
					-2N\prn{\csp{\beta}\ct{\delta}{^E_B}+\csp{\lambda}\ctc{\epsilon}{_{B}^E}}\ctp{C}{_E}&=\cdc{_C}\ctp{n}{_B^C} \label{eq:lb-CApNewsp}\spacef,\\
					-2N\prn{\csm{\beta}\ct{\delta}{^E_B}+\csm{\lambda}\ctc{\epsilon}{_{B}^E}}\ctm{C}{_E}&=\frac{1}{2}\cdc{_C}\ctm{n}{_B^C} \label{eq:lb-CAmNewsm}\spacef,
					\end{align}
					for $ \ctpm{n}{_{AB}} $ symmetric traceless gauge invariant tensor fields on $ \Sc $. Notice that for $ \cspm{\lambda}=1 $ and $ \cspm{\beta}=0 $ we are in the situation described above for \cref{eq:CApNewsp,eq:CAmNewsm}. Now, one has to define 
					\begin{equation}
					\ctp{Y}{_B}\defeq N\prn{\csp{\beta}\ct{\delta}{^E_B}+\csp{\lambda}\ctc{\epsilon}{_{B}^E}}\ctp{C}{_E},\quad
					\ctm{Y}{_B}\defeq  N\prn{\csm{\beta}\ct{\delta}{^E_B}+\csm{\lambda}\ctc{\epsilon}{_{B}^E}}\ctm{C}{_E}.
					\end{equation}
					Again, one expects the \emph{second components} $ \ctpm{X}{_{AB}} $  coming from an equation for $ \ctc{D}{_{ab}} $ to be part of $ \ctpm{n}{_{AB}} $, together with $ \ct{V}{_{AB}} $:
					\begin{align}
					\ctp{n}{_{AB}}\defeq \ct{V}{_{AB}}+\ctp{X}{_{AB}}\spacef,\label{eq:lb-nABplus}\\
					\ctm{n}{_{AB}}\defeq \ct{V}{_{AB}}+\ctm{X}{_{AB}}\spacef.\label{eq:lb-nABminus}
					\end{align}
					It can be checked by direct computation that the necessary and sufficient conditions on $ \cspm{\lambda} $ and $ \cspm{\beta} $ for \cref{eq:lb-CApNewsp,eq:lb-CAmNewsm} to hold are
					\begin{align}
					-\frac{1}{2}N\ctc{D}{_B}&= \ct{T}{_B}+\prn{\csp{\lambda}-1}N\ctc{\epsilon}{_{BC}}\ctcp{C}{^C}+\csp{\beta}N\ctcp{C}{_B}+\frac{1}{2}\cdc{_C}\ctp{X}{_B^C}\spacef,\label{eq:lb-conditionPlus}\\
					\frac{1}{2}N\ctc{D}{_B}&= \ct{T}{_B}+\prn{\csm{\lambda}-1}N\ctc{\epsilon}{_{BC}}\ctcm{C}{^C}+\csm{\beta}N\ctcm{C}{_B}+\frac{1}{2}\cdc{_C}\ctm{X}{_B^C}\spacef,\label{eq:lb-conditionMinus}
					\end{align}
					which are satisfied if and only if 	for any CKVF $ \ct{\chi}{^B} $ on $ \Sc $
					\begin{align}
					\int_\Sc \ct{\chi}{^B} \prn{\csp{\beta}\ct{\delta}{^E_B}+\csp{\lambda}\ctc{\epsilon}{_{B}^E}}\ctp{C}{_E}\csC{\epsilon} = 0\spacef,\label{eq:lb-conditionPlusChi}\\
					\int_\Sc \ct{\chi}{^B}\prn{\csm{\beta}\ct{\delta}{^E_B}+\csm{\lambda}\ctc{\epsilon}{_{B}^E}}\ctm{C}{_E}\csC{\epsilon} = 0\spacef,\label{eq:lb-conditionMinusChi}
					\end{align}
				which again meets the general construction of \cref{ssec:general-approach} --in particular, see \cref{rmk:existence-solutions-NAB}. Notice that equivalent expressions to  \cref{eq:lb-conditionMinus,eq:lb-conditionPlus}  in terms of $ \ctc{C}{_A} $ are
					\begin{align}
					-N\frac{1}{2}\prn{\csp{\lambda}\ct{\delta}{^D_B}-\csp{\beta}\ctc{\epsilon}{_B^D}}\ctc{D}{_D}&= \ct{T}{_B}+N\frac{1}{2}\brkt{\csp{\beta}\ct{\delta}{^D_B}+\prn{\csp{\lambda}-1}\ctc{\epsilon}{_B^D}}\ctc{C}{_D}+\frac{1}{2}\cdc{_C}\ctp{X}{_B^C}\spacef,\\
					N\frac{1}{2}\prn{\csm{\lambda}\ct{\delta}{^D_B}-\csm{\beta}\ctc{\epsilon}{_B^D}}\ctc{D}{_D}&= \ct{T}{_B}+N\frac{1}{2}\brkt{\csm{\beta}\ct{\delta}{^D_B}+\prn{\csm{\lambda}-1}\ctc{\epsilon}{_B^D}}\ctc{C}{_D}+\frac{1}{2}\cdc{_C}\ctm{X}{_B^C}\spacef.
					\end{align}
				This time, the conclusion is that if  \cref{eq:lb-conditionMinus,eq:lb-conditionPlus}  hold,  \cref{eq:lb-CApNewsp,eq:lb-CAmNewsm} are satisfied for the symmetric traceless gauge-invariant tensor fields $ \ctpm{n}{_{AB}} $ of \cref{eq:lb-nABminus,eq:lb-nABplus}. In that case, one has:
					\begin{align}
					N^2\prn{\csp{\beta}^2+\csp{\lambda}^2}\csp{\Z}&= \cdc{_C}\prn{\ctp{n}{_B^C}}\cdc{_D}\prn{\ctp{n}{^{BD}}}\spacef,\\
					N^2\prn{\csm{\beta}^2+\csm{\lambda}^2}\csm{\Z}&= \cdc{_C}\prn{\ctm{n}{_B^C}}\cdc{_D}\prn{\ctm{n}{^{BD}}}\spacef.
					\end{align}				
				 \Cref{thm:theTwoNewsProp} can be generalised straightforwardly:
					\begin{prop}[Generalized radiant news]\label{thm:lb-theTwoNewsProp}
						If the condition of \cref{eq:lb-conditionPlus} (\cref{eq:lb-conditionMinus}) holds on a cut $ \Sc $ with $ \mathbb{S}^2 $-topology, then
						\begin{align}
						\ctp{n}{_{AB}}= 0 \iff \csp{\Z}= 0\spacef.\quad\\
						\prn{\ctm{n}{_{AB}}= 0 \iff \csm{\Z}= 0}\quad  	
						\end{align}
						and $ \ctp{n}{_{AB}} $ ($ \ctm{n}{_{AB}} $) fulfils \cref{it:propertyNewsCandF,it:propertyNewsGaugeInvariant,it:propertyNewsSymmetric,it:propertyNewsTraceless,it:propertyNewsTwoDim,it:propertyNewsZ} on page \pageref{it:propertyNewsCandF}. 
						\begin{proof}
							The proof is very much the same as the one of \cref{thm:theTwoNewsProp}, only that now one uses \cref{eq:lb-nABplus,eq:lb-conditionPlus,eq:lb-CApNewsp} (\cref{eq:lb-nABminus,eq:lb-conditionMinus,eq:lb-CAmNewsm}) instead of \cref{eq:nABplus,eq:conditionPlus,eq:CApNewsp} (\cref{eq:nABminus,eq:conditionMinus,eq:CAmNewsm}).
						\end{proof}
					\end{prop}
					One can generalise \cref{thm:oneNewsUmbilical} too. In fact, the following result serves to exemplify the role of the coefficients $ \cspm{\lambda} $ and $ \cspm{\beta} $:
					\begin{lemma}\label{thm:lb-oneNewsUmbilical}
						Consider any \emph{umbilical} cut $ \Sc $ with $ \mathbb{S}^2 $-topology on $ \scri $ such that $ \ct{r}{_a} $ defines a strong orientation on $ \Sc $, i. e., $ \csm{\Z}\eqc 0 $. Then, there always exists the radiant news $ \ctp{n}{_{AB}} $ for $ \csp{\lambda}=\text{constant} $ and $ \csp{\beta}=0 $ given by
						\begin{equation}
						\ctp{n}{_{AB}}=2\csp{\lambda}\ct{V}{_{AB}}\spacef.
						\end{equation}
					\end{lemma}
					\begin{proof}
						One follows the same steps as in the proof of \cref{thm:oneNewsUmbilical}, this time using \cref{eq:lb-CApNewsp} instead of \cref{eq:CApNewsp}, arriving at:
						\begin{equation}\label{eq:lb-exampleBetaLambda}
						-\csp{\lambda}\cdc{_E}\ct{V}{_B^E}+\csp{\beta}\ctc{\epsilon}{^{CD}}\cdc{_{[C}}\prn{\ct{V}{_{D]B}}}=-\frac{1}{2}\cdc{_C}\ctp{n}{_B^C}\spacef.
						\end{equation}	
						Setting $ \csp{\lambda}=\text{constant} $ and $ \csp{\beta}=0 $, one finds $ \cdc{_{E}}\prn{\ctcp{n}{_{B}^E}-2\csp{\lambda}\ct{V}{_{B}^E}} =0$, which holds true if and only if $ \ctp{n}{_{AB}}=2 \csp{\lambda}\ct{V}{_{AB}} $.	
					\end{proof}
					\begin{remark}
						Given the assumptions of \cref{thm:lb-oneNewsUmbilical}, other solutions exist, for instance $ \csp{\beta}=\text{constant} $ and $ \csp{\lambda}=0 $. The role of $ \cspm{\beta} $ and $ \cspm{\lambda} $ is nothing more but to find solutions, i.e., $ \ctpm{X}{_{AB}} $ tensors, to \cref{eq:lb-conditionMinus,eq:lb-conditionPlus}. However, for particular cases --as the one described in \cref{thm:lb-oneNewsUmbilical}-- they are pure gauge, in the sense that setting them to one constant value or another provides a combination of the \emph{same} gauge-invariant symmetric traceless tensor field's divergence and its dual, as it is the case of \cref{eq:lb-exampleBetaLambda} where the (two) fundamental degrees of freedom are encoded in $ \ct{V}{_{AB}} $. Still in that case, if one considers $ \cspm{\lambda} $ and $ \cspm{\beta} $ as functions it does not affect the fact that the vanishing of $ \cspm{\Z}$ is equivalent to the vanishing of $ \ct{V}{_{AB}} $. This gauge freedom arises as a consequence of asking $ \cspm{\Z} $, a scalar function, to vanish if and only if the divergence of some radiant news tensor does, which allows one to construct combinations of $ \ctcupm{C}{_A} $ and its dual, as in \cref{eq:lb-CApNewsp,eq:lb-CAmNewsm}.
					\end{remark}

	\section{Equipped $ \scri $ and the problem of incoming radiation}\label{sec:additional-structure}

			If one wishes to generalise the concept of  radiant news presented in \cref{sec:news} for arbitrary cuts of $\scri$ to a tensor field on $ \scri $, study some sort of evolution of the physical fields intrinsically to $ \scri $ or carry out a close analogy with the $ \Lambda=0 $ scenario, one needs to endow $ \scri $ with further structure. In particular, we are going to introduce a selected family of curves, trying to keep it as general as possible. In \cref{ssec:noincomingrad} we will see that such kind of additional structures can be well motivated by physical conditions. Hence, in this section we will use the formalism presented in \cref{app:congruences}, where the necessary notation and definitions for the congruences of curves associated to a unit vector field $ \ct{m}{^a} $ on $ \scri $ can be found. Also, we follow the same notation as in \cref{sec:news} for objects associated to the decomposition of fields with respect to $ \ct{m}{^a} $, only that now underbars will be used instead of over-rings so that they become distinguishable from quantities resulting from the decomposition with respect to $ \ct{r}{^a} $. For instance, for the intrinsic Schouten tensor we write
			\begin{equation}\label{eq:decomp-Schouten-m}
			\cts{S}{_{ab}}\eqc \csS{\ubar{S}}\ct{m}{_a}\ct{m}{_b}+2\ctcn{S}{_B}\ct{m}{_{(a}}\ctcn{W}{_{b)}^B}+\ctcn{S}{_{AB}}\ctcn{W}{_a^A}\ctcn{W}{_b^B}\quad
			\end{equation}
		and, in general, for any symmetric tensor $ \ct{B}{_{\mu\nu}}$,
			\begin{equation}	\ct{B}{_{\alpha\beta}}\eqs\ct{n}{^{\mu}}\ct{n}{^{\nu}}\ct{B}{_{\mu\nu}}\ct{n}{_{\alpha}}\ct{n}{_{\beta}}+\ct{n}{^{\mu}}\ctcn{P}{^{\nu}_{(\alpha}}\ct{n}{_{\beta)}} \ct{B}{_{\mu\nu}}+2\ct{B}{_{\mu}}\ct{n}{^\mu}\ct{m}{_{(\alpha}}\ct{n}{_{\beta)}}+\ct{m}{_\alpha}\ct{m}{_\beta}\cs{B}+2\ctcn{B}{_{(\alpha}}\ct{m}{_{\beta)}}+\ctcn{B}{_{\alpha\beta}}\spacef,
			\end{equation}
		and 
			\begin{equation}
			\ctt{B}{_{\alpha\beta}}\defeq \ctcn{B}{_{\alpha\beta}}-\frac{1}{2}\ctcn{P}{_{\alpha\beta}}\ctcn{P}{^{\mu\nu}}\ctcn{B}{_{\mu\nu}}\spacef,
			\end{equation}
		with
			\begin{equation}\label{eq:notation-symmetric-tensor-m}
			\ctcn{B}{_{\alpha\beta}}\defeq \ctcn{P}{^\mu_\alpha}\ctcn{P}{^\nu_\beta}\ct{B}{_{\mu\nu}},\quad\ct{B}{_{\alpha}}\defeq \ct{m}{^\mu}\ct{B}{_{\mu\alpha}},\quad	\ctcn{B}{_{\alpha}}\defeq \ctcn{P}{^\nu_\alpha}\ct{m}{^\mu}\ct{B}{_{\mu\nu}},\quad\cs{B}\defeq \ct{m}{^\mu}\ct{m}{^\nu}\ct{B}{_{\mu\nu}},
			\end{equation}
		and with capital indices too,
			\begin{equation}\label{eq:notation-symmetric-tensorLatin-m}
			\ctcn{B}{_{AB}}\defeq \ctcn{E}{^\alpha_A}\ctcn{E}{^\beta_B}\ctcn{B}{_{\alpha\beta}} ,\quad	\cttcn{B}{_{AB}}\defeq \ctcn{B}{_{AB}}-\frac{1}{2}\mcn{_{AB}}\ctcn{B}{^C_{C}},\quad\ctcn{B}{_{A}}\defeq \ctcn{E}{^A_\alpha}\ct{m}{^\mu}\ct{B}{_{\mu\alpha}}\spacef.
			\end{equation}
		 Also, we define a couple of lightlike vector fields on $ \scri $ as
			\begin{align}
			\ctp{k}{^\alpha}&\defeq\frac{1}{\sqrt{2}}\left(\ct{n}{^\alpha}+\ct{m}{^\alpha}\right)\spacef,\label{eq:kp-def-m}\\
			\ctm{k}{^\alpha}&\defeq\frac{1}{\sqrt{2}}\left(\ct{n}{^\alpha}-\ct{m}{^\alpha}\right)\spacef.\label{eq:km-def-m}
			\end{align}
		This notation applies to all the objects coming from the orthogonal and lightlike decomposition of the rescaled-Weyl tensor (see \cref{ssec:lightlike-projections}).
		\\
		
		Let us  start by stating what we mean by additional structure on $ \scri $.
			 \begin{deff}[Equipped $ \scri $]\label{def:additional-structure}
		 		We say  that an open, connected, subset $ \Delta\subset \scri $ with the same topology than $ \scri $ is {\em equipped}  when it is endowed with a congruence $ \C $ of curves characterised by a unit vector field $ \ct{m}{^a} $. The projected surface $ \Scn\defeq \Delta/\C $ and $ \C $ are characterised by the conformal family of pairs
		 		\begin{equation}\label{eq:pairs-additional-structure}
		 		\prn{\ctcn{P}{_{ab}},\ct{m}{_a}}\spacef,
		 		\end{equation}
		 		where $ \ctcn{P}{_{ab}} $ is the projector to $ \Scn $. Two members belong to the same family if and only if $ \prn{\ctcn{P'}{_{ab}},\ct{m'}{_a}}=\prn{\Psi^2\ctcn{P}{_{ab}},\Psi\ct{m}{_a}} $, where $ \Psi $ is a positive function on $ \scri $.
		 	\end{deff}
		  We will usually assume that $\Delta$ is actually one entire connected component of $\scri$.
		 The curves of $ \C $ are parametrised by any scalar function $ v $ such that $ \lied_{\vec{m}}v\neq 0 $, and thus $v$  it is only defined up to the following changes:
		 	\begin{equation}\label{eq:change-v-congruences}
		 			v \rightarrow v'\prn{v,\zeta^A},\quad\frac{\partial v'}{\partial v}\neq 0\spacef,
		 	\end{equation}
		 where $ \zeta^A $ 
		  label each curve --see \cref{app:congruences} for further details. One can always choose adapted coordinates such that
		 	\begin{equation}\label{eq:parameter-along-curves}
		 		\ct{m}{^a}=A\delta^a_v
		 	\end{equation}
		 for some function $ A $. This form is preserved by \eqref{eq:change-v-congruences} as long as \eqref{eq:coordinate-change-projectiveS} is enforced for the $ \ct{\zeta}{^A} $.\\
		 
		 The orthogonal decomposition with respect to $ \ct{m}{^a} $ of $ \cts{S}{_{ab}} $ and $ \ct{C}{_{ab}} $ gives among other quantities
			\begin{align}
				\ctcn{S}{_{B}}&= \cdcn{_C}\prn{\ctcn{\kappa}{_B^C}+\ctcn{\omega}{_B^C}}-\cdcn{_B}\cscn{\kappa}-2\ctcn{a}{^C}\ctcn{\omega}{_{BC}}\spacef,\label{eq:SAkappa-congruence}	\\
				\ctcn{S}{^E_E}&\eqc \cscn{K}+\frac{1}{2}\cscn{\Sigma}^2-\frac{1}{4}\cscn{\kappa}^2-\frac{3}{2}\cscn{\omega}^2\label{eq:traceSchouten-Congruence}\spacef,\\
				N\ctcn{C}{^A}&= \ctcn{\epsilon}{^{CD}}\prn{\cdcn{_{[C}}\ctcn{S}{_{D]}^A}+\ctcn{\kappa}{_{[C}^A}\ctcn{S}{_{D]}}+\frac{3}{2}\ctcn{S}{^{A}}\ctcn{\omega}{_{CD}}}\spacef.\label{eq:CA-schouten-congruence}
			\end{align}
		Observe that $ \cdcn{_A} $, $ \mcn{_{AB}} $ and $ \ctcn{\epsilon}{_{AB}} $ represent a  one-parameter  family of connections, metrics and volume forms on $ \Scn $, because they also `depend on $ v $ ' ---see \cref{app:congruences}. Here, $ \cscn{K} $ is the function appearing in \cref{eq:curvature-congruences}. Taking these remarks into account, the formulae look aesthetically the same as for any single cut $ \Sc $ if  $ \C $ is a foliation ($ \iff \ctcn{\omega}{_{AB}}=0 $)  in which case we use a different name according to the following definition
		\begin{deff}[Strictly equipped $ \scri $]\label{def:additional-structure1}
		 		We say  that $ \scri $ is {\em strictly equipped} when it is equipped and the unit vector field $ \ct{m}{^a} $ is surface-orthognal, providing a foliation by cuts of $\scri$. 
		\end{deff}
		
		Indeed, many of the forthcoming results are considered if this happens, however one has to notice that even when the equations resemble the ones for cuts, they are different. Some insights into the case of general $ \C $ will be given in \cref{ssec:symmetries,ssec:news-congruences} as well. 
			
				 There is a third level of equipment, the highest one, given by
		\begin{deff}[Strongly equipped $ \scri $]\label{def:strong-structure}
					We say that $ \scri $ is strongly equipped when it is strictly equipped and $\ct{m}{^a}$ is shear-free, so that it defines a \emph{foliation}  by \emph{umbilical} cuts $ \Sc_v $ for $ v= $constant , that is
					$$\ct{m}{_a}=F\cds{_a}v\quad $$ for some scalar function $ F $. 
				\end{deff}	
				
				\begin{remark}
					This definition is the particular case of  \cref{def:additional-structure} with $ \ctcn{\omega}{_{ab}}=0 $ (i.e., $ \ct{m}{^a}  $ orthogonal to cuts) and $ \ctcn{\Sigma}{_{ab}}=0 $ (i.e., umbilical cuts).
				\end{remark}	

			\subsection{Decomposition of the Schouten tensor: kinematic expression}
			We are going to deduce an expression for $ \ctcn{S}{_{ab}} $ in terms of the kinematic quantities of $ \ct{m}{^a} $. To begin with, note that  the combination
				\begin{equation}
				-\ct{f}{_{ab}}= \lied_{\vec{m}}\ctcn{\Sigma}{_{ab}}-2\ctcn{\Sigma}{_{ad}}\ctcn{\Sigma}{_b^{d}}-\frac{1}{2}\cs{\kappa}\ctcn{\Sigma}{_{ab}}
				\end{equation}
			satisfies \cref{it:propertyNewsGaugeInvariant,it:propertyNewsSymmetric,it:propertyNewsTraceless,it:propertyNewsTwoDim} on page \pageref{it:propertyNewsGaugeInvariant}. Consider its pullback to $ \Scn $ --and use the identity $ 2\ctcn{\Sigma}{_{AD}}\ctcn{\Sigma}{_B^{D}}= \mcn{_{AB}}\cscn{\Sigma}^2 $:
					\begin{equation}\label{eq:fakenewsdef}
					-\ct{f}{_{AB}}\defeqcn \ctcn{\Sigma'}{_{AB}}-\mcn{_{AB}}\cscn{\Sigma}^2-\frac{1}{2}\cs{\kappa}\ctcn{\Sigma}{_{AB}}\spacef,
					\end{equation}
			with 
				\begin{equation}
					\ctcn{\Sigma'}{_{AB}}\defeq\ctcn{E}{^a_A}\ctcn{E}{^b_B}\lied_{\vec{m}}\ctcn{\Sigma}{_{ab}}=\lied_{\vec{m}}\ctcn{\Sigma}{_{AB}}\spacef,
				\end{equation}
			where the right-hand side follows from \cref{eq:liedEcongruence}. This term can be replaced  using the projection of  the Schouten tensor noticing that
					\begin{align}
					\ctcn{P}{^b_f}\ctcn{P}{^c_d}\ct{m}{^e}\ct{m}{_d}\cts{R}{_{ebc}^d}&=\ctcn{P}{^b_f}\ctcn{P}{^c_d}\ct{m}{^e}\prn{\cds{_e}\cds{_b}\ct{m}{_c}-\cds{_b}\cds{_e}\ct{m}{_c}}\nonumber\\
					&= \lied_{\vec{m}}\ctcn{\Sigma}{_{fd}}-2\ctcn{\Sigma}{_{e(d}}\cds{_{d)}}\ct{m}{^e}-\cscn{\kappa}\ct{m}{_{(f}}\ctcn{a}{_{d)}}+2\ct{m}{_{(d}}\ctcn{\Sigma}{_{f)}^e}\ctcn{a}{_e}+\ctcn{a}{_f}\ctcn{a}{_d}-\ctcn{P}{^b_f}\ctcn{P}{^c_d}\cds{_b}\ctcn{a}{_c}\nonumber\\
					&+\cscn{\kappa}\ct{m}{_{(d}}\ctcn{a}{_{f)}}+\cscn{\kappa}\ctcn{\Sigma}{_{df}}+\ctcn{\Sigma}{_d^e}\ctcn{\Sigma}{_{fe}}+\prn{\frac{1}{4}\cscn{\kappa}^2+\frac{1}{2}\ct{m}{^e}\cds{_e}\cscn{\kappa}}\mcn{_{df}}\spacef,
					\end{align}
			or equivalently
					\begin{equation}\label{eq:aux14}
					\ctcn{E}{^a_A}\ctcn{E}{^b_B}\ct{m}{^d}\ct{m}{^e}\cts{R}{_{dabe}}=\ctcn{\Sigma'}{_{AB}}+\ctcn{a}{_A}\ctcn{a}{_B}-\cdcn{_A}\ctcn{a}{_B}+\ctcn{q}{_{AB}}\prn{\frac{1}{2}\lied_{\vec{m}}\cscn{\kappa}-\frac{1}{2}\cscn{\Sigma}^2+\frac{1}{4}\cscn{\kappa}^2}\spacef.
					\end{equation}
		 	Next, use \cref{eq:intrinsicriemannScriSchoutenRelation} to get
					\begin{equation}\label{eq:aux18}
					\ctcn{P}{^b_f}\ctcn{P}{^c_d}\ct{m}{^e}\ct{m}{_d}\cts{R}{_{ebc}^d}=-\ctcn{P}{_{df}}\csS{\ubar{S}}-\ctcn{S}{_{df}}\spacef.
					\end{equation}						
			Then, take the trace of \cref{eq:aux14} and replace \cref{eq:traceSchouten-Congruence} in the resulting expression,
					\begin{equation}
					-\csS{\ubar{S}}=\frac{1}{2} \ctcn{\Sigma'}{^C_C}+ \frac{1}{2}\cscn{K}-\frac{1}{4}\cscn{\Sigma}^2+\frac{1}{8}\cscn{\kappa}^2+\frac{1}{2}\prn{\ctcn{a}{^E}\ctcn{a}{_E}-\cdcn{_E}\ctcn{a}{^E}+\lied_{\vec{m}}\cscn{\kappa}}\spacef.\label{eq:aux17}
					\end{equation}
			After that, use \cref{eq:aux14,eq:aux17,eq:aux18} to derive a formula for the projection of the intrinsic Schouten tensor
					\begin{align}
					\ctcn{S}{_{AB}}&= -\ctcn{\Sigma'}{_{AB}}+\cdcn{_A}\ctcn{a}{_B}-\ctcn{a}{_A}\ctcn{a}{_B}-\mcn{_{AB}}\brkt{\frac{1}{2}\prn{\cdcn{_E}\ctcn{a}{^E}-\ctcn{a}{_E}\ctcn{a}{^E}}-\frac{5}{4}\cscn{\Sigma^2}-\frac{1}{2}\cscn{K}+\frac{1}{8}\cscn{\kappa}^2}\spacef,\label{eq:schouten-acceleration}
					\end{align}
				or more compactly,
					\begin{align}
					\ctcn{S}{_{AB}}&= -\ctcn{\Sigma'}{_{AB}}+\cdcn{_A}\ctcn{a}{_B}-\ctcn{a}{_A}\ctcn{a}{_B}-\mcn{_{AB}}\brkt{\frac{1}{2}\prn{\cdcn{_E}\ctcn{a}{^E}-\ctcn{a}{_E}\ctcn{a}{^E}}-\frac{1}{2}\ctcn{\Sigma'}{^C_C}-\frac{1}{2}\ctcn{S}{^C_C}}\spacef.
					\end{align}	
				These formulae are interesting on their own and valid for a general foliation on $ \scri $. It is clear that they have the correct trace and, using the formulae in  \cref{app:gauge-transformations}, they give the right gauge behaviour ---compare with \cref{eq:gaugeSAB} in that same appendix.
				In terms of the gauge invariant quantity $ \ct{f}{_{AB}} $, we have
					\begin{equation}\label{eq:SAB-acceleration}
					\ctcn{S}{_{AB}}= \ct{f}{_{AB}}+\cdcn{_A}\ctcn{a}{_B}-\ctcn{a}{_A}\ctcn{a}{_B}-\frac{1}{2}\cscn{\kappa}\ctcn{\Sigma}{_{AB}}-\mcn{_{AB}}\brkt{\frac{1}{2}\prn{\cdcn{_E}\ctcn{a}{^E}-\ctcn{a}{_E}\ctcn{a}{^E}}-\frac{1}{4}\cscn{\Sigma^2}-\frac{1}{2}\cscn{K}+\frac{1}{8}\cscn{\kappa}^2}.
					\end{equation}

		\subsection{ Radiant news tensor field on equipped $\scri$}\label{ssec:news-congruences}
			In \cref{sec:news} we have shown that a gauge-invariant traceless symmetric tensor field $ \ct{V}{_{AB}} $  exists on any two-dimensional surface (with $ \mathbb{S}^2 $-topology or, with further assumptions, non-$ \mathbb{S}^2 $). This applies to cuts with $ \mathbb{S}^2$-topology
			on $ \scri $ where it represents \emph{a first component of news-like tensors} and  actually defines, under suitable conditions --\cref{thm:theTwoNewsProp}--, radiant news for a radiant supermomentum. However, the question of a news-like tensor field \emph{on} $ \scri $ is still open. To address it, first we present some geometrical results:
				\begin{lemma}\label{thm:no-general-scri-news}
					Let $ \prn{\I,\ms{_{ab}}} $  be any spacelike hypersurface $ \I $ with metric $ \ms{_{ab}} $ and define $ \ctc{P}{^c_a}\defeq \delta^c_a-\ct{m}{^c}\ct{m}{_a} $ for any unit vector field $ \ct{m}{^a} $ on $ \I $. Then, there are no tensor fields $ \ct{M}{_{ab}} $ on $ \I $ such that $ \ctc{M}{_{ab}}\defeq \ctc{P}{^c_a}\ctc{P}{^d_b}\ct{M}{_{cd}} $ is symmetric and traceless ($ \ctc{P}{^{cd}}\ctc{M}{_{cd}}=0 $) for all possible $ \ct{m}{^a} $,
						\begin{equation}
							\cbrkt{\nexists \ct{M}{_{ab}}\quad\big/\quad\ctc{M}{_{ab}}=\ctc{M}{_{ba}},\quad\ctc{P}{^{ab}}\ctc{M}{_{ab}}=0\quad\forall\ct{m}{^a}}.
						\end{equation}
				\end{lemma}
				\begin{proof}
					Given $ \ct{M}{_{ab}} $, assume that two different vector fields $ \ct{m}{^a} $, $ \ct{m'}{^a} $ exist such that
						\begin{align}
							0=\ctc{M'}{^c_c}=\ct{M}{^c_c}-\ct{m'}{^c}\ct{m'}{^d}\ct{M}{_{cd}}\spacef,\\
							0=\ctc{M}{^c_c}=\ct{M}{^c_c}-\ct{m}{^c}\ct{m}{^d}\ct{M}{_{cd}}\spacef,
						\end{align}
					Then, the only possibility for this to happen $ \forall \ct{m}{^a} $ is
						\begin{equation}
							\ct{m}{^c}\ct{m}{^d}\ct{M}{_{cd}}=\ct{m'}{^c}\ct{m'}{^d}\ct{M}{_{cd}}\quad\forall \ct{m'}{^a}\neq \ct{m}{^a}.
						\end{equation}	
					Thus, either $ \ct{M}{_{ab}}=-\ct{M}{_{ba}} $ (which cannot give rise to a symmetric tensor $ \ctc{M}{_{ab}} $) or $ \ct{M}{_{ab}}=0 $.
				\end{proof}
			This is in contrast with what happens at the conformal boundary for a vanishing cosmological constant, where any symmetric and traceless tensor field orthogonal to $ \ct{N}{^\alpha}\evalat{_{\Lambda=0}}$ on $ \scri $\footnote{Recall that $ \scri $ is lightlike for $ \Lambda=0 $, and $ \ct{N}{^\alpha} $ is the vector field tangent to the generators.} is a symmetric and traceless tensor field \emph{on any cut}. Precisely, this applies to the news tensor $ \ct{N}{_{ab}} $ on $ \scri $ for $ \Lambda=0 $: its pullback $ \ct{N}{_{AB}} $ to \emph{any cut} is symmetric and traceless there (see the companion paper \cite{Fernandez-Alvarez_Senovilla-afs}). In any case, in general
 one has
				\begin{lemma}\label{thm:linear-combination-news}
				Let $ \prn{\I,\ms{_{ab}}} $  be any spacelike hypersurface $ \I $ with metric $ \ms{_{ab}} $ and $ \ct{M}{_{ab}} $ and $ \ct{M'}{_{ab}} $ any couple of symmetric tensor fields on $ \I $, each one orthogonal to a unit vector field $ \ct{m}{^a} $ and $ \ct{m'}{^a} $, respectively. Assume that $ \ctc{P}{^{cd}}\ct{M}{_{cd}}=0 $ and $ \ctc{P'}{^{cd}}\ct{M}{_{cd}}=0 $,  where $ \ctc{P}{^c_a}$ and $ \ctc{P'}{^c_a} $ are the orthogonal projectors associated to $ \ct{m}{^a} $ and $ \ct{m'}{^a} $. Then, the tensor field $ \ct{B}{_{ab}}\defeq \lambda \ct{M}{_{ab}}+\delta \ct{M'}{_{ab}}$, for arbitrary coefficients $ \lambda $ and $ \beta $, is symmetric and traceless, $ \ms{^{cd}}\ct{B}{_{cd}}=0 $.
				\end{lemma}
				\begin{proof}
				The tensor field $ \ct{B}{_{ab}} $ is symmetric and notice that
				\begin{align}
				\ms{^{cd}}\ct{M}{_{cd}}&=\ct{m}{^c}\ct{m}{^d}\ct{M}{_{cd}}+\ctc{P}{^{cd}}\ct{M}{_{cd}}=0\spacef,\\
				\ms{^{cd}}\ct{M'}{_{cd}}&=\ct{m'}{^c}\ct{m'}{^d}\ct{M'}{_{cd}}+\ctc{P'}{^{cd}}\ct{M'}{_{cd}}=0\spacef.
				\end{align}
			 Therefore, $ \ms{^{cd}}\ct{B}{_{cd}}=0 $.
				\end{proof}
		
			From \cref{thm:no-general-scri-news} it follows that a unique tensor field on $ \scri $ cannot generate a would-be news tensor field assigned to every possible cut $ \Sc $ on $ \scri $. Also, \cref{thm:linear-combination-news} shows that a linear combination of would-be news tensor fields, associated each one to a  different family of cuts, gives rise to a gauge-invariant traceless symmetric tensor field on $ \scri $. Such a combination will have in general more than two degrees of freedom. All in all, we are led to search for a tensor field on $ \scri $ associated to the congruence $ \C $ of \cref{def:additional-structure} equipping $\scri$. 
			
			Having presented the above general results, let us come back to the asymptotic geometry. First, consider the case of a general $ \C $ ($ \ctcn{\omega}{_{AB}} \neq0 $). \Cref{eq:CA-schouten-congruence} can be written using \cref{eq:SAkappa-congruence} as
				\begin{equation}\label{eq:CA-U-congruences}
					N\ctcn{C}{^A}= \ctcn{\epsilon}{^{CD}}\prn{\cdcn{_{[C}}\ctcn{U}{_{D]}^A}+\ct{W}{_{CD}^A}-\ct{S}{_{CD}^A}}\spacef,
				\end{equation}
			where
				\begin{align}
					\ctcn{U}{_{AB}}&\defeq \ctcn{S}{_{AB}}+\frac{1}{2}\cscn{\kappa}\ctcn{\Sigma}{_{AB}}+\ct{L}{_{AB}}\spacef,\label{eq:UABdef-congruences}\\
					\ct{L}{_{AB}}&\defeq \prn{\frac{1}{8}\cscn{\kappa}^2-\frac{1}{4}\cscn{\Sigma}^2+\frac{3}{4}\cscn{\omega}^2}\mcn{_{AB}}\spacef,\label{eq:LAB-congruences}\\
					\ct{S}{_{CAB}}&\defeq \ctcn{T}{_{CAB}}+3\brkt{\cdcn{_D}\prn{\ctcn{\omega}{_B^D}}-\ctcn{a}{^D}\ctcn{\omega}{_{BD}}\ctcn{\omega}{_{CA}}}\label{eq:SABC-congruences}\spacef,\\
					\ctcn{T}{_{CAB}} &\defeq \frac{1}{2}\brkt{\mcn{_{B[C}}\cdcn{_{A]}}\cscn{\Sigma}^2-\cdcn{_B}\prn{\ctcn{\Sigma}{^D_{[A}}}\ctcn{\Sigma}{_{C]D}}}\spacef,\label{eq:TABC-congruences}\\
					\ct{W}{_{CAB}}&\defeq -\frac{1}{2}\ctcn{\kappa}{^D_B}\cdcn{_D}\ctcn{\omega}{_{CA}}+\ctcn{a}{^D}\ctcn{\kappa}{_{DB}}\ctcn{\omega}{_{CA}}+\frac{3}{2}\ctcn{\omega}{_{CA}}\cdcn{_D}\ctcn{\kappa}{_B^D}-\frac{3}{2}\ctcn{\omega}{_{CA}}\cdcn{_B}\cscn{\kappa}\spacef,\label{eq:WABC-congruences}
				\end{align}
			and it will be convenient to define
				\begin{equation}
					\ctcn{T}{_A}\defeq\ctcn{T}{_{CA}^C}\spacef.\label{eq:TA-congruences}
				\end{equation}
			The gauge behaviour of these fields follows by direct computation, using the formulae of \cref{app:gauge-transformations},
				\begin{align}
					\ctcng{U}{_{AB}}&=\ctcn{U}{_{AB}}-\frac{1}{\omega}\cdcn{_{(A}}\ctcn{\omega}{_{B)}}+\frac{2}{\omega^2}\ctcn{\omega}{_A}\ctcn{\omega}{_B}-\frac{1}{2\omega^2}\ctcn{\omega}{_C}\ctcn{\omega}{^C}\mc{_{AB}}\spacef,\label{eq:gauge-UAB-congruences}\\
					\ctcng{S}{_{CAB}}&=\ctcn{S}{_{CAB}}\spacef,\\
					\ctcng{T}{_{CAB}}&=\ctcn{T}{_{CAB}}\spacef,\\
					\ctcng{W}{_{CAB}}&=\ctcn{W}{_{CAB}}-\frac{1}{2\omega}\lied_{\vec{m}}\omega\cdcn{_B}\ctcn{\omega}{_{CA}}+\frac{1}{\omega}\ctcn{\omega}{_{CA}}\brkt{\ctcn{\kappa}{^D_B}\ctcn{\omega}{_D}-\frac{1}{2}\cdcn{_B}\lied_{\vec{m}}\omega-\lied_{\vec{m}}\ctcn{\omega}{_{B}}+\frac{5}{2\omega}\lied_{\vec{m}}\omega\ctcn{\omega}{_B}}.
				\end{align} 	
			Precisely, the interest of these definitions is that the combination $ \cdcn{_{[C}}\ctcn{U}{_{D]A}}+\ct{W}{_{CDA}} $ is gauge invariant,
				\begin{equation}
					\cdcng{_{[C}}\ctg{U}{_{D]A}}+\ctg{W}{_{CDA}}=\cdcn{_{[C}}\ctcn{U}{_{D]A}}+\ct{W}{_{CDA}}\quad
				\end{equation}
			 and
				\begin{equation}\label{eq:UAB-trace-congruences}
					\ctcn{U}{^E_E}=\cscn{K}\spacef.
				\end{equation}
			In order to recover a result of the kind of \cref{thm:onepiece-news}, it is reasonable to consider the splitting
				\begin{equation}
					\ctcn{V}{_{AB}}\defeq \ctcn{U}{_{AB}}-\ctcn{\rho}{_{AB}}\spacef,
				\end{equation}
			for some $ \ctcn{\rho}{_{AB}} $ fulfilling the gauge-invariant equation
				\begin{equation}\label{eq:rho-diffeq-congruences}
					\cdcn{_{[A}}\ct{\rho}{_{B]C}}+\ct{W}{_{ABC}}=0\spacef,
				\end{equation}
			and $ \ct{V}{_{AB}} $ a two-dimensional gauge-invariant symmetric traceless tensor field on $ \scri $. This would constitute  the \emph{first component of news-like objects} when $\scri$ is equipped. However, while existence of general solutions to \cref{eq:rho-diffeq-congruences} may be provable, uniqueness is in principle a non-trivial task. Indeed, this is an open problem which should be studied carefully. \\
		
			Now we focus on the case of a strictly equipped $\scri$, so that $ \C $ defines also a foliation ($ \ctcn{\omega}{_{ab}}=0 $). In this case, it is always possible to write \cref{eq:m-foliations}, using the freedom \eqref{eq:change-v-congruences} if needed, as
				\begin{equation}\label{eq:m-congruences-sec}
				\ct{m}{_a}=F\cds{_a}v\quad\text{with }	\frac{1}{F}=\lied_{_{\vec{m}}}v,
				\end{equation}
			and each leaf of the foliation $ \C $ is a cut $ \Sc_{v}$, labelled by a constant value of the parameter along the curves, $ v=\hat{v}=\text{constant} $, and with basis $ \cbrkt{\ct{E}{^a_A}}_v $, $ \cbrkt{\ct{W}{_a^A}}_v $. Therefore, on each leaf we are in the situation described in \cref{sec:news} for any single cut. In other words, \cref{thm:onepiece-news,thm:rho-tensor-noncompact,thm:rho-tensor} apply on each leaf. Hence, one has on each cut
				\begin{equation}
				N\ctc{C}{^A} \eqSv{v} \ctc{\epsilon}{^{CD}}\prn{\cdc{_{[C}}\ct{V}{_{D]}^A}-\ct{T}{_{CD}^A}}\spacef,\label{eq:CA-V-cut-foliation}
				\end{equation}
			or, by means of $ \ctcnupm{C}{_{A}} $,	
				\begin{align}
				2N\ctcp{C}{^A}+N\ctc{\epsilon}{^{AC}}\ctc{D}{_C}&\eqSv{v} \ctc{\epsilon}{^{CD}}\prn{\cdcn{_{[C}}\ct{V}{_{D]}^A}-\ct{T}{_{CD}^A}}\spacef, \label{eq:CApV-cut-foliation}\\
				2N\ctcm{C}{^A}-N\ctc{\epsilon}{^{AC}}\ctc{D}{_C}&\eqSv{v}\ctc{\epsilon}{^{CD}}\prn{\cdcn{_{[C}}\ct{V}{_{D]}^A}-\ct{T}{_{CD}^A}}\spacef, \label{eq:CAmV-cut-foliation}
				\end{align}
			Moreover, because $ \ctcn{\omega}{_{AB}}=0 $,
				\begin{equation}\label{eq:omega-null}
				\ct{S}{_{ABC}}=\ctcn{T}{_{ABC}},\quad\ct{W}{_{ABC}}=0\spacef,
				\end{equation}
			which makes \cref{eq:CA-U-congruences} read
				\begin{equation}\label{eq:CA-U-foliation}
					N\ctcn{C}{^A}=\ctcn{\epsilon}{^{CD}}\prn{\cdcn{_{[C}}\ctcn{U}{_{D]}^A}-\ctcn{T}{_{CD}^A}}\spacef.
				\end{equation}
			Inserting $ \ctcn{\omega}{_{AB}}=0 $ in \cref{eq:gauge-UAB-congruences} too, noting \cref{eq:commutator-Es}, it turns out that $ \ctcn{U}{_{AB}} $ has a recognisable gauge behaviour,
				\begin{equation}
					\ctg{U}{_{AB}}=\ct{U}{_{AB}}-a\frac{1}{\omega}\cdcn{_A}\ctcn{\omega}{_B}+\frac{2a}{\omega^2}\ctcn{\omega}{_A}\ctcn{\omega}{_B}-\frac{a}{2\omega^2}\ctcn{\omega}{_C}\ctcn{\omega}{^C}\mcn{_{AB}}\spacef.
				\end{equation}
			Then, one can show some important results for  strictly equipped $\scri$  (the first two are general i.e., not only for $ \scri $ but for any Riemannian 3-dimensional hypersurface $ \I $ equipped with a foliation),
				\begin{lemma}\label{thm:general-gauge-trace-foliation}
				 Let $\scri$ be strictly equipped and $ \ct{t}{_{AB}} $ be any symmetric tensor field on $ \Scn $ whose behaviour under conformal rescalings \eqref{eq:gauge-mcn} is
				\begin{equation}\label{eq:gauge-behaviour-appropriate-foliations}
				\ctg{t}{_{AB}}=\ct{t}{_{AB}}-a\frac{1}{\omega}\cdcn{_A}\ctcn{\omega}{_B}+\frac{2a}{\omega^2}\ctcn{\omega}{_A}\ctcn{\omega}{_B}-\frac{a}{2\omega^2}\ctcn{\omega}{_C}\ctcn{\omega}{^C}\mcn{_{AB}}
				\end{equation}
				for some fixed constant $ a\in\mathbb{R} $. Then,
				\begin{equation}\label{eq:gauge-derivative-appropiate-behaviour-foliations}
				\cdcng{_{[C}}\ctg{t}{_{A]B}}=\cdcn{_{[C}}\ct{t}{_{A]B}}+\frac{1}{\omega}\prn{a\cscn{K}-\ct{t}{^E_E}}\ctcn{\omega}{_{[C}}\mcn{_{A]B}}\spacef,
				\end{equation}						
				where now $ \cscn{K} $ coincides with the Gaussian curvature $ \cs{K} $ of each cut at $ v=$constant. In particular, for any symmetric gauge-invariant tensor field $ \ct{B}{_{AB}} $ on $ \Sc $,
				\begin{equation}\label{eq:aux23-foliations}
				\cdcng{_{[C}}\ctg{B}{_{A]B}}=\cdcn{_{[C}}\ct{B}{_{A]B}}-\frac{1}{\omega}\ct{B}{^E_E}\ctcn{\omega}{_{[C}}\mcn{_{A]B}}\quad
				\end{equation}
			\end{lemma}
			\begin{proof}
				One proceeds as in the proof of \cref{thm:general-gauge-trace}.
			\end{proof}
			\begin{corollary}\label{thm:gauge-invariant-diff-foliation}
				Under the same assumptions, a symmetric gauge-invariant tensor field $ \ct{B}{_{AB}} $ on $ \Scn $ satisfies
				\begin{equation}
				\cdcng{_{[C}}\ctg{B}{_{B]A}}=\cdcn{_{[C}}\ct{B}{_{B]A}}
				\end{equation}
				if and only if $ \ct{B}{^E_E}=0 $.
			\end{corollary}
			Now one can prove the following two results:
				\begin{corollary}[The tensor field $ \rho $ for strictly equipped $\scri$ with $\mathbb{S}^2$ leaves]\label{thm:rho-tensor-foliation}
				Assume $\scri$ is strictly equipped
				with $ v $ the parameter along the curves \eqref{eq:parameter-along-curves} and such that \cref{eq:m-congruences-sec} holds. If the leaves have $ \mathbb{S}^2 $-topology, there is a unique tensor field $ \ctcn{\rho}{_{ab}} $  on $ \scri $ orthogonal to $ \ct{m}{^a} $ (equivalently, a one-parameter family of symmetric tensor fields $ \ctcn{\rho}{_{AB}}\prn{v}\defeq \ctcn{E}{^a_A}\ctcn{E}{^b_B}\ctcn{\rho}{_{ab}} $ on the projected surface $ \Scn $) whose behaviour under conformal rescalings \eqref{eq:gauge-mcn} is as in \cref{eq:gauge-behaviour-appropriate-foliations} and satisfies the equation
				 	\begin{equation}\label{eq:rho-diff-eq-foliation}
					\ctcn{P}{^d_a}\ctcn{P}{^e_b}\ctcn{P}{^f_c}\cds{_{[f}}\ctcn{\rho}{_{d]e}}= 0
					\end{equation}
				in any conformal frame. This tensor field must have a trace $ \ctcn{\rho}{^e_e}\defeq \ctcn{P}{^{ae}}\ctcn{\rho}{_{ae}}=a\cscn{K} $ and reduces, at each leaf, to the corresponding tensor of \cref{thm:rho-tensor} with all its properties.
In particular, it is given for the round-sphere one-parameter family of metrics by $ \ctcn{\rho}{_{ab}}=\ctcn{P}{_{ab}}a\cscn{K}/2 $.
				\end{corollary}
				\begin{proof}
				Let $ \Sc_{\hat{v}} $ represent a leaf of the foliation for constant $ v=\hat{v} $.
				  If we evaluate \cref{eq:rho-diff-eq-foliation} at $ v=\hat{v} $, i.e., on the leaf $ \Sc_{\hat{v}} $ and contract all the indices with the basis on $ \Sc_{\hat{v}} $, $ \cbrkt{ \ct{E}{^a_A} }$, we obtain the following equation there
				 	\begin{equation}
				 		\cdc{_{[C}}\prn{\ctcn{\rho}{_{A]B}}}\eqSv{\hat{v}}0\spacef,
				 	\end{equation}
				 where $ \cdc{_A} $ is the canonical covariant derivative on $ \Sc_{\hat{v}} $. But the solution to this equation exists and is unique
				 	\begin{equation}
				 		\ctcn{\rho}{_{AB}}\eqSv{\hat{v}} \ctrd{\hat{v}}{\rho}{_{AB}}\spacef,
				 	\end{equation} 
				 with $ \ctrd{\hat{v}}{\rho}{_{AB}} $ the tensor field of \cref{thm:rho-tensor} corresponding to $ \Sc_{\hat{v}} $. 
				 Then, one can define $ \ctcn{\rho}{_{ab}} $ at any leaf simply by
				 	\begin{equation}\label{eq:aux33}
					 	\ctcn{\rho}{_{ab}}\eqSv{\hat{v}}\ct{W}{_a^A}\ct{W}{_b^B}\ctrd{\hat{v}}{\rho}{_{AB}}\quad
				 	\end{equation}
				 and this holds on each leaf --i.e., at any value of $ v $. Since  $ \scri=\underset{v}{\cup} \Sc_{v} $ and $ \Sc_{\hat{v}_1}\cap\Sc_{\hat{v}_2}=\emptyset $  for $ \hat{v}_1\neq\hat{v}_2 $, at every point on $ \scri$  \cref{eq:rho-diff-eq-foliation} has a unique solution given by
				 	\begin{equation}
				 	 \ctcn{\rho}{_{ab}}=\ctcn{W}{_a^A}\ctcn{W}{_b^B}\ctrd{v}{\rho}{_{AB}},\quad\ct{m}{^a}\ctcn{\rho}{_{ab}}=0\spacef.\label{eq:rho-foliation}
				 	\end{equation}
				 Note that contraction of \cref{eq:rho-diff-eq-foliation} with $ \cbrkt{\ctcn{E}{^a_A}} $ gives the equivalent equation on $ \Scn $
 					\begin{equation}\label{eq:rho-diff-eq-foliation-scn}
 						\cdcn{_{[C}}\ctcn{\rho}{_{A]B}}= 0.
 					\end{equation}	
					  According to \cref{thm:general-gauge-trace-foliation}, \cref{eq:rho-diff-eq-foliation-scn} is satisfied in any conformal frame if and only if $ \ctcn{\rho}{^E_E}=a\cscn{K} $, which using the definition \eqref{eq:projector-foliation} of the projector $ \ctcn{P}{^a_b} $ can be recast as $ \ctcn{\rho}{^e_e}\defeq \ctcn{P}{^{ae}}\ctcn{\rho}{_{ae}}=a\cscn{K} $. Finally, notice that by \eqref{eq:rho-foliation} and according to \cref{thm:rho-tensor}, the solution $ \ctcn{\rho}{_{AB}} $ on each leaf is given by  $ \ctcn{\rho}{_{AB}}\eqSv{v}\mc{_{AB}}aK/2$ for every cut with a round metric and Gaussian constant curvature $ \cs{K} $. But $ \cscn{K} $ coincides on each cut with $ \cs{K} $, and then contraction with $ \cbrkt{\ctcn{W}{_a^A}} $ gives $ \ctcn{\rho}{_{ab}}=\ctcn{P}{_{ab}}a\cscn{K}/2 $.
					\end{proof}

					\begin{corollary}[The tensor $ \rho $ for strictly equipped $\scri$ with non-$ \mathbb{S}^2 $ leaves]\label{thm:rho-tensor-noncompact-foliation}
					Assume $\scri$ is strictly equipped
					with $ v $ the parameter along the curves \eqref{eq:parameter-along-curves} and such that \cref{eq:m-congruences-sec} holds. 
					 Assume that the leaves $ \prn{\Sc_v,\mc{_{AB}}} $ are non-necessarily topological-$\mathbb{S}^2$ and that there is a vector field $ \ct{\chi}{^a} $ such that $ \ctcn{\chi}{^A}\defeq \ctcn{W}{_a^A}\ct{\chi}{^a} $ is a CKVF, and has a fixed point, on each leaf.
					 Then, there is a unique tensor field $ \ctcn{\rho}{_{ab}} $  on $ \scri $ orthogonal to $ \ct{m}{^a} $ (equivalently, a one-parameter family of symmetric tensor fields $ \ctcn{\rho}{_{AB}}\prn{v}\defeq \ctcn{E}{^a_A}\ctcn{E}{^b_B}\ctcn{\rho}{_{ab}} $ on the projected surface $ \Scn $) whose behaviour under conformal rescalings \eqref{eq:gauge-mcn} is as in \cref{eq:gauge-behaviour-appropriate-foliations} and satisfies the equations
						\begin{align}
						\ctcn{P}{^d_a}\ctcn{P}{^e_b}\ctcn{P}{^f_c}\cds{_{[f}}\ctcn{\rho}{_{d]e}}&= 0\spacef,\\
						\lied_{\vec{\cscn{\chi}}}\ctcn{\rho}{_{AB}}+a\cdcn{_{A}}\cdcn{_{B}}\phi &=0\spacef, \label{eq:lie-rho-foliations-nonS}
						\end{align}
					in any conformal frame, where $\phi\defeq\cdcn{_C}\ctcn{\chi}{^C}/2 $. Furthermore, this tensor field must have a trace $ \ctcn{\rho}{^e_e}\defeq \ctcn{P}{^{ae}}\ctcn{\rho}{_{ae}}=a\cscn{K} $ and coincides, at each leaf, with the corresponding tensor of \cref{thm:rho-tensor-noncompact} with all its properties.
				\end{corollary}
				\begin{remark}\label{rem:axial}
				An outstanding case for the existence of the vector field  $ \ct{\chi}{^a} $ is when this is an {\em axial} CKVF of $\ct{h}{_{ab}}$ orthogonal to $\ct{m}{^a}$, that is, tangent to the leaves, and such that it leaves the equipping congruence conformally invariant ($\lied_{\vec\chi}\ct{m}{_a} \propto \ct{m}{_a}$). 
				\end{remark}
				
				\begin{proof}
					One follows the same reasoning as in the proof of \cref{thm:rho-tensor-foliation}, this time using \cref{thm:rho-tensor-noncompact} instead of \cref{thm:rho-tensor}. After the first steps, one finds
						\begin{equation}
						\cdc{_{[C}}\prn{\ctcn{\rho}{_{A]B}}}\eqSv{\hat{v}}0\spacef,
						\end{equation}
					on each cut. Taking into account that $ \ctcn{\chi}{^A} $ has a fixed point on every $ \Sc_{v} $ and \cref{eq:lie-rho-foliations-nonS}, the solution to this equation exists and is unique on each cut
						\begin{equation}
						\ctcn{\rho}{_{AB}}\eqSv{\hat{v}} \ctrd{\hat{v}}{\rho}{_{AB}}\spacef,
						\end{equation} 
					given by the tensor $ \ct{\rho}{_{AB}}$ of \cref{thm:rho-tensor-noncompact}. The rest of the proof follows the same steps as in \cref{thm:rho-tensor-foliation}.
				\end{proof}
					\begin{lemma}[No traceless Codazzi tensor fields on $ \Scn $ for foliations]\label{thm:codten-foliation}
					Let $ \ctcn{M}{_{AB}} $ be any one-parameter family of traceless and symmetric tensor fields on $ \Scn $ associated to a congruence of curves $ \C $ orthogonal to a family of surfaces foliating a 3-dimensional space-like hypersurface $ \I $ with topological-$ \mathbb{S}^2 $ leaves. Then
					\begin{equation}
					\ctcn{M}{_{AB}}=0\iff \cdcn{_{[C}}\ctcn{M}{_{A]B}}=0.
					\end{equation}
				\end{lemma}
				\begin{proof}
					Defining on $ \I $ $ \ctcn{M}{_{ab}}\defeq  \ctcn{W}{_a^A}\ctcn{W}{_b^B}\ctcn{M}{_{AB}}$, it vanishes if and only if $ \ctcn{M}{_{AB}} $ so does, and satisfies $  \ctcn{P}{^c_d}\ctcn{P}{^a_e}\ctcn{P}{^b_f}\cds{_{[c}} \ctcn{M}{_{a]b}} = 0 $ if and only if $ \cdcn{_{[C}}\ctcn{M}{_{A]B}}=0 $. Evaluating on each leaf $ \Sc_{v} $ and contracting the equation $  \ctcn{P}{^c_d}\ctcn{P}{^a_e}\ctcn{P}{^b_f} \cds{_{[c}}\ctcn{M}{_{a]b}} = 0 $ with $ \cbrkt{\ct{E}{^a_A}} $, one finds  $ \cdcn{_{[C}}\prn{\ctcn{M}{_{A]B}}}\eqSv{v}0 $ $ \forall v $ which using \eqref{eq:dimensional-identity} is equivalent to its trace and holds if and only if $ \ctcn{M}{_{AB}}\eqSv{v}0  $ $ \forall v $  because $ \ctcn{M}{_{AB}} $ is a symmetric and traceless Codazzi tensor on each compact, two-dimensional cut $\Sc_v$ (see e.g. \cite{Liu1998} and references therein). This is equivalent to the vanishing of $ \ctcn{M}{_{ab}} $ on each $ \Sc_v $ and, since  $ \I=\underset{v}{\cup} \Sc_{v} $ and $ \Sc_{\hat{v}_1}\cap\Sc_{\hat{v}_2}=\emptyset $  for $ \hat{v}_1\neq\hat{v}_2 $, to the vanishing of $ \ctcn{M}{_{ab}} $ at every point on $ \I $ too.
				\end{proof}
			Let us continue by showing the existence and uniqueness of a \emph{first component of news} on strictly equipped $ \scri$ with topological $\mathbb{S}^2$ leaves:
				\begin{prop}[The first component of news on strictly equipped $\scri$ with $\mathbb{S}^2$ leaves]\label{thm:onepiece-news-foliations}
				Assume $\scri$ is strictly equipped
					with $ v $ the parameter along the curves \eqref{eq:parameter-along-curves} and such that \cref{eq:m-congruences-sec} holds. If the leaves have $ \mathbb{S}^2 $-topology, there is a one-parameter family of symmetric traceless gauge-invariant tensor fields
					\begin{equation}
					\ctcn{V}{_{AB}}\defeq \ctcn{U}{_{AB}}-\ctcn{\rho}{_{AB}}\spacef,
					\end{equation}
					that satisfies the gauge-invariant equation
					\begin{equation}\label{eq:diffUAB-foliations}
					\cdcn{_{[A}}\ctcn{U}{_{B]C}}=\cdcn{_{[A}}\ctcn{V}{_{B]C}}\spacef,
					\end{equation}
					where $ \ctcn{\rho}{_{AB}} $ is the family of tensor fields of \cref{thm:rho-tensor-foliation} (for $ a=1 $). Besides, $ \ctcn{V}{_{AB}} $ is unique with these properties.
				\end{prop}
				\begin{proof}
					The one-parameter family of tensor fields $ \ctcn{V}{_{AB}} $ is symmetric, traceless and gauge invariant as a consequence of \cref{eq:UABdef-congruences,eq:gauge-UAB-congruences,eq:UAB-trace-congruences}, recalling $ \ctcn{\omega}{_{AB}}=0 $, and \cref{thm:rho-tensor-foliation}. The uniqueness of $ \ctcn{V}{_{AB}} $ follows from \cref{thm:rho-tensor-foliation} too and \Cref{eq:diffUAB-foliations}.
				\end{proof}
				\begin{corollary}[The first component of news on strictly equipped $\scri$ with non-$\mathbb{S}^2$ leaves]\label{thm:onepiece-news-non-compact-foliations}
							Assume $\scri$ is strictly equipped
							with $ v $ the parameter along the curves \eqref{eq:parameter-along-curves} and such that \cref{eq:m-congruences-sec} holds.
							Assume that the leaves $ \prn{\Sc_v,\mc{_{AB}}} $ are non-necessarily topological-$\mathbb{S}^2$ and that there is a vector field $ \ct{\chi}{^a} $ such that $ \ctcn{\chi}{^A}\defeq \ctcn{W}{_a^A}\ct{\chi}{^a} $ is a CKVF of the metric $ \mc{_{AB}} $ and has a fixed point on each leaf.
							  Then, there is a one-parameter family of symmetric traceless gauge invariant tensor fields
						\begin{equation}\label{eq:VdefFoliations}
						\ctcn{V}{_{AB}}\defeq \ctcn{U}{_{AB}}-\ctcn{\rho}{_{AB}}\spacef,
						\end{equation}
						that satisfies the gauge-invariant equation
						\begin{equation}
						\cdcn{_{[A}}\ctcn{U}{_{B]C}}=\cdcn{_{[A}}\ctcn{V}{_{B]C}}\spacef,
						\end{equation}
						where $ \ctcn{\rho}{_{AB}} $ is the tensor field of \cref{thm:rho-tensor-noncompact-foliation} (for $ a=1 $). Besides, $ \ctcn{V}{_{AB}} $ is unique with these properties.
					\end{corollary}
					\begin{proof}
						The proof proceeds as the one of \cref{thm:onepiece-news-foliations}, only that this time one uses \cref{thm:rho-tensor-noncompact-foliation} instead of \cref{thm:rho-tensor-foliation}.
				\end{proof}
			Then, under assumptions of \cref{thm:onepiece-news-foliations} or \cref{thm:onepiece-news-non-compact-foliations}, one has on $ \Scn $ (equivalently on $ \scri $ by taking the pullback)
				\begin{equation}
					N\ctcn{C}{^A} \eqcn \ctcn{\epsilon}{^{CD}}\prn{\cdcn{_{[C}}\ctcn{V}{_{D]}^A}-\ctcn{T}{_{CD}^A}}\spacef,\label{eq:CA-V-foliation}
				\end{equation}
				\begin{corollary}\label{thm:VcnVc}
					The tensor field on $ \scri $
						 \begin{equation}
						 	\ctcn{V}{_{ab}}(v)\defeq \ctcn{W}{_a^A} \ctcn{W}{_b^B}\ctcn{V}{_{AB}}
						 \end{equation}
					satisfies
						\begin{equation}
						 \ctcn{V}{_{ab}}=\ctcn{W}{_a^A}\ctcn{W}{_b^B}\ctrd{v}{V}{_{AB}},\quad\ct{m}{^a}\ctcn{V}{_{ab}}=0\spacef,
						\end{equation}
					where $ \ctrd{_v}{V}{_{AB}} $ is the first component of news associated to each leaf $ \Sc_{v} $ defined in \cref{thm:onepiece-news}.
				\end{corollary}
				\begin{proof}
					One can take the pullback to $ \scri $ with $ \cbrkt{\ctcn{W}{_a^A}} $ of \cref{eq:VdefFoliations},
						\begin{equation}
							\ctcn{V}{_{ab}}=\ctcn{U}{_{ab}}-\ctcn{\rho}{_{ab}}\spacef,
						\end{equation}
					and see that 
						\begin{equation}
							\ctcn{U}{_{ab}}=\ctcn{W}{_a^A}\ctcn{W}{_b^B}\ctrd{v}{U}{_{AB}},\quad\ct{m}{^a}\ctcn{U}{_{ab}}=0
						\end{equation}
						where $ \ctrd{v}{U}{_{AB}} $ is \eqref{eq:UABdef} for each leaf $ \Sc_v $ --using that $ \ct{m}{^a} $ and $ \ctcn{P}{^a_b} $ are the normal and the projector to each cut for constant $ v $, respectively. Now, we have already shown that (see \cref{thm:rho-tensor-foliation,eq:rho-foliation}) $ \ct{E}{^a_A}\ct{E}{^b_B}\ctcn{\rho}{_{ab}}\eqSv{v}\ctrd{v}{\rho}{_{AB}} $ where $ \ctrd{v}{\rho}{_{AB}} $ is the tensor of  \cref{thm:rho-tensor} or \cref{thm:rho-tensor-noncompact}  for each cut $ \Sc_{v} $. Hence, one deduces that $  \ct{E}{^a_A}\ct{E}{^b_B}\ctcn{V}{_{ab}}\eqSv{v}\ctrd{v}{V}{_{AB}} $ with $ \ctrd{v}{V}{_{AB}} $ the first component of news of \cref{thm:onepiece-news}  for each leaf $ \Sc_{v} $.
				\end{proof}
			 By means of \cref{eq:SAB-acceleration}, a formula for $ \ctcn{V}{_{AB}} $ in terms of the acceleration $ \ctcn{a}{_A} $, $ \cscn{K} $, $ \ctcn{\rho}{_{AB}} $ and the gauge invariant tensor field $ \ct{f}{_{AB}} $ is obtained for a general foliation:
				\begin{equation}\label{eq:VAB-acceleration}
					\ctcn{V}{_{AB}}=\ct{f}{_{AB}}-\ctcn{\rho}{_{AB}}+\cdcn{_A}\ctcn{a}{_B}-\ctcn{a}{_A}\ctcn{a}{_B}-\frac{1}{2}\mcn{_{AB}}\prn{\cdcn{_E}\ctcn{a}{^E}-\ctcn{a}{_E}\ctcn{a}{^E}-\cscn{K}}\spacef.
				\end{equation}
			 It is convenient to define the one-parameter family of tensor fields on $  \Scn $
				\begin{equation}
					\ct{\tau}{_{AB}}\defeq \cdcn{_A}\ctcn{a}{_B}-\ctcn{a}{_A}\ctcn{a}{_B}-\frac{1}{2}\mcn{_{AB}}\prn{\cdcn{_E}\ctcn{a}{^E}-\ctcn{a}{_E}\ctcn{a}{^E}-\cscn{K}}\spacef.
				\end{equation}
			Interestingly, its gauge change is the same as the one of $ \ctcn{U}{_{AB}} $ and $ \ctcn{\rho}{_{AB}} $ --see \cref{eq:gauge-behaviour-appropriate-foliations}-- and its trace coincides with the trace of $ \ctcn{\rho}{_{AB}} $, i.e., $ \ct{\tau}{^E_E}=\cscn{K} $. Furthermore, taking the covariant derivative of \cref{eq:VAB-acceleration}, antisymmetrising and using \cref{eq:rho-diff-eq-foliation}  one has
				\begin{equation}\label{eq:taudiffeq}
					\cdcn{_{[C}}\ct{\tau}{_{A]B}}=\cdcn{_{[C}}\prn{\ctcn{V}{_{A]B}}-\ct{f}{_{A]B}}}\spacef,
				\end{equation}
			which can be checked to be gauge-invariant, noting that $ \ctcn{V}{_{AB}}-\ct{f}{_{AB}} $ is a symmetric, traceless and gauge-invariant tensor that fulfils \cref{thm:gauge-invariant-diff-foliation}, and that $ \ctcn{\tau}{_{AB}} $ satisfies \cref{thm:general-gauge-trace-foliation} for $ a=1 $.
				\begin{lemma}
					The vanishing of the \emph{first component of news} $ \ctcn{V}{_{AB}} $ of \cref{thm:onepiece-news-foliations} on $ \scri $ can be written as a relation between the kinematical quantities of $ \ct{m}{^a} $ (shear, expansion and acceleration) and the curvature $ \cscn{K} $,
						\begin{equation}
							\ctcn{V}{_{AB}}=0 \iff \cdcn{_{[C}}\ct{f}{_{A]B}}=\cdcn{_{[C}}\prn{\cdcn{_{A]}}\ctcn{a}{_B}-\ctcn{a}{_{A]}}\ctcn{a}{_B}}-\frac{1}{2}\ctcn{q}{_{B[A}}\cdcn{_{C]}}\prn{\cdcn{_E}\ctcn{a}{^E}-\ctcn{a}{_E}\ctcn{a}{^E}-\cscn{K}}.
						\end{equation}
					Equivalently,
						\begin{equation}
							\ctcn{V}{_{AB}}=0 \iff \cdcn{_{[C}}\prn{\ct{\tau}{_{A]B}}-\ct{f}{_{A]B}}}=0\spacef.
						\end{equation}
				\end{lemma}
				\begin{corollary}
					If $\scri$ is strongly equipped, ergo the leaves of the foliation are umbilical ($ \ctcn{\Sigma}{_{AB}} $ vanishes on $ \scri $) then
						\begin{equation}
							\ctcn{V}{_{AB}}=0 \iff \cdcn{_{[C}}\prn{\cdcn{_{A]}}\ctcn{a}{_B}-\ctcn{a}{_{A]}}\ctcn{a}{_B}}-\frac{1}{2}\ctcn{q}{_{B[A}}\cdcn{_{C]}}\prn{\cdcn{_E}\ctcn{a}{^E}-\ctcn{a}{_E}\ctcn{a}{^E}-\frac{1}{2}\cscn{K}}=0.
						\end{equation}
					Equivalently,
						\begin{equation}
							\ctcn{V}{_{AB}}=0 \iff \cdcn{_{[C}}\ct{\tau}{_{A]B}}=0\spacef.
						\end{equation}
				\end{corollary}
				\begin{remark}
					The dependence on $ \cscn{\kappa} $ and $ \ctcn{\Sigma}{_{AB}} $ is encoded in $ \ct{f}{_{AB}} $, see \cref{eq:fakenewsdef}.
				\end{remark}
				\begin{proof}
					Firstly, take the derivative of \cref{eq:VAB-acceleration} with $ \cdcn{_C} $ and then antisymmetrise. Secondly, apply \cref{thm:codten-foliation}.
				\end{proof}
				
				Following the programme developed in \cref{sec:news}, we look now for the second components of news and construct a couple of traceless gauge invariant symmetric families of tensor fields $ \ctcnupm{X}{_{AB}} $ on $ \Scn $, such that the pullback to $ \scri $, $ \ctcnupm{X}{_{ab}}(v)\defeq \ctcn{W}{_a^A}\ctcn{W}{_b^B}\ctcnupm{X}{_{AB}} $, satisfies
					\begin{equation}\label{eq:pullback-XAB}
					 \ct{E}{^a_A}\ct{E}{^b_B}\ctcnupm{X}{_{ab}}\eqSv{v}\ctrdupm{v}{X}{_{AB}} \spacef,
					\end{equation}
				where $ \ctrdupm{v}{X}{_{AB}} $ are the (undetermined) second components of news defined in  \cref{sec:news} fulfilling \cref{eq:conditionPlus,eq:conditionMinus} on each cut $ \Sc_{v} $. The pullback of this pair of equations to $ \scri $, taken with respect to $ \cbrkt{\ct{W}{_b^B}}_{v} $, reads
					\begin{align}
						-\frac{1}{2}N\ctcn{P}{^e_b}\ct{D}{_e}&\eqSv{v} \ctcn{T}{_b}+\frac{1}{2}\ctcn{P}{^d_c}\cds{_d}\ctcnp{X}{_b^c}\spacef,\label{eq:conditionPlus-foliation-pre}\\
						\frac{1}{2}N\ctcn{P}{^e_b}\ct{D}{_e}&\eqSv{v} \ctcn{T}{_b}+\frac{1}{2}\ctcn{P}{^d_c}\cds{_d}\ctcnm{X}{_b^c}\spacef.\label{eq:conditionMinus-foliation-pre}
					\end{align}
				Because $ \ct{m}{^a} $ is orthogonal to each cut, observe that $ \ctcn{P}{^a_b}\eqSv{v}\ctc{P}{^a_b} $, $ \ctcn{\epsilon}{_{ab}}=\ct{W}{_a^A}\ct{W}{_b^B}\ctc{\epsilon}{_{AB}} $ and $ \ctcn{T}{_b}\defeq\ctcn{W}{_b^B}\ctcn{T}{_B}=\ct{W}{_b^B}\ct{T}{_B} $. Thus, remarking that $ \scri=\underset{v}{\cup} \Sc_{v} $ and $ \Sc_{\hat{v}_1}\cap\Sc_{\hat{v}_2}=\emptyset $  for $ \hat{v}_1\neq\hat{v}_2 $, \cref{eq:conditionPlus-foliation-pre,eq:conditionMinus-foliation-pre} hold everywhere on $ \scri $ and one can take the push-forward to $ \Scn $ using $ \cbrkt{\ctcn{E}{^a_A}} $,
					\begin{align}
						-\frac{1}{2}N\ctcn{D}{_B}&= \ctcn{T}{_B}+\frac{1}{2}\cdcn{_C}\ctcnp{X}{_B^C}\spacef,\label{eq:conditionPlus-foliation}\\
						\frac{1}{2}N\ctcn{D}{_B}&= \ctcn{T}{_B}+\frac{1}{2}\cdcn{_C}\ctcnm{X}{_B^C}\spacef,\label{eq:conditionMinus-foliation}
					\end{align}
				Then, one has on $ \Scn $ (equivalently, on $ \scri $ after taking the pull-back)
					\begin{align}
						N^2\csp{\Z}&= \cdcn{_C}\prn{\ctcnp{n}{_B^C}}\cdcn{_D}\prn{\ctcnp{n}{^{BD}}}\spacef,\label{eq:divNewsZp-foliation}\\
						N^2\csm{\Z}&= \cdcn{_C}\prn{\ctcnm{n}{_B^C}}\cdcn{_D}\prn{\ctcnm{n}{^{BD}}}\spacef,\label{eq:divNewsZm-foliation}
					\end{align}		
				with 
					\begin{align}
						\ctcnp{n}{_{AB}}\defeq \ct{V}{_{AB}}+\ctcnp{X}{_{AB}}\spacef,\label{eq:nABplus-foliations}\\
						\ctcnm{n}{_{AB}}\defeq \ct{V}{_{AB}}+\ctcnm{X}{_{AB}}\spacef,\label{eq:nABminus-foliations}
					\end{align}
				such that $ \ctcnupm{n}{_{ab}}\defeq\ctcn{W}{_a^A}\ctcn{W}{_b^B}\ctcnupm{n}{_{AB}} $ fulfils 
					\begin{equation}\label{eq:news-foliation-cut}
						\ct{E}{^a_A}\ct{E}{^b_B}\ctcnupm{n}{_{ab}}\eqSv{v}\ctpm{n}{_{AB}}\quad\forall v\spacef.
					\end{equation}
				A  generalisation of \cref{thm:theTwoNewsProp} can be written for strictly equipped $\scri$:
					\begin{prop}[Radiant news on strictly equipped $ \scri $ with $\mathbb{S}^2$ leaves]\label{thm:theTwoNewsProp-foliation}
						Assume that $\scri$ is strictly equipped
						with $ v $ the parameter along the curves \eqref{eq:parameter-along-curves} and that the leaves have $ \mathbb{S}^2 $-topology. Then, if the condition of \cref{eq:conditionPlus-foliation} (\cref{eq:conditionMinus-foliation}) holds,
						\begin{align}
						\ctcnp{n}{_{AB}}= 0 \iff \csp{\Z}= 0\spacef,\quad\\
						\prn{\ctcnm{n}{_{AB}}= 0 \iff \csm{\Z}= 0}\quad
						\end{align}
						where $ \ctcnp{n}{_{AB}} $ ($ \ctcnm{n}{_{AB}} $) is the one-parameter family of tensor fields on $ \Scn $ given by \cref{eq:nABplus-foliations} (\cref{eq:nABminus-foliations}) that 
						fulfils \cref{it:propertyNewsCandF,it:propertyNewsGaugeInvariant,it:propertyNewsSymmetric,it:propertyNewsTraceless,it:propertyNewsTwoDim,it:propertyNewsZ} on page \pageref{it:propertyNewsCandF}. Its pullback to $ \scri $ as
							\begin{equation}
								\ctcnp{n}{_{ab}} \defeq \ctcn{W}{_a^A}\ctcn{W}{_b^B}\ctcnp{n}{_{AB}},\quad\ct{m}{^a}\ctcnp{n}{_{ab}}=0\quad
							\end{equation}
							analogously for $ \ctcnm{n}{_{ab}}  $
						is a $ v $-dependent tensor field on $ \scri $ fulfilling \cref{eq:news-foliation-cut}. Hence, we call it {\em radiant news}  on $ \scri $ for the radiant $ \ctps{\Q}{^a} $ ($ \ctms{\Q}{^k}  $).
				\end{prop}
				\begin{proof}
					By definition \cref{eq:nABplus-foliations} (\cref{eq:nABminus-foliations}) the one-parameter family of tensor fields $ \ctcnp{n}{_{AB}} $ ($ \ctcnm{n}{_{AB}} $)  satisfies \cref{it:propertyNewsTwoDim,it:propertyNewsSymmetric,it:propertyNewsTraceless,it:propertyNewsGaugeInvariant} on page \pageref{it:propertyNewsTwoDim}.  \Cref{it:propertyNewsCandF} is fulfilled as well, which can be checked by inspection of \cref{eq:conditionPlus-foliation,eq:CA-V-foliation} (\cref{eq:conditionMinus-foliation,eq:CA-V-foliation}). Now, $ \ctcnp{n}{_{AB}}$ ($ \ctcnm{n}{_{AB}} $) is symmetric and traceless on $ \Scn $ and by \cref{thm:codten-foliation} $ \ctcnp{n}{_{AB}}= 0\iff \cdcn{_C}\ctcnp{n}{_B^C}= 0 $ ($ \ctcnm{n}{_{AB}}= 0\iff \cdcn{_C}\ctcnm{n}{_B^C}= 0 $). But this vanishes if and only if  $ \csp{\Z}=0 $ ($ \csm{\Z}=0 $) because of \cref{eq:divNewsZp-foliation} (\cref{eq:divNewsZm-foliation}).
				\end{proof}
				
					\begin{prop}[Radiant pseudo-news on strictly equipped $ \scri $ with non-$\mathbb{S}^2$ leaves]\label{thm:theTwoNewsProp-foliation-non-S2}
					Assume $\scri$ is strictly equipped
					with $ v $ the parameter along the curves \eqref{eq:parameter-along-curves} and assume the conditions of \cref{thm:onepiece-news-non-compact-foliations}. Then, if the condition of \cref{eq:conditionPlus-foliation} (\cref{eq:conditionMinus-foliation}) holds,
					\begin{align}
					\ctcnp{n}{_{AB}}= 0 \implies \csp{\Z}= 0\spacef,\\
					(\ctcnm{n}{_{AB}}= 0 \implies \csm{\Z}= 0)\spacef,
					\end{align}
					where $ \ctcnp{n}{_{AB}} $ ($ \ctcnm{n}{_{AB}} $) is the one-parameter family of tensor fields on $ \Scn $ given by \cref{eq:nABplus-foliations} (\cref{eq:nABminus-foliations}) that has the properties \cref{it:propertyNewsCandF,it:propertyNewsGaugeInvariant,it:propertyNewsSymmetric,it:propertyNewsTraceless,it:propertyNewsTwoDim} on page \pageref{it:propertyNewsCandF}, but in general it does not fulfils \cref{it:propertyNewsZ}. One defines its pullback to $ \scri $ as
					\begin{equation}
					\ctcnp{n}{_{ab}} \defeq \ctcn{W}{_a^A}\ctcn{W}{_b^B}\ctcnp{n}{_{AB}},\quad\ct{m}{^a}\ctcnp{n}{_{ab}}=0\quad
					\end{equation}
					and analogously for $ \ctcnm{n}{_{ab}} $. 
					The tensor field $ \ctcnp{n}{_{ab}}  $ ($ \ctcnm{n}{_{ab}} $) is a $ v $-dependent tensor field on $ \scri $ fulfilling \cref{eq:news-foliation-cut}. 
				\end{prop}
				\begin{proof}
					The proof is very much as the one in \cref{thm:theTwoNewsProp-foliation}, except for that now the tensor $ \ctcn{V}{_{AB}} $ in \cref{eq:CA-V-foliation} corresponds to the one of \cref{thm:onepiece-news-non-compact-foliations}. Then, by \cref{eq:divNewsZp-foliation} one has $ \csp{\Z}=0 \implies \ctcnp{n}{_{AB}}=0$. Due to the non-$ \mathbb{S}^2 $ topology of the cuts, \cref{thm:codten-foliation} does not apply and the inverse implications does not follow.
				\end{proof}
			
				\subsubsection{Relation to the radiation condition}
				We have shown that under appropriate conditions radiant news  $ \ctcnupm{n}{_{ab}} $ for $ \ctpms{\Q}{^\alpha} $ exist as tensor fields on strictly equipped $ \scri $. 
			  Next task of our programme is to find equations for the derivatives along $ \ct{m}{^a} $ of these objects. In principle, guiding ourselves by the $ \Lambda=0 $ case, the derivative along the `evolution' direction of a radiant news-like object should be related to $ \cspm{\W} $. The approach that we will follow is similar to the one in \cref{ssec:conditions-news}.\\
				
				Begin contracting \cref{eq:magneticSchouten} with $ \ctcn{E}{^a_A}\ctcn{E}{^b_B} $ and symmetrising
					\begin{equation}\label{eq:CtAB-schouten}
						N\ctcn{\epsilon}{_{A}^E}\ctt{C}{_{BE}}\eqc \ctcn{S'}{_{AB}}-\ctcn{\kappa}{_{(A}^D}\ctcn{S}{_{B)D}}-\cdcn{_{(A}}\ctcn{S}{_{B)}}-\csS{\ubar{S}}\ctcn{\kappa}{_{AB}}+2\ctcn{S}{_{(B}}\ctcn{a}{_{A)}}\spacef,
					\end{equation}
				where we have defined
					\begin{equation}
						\ctcn{S'}{_{AB}}\defeq \ctcn{E}{^a_A}\ctcn{E}{^b_B}\lied_{\vec{m}}\cts{S}{_{ab}}\spacef.
					\end{equation}
				Note that $ \ctcn{\epsilon}{_{(A}^E}\ctcn{C}{_{B)E}}= \ctcn{\epsilon}{_{(A}^E}\ctt{C}{_{B)E}}= \ctcn{\epsilon}{_{A}^E}\ctt{C}{_{BE}} $ and also that $ \ctcn{S'}{_{AB}} =\lied_{\vec{m}}\ctcn{S}{_{AB}}$ --see \cref{eq:liedEcongruence}. \Cref{eq:CtAB-schouten} can be expressed in terms of $ \ctcnupm{C}{_{AB}} $ using \cref{it:nullDecompProp11,it:nullDecompProp11F} on page \pageref{it:nullDecompProp11} of \cref{app:lightlike-projections} as
					\begin{equation}\label{eq:CpmAB-schouten}
						N\ctcn{\epsilon}{_B^E}\ctcnupm{C}{_{AE}}= \ctcn{S'}{_{AB}}-\ctcn{\kappa}{_{(A}^D}\ctcn{S}{_{B)D}}-\cdcn{_{(A}}\ctcn{S}{_{B)}}-\csS{\ubar{S}}\ctcn{\kappa}{_{AB}}+2\ctcn{S}{_{(B}}\ctcn{a}{_{A)}}\pm N\ctt{D}{_{AB}}\spacef.
					\end{equation} 
				We can write this equation in terms of $ \ctcn{V}{_{AB}} $, $ \ctcn{\rho}{_{AB}} $, $ \ct{L}{_{AB}} $,
					\begin{align}\label{eq:CpmABdual-V}
						N\ctcn{\epsilon}{_B^E}\ctcnupm{C}{_{AE}}&= \ct{V'}{_{AB}}+\ctcn{\rho'}{_{AB}}-\ctcn{L'}{_{AB}}-\ctcn{\Sigma}{_{(A}^D}\prn{\ct{V}{_{B)D}}+\ctcn{\rho}{_{B)D}}-\ctcn{L}{_{B)D}}}\nonumber\\
						-&\frac{1}{2}\cscn{\kappa}\prn{\ct{V}{_{AB}}+\ctcn{\rho}{_{AB}}-\ctcn{L}{_{AB}}}-\cdcn{_{(A}}\ctcn{S}{_{B)}}-\csS{\ubar{S}}\ctcn{\kappa}{_{AB}}+2\ctcn{S}{_{(B}}\ctcn{a}{_{A)}}\pm N\ctt{D}{_{AB}}\spacef,
					\end{align} 
				where, in addition, we have expanded $ \ctcn{\kappa}{_{AB}} $ in terms of $ \ctcn{\Sigma}{_{AB}} $ and $ \cscn{\kappa} $ and defined
					\begin{align}
						\ctcn{\rho'}{_{AB}}&\defeq \lied_{\vec{m}}\ctcn{\rho}{_{AB}}\spacef,\\
						\ctcnupm{n'}{_{AB}}&\defeq \lied_{\vec{m}}\ctcnupm{n}{_{AB}}\spacef.
					\end{align}
				A similar expression follows for $ \ctcnupm{C}{_{AB}} $,
					\begin{align}\label{eq:CpmAB-V}
						N\ctcnupm{C}{_{AB}}&=\ctcn{\epsilon}{_B^E}\bbrkt{ \ctcn{V'}{_{AE}}+\ctcn{\rho'}{_{AE}}-\ctcn{L'}{_{AE}}-\ctcn{\Sigma}{_{(A}^D}\prn{\ctcn{V}{_{E)D}}+\ctcn{\rho}{_{E)D}}-\ctcn{L}{_{E)D}}}\nonumber\\
						-&\frac{1}{2}\cscn{\kappa}\prn{\ctcn{V}{_{AE}}+\ctcn{\rho}{_{AE}}-\ctcn{L}{_{AE}}}-\cdcn{_{(A}}\ctcn{S}{_{E)}}-\csS{\ubar{S}}\ctcn{\kappa}{_{AE}}+2\ctcn{S}{_{(E}}\ctcn{a}{_{A)}}\pm N\ctt{D}{_{AE}}}\spacef.
					\end{align} 
				Now, we propose the following `transport' equations for $ \ctcnupm{n}{_{AB}} $:
					\begin{align}
						N\ctcn{\epsilon}{_B^E}\ctp{C}{_{AE}}&= \ctp{n'}{_{AB}}-\ctcn{\Sigma}{_{(A}^C}\ctp{n}{_{B)C}}\spacef,\label{eq:CpAB-nAB}\\
						N\ctcn{\epsilon}{_B^E}\ctm{C}{_{AE}}&= \ctm{n'}{_{AB}}-\ctcn{\Sigma}{_{(A}^C}\ctm{n}{_{B)C}}\spacef,\label{eq:CmAB-nAB}
					\end{align}
				with $ \ctpm{n}{_{AB}} $ defined as in  \cref{eq:nABplus,eq:nABminus}. The square of this expressions reads
					\begin{align}
						N^2\csp{\W}&=\prn{\ctp{n'}{_{AB}}-\ctcn{\Sigma}{_{(A}^C}\ctp{n}{_{B)C}}}\prn{\ctp{n'}{^{AB}}-\ctcn{\Sigma}{^{(A}_C}\ctp{n}{^{B)C}}}\spacef,\label{eq:WpNewsp}\\
						N^2\csm{\W}&=\prn{\ctm{n'}{_{AB}}-\ctcn{\Sigma}{_{(A}^C}\ctm{n}{_{B)C}}}\prn{\ctm{n'}{^{AB}}-\ctcn{\Sigma}{^{(A}_C}\ctm{n}{^{B)C}}}\spacef.\label{eq:WmNewsm}
					\end{align} 
				Let us remark that \cref{eq:CpAB-nAB,eq:CmAB-nAB} are gauge invariant, which follows from the gauge transformations presented in \cref{app:gauge-transformations}, from where the next result is derived as well:
					\begin{lemma}
						Let $ \ct{j}{_{ab}} $ be any symmetric gauge invariant tensor field on equipped $ \scri $ orthogonal to $ \ct{m}{^a} $, i.e., $ \ct{m}{^a}\ct{j}{_{ab}} =0$. Then,
							\begin{equation}
							\ctg{j'}{_{ab}}=\frac{1}{\omega}\ct{j'}{_{ab}}\spacef,
							\end{equation}
						where $ \ctcn{j'}{_{ab}}\defeq \lied_{\vec{m}}\ctcn{j}{_{ab}} $.
					\end{lemma}	
				The sufficient and necessary conditions for \cref{eq:CpAB-nAB,eq:CmAB-nAB} to hold are, respectively:
					\begin{align}
						-N\ctt{D}{_{AB}}&= -\ctp{X'}{_{AB}}+\ctp{X}{_{C(B}}\ctcn{\Sigma}{_{A)}^C}-\ctcn{\kappa}{_{(A}^D}\prn{\ctcn{\rho}{_{B)D}}-\ctcn{L}{_{B)D}}}-\frac{1}{2}\cscn{\kappa}\prn{\ct{V}{_{AB}}}-\ct{L'}{_{AB}}\nonumber\\
						&+\ct{\rho'}{_{AB}}-\cdcn{_{(A}}\ctcn{S}{_{B)}}-\csS{\ubar{S}}\ctcn{\kappa}{_{AB}}+2\ctcn{S}{_{(B}}\ctcn{a}{_{A)}}\spacef,\label{eq:conditionWPlus}\\
						N\ctt{D}{_{AB}}&= -\ctm{X'}{_{AB}}+\ctm{X}{_{C(B}}\ctcn{\Sigma}{_{A)}^C}-\ctcn{\kappa}{_{(A}^D}\prn{\ctcn{\rho}{_{B)D}}-\ctcn{L}{_{B)D}}}-\frac{1}{2}\cscn{\kappa}\prn{\ct{V}{_{AB}}}-\ct{L'}{_{AB}}\nonumber\\
						&+\ct{\rho'}{_{AB}}-\cdcn{_{(A}}\ctcn{S}{_{B)}}-\csS{\ubar{S}}\ctcn{\kappa}{_{AB}}+2\ctcn{S}{_{(B}}\ctcn{a}{_{A)}}\spacef.\label{eq:conditionWMinus}
					\end{align}
	
				Therefore, by \cref{eq:WpNewsp,eq:WmNewsm}, one has
					\begin{lemma}\label{thm:noNewsnoW}
				Assume $\scri$ is strictly equipped 
				with $ v $ the parameter along the curves \eqref{eq:parameter-along-curves} and such that \cref{eq:m-congruences-sec} holds. Assume that the leaves have $ \mathbb{S}^2 $-topology and \cref{eq:conditionPlus-foliation,eq:conditionWPlus} (\cref{eq:conditionWMinus,eq:conditionMinus-foliation}) hold there. Then,
						\begin{align}
						\ctcnp{n}{_{ab}}= 0 &\implies \csp{\W}= 0\spacef.\\
						\bprn{\ctcnm{n}{_{ab}}= 0 &\implies \csm{\W}= 0} 	
						\end{align}
					\end{lemma}
				To see the effects of a vanishing $ \ctcnpm{n}{_{ab}} $ on the presence of radiation at $ \scri $, it is easier if one studies the relation with the radiant supermomenta first
					\begin{prop}[Radiant news and radiant supermomenta]\label{thm:noNewsiffnoQ}
						Under the same assumtptions of the previous Lemma
							\begin{align}
							\ctcnp{n}{_{ab}}= 0 \iff \ctp{\Q}{^\alpha}= 0\spacef.\quad\\
							\prn{\ctcnm{n}{_{ab}}= 0 \iff \ctm{\Q}{^\alpha}= 0}\spacef.\ 	
							\end{align}
					\end{prop}
					\begin{proof}
						We give the proof for $ \ctp{\Q}{^\alpha} $. By \cref{thm:theTwoNewsProp-foliation}, one has that $ \csp{\Z}=0 \iff \ctcnp{n}{_{AB}}=0$ and, by \cref{thm:noNewsnoW}, that $ \ctcnp{n}{_{AB}}=0\implies \csp{\W}=0$, therefore $ \ctp{n}{_{AB}}=0 \implies \ctp{\Q}{^\alpha}=0 $ --see \cref{it:noQnoWnoZ} on page \pageref{it:noQnoWnoZ}. For the converse, $ \ctp{\Q}{^\alpha}=0 \implies \csp{\W}=0=\csp{\Z} $ and, by \cref{thm:theTwoNewsProp-foliation} again, $  \csp{\Z}=0\implies \ctcnp{n}{_{AB}}=0 $.
					\end{proof}
				With this intermediate result, we are able to write a theorem for the asymptotic canonical super-Poynting vector field
					\begin{thm}[Asymptotic super-Poynting vector and radiant news]\label{thm:noNewspm-noPoynting}
					Assume $\scri$ strictly equipped such 
					 that the leaves have $ \mathbb{S}^2 $-topology and that \cref{eq:conditionPlus-foliation,eq:conditionWPlus,eq:conditionWMinus,eq:conditionMinus-foliation} hold there. Then,		
							\begin{equation}
							\ctcnp{n}{_{ab}}= 0 = \ctcnm{n}{_{ab}} \implies \cts{\P}{^a}= 0 \spacef.
							\end{equation}
					\end{thm}
					\begin{remark}
						According to \cref{def:criterionGlobal}, the result states that $ 	\ctcnp{n}{_{ab}}= 0 = \ctcnm{n}{_{ab}} \implies $ \emph{no radiation at $ \scri $}.
					\end{remark}
					\begin{proof}
						The proof follows directly by \cref{thm:noNewsiffnoQ} and \cref{thm:oneQnull-then}.
					\end{proof}
				
					\subsubsection{Possible generalisation}\label{sssec:possible-generalisation-congruences}
					We proceed to generalise the above results using the same technique of \cref{sssec:possible-generalisation}.  As before, we ask for a couple of families of traceless gauge invariant symmetric tensor fields $ \ctcnupm{X}{_{AB}} $ on $ \Scn $ satisfying \cref{eq:pullback-XAB} where this time  $ \ctrdupm{v}{X}{_{AB}} $ is the unknown tensor field of  \cref{sec:news} appearing in \cref{eq:lb-conditionPlus,eq:lb-conditionMinus} (instead of \cref{eq:conditionPlus,eq:conditionMinus}). The pullback of these equations to $ \scri $ reads
					\begin{align}
					-\frac{1}{2}N\ctcn{P}{^e_b}\ct{D}{_e}&\eqSv{v} \ctcn{T}{_b}+\prn{\csp{\lambda}-1}N\ctcn{\epsilon}{_{bc}}\ctcn{P}{^c_d}\ctp{C}{^d}+\csp{\beta}N\ctcn{P}{^e_b}\ctp{C}{_e}+\frac{1}{2}\ctcn{P}{^d_c}\cds{_d}\ctcnp{X}{_b^c}\spacef,\label{eq:lb-conditionPlus-foliation-pre}\\
					\frac{1}{2}N\ctcn{P}{^e_b}\ct{D}{_e}&\eqSv{v} \ctcn{T}{_b}+\prn{\csm{\lambda}-1}N\ctcn{\epsilon}{_{bc}}\ctcn{P}{^c_d}\ctm{C}{^d}+\csm{\beta}N\ctcn{P}{^e_b}\ctm{C}{_e}+\frac{1}{2}\ctcn{P}{^d_c}\cds{_d}\ctcnm{X}{_b^c}\spacef.\label{eq:lb-conditionMinus-foliation-pre}
					\end{align}
					On $ \Scn $, contracting with $ \cbrkt{\ctcn{E}{^a_A}} $,
					\begin{align}
					-\frac{1}{2}N\ctcn{D}{_B}&= \ctcn{T}{_B}+\prn{\cscnp{\lambda}-1}N\ctcn{\epsilon}{_{BC}}\ctcnp{C}{^C}+\cscnp{\beta}N\ctcnp{C}{_B}+\frac{1}{2}\cdcn{_C}\ctcnp{X}{_B^C}\spacef,\label{eq:lb-conditionPlus-foliation}\\
					\frac{1}{2}N\ctcn{D}{_B}&= \ctcn{T}{_B}+\prn{\cscnm{\lambda}-1}N\ctcn{\epsilon}{_{BC}}\ctcnm{C}{^C}+\cscnm{\beta}N\ctcnm{C}{_B}+\frac{1}{2}\cdcn{_C}\ctcnm{X}{_B^C}\spacef,\label{eq:lb-conditionMinus-foliation}
					\end{align}
					where
					\begin{equation}
					\cbrkt{\cscnpm{\lambda}\quad\Big/\quad \cscnpm{\lambda}\eqSv{v}\cspm{\lambda}\quad\forall v},\quad
					\cbrkt{\cscnpm{\beta}\quad\Big/\quad \cscnpm{\beta}\eqSv{v}\cspm{\beta}\quad\forall v}\quad
					\end{equation}	
					and we assume $ \cscnpm{\lambda} $ and $ \cscnpm{\beta} $ differentiable enough. A direct calculation provides on $ \Scn $ (equivalently, on $ \scri $ after taking the pull-back)
					\begin{align}
					N^2\prn{\cscnp{\beta}^2+\cscnp{\lambda}^2}\csp{\Z}&= \cdcn{_C}\prn{\ctcnp{n}{_B^C}}\cdcn{_D}\prn{\ctcnp{n}{^{BD}}}\spacef,\label{eq:lb-divNewsZp-foliation}\\
					N^2\prn{\cscnm{\beta}^2+\cscnm{\lambda}^2}\csm{\Z}&= \cdcn{_C}\prn{\ctcnm{n}{_B^C}}\cdcn{_D}\prn{\ctcnm{n}{^{BD}}}\spacef,\label{eq:lb-divNewsZm-foliation}
					\end{align}		
					where the definitions 
					\begin{align}
					\ctcnp{n}{_{AB}}\defeq \ct{V}{_{AB}}+\ctcnp{X}{_{AB}}\spacef,\label{eq:lb-nABplus-foliations}\\
					\ctcnm{n}{_{AB}}\defeq \ct{V}{_{AB}}+\ctcnm{X}{_{AB}}\spacef,\label{eq:lb-nABminus-foliations}
					\end{align}
					were introduced. In a similar fashion, recalling that $ \cspm{\W} $ vanishes if and only if $ \ctcnupm{C}{_{AB}} $ does so, we consider combinations of the form
					\begin{equation}\label{eq:lb-combinationCpmAB}
					\cspm{\delta}\ctcnupm{C}{_{AB}} + \cspm{\gamma}\ctcn{\epsilon}{_B^C}\ctcnupm{C}{_{AC}}\spacef,
					\end{equation}
					with $ \delta $, $ \gamma $ gauge-invariant, dimensionless, scalar functions obeying
					\begin{align}
					\cspm{\delta}= 0 \implies \cspm{\gamma}\neq 0\spacef,\\
					\cspm{\gamma}= 0 \implies \cspm{\delta}\neq 0\spacef.
					\end{align} 
					Now, we propose the following `transport' equations for $ \ctcnupm{n}{_{AB}} $:
					\begin{align}
					N\prn{\csp{\delta}\ctp{C}{_{AB}}+\csp{\gamma}\ctcn{\epsilon}{_B^E}\ctp{C}{_{AE}}}&= \ctcnp{n'}{_{AB}}-\ctcn{\Sigma}{_{(A}^C}\ctcnp{n}{_{B)C}}\spacef,\label{eq:lb-CpAB-nAB}\\
					N\prn{\csm{\delta}\ctm{C}{_{AB}}+\csm{\gamma}\ctcn{\epsilon}{_B^E}\ctm{C}{_{AE}}}&= \ctcnm{n'}{_{AB}}-\ctcn{\Sigma}{_{(A}^C}\ctcnm{n}{_{B)C}}\spacef,\label{eq:lb-CmAB-nAB}
					\end{align}
					The square of this expressions reads
					\begin{align}
					N^2\csp{\W}\prn{\csp{\delta}^2+\csp{\gamma}^2}&=\prn{\ctcnp{n'}{_{AB}}-\ctcn{\Sigma}{_{(A}^C}\ctcnp{n}{_{B)C}}}\prn{\ctcnp{n'}{^{AB}}-\ctcn{\Sigma}{^{(A}_C}\ctcnp{n}{^{B)C}}}\spacef,\label{eq:lb-WpNewsp}\\
					N^2\csm{\W}\prn{\csm{\delta}^2+\csm{\gamma}^2}&=\prn{\ctcnm{n'}{_{AB}}-\ctcn{\Sigma}{_{(A}^C}\ctcnm{n}{_{B)C}}}\prn{\ctcnm{n'}{^{AB}}-\ctcn{\Sigma}{^{(A}_C}\ctcnm{n}{^{B)C}}}\spacef.\label{eq:lb-WmNewsm}
					\end{align} 
					This time, the sufficient and necessary conditions for \cref{eq:lb-CpAB-nAB,eq:lb-CmAB-nAB} to hold are, respectively:
					\begin{align}
					-N\ctt{D}{_{AB}}&= -\ctp{X'}{_{AB}}+\ctp{X}{_{C(B}}\ctcn{\Sigma}{_{A)}^C}-\ctcn{\kappa}{_{(A}^D}\prn{\ctcn{\rho}{_{B)D}}-\ctcn{L}{_{B)D}}}-\frac{1}{2}\cscn{\kappa}\prn{\ct{V}{_{AB}}}-\ct{L'}{_{AB}}+\ct{\rho'}{_{AB}}\nonumber\\
					&-\cdcn{_{(A}}\ctcn{S}{_{B)}}-\csS{\ubar{S}}\ctcn{\kappa}{_{AB}}+2\ctcn{S}{_{(B}}\ctcn{a}{_{A)}}+N\prn{\csp{\gamma}-1}\ctcn{\epsilon}{_B^E}\ctp{C}{_{AE}}+N\csp{\delta}\ctp{C}{_{AB}}\spacef,\label{eq:lb-conditionWPlus}\\
					N\ctt{D}{_{AB}}&= -\ctm{X'}{_{AB}}+\ctm{X}{_{C(B}}\ctcn{\Sigma}{_{A)}^C}-\ctcn{\kappa}{_{(A}^D}\prn{\ctcn{\rho}{_{B)D}}-\ctcn{L}{_{B)D}}}-\frac{1}{2}\cscn{\kappa}\prn{\ct{V}{_{AB}}}-\ct{L'}{_{AB}}+\ct{\rho'}{_{AB}}\nonumber\\
					&-\cdcn{_{(A}}\ctcn{S}{_{B)}}-\csS{\ubar{S}}\ctcn{\kappa}{_{AB}}+2\ctcn{S}{_{(B}}\ctcn{a}{_{A)}}+N\prn{\csm{\gamma}-1}\ctcn{\epsilon}{_B^E}\ctm{C}{_{AE}}+N\csm{\delta}\ctm{C}{_{AB}}\spacef.\label{eq:lb-conditionWMinus}
					\end{align}
					
					Finally, one can write a generalised version of \cref{thm:noNewsnoW}, \cref{thm:noNewspm-noPoynting}, \cref{thm:theTwoNewsProp-foliation} and \cref{thm:noNewsiffnoQ} by means of  \cref{eq:lb-conditionWMinus,eq:lb-conditionWPlus,eq:lb-conditionMinus-foliation,eq:lb-conditionPlus-foliation} instead of \cref{eq:conditionWMinus,eq:conditionWPlus,eq:conditionMinus-foliation,eq:conditionPlus-foliation}.
					
		\subsection{Dealing with incoming radiation}\label{ssec:noincomingrad}
			We turn now to investigate possible ways of isolating \emph{outgoing} radiation from \emph{incoming} components at $ \scri $. This issue is relevant for characterising isolated sources which on physical grounds one expects to contain no incoming contributions but only to emit gravitational radiation that constitutes the outgoing component. In this section we will consider radiation arriving at the future component of the conformal boundary, $ \scri^+ $. The case of $ \scri^- $ can be treated similarly. Let us point out that the asymptotically flat scenario automatically has a structure adapted to the outgoing radiation due to the lightlike character of $ \scri^+ $. In simple words, when $ \Lambda=0 $ the radiation arriving at infinity and escaping from the space-time necessarily follows lightlike directions transversal to the conformal boundary. Therefore, the generators of $ \scri^+ $ can be considered to point along the direction of propagation of incoming radiation or, from another point of view, incoming radiation never propagates transversally to $ \scri^+ $. In contrast, the $ \Lambda>0 $ case presents the following difficulty: \emph{every radiation component, incoming or outgoing, crosses $ \scri^+ $ and escapes from the space-time}. Hence, one is left with the problem of specifying physically reasonable conditions capable of ruling out one of the radiative components --in our setting the incoming one, by definition. This sort of constraints sometimes receives the name of \emph{no incoming radiation conditions}. There is already a proposal \cite{Ashtekar2019} in the literature which  requires information from the physical space-time. Since according to \cref{def:criterionGlobal} the presence of radiation at $ \scri^+ $ is determined by the information encoded in $\prn{\scri^+, \ms{_{ab}}, \ct{D}{_{ab}}}$ --see  \cref{it:crit-prop4} on page \pageref{it:crit-prop4} and \cref{rmk:criterion}--, we believe that absence of incoming radiation \emph{\fblue{at $ \scri^+$}} should be encoded upon this same data.\\
			
\begin{figure}[h]
				\begin{subfigure}{0.5\textwidth}
					\begin{flushleft}
						\includegraphics[width=0.93\textwidth]{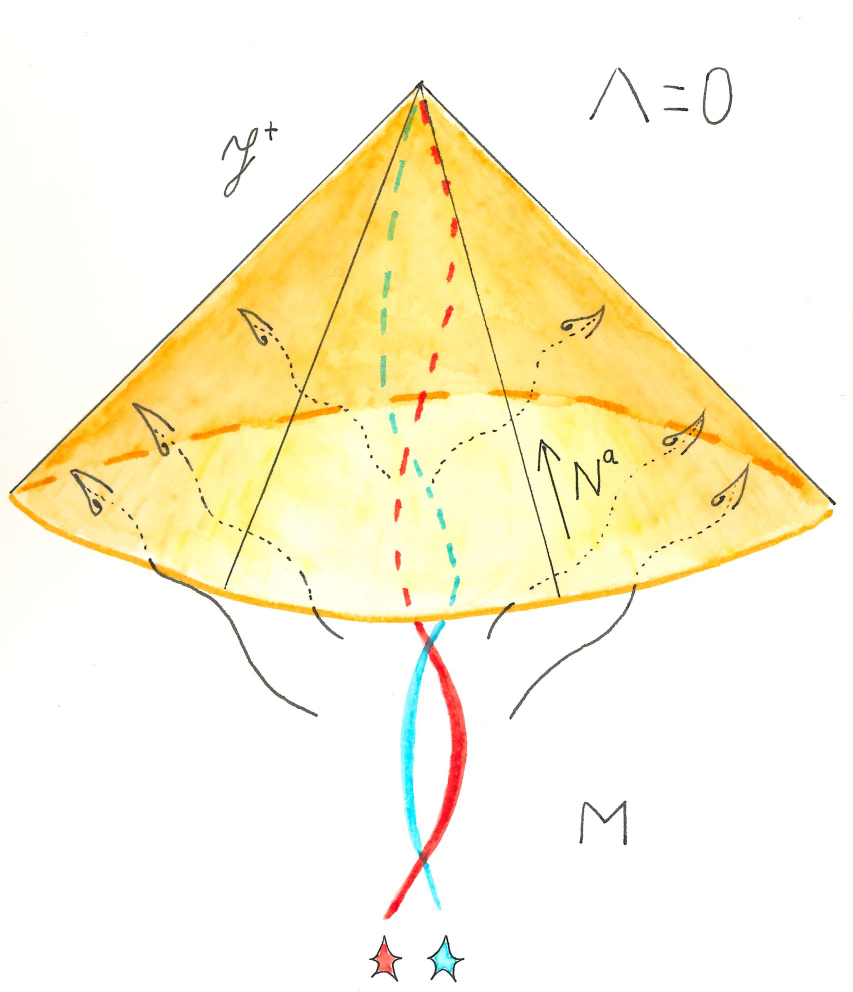}
					\end{flushleft}
				 \end{subfigure}%
 				\begin{subfigure}{0.5\textwidth}
					\begin{flushright}
						\includegraphics[width=0.93\textwidth]{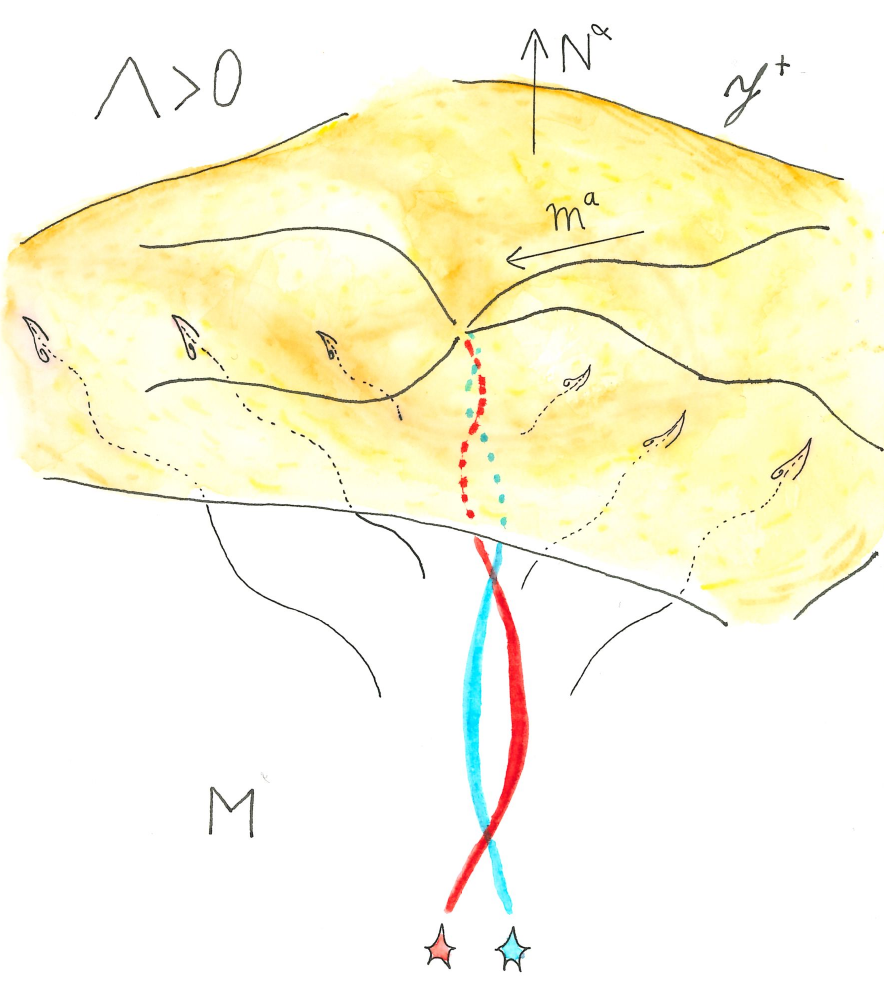}
					\end{flushright}
				\end{subfigure}
				\caption[Comparison between $ \Lambda>0$ and $ \Lambda=0 $-scenarios]{On the left: the asymptotically flat case, where the generators of $ \scri $ rule the natural evolution direction and outgoing radiation crosses $ \scri^+ $ transversally. On the right: the  $ \Lambda>0 $ scenario, where any direction of propagation of gravitational radiation is transversal to $ \scri^+ $ and \cref{def:noIncomingRadDefOpen} selects an intrinsic `evolution' direction given by  $ \ct{m}{^a} $, which points towards the region where the source meets $ \scri^+ $.}\label{fig:scri-ds-afs}
			\end{figure}

			Motivated by the $ \Lambda=0 $ case \cite{Fernandez-Alvarez_Senovilla20,Fernandez-Alvarez_Senovilla-afs}, it is reasonable to think that the vanishing of a radiant supermomentum $ \ctl{\Q}{^\alpha} $ is related to the absence of radiation propagating transversally to the null direction $ \ct{\ell}{^\alpha} $. This suggests that in our setup the vanishing of one radiant supermomenta, say $ \ctm{\Q}{^\alpha} $, could suppress the radiation travelling along transversal directions, in particular along $ \ctp{k}{^\alpha} $. Looking at the definition in \cref{eq:kp-def-m}, this restriction automatically turns $ \ct{m}{^a} $ into an intrinsic incoming direction field which in particular equips
			$ \scri^+ $  --or the open portion $ \Delta\subset\scri^+ $ with the same topology where $ \ctm{\Q}{^\alpha} $ vanishes. \emph{Observe that the `incoming direction' nomenclature here simply means opposite to $ -\ct{m}{^a} $}. In view of these properties, it makes sense to consider $ \ct{m}{^a} $ as an intrinsic `evolution direction' on $ \scri^+ $: if we compare with the $ \Lambda=0 $ case, the incoming direction given by the generators of $ \scri^+ $ defines the evolution direction; the analogy goes further if we notice that the vector field $ \ct{m}{^a} $ points towards the region where the worldlines of the isolated sources meet $ \scri^+ $ --see \cref{fig:scri-ds-afs}. 
			As a further positive property, the restriction $ \ctm{\Q}{^\alpha}=0 $ can be expressed entirely with the information available in $ \prn{\scri^+, \ms{_{ab}}, \ct{D}{_{ab}}} $:
			
				\begin{lemma}\label{thm:nircCD}
					Let $ \scri$ (or an open portion thereof) be equipped as in \cref{def:additional-structure} and define $ \ctm{\Q}{^\alpha} $ according to definitions of \cref{eq:Qmdef,eq:km-def-m}. Then
						\begin{equation}
							\ctm{\Q}{^\alpha}=0 \iff 	\ct{D}{_{ab}}-\frac{1}{2}\ct{D}{_{ef}}\ct{m}{^e} \ct{m}{^f} \prn{3\ct{m}{_a} \ct{m}{_b} - \ms{_{ab}} }= \ct{m}{^d}\ct{\epsilon}{_{ed(a}} \prn{\ct{C}{_{b)}^e} +\ct{m}{_{b)} }\ct{m}{^f} \ct{C}{_f^e}}\spacef.
						\end{equation}
				\end{lemma}
				\begin{proof}
					By \cref{eq:Wmdef,eq:Zmdef} and \cref{it:noQnoWnoZ} on page \pageref{it:noQnoWnoZ}, $ \ctm{\Q}{^\alpha}=0\iff \ctcnm{D}{_{AB}}=\ctcnm{C}{_{AB}}=0=\ctcnm{D}{_A}=\ctcnm{C}{_A} $. The result is obtained setting these values into \cref{eq:Dab-decomposition} and using \cref{it:nullDecompProp18} on page \pageref{it:nullDecompProp18}.
				\end{proof}
			 Then, our proposal 
			 reads as follows:
				\begin{crit}[No incoming radiation on $ \scri^+ $]\label{def:noIncomingRadDefOpen}
					We say that there is no incoming radiation 
					(or on an open portion $ \Delta\subset\scri^+ $ with the same topology) propagating along a vector field $  \ct{m}{^a}  $ on $ \scri^+ $, if $ \ct{m}{^a} $  is such that, according to definitions \lmultieqref{eq:Qmdef}\rmultieqref{eq:km-def-m},
					\begin{equation}\label{eq:no-ir}
					\ctm{\Q}{^\alpha}= 0 \quad
					\end{equation}
					or equivalently
					\begin{equation}\label{eq:no-ir-C-D}
					\ct{D}{_{ab}}-\frac{1}{2}\ct{D}{_{ef}}\ct{m}{^e} \ct{m}{^f} \prn{3\ct{m}{_a} \ct{m}{_b} - \ms{_{ab}} }= \ct{m}{^d}\ct{\epsilon}{_{ed(a}} \prn{\ct{C}{_{b)}^e} +\ct{m}{_{b)} }\ct{m}{^f} \ct{C}{_f^e}}\spacef.
					\end{equation}
				\end{crit}
				\begin{remark}
					It has to be stressed that the `incoming' terminology refers to components of the gravitational field at $ \scri^+ $.
				\end{remark}
				\begin{remark}
					Equivalently, there is no incoming radiation propagating along $ \ct{m}{^a} $ on $ (\Delta\subset)\scri^+ $ when $ \ct{m}{^a} $ defines a strong orientation there --see \cref{def:orientationStrong}.
				\end{remark}
				\begin{remark}
					If criterion \ref{def:noIncomingRadDefOpen} holds, all the components of $ \ct{D}{_{ab}} $ except $ \ct{m}{^a}\ct{m}{^b}\ct{D}{_{ab}} $ are determined by $ \ct{C}{_{ab}} $. This is in close analogy to what happens at the conformal boundary for $ \Lambda=0 $ where the `electric' part of the rescaled Weyl tensor defined with respect to the null normal $ \ct{N}{^\alpha} $ (which algebraically is of the kind \eqref{eq:DpDef}), is determined by the `magnetic' part (which is of the sort \eqref{eq:CpDef}) except for the $ \ct{N}{^\alpha}\ct{N}{^\beta} $ component. In both scenarios, this free component carries the information related to the Coulomb part of the gravitational field (see \cref{eq:np-coulomb}). This evinces that \cref{def:noIncomingRadDefOpen} is a constraint that affects the radiative degrees of freedom.
				\end{remark}
				
				In other words, criterion \ref{def:noIncomingRadDefOpen} identifies a class of space-times which can be safely considered to describe situations with only outgoing gravitational radiation arriving at $\scri^+$: those where the free data $\ct{D}{_{ab}}$ are determined by the intrinsic geometry of $(\scri^+,\ct{h}{_{ab}})$ according to \eqref{eq:no-ir-C-D} (for unit  $ \ct{m}{^a} $) except for the one component $ \ct{m}{^a}\ct{m}{^b}\ct{D}{_{ab}} $ which remains as the only extra free data independent of $(\scri^+,\ct{h}{_{ab}})$. It seems  interesting to study in deep this class of space-times.
				
			As a consequence of \cref{thm:oneQnull-then} one has
				\begin{corollary}\label{thm:oneQnull-nirc}
					Assume that the condition of \cref{def:noIncomingRadDefOpen} holds. Then,
						\begin{equation}
							\ctp{\Q}{^\alpha}=0 \iff \cts{\P}{^a}=0.
						\end{equation}
				\end{corollary}
				\begin{remark}
					This provides further support to \cref{def:noIncomingRadDefOpen} because if condition \eqref{eq:no-ir} holds, the presence of gravitational waves at $ \scri^+ $ (or on an open portion $ \Delta\in\scri^+ $) is completely determined, according to our \cref{def:criterionGlobal}, by the \emph{outgoing} components of the radiation ---which are associated to $ \ctp{\Q}{^\alpha} $. 
				\end{remark}
			Of especial interest is the case of a strictly equipped $\scri^+$, so that $ \ct{m}{^a} $ defines a foliation ($ \iff \ctcn{\omega}{_{ab}}=0 $). In particular,
				\begin{lemma}
					Let $ \scri^+ $ by strictly equipped such that $ \ct{m}{^a} $ satisfies condition \eqref{eq:no-ir} of \cref{def:noIncomingRadDefOpen}. Assume that conditions in \cref{thm:theTwoNewsProp-foliation} and  \cref{eq:conditionMinus-foliation} hold on $ \scri^+ $. Then,
					\begin{equation}
					\ctcnm{n}{_{AB}}= 0\quad
					\end{equation}
					with $ \ctcnm{n}{_{AB}} $ one of the news tensor fields of \cref{thm:theTwoNewsProp-foliation}.
					\begin{proof}
						On the one hand, condition in \cref{def:noIncomingRadDefOpen} is saying that $ \ctm{\Q}{^\alpha}=0 $, which implies $ \csm{\Z}=0 $. On the other hand, because of \cref{eq:conditionMinus-foliation}, \cref{thm:theTwoNewsProp-foliation} tells us $ \csm{\Z}=0\iff \ctcnm{n}{_{AB}}=0 $.
					\end{proof}
				\end{lemma}
			And for the particular case with $\scri^+$ strongly equipped: 
					\begin{lemma}\label{thm:noincoming-np}
					If $\scri^+$ is strongly equipped 
					($ \ctcn{\Sigma}{_{ab}}=0 $) with $\mathbb{S}^2$ leaves and the condition \eqref{eq:no-ir} of \cref{def:noIncomingRadDefOpen} is satisfied, then there always exists the radiant news $ \ctcnp{n}{_{AB}} $ of \cref{thm:theTwoNewsProp-foliation} and is given by
					\begin{equation}
					\ctcnp{n}{_{AB}}=2\ctcn{V}{_{AB}}\spacef,
					\end{equation}
					where $ \ctcn{V}{_{AB}} $ is the \emph{first component of news} of \cref{thm:onepiece-news-foliations}.
					\begin{proof}
						Note that \cref{eq:no-ir} imposes $ \ctcnp{C}{^A}=\ctcn{C}{^A} $ (and $ \ctcn{D}{_A}\ctcn{\epsilon}{^{AB}}=-\ctcn{C}{^B} $, see \cref{thm:oneQvanishes}) and umbilicity implies $ \ctcn{T}{_A}=0 $ (see \eqref{eq:TA-congruences}). It is easy to see that \cref{eq:conditionPlus-foliation} is satisfied with $ \ctcnp{X}{_{AB}}=\ctcn{V}{_{AB}} $. The result follows then by \cref{thm:theTwoNewsProp-foliation}.
					\end{proof}
				\end{lemma}
				\begin{lemma}\label{thm:noincoming-np-non-compact}
						Assume $\scri^+$ is strongly equipped with leaves that are non-necessarily topological-$\mathbb{S}^2$ and
						such that condition \eqref{eq:no-ir} of \cref{def:noIncomingRadDefOpen} is satisfied.  
						Assume also that there is a vector field $ \ct{\chi}{^a} $ such that $ \ctcn{\chi}{^A}\defeq \ctcn{W}{_a^A}\ct{\chi}{^a} $ is a CKVF with a fixed point on each leaf
						 Then, there always exists the radiant pseudo-news $ \ctcnp{n}{_{AB}} $ of \cref{thm:theTwoNewsProp-foliation-non-S2} and is given by
					\begin{equation}
					\ctcnp{n}{_{AB}}=2\ctcn{V}{_{AB}}\spacef,
					\end{equation}
					where $ \ctcn{V}{_{AB}} $ is the \emph{first component of news} of \cref{thm:onepiece-news-non-compact-foliations}.
					\begin{proof}
						The proof follows as in \cref{thm:noincoming-np}, but now one uses \cref{thm:theTwoNewsProp-foliation-non-S2} instead of \cref{thm:theTwoNewsProp-foliation}.
					\end{proof}
				\end{lemma}
				\begin{remark}
					If instead one uses the generalised approach of \cref{sssec:possible-generalisation-congruences}, it is possible to show in a similar fashion that solutions 
						\begin{equation}
						\ctcnp{n}{_{AB}}=2\csp{\lambda}\ctcn{V}{_{AB}}\quad
						\end{equation}
					always exist, where the values $ \csp{\lambda}=\text{constant} $ and $ \csp{\beta}=0 $ are fixed. It may be the case that the value of $ \csp{\lambda}$ can be fixed by physical arguments.
				\end{remark}

			If we use \cref{eq:conditionPlus-foliation,eq:conditionWPlus}, we end up with a theorem on the presence of radiation,
				\begin{thm}[Asymptotic super-Poynting and radiant news under \Cref{def:noIncomingRadDefOpen}]\label{thm:noradIncoming} 
					Assume that $ \scri^+ $ is strictly equipped with $ \mathbb{S}^2 $ leaves and that condition \eqref{eq:no-ir} of \cref{def:noIncomingRadDefOpen} holds. Assume also that \cref{eq:conditionPlus-foliation,eq:conditionWPlus} hold on $ \scri^+ $. Then, radiant news $ \ctcnp{n}{_{AB}} $ exists such that
					\begin{equation}
					\ctcnp{n}{_{AB}}= 0  \iff \cts{\P}{^a}= 0 \iff \text{There is no radiation at $ \scri^+ $ }\spacef.
					\end{equation}
					\begin{proof}
						\Cref{def:noIncomingRadDefOpen} implies that $ \ctm{\Q}{^\alpha}=0 $, and from \cref{thm:oneQnull-nirc} 
						we have $ \ctp{\Q}{^\alpha}=0\iff\cts{\P}{^a}=0 $ (which according to \cref{def:criterionGlobal} occurs if and only if there is no radiation at $ \scri^+ $). Then, \cref{thm:noNewsiffnoQ} shows that $\ctp{\Q}{^\alpha}=0\iff\ctcnp{n}{_{AB}}=0 $ --the existence of $ \ctcnp{n}{_{AB}} $ follows from \cref{thm:theTwoNewsProp-foliation}.
					\end{proof}
				\end{thm}
				\begin{corollary}\label{thm:noradV}
					Let the assumptions of \cref{thm:noradIncoming} hold but now with a strongly equipped $\scri^+$.  Then
						\begin{equation}
						\ctcn{V}{_{AB}}= 0  \iff \cts{\P}{^a}= 0 \iff \text{There is no radiation at $ \scri $ }\spacef.
						\end{equation}
				\end{corollary}
				\begin{proof}
					By \cref{thm:noincoming-np} $ \ctcnp{n}{_{AB}} $ exists such that $ \ctcn{V}{_{AB}}= 0 \iff \ctcnp{n}{_{AB}}= 0  $. The result follows then by \cref{thm:noradIncoming}.
				\end{proof}
						

	\section{Symmetries}\label{ssec:symmetries}

			Another field of study is that of symmetries at infinity. Although the asymptotically flat scenario is well understood in this respect, such is not the case for the $ \Lambda>0 $ case. For a vanishing cosmological constant, a universal group of symmetries at $ \scri $ --the so called BMS group-- emerges following different approaches. One possibility is to work in the physical space-time and define the symmetries as those transformations preserving some coordinate boundary condition, as in the original work of Bondi, Metzner and Sachs \cite{Bondi1962,Sachs1962,Penrose1966}--after whom the symmetry group inherits its name-- or by defining `approximate asymptotic symmetries' \cite{Geroch81,Wald1984}. Alternatively one can work in the conformal space-time and define the asymptotic symmetry group as those mappings that leave invariant a particular conformal-gauge fixing, sometimes called `Bondi systems' \cite{Stewart1991}, or as those transformations which leave invariant certain structure consisting on the degenerate metric and the generators of $ \scri $ \cite{Ashtekar81,Geroch1977}. Moreover, there is an alternative definition of asymptotic symmetries as those which leave unchanged some  gauge-invariant tensorial quantity constructed with the elementary objects on $ \scri $ --an `asymptotic geometry', to put it in Geroch's words \cite{Geroch1977}. Indeed, this is the first approach we will consider for $ \Lambda>0 $ and, as we will see, it does not lead to the type of enhanced group of symmetries --analogous to the BMS in a broad sense-- that one may wish; for this reason we will explore other different paths too, eventually arriving at a proposal providing an infinite-dimensional Lie algebra. \\
				
			Consider the gauge invariant object
				\begin{equation}\label{eq:symmDef1}
					\ct{\Upsilon}{_{abcdef}} \defeq \ms{_{ab}}\ct{D}{_{cd}}\ct{D}{_{ef}}\quad
				\end{equation}
				which contains the elements of the fundamental structure $(\scri^+,\ct{h}{_{ab}},\ct{D}{_{ab}})$
			and define the generators of infinitesimal symmetries $ \ct{\xi}{^a} $ as
				\begin{equation}
					\lied_{\vec{\xi}}\ct{\Upsilon}{_{abcde}}=0\spacef.
				\end{equation}
			Expanding this equation we find
				\begin{equation}\label{eq:aux19}
					2\cds{_{(a}}\ct{\xi}{_{b)}}\ct{D}{_{cd}}\ct{D}{_{ef}}=-\ms{_{ab}}\brkt{\ct{D}{_{cd}}\lied_{\vec{\xi}}\ct{D}{_{ef}}+\ct{D}{_{ef}}\lied_{\vec{\xi}}\ct{D}{_{cd}}}\spacef,
				\end{equation}
			from where
				\begin{equation}\label{eq:aux20}
					\lied_{\vec{\xi}}\ms{_{ab}}=2\cds{_{(a}}\ct{\xi}{_{b)}}=2\psi\ms{_{ab}}\quad\text{with }\psi\defeq \frac{1}{3}\cds{_c}\ct{\xi}{^c}\spacef.
				\end{equation}
			Using this back into \cref{eq:aux19} one gets
				\begin{equation}\label{eq:aux21}
					\lied_{\vec{\xi}}\ct{D}{_{cd}}=-\psi\ct{D}{_{cd}}\spacef.
				\end{equation}
			\Cref{eq:aux20} implies that $ \ct{\xi}{^a} $ is a CKVF of the metric $ \ms{_{ab}} $. A result in \cite{Paetz2016} states that a Killing vector field of a $ \Lambda>0 $-vacuum space-time induces a vector field on $ \scri $ that satisfies precisely \cref{eq:aux19,eq:aux21} and, conversely, a vector field on $ \scri $ satisfying \cref{eq:aux19,eq:aux21} gives rise, via an initial value problem, to a KVF of the physical space-time. From this point of view, the proposal of preserving \eqref{eq:symmDef1} is fully justified. Importantly, this definition does not require fixing the topology of $ \scri $ nor requires the metric to be conformally flat --with the high-restrictive aftermath this implies \cite{Ashtekar2014}. Also, it includes $ \ct{D}{_{ab}} $ as a fundamental ingredient, in accordance with our repeated claim that one has to bring $ \ct{D}{_{ab}} $ into the picture.  These are the basic asymptotic symmetries 
				\begin{deff}[Basic infinitesimal asymptotic symmetries]\label{def:basic-symmetries}
					We define the basic infinitesimal asymptotic symmetries as those CKVF $ \ct{\xi}{^a} $ of $ \prn{\scri ,\ms{_{ab}}} $ which satisfy
						\begin{equation}
							\lied_{\vec{\xi}}\ct{D}{_{cd}}=-\frac{1}{3}\cds{_a}\ct{\xi}{^a} \ct{D}{_{cd}}\spacef.
						\end{equation}
				\end{deff}
				Nevertheless, definition \ref{def:basic-symmetries} is not completely satisfactory as there may be cases in which no such basic asymptotic symmetries exist.
			Alternatively, we can define other asymptotic symmetries as those which preserve the structure of \cref{def:additional-structure} in the following sense:
				 \begin{deff}[Equipped $ \scri $ symmetries.]\label{def:asymptotic-symmetries}
					Consider $ \scri $ equipped according to \cref{def:additional-structure}. The extended asymptotic symmetries are those preserving the conformal class of the one-parameter family of projectors to $ \Scn $, and the direction of the congruence $ \C $ on $ \scri $. In other words, these symmetries are the transformations acting on the pairs $ \prn{\ctcn{P}{_{ab}},\ct{m}{_a}} $ as
					\begin{equation*}
					\prn{\ctcn{P}{_{ab}},\ct{m}{_a}}\longrightarrow\prn{\Psi^2\ctcn{P}{_{ab}},\Phi\ct{m}{_a}}\spacef.
					\end{equation*}
				\end{deff}
				\begin{remark}
					The infinitesimal version $ \ct{\xi}{^a} $ of these transformations is
					\begin{align}
					\lied_{\vec{\xi}}\ctcn{P}{_{ab}}&= 2\psi\ctcn{P}{_{ab}}\spacef,\label{eq:biconformal-P}\\
					\lied_{\vec{\xi}}\ct{m}{_{a}}&= \phi\ct{m}{_{a}}\spacef,\label{eq:biconformal-m}
					\end{align}
					where $ \ct{\xi}{^a} $ generates a one-parameter ($ \epsilon $) family of transformations of the type of \cref{def:asymptotic-symmetries}, with $ \phi\defeq \partial_\epsilon \Phi\prn{\epsilon} \vert_{\epsilon=0} $, $ \psi\defeq \partial_\epsilon \Psi\prn{\epsilon} \vert_{\epsilon=0} $. Note that from these equations it also follows 
					\begin{equation}
					\lied_{\vec{\xi}}\ct{m}{^a}= -\cs{\phi}\ct{m}{^a}\quad\label{eq:biconformal-mform}
					\end{equation}
					and
					\begin{equation}
					\lied_{\vec{\xi}}\ms{_{ab}}=2\phi\ct{m}{_a}\ct{m}{_b}+2\psi\ctcn{P}{_{ab}}\spacef.
					\end{equation}
					The following gauge-changes follow from \cref{eq:biconformal-m,eq:biconformal-P}:
						\begin{align}
							\csg{\psi}&=\psi+\lied_{\vec{\xi}}\prn{\ln \omega}\spacef,\\
							\csg{\phi}&=\phi+\lied_{\vec{\xi}}\prn{\ln \omega}\spacef.
						\end{align} 
					Observe  that $\csg{\psi}-\csg{\phi}=\cs{\psi}-\cs{\phi} $ is gauge invariant.
				\end{remark}
			The group of symmetries of \cref{def:asymptotic-symmetries} will be denoted by $ \mathsf{B}$ and it constitutes a case of the so called bi-conformal transformations  \cite{GarciaParrado2004}. Taking into account that the Lie derivative acts linearly and using the property $ \lied_{\brkt{\vec{\xi_1},\vec{\xi_2}}}=\lied_{\vec{\xi_1}}\lied_{\vec{\xi_2}}-\lied_{\vec{\xi_2}}\lied_{\vec{\xi_1}} $, it can be easily shown  that these infinitesimal transformations form a Lie algebra which we denote by $ \fb $ and that for $ \ctrd{3}{\xi}{^a}=\commute{\csrd{1}{\xi}}{\csrd{2}{\xi}}^a $ one has
				\begin{align}
					\csrd{3}{\psi}=\lied_{\vec{\csrd{1}{\xi}}}\cs{\csrd{2}{\psi}}-\lied_{\vec{\csrd{2}{\xi}}}\cs{\csrd{1}{\psi}}\spacef,\\
					\csrd{3}{\phi}=\lied_{\vec{\csrd{1}{\xi}}}\cs{\csrd{2}{\phi}}-\lied_{\vec{\csrd{2}{\xi}}}\cs{\csrd{1}{\phi}}\spacef.
				\end{align}
			\\
			
			Consider the general decomposition 
				\begin{equation}
				\ct{\xi}{^a}= \beta \ct{m}{^a}+ \ct{\chi}{^a}\spacef,\quad\ct{\chi}{^a}\ct{m}{_a}= 0.
				\end{equation}
			We can obtain the necessary and sufficient conditions that $ \beta $ and $ \ct{\chi}{^a} $ have to satisfy so that $ \ct{\xi}{^a} \in \fb$ by decomposing into tangent and orthogonal parts \cref{eq:biconformal-P,eq:biconformal-m,eq:biconformal-mform},
				\begin{empheq}
					[left={	\begin{cases}
					\lied_{\vec{\xi}}\ct{P}{_{ab}}= 2\psi\ctcn{P}{_{ab}}\spacef,\\
					\lied_{\vec{\xi}}\ct{m}{_{a}}= \phi\ct{m}{_{a}}\spacef,
					\end{cases}
					\iff\quad}\empheqlbrace]
				{align}
					&\lied_{\vec{m}}\beta-\ctcn{a}{^e}\ct{\chi}{_e}= \phi\spacef,\label{eq:biconformal-mbeta}\\
					&\beta\ctcn{a}{_b}+\cdcn{_b}\beta+2\ctcn{\omega}{_{eb}}\ct{\chi}{^e}= 0 \spacef,\label{eq:biconformal-Dbeta}\\
					&\lied_{\vec{m}}\ct{\chi}{^a}+\ct{\chi}{^e}\ctcn{a}{_e}\ct{m}{^a}= 0\spacef,\label{eq:biconformal-m-chi}\\
					&2\cdcn{_{(d}}\ct{\chi}{_{c)}}+2\beta\ctcn{\kappa}{_{cb}}-2\psi\ctcn{P}{_{cd}}= 0\spacef.\label{eq:biconformal-beta-psi-chi}
				\end{empheq}
			 In order to identify some sort of translational subgroup, it seems natural to ask for the existence of a particular class of generators  $ \ct{\tau}{^a}\defeqs \alpha \ct{m}{^a} $ completely tangent to $ \vec{m} $,
					\begin{empheq}	
						[left={\begin{cases}
									\lied_{\vec{\tau}}\ct{P}{_{ab}}= 	2\theta\ctcn{P}{_{ab}}\spacef,\\
								\lied_{\vec{\tau}}\ct{m}{_{a}}= \lambda\ct{m}{_{a}}\spacef,
							\end{cases}
							\iff 
						\quad}\empheqlbrace]
					{align}
								&\lied_{\vec{m}}\alpha= \lambda\spacef,\label{eq:bitranslation-ralpha}\\
								&\alpha\ctcn{a}{_b}+\cdcn{_b}\alpha= 0\spacef,\label{eq:bitranslation-Dalpha}\\
								&\theta\ctcn{P}{_{cd}}-\alpha\ctcn{\kappa}{_{cd}}= 0 \spacef.\label{eq:bitranslation-alpha-theta}
					\end{empheq}
			Notice that \eqfref{eq:bitranslation-alpha-theta} requires $ \ct{m}{^a} $ to be shear-free. However, this is not an assumption in \cref{def:additional-structure} and in general one has $ \ctcn{\Sigma}{_{ab}}\neq 0 $.\\
			
			Furthermore, asking for $ \ct{\eta}{^a} $ to be a symmetry orthogonal to $ \ct{m}{_a} $ ($ \ct{\eta}{^e}\ct{m}{_e}=0 $) produces the following set of conditions:
				\begin{empheq}[left=
					{\begin{cases}
					\lied_{\vec{\eta}}\ct{P}{_{ab}}= 2\varphi\ctcn{P}{_{ab}}\spacef,\label{biconformaltangent-P}\\
					\lied_{\vec{\eta}}\ct{m}{_{a}}= \mu \ct{m}{_{a}}\spacef,\label{biconformaltangent-r}
					\end{cases}
					\iff\quad}\empheqlbrace]
					{align}
					&\ctcn{a}{^e}\ct{\eta}{_e}=-\mu\spacef,\label{eq:conformalS-aeta}\\
					&\ct{\eta}{^e}\ctcn{\omega}{_{eb}}= 0\spacef,\label{eq:conformalS-omega}\\
					&\lied_{\vec{m}}\ct{\eta}{^a}+\ct{\eta}{^e}\ctcn{a}{_e}\ct{m}{^a}= 0\spacef,\label{eq:conformalS-reta}\\
					&\cdcn{_{(a}}\ct{\eta}{_{b)}}-\varphi\ctcn{P}{_{ab}}= 0\spacef.\label{eq:conformalS-Deta}
				\end{empheq}
			In this case, \eqfref{eq:conformalS-omega} requires $ \ctcn{\omega}{_{ab}}=0 $ ---to prove this, note that $ 2\ctcn{\omega}{_{ab}}=\ctcn{\omega}{_{cd}}\ctcn{\epsilon}{^{cd}}\ctcn{\epsilon}{_{ab}} $. However, according to \cref{def:additional-structure}, the vector field $ \vec{m} $ has non-vanishing vorticity, in general. Importantly, to account for the existence of symmetries one has to study  integrability conditions too. Then, (multiple) solutions to the above equations may exist or not. The general form of such conditions are out of the scope of this work, but one can study them for each particular metric.\\
			
			What we have seen is that, in general, \cref{def:additional-structure} is too weak in order to get a notion of translations within $ \scri $ and along $ \ct{m}{^a} $. We will explore the particular case in which this kind of symmetries are present in \cref{ssec:strong-structure}. Before that, we present a derivation of the transformations of \cref{def:asymptotic-symmetries} without further constraints using a different approach which also partly justifies the definition.
				\subsection{Derivation from approximate space-time symmetries}
					We are about to show that a particular sort of approximate space-time symmetries can lead at infinity to the equipped-$\scri$ symmetries of definition \ref{def:asymptotic-symmetries}.
					
					Begin by considering a vector field $ \pt{\xi}{^\alpha} $ on the physical space-time $ \prn{\ps{M},\pt{g}{_{\alpha\beta}}} $ with a smooth extension to $ \scri $ on the unphysical space-time $ \prn{\cs{M},\ct{g}{_{\alpha\beta}}} $, which in this subsection we consider foliated by $ \Omega= $constant-hypersurfaces near $ \scri $. On $ \ps{M} $ one has
						\begin{equation}\label{eq:aux35}
							\lied_{\vec{\ps{\xi}}}\ct{g}{_{\alpha\beta}}=\Omega^2\lied_{\vec{\ps{\xi}}}\pt{g}{_{\alpha\beta}}+\frac{2}{\Omega}\lied_{\vec{\ps{\xi}}}\prn{\Omega}\pt{g}{_{\alpha\beta}}\spacef.
						\end{equation}
					We will require that
						\begin{equation}\label{eq:aux36}
							\Omega^2\lied_{\vec{\ps{\xi}}}\pt{g}{_{\alpha\beta}}=\ct{H}{_{\alpha\beta}}
						\end{equation}
					for some symmetric tensor field $ \ct{H}{_{\alpha\beta}} $ regular at $ \scri $.  Then, the idea is to ask $ \ct{H}{_{\alpha\beta}} $ to fulfil certain conditions such that $ \pt{\xi}{^\alpha} $ `approximates' a symmetry near $ \scri $. Some obvious examples are:
					\begin{itemize}
						\item If $ \pt{\xi}{^\alpha} $ is a KVF for $ \pt{g}{_{\alpha\beta}} $, then $ \ct{H}{_{\alpha \beta}}=0 $.
						\item If $ \pt{\xi}{^\alpha} $ is a CKVF for $ \pt{g}{_{\alpha\beta}} $, then $ \ct{H}{_{\alpha \beta}}\propto \ct{g}{_{\alpha\beta}} $.
					\end{itemize}
					Observe that assumption \eqref{eq:aux36} and regularity of \cref{eq:aux35} at $ \scri $ require
						\begin{equation}\label{eq:aux42}
							\lied_{\vec{\ps{\xi}}}\prn{\Omega}=\Omega J
						\end{equation}
					for some scalar function $ J $ regular at $ \scri $. Of course, this implies that
						\begin{equation}
							\pt{\xi}{^\alpha}\ct{N}{_\alpha}\eqs 0\spacef.
						\end{equation}
					Hence, $ \pt{\xi}{^\alpha} $ has to be tangent to $ \scri $. For later convenience let us define
						\begin{equation}
							\ct{\xi}{^\alpha}\defeqs \pt{\xi}{^\alpha}\eqs \ct{e}{^\alpha_a}\ct{\xi}{^a},
						\end{equation}
					where $ \cbrkt{\ct{e}{^a_\alpha}} $ is a basis on $ \scri $. Eq.(\ref{eq:aux35}) reads 
						\begin{equation}\label{eq:aux41}
								\lied_{\vec{\ps{\xi}}}\ct{g}{_{\alpha\beta}}=2J\ct{g}{_{\alpha\beta}}+\ct{H}{_{\alpha\beta}}\spacef.
						\end{equation}
					It is easy to obtain
						\begin{equation}
							\lied_{\vec{\ps{\xi}}}\ct{N}{_{\alpha}}=\ct{N}{_\alpha}J+\Omega\cd{_\alpha}J,
						\end{equation}
					from where at $ \scri $ (recall that $ \ct{P}{^\alpha_\beta} $ is the projector to $ \scri $ \eqref{eq:projectorScri})
						\begin{equation}\label{eq:app-sym-lie-projector}
							\lied_{\vec{\ps{\xi}}}\ct{P}{_{\alpha\beta}}\eqs 2J\ct{P}{_{\alpha\beta}}+\ct{H}{_{\alpha\beta}}\spacef.
						\end{equation}
					Next, we are going to see whether this equation contains components along $ \ct{N}{_\alpha} $ or not. Expanding the left-hand side of \cref{eq:app-sym-lie-projector} and contracting with $ \ct{N}{^\alpha} $,
						\begin{equation}\label{eq:aux38}
							\lied_{\vec{\ps{\xi}}}\ct{N}{^\alpha}=-J\ct{N}{^\alpha}-\ct{H}{^\alpha}+\Omega \cd{^\alpha}J
						\end{equation}
					with
						\begin{equation}
							\ct{H}{^\alpha}\defeqs \ct{N}{^\mu}\ct{H}{_\mu^\alpha}\spacef,
						\end{equation} 	
					from where,
						\begin{equation}\label{eq:aux37}
							\ct{N}{^\mu}\lied_{\vec{\ps{\xi}}}\ct{P}{_{\alpha\beta}}=\ct{H}{_\beta}+ \Omega\brkt{-\cd{_\beta}J+\ct{n}{_\beta}\ct{n}{^\alpha}\cd{_\alpha}J-2\ct{n}{_\beta}\prn{\lied_{\vec{\ps{\xi}}}f+fJ}},
						\end{equation}
					where $ f $ is the scalar \eqref{eq:friedrichscalar}. Then, contraction of \cref{eq:app-sym-lie-projector} with $ \ct{N}{^\alpha} $ gives
						\begin{equation}
							\ct{N}{^\alpha}\lied_{\vec{\xi}}\ct{P}{_{\alpha\beta}}\eqs \ct{H}{_\beta}
						\end{equation}
					and contraction of \cref{eq:aux37} with $ \ct{N}{^\beta} $,
						\begin{equation}\label{eq:aux40}
							\ct{N}{^\alpha}\ct{N}{^\beta}\ct{H}{_{\alpha\beta}}=\ct{N}{^\beta}\ct{H}{_\beta}\eqs 0.
						\end{equation}
					Then, $ \ct{H}{^\alpha}\eqs \ct{e}{^\alpha_a}\ct{H}{^a}  $ and by \cref{eq:aux38}
						\begin{equation}\label{eq:app-sym-Hvector-scri}
							\ct{H}{^\alpha}\eqs -\prn{\lied_{\vec{\ps{\xi}}}\ct{N}{^\alpha}+J\ct{N}{^\alpha}}\spacef.
						\end{equation}
					To see if there is any consistency condition for $ \ct{H}{_{\alpha\beta}} $ compute the following:
						\begin{align}
							\lied_{\vec{\ps{\xi}}}\lied_{\vec{N}}\ct{g}{_{\alpha\beta}}&=2\lied_{\vec{\ps{\xi}}}f\ct{g}{_{\alpha\beta}}+2f\prn{2J\ct{g}{_{\alpha\beta}}+\ct{H}{_{\alpha\beta}}}-\Omega J \ct{S}{_{\alpha\beta}}-\Omega \lied_{\vec{\ps{\xi}}}\ct{S}{_{\alpha\beta}}\spacef,\\
							\lied_{\vec{N}}\lied_{\vec{\ps{\xi}}}\ct{g}{_{\alpha\beta}}&=2\lied_{\vec{N}}J \ct{g}{_{\alpha\beta}}+2J\prn{2f\ct{g}{_{\alpha\beta}}-\Omega\ct{S}{_{\alpha\beta}}}+\lied_{\vec{N}} \ct{H}{_{\alpha\beta}}\spacef,\\
							\lied_{\commute{\vec{\ps{\xi}}}{\vec{N}}}\ct{g}{_{\alpha\beta}}&=\lied_{\prn{-J\vec{N}-\vec{H}-\Omega \vec{\nabla J}}}\ct{g}{_{\alpha\beta}}=-J\prn{2f\ct{g}{_{\alpha\beta}}-\Omega\ct{S}{_{\alpha\beta}}}-\cd{_\alpha} \ct{H}{_\beta}-\cd{_\beta}\ct{H}{_\alpha}+2\Omega \cd{_\alpha}\cd{_\beta} J
						\end{align}
					and then, use the identity $ \lied_{\commute{\vec{\ps{\xi}}}{\vec{N}}}= \lied_{\vec{\ps{\xi}}}\lied_{\vec{N}}-	\lied_{\vec{N}}\lied_{\vec{\ps{\xi}}}$ to get
						\begin{equation}\label{eq:aux39}
							0=2\prn{\lied_{\vec{\ps{\xi}}} f -\lied_{\vec{N}} J +fJ}\ct{g}{_{\alpha\beta}}+2f\ct{H}{_{\alpha\beta}}-\lied_{\vec{N}}\ct{H}{_{\alpha\beta}}+\cd{_\alpha}\ct{H}{_\beta}+\cd{_\beta}\ct{H}{_\alpha}-\Omega\prn{2\cd{_\alpha}\cd{_\beta}J+\lied_{\vec{\ps{\xi}}}\ct{S}{_{\alpha\beta}}}.
						\end{equation}
					After some computation, it can be checked that the right-hand side of \cref{eq:aux39} does not have components along $ \ct{N}{^\alpha} $, therefore this equation contains no information orthogonal to the $ \Omega=$constant hypersurfaces. Expanding the Lie derivative of $ \ct{H}{_{\alpha\beta}} $, \cref{eq:aux39} turns into an expression for the derivative of this tensor along $ \ct{N}{^\alpha} $, $ \ctd{H}{_{\alpha\beta}}\defeq\ct{N}{^\mu}\cd{_\mu}\ct{H}{_{\alpha\beta}} $,
						\begin{equation}
							\ctd{H}{_{\alpha\beta}}=\cd{_\alpha}\ct{H}{_\beta}+\cd{_\beta}\ct{H}{_\alpha}+\prn{\lied_{\vec{\ps{\xi}}} f -\lied_{\vec{N}}J+fJ}\ct{g}{_{\alpha\beta}}+\Omega\prn{\ct{H}{_{\mu(\alpha}}\ct{S}{_{\beta)}^\mu}-2\cd{_\alpha}\cd{_\beta}J-\lied_{\vec{\ps{\xi}}}\ct{S}{_{\alpha\beta}}}\spacef.
						\end{equation}
					If one projects this equation to $ \scri $ with $ \cbrkt{\ct{e}{^\alpha_a}} $ and uses \cref{eq:aux40}, it reads
						\begin{equation}
							N \ctd{H}{_{ab}}\eqs 2\cds{_{(a}}\ct{H}{_{b)}} -\lied_{\vec{N}}\csS{J}\ms{_{ab}}\spacef,
						\end{equation}
					where $ \csS{J}\eqs J $ and we have used $ f\eqs0 $ (see \cref{eq:fscalarScri}) and
						\begin{equation}
							\ct{\xi}{^\alpha}\cd{_\alpha}f\eqs 0\spacef.
						\end{equation}
					Next, take the pullback of \cref{eq:app-sym-lie-projector} to $ \scri $,
						\begin{equation}\label{eq:app-sym-lie-ms}
							\lied_{\vec{\xi}} \ms{_{ab}}\eqs 2\csS{J}\ms{_{ab}}+\ct{H}{_{ab}}\spacef,
						\end{equation}
					where it is evident that only the tangent part of $ \ct{H}{_{\alpha\beta}} $ intervenes. \Cref{eq:app-sym-lie-ms} is important, as the meaning of $ \ct{H}{_{ab}} $ on $ \scri $ it is clear here: how we choose $ \ct{H}{_{ab}} $ defines how we define $ \ct{\xi}{^a} $ as an asymptotic-symmetry. Our goal is to choose $ \ct{H}{_{ab}} $ such that one can say that \cref{eq:app-sym-lie-ms} comes from an approximate space-time symmetry --as much as possible. Before entering into this task, let us remark that $ \ct{\xi}{^a} $ is in one-to-one correspondence with the equivalence class
						\begin{equation}
							\cbrkt{\ctt{\ps{\xi}}{^\alpha}\in \brkt{\ct{\ps{\xi}}{^\alpha}}\quad \iff\quad\ctt{\ps{\xi}}{^\alpha}-\ct{\ps{\xi}}{^\alpha}=\Omega \ct{v}{^\alpha}},
						\end{equation}
					where $ \ct{v}{^\alpha}$ is any vector field on $\cs{M} $. However, if we want any element of the equivalence class to generate an asymptotic symmetry of the kind \eqref{eq:app-sym-lie-ms}, $ \Omega \ct{v}{^\alpha} $ itself has to satisfy all the equations so far. Calling $ \ctrd{0}{H}{_{\alpha\beta}} $ and $ \csrd{0}{J} $ the $ \ct{H}{_{\alpha\beta}} $ and $ J $ associated to $ \Omega\ct{v}{^\alpha} $, respectively, from \cref{eq:aux41,eq:aux42}  we have
						\begin{align}
							\lied_{\Omega\vec{v}}\Omega&=\Omega \ct{v}{^\mu}\ct{N}{_\mu}=\Omega\csrd{0}{J}\spacef,\label{eq:app-symm-Jo}\\
							\lied_{\Omega\vec{v}}\ct{g}{_{\alpha\beta}}&=2\Omega\cd{_{{(\alpha}}}\ct{v}{_{\beta)}}+2\ct{v}{_{(\alpha}}\ct{N}{_{\beta)}}=2\csrd{0}{J}\ct{g}{_{\alpha\beta}}+\ctrd{0}{H}{_{\alpha\beta}}\spacef.
						\end{align}
					Then, putting together these two equations we get a formula for $ \ctru{0}{H}{_{\alpha\beta}} $:
						\begin{equation}\label{eq:aux44}
							\ctrd{0}{H}{_{\alpha\beta}}=2\Omega\cd{_{{(\alpha}}}\ct{v}{_{\beta)}}+2\ct{v}{_{(\alpha}}\ct{N}{_{\beta)}}-2\ct{v}{^\mu}\ct{N}{_\mu}\ct{g}{_{\alpha\beta}}\spacef.
						\end{equation}
					It can also be shown that $ \ctrd{0}{H}{_\alpha}\defeq \ct{N}{^\mu}\ctrd{0}{H}{_{\alpha\mu}} $ is tangent at $ \scri $ and satisfies 
						\begin{align}\label{eq:app-symm-redundacy-terms}
							\ctrd{0}{H}{_{\alpha}}\eqs -\ct{v}{^\mu}\ct{N}{_\mu}-N^2\ct{v}{_\alpha}\eqs -\ct{P}{^\mu_\alpha}\ct{v}{_{\mu}}\spacef,\nonumber\\
							=-\prn{\lied_{\Omega\vec{v}}\ct{N}{_\alpha}+\csrd{0}{J}\ct{N}{_\alpha}}
						\end{align}
					---compare with \cref{eq:app-sym-Hvector-scri}. The combination $ \ctrd{0}{H}{_{\alpha\beta}} $ for arbitrary $ \ct{v}{^\alpha} $ has no relevance for $ \ct{\xi}{^a} $, then, it  can be considered as a gauge part in $ \ct{H}{_{\alpha\beta}} $. Hence, for any $ \ctt{\hat{\xi}}{^\alpha}\in\brkt{\pt{\xi}{^\alpha}} $ one uses
						\begin{equation}\label{eq:app-symm-gauge-corrections}
							\ctt{H}{_{\alpha\beta}}=\ct{H}{_{\alpha\beta}}+2\ct{N}{_{(\alpha}}\cts{v}{_{\beta)}}-2\ct{v}{^\mu}\ct{N}{_\mu}\ct{P}{_{\alpha\beta}}+2\Omega\cd{_{{(\alpha}}}\ct{v}{_{\beta)}}\spacef,
						\end{equation}
					where we have defined $ \cts{v}{_\alpha}\defeq \ct{P}{^\mu_\alpha}\ct{v}{_\mu} $. Note that any term of type $ \ct{N}{_{\alpha}}\ct{v}{_{\beta}} + \ct{N}{_{\beta}}\ct{v}{_{\alpha}} $ is pure gauge in $ \ct{H}{_{\alpha\beta}} $, and the term in $ \ct{P}{_{\alpha\beta}} $ is the one that makes the definition of $ \ct{\xi}{^a} $ unambiguous. This is clearly seen projecting to $ \scri $,
						\begin{equation}\label{eq:app-H-Hprimed}
							\ctt{H}{_{ab}}=\ct{H}{_{ab}}-2\prn{\ct{v}{^\mu}\ct{N}{_\mu}}\evalat{\scri}\ms{_{ab}}\spacef.
						\end{equation}
					Therefore, within $ \scri $ one gets
						\begin{equation}
							\lied_{\vec{\grave{\xi}}}\ms{_{ab}}=2\csS{\grave{J}}\ms{_{ab}}+\ctt{H}{_{ab}}=2\csS{J}\ms{_{ab}}+\ct{H}{_{ab}}=\lied_{\vec{\xi}}\ms{_{ab}}\spacef,
						\end{equation}
					where we have used \cref{eq:app-sym-lie-ms,eq:app-H-Hprimed,eq:app-symm-Jo} together with $ \csS{\grave{J}}=\csS{J}+\csrd{0}{\csS{J}}$. By typical calculations, it can be proven that the set of such vector fields $ \pt{\xi}{^\alpha}$ on $ \prn{\ps{M},\pt{g}{_{\alpha\beta}}} $ form a Lie algebra, as well as their equivalence classes. One should not forget the conformal gauge freedom \eqref{eq:conformal-gauge}. Under such rescalings,
						\begin{align}
							\csg{J}&=\cs{J}+\frac{1}{\omega}\lied_{\vec{\ps{\xi}}}\omega\spacef,\label{eq:gauge-J}\\
							\ctg{H}{_{\alpha\beta}}&=\omega^2 \ct{H}{_{\alpha\beta}}\spacef,\\
							\ctg{H}{_{\alpha}}&=\omega\ct{H}{_\alpha}+\Omega \ct{\omega}{^\alpha}\ct{H}{_{\alpha\beta}},\label{eq:gauge-H}
						\end{align}
					which follow form \cref{eq:aux41,eq:aux42} and the gauge transformations of \cref{app:gauge-transformations}.\\
					
					Now we have to make a choice for $ \ct{H}{_{\alpha\beta}} $. Should we follow Geroch and Winicour \cite{Geroch81}, we would have to set $ \ct{H}{_{\alpha\beta}}\eqs0 $. This fixing makes $ \ct{\xi}{^a} $ a CKVF of $ \ms{_{ab}} $, hence one probably, at the best, recovers the basic symmetries preserving \eqref{eq:symmDef1} of definition \ref{def:basic-symmetries}. However, as we have argued, these kind of symmetries are not satisfactory. Thus, one is left with the problem of specifying a different kind of $ \ct{H}{_{\alpha\beta}} $. It makes sense to think that $ \ct{H}{_{\alpha\beta}} $ should be a rank-1 matrix --at least on $ \scri $-- , up to redundancy-correction terms, that is 
						\begin{equation}\label{eq:aux43}
							A\ct{m}{_\alpha}\ct{m}{_\beta}+\Omega \ct{x}{_{\alpha\beta}}
						\end{equation}
					for some scalar function $ A $ and tensor field $ \ct{x}{_{\alpha\beta}} $. Moreover, the one-form $ \ct{m}{_{\alpha}} $ has to be tangent to $ \scri $, i.e. $ \ct{m}{_\alpha}\ct{N}{^\alpha}\eqs 0 $, so that it fulfils \cref{eq:aux40}. Still, in order to use \eqref{eq:aux43} as $ \ct{H}{_{\alpha\beta}} $, one has to add the redundancy-correction terms \eqref{eq:aux44}; the resulting expression reads
						\begin{equation}
							\ct{H}{_{\alpha\beta}}=	A\ct{m}{_\alpha}\ct{m}{_\beta}+2\ct{N}{_{(\alpha}}\cts{v}{_{\beta)}}+C\ct{P}{_{\alpha\beta}}+\Omega\ct{x}{_{\alpha\beta}},
						\end{equation}
					where we have set $ C\defeq -2\ct{v}{_\mu}\ct{N}{^\mu} $. The parameters $ A $ and $ C $ are general and should not be fixed beforehand, as doing so would restrict the available $ \ct{\xi}{^a} $. The pullback to $ \scri $ is
						\begin{equation}
							\ct{H}{_{ab}}= A\ct{m}{_a}\ct{m}{_b}+C\ms{_{ab}}\spacef,
						\end{equation}
					by means of which we can write \cref{eq:app-sym-lie-ms} as
						\begin{equation}
							\lied_{\vec{\xi}} \ms{_{ab}}= \prn{2\csS{J}+C}\ms{_{ab}}+A\ct{m}{_a}\ct{m}{_b}\spacef.
						\end{equation}
					Observe that \cref{eq:gauge-H,eq:gauge-mcn,eq:gauge-m-vector} impose the following gauge behaviour
						\begin{equation}\label{eq:gauge-A-C}
							\csg{C}\eqs C,\quad\csg{A}\eqs A\spacef.
						\end{equation}
					Let us define $ \ctcn{P}{_{ab}}\defeq \ms{_{ab}}-\ct{m}{_a}\ct{m}{_b} $, as in \cref{eq:projector-congruences}, to write the last formula as
						\begin{equation}\label{eq:app-symm-biconformal}
							\lied_{\vec{\xi}} \ms{_{ab}}= \prn{2\csS{J}+C}\ctcn{P}{_{ab}}+\prn{A+2\csS{J}+C}\ct{m}{_a}\ct{m}{_b}\spacef.
						\end{equation}
					From this expression it becomes manifest that the resulting $ \ct{\xi}{^a} $ are \emph{biconformal vector fields on $ \scri $}. The Lie-algebra structure of these infinitesimal transformations and \cref{eq:app-symm-biconformal} require
						\begin{align}
						\lied_{\vec{\xi}}\ctcn{P}{_{ab}}&= 2\psi\ctcn{P}{_{ab}}\quad\text{with}\quad \psi\defeq 2\csS{J}+C,\\
						\lied_{\vec{\xi}}\ct{m}{_{a}}&= \phi\ct{m}{_{a}}\quad\text{with}\quad\phi\defeq A+2\csS{J}+C,
						\end{align}
					which are actually \cref{eq:biconformal-P,eq:biconformal-m} for $ \ct{m}{^a} $ the vector field of \cref{def:additional-structure}. Observe that from \cref{eq:gauge-A-C,eq:gauge-J} one can deduce the gauge transformation of $ \cs{\psi} $ and $ \cs{\phi} $:
						\begin{equation}
						 	\csg{\psi}=\cs{\psi}+\frac{2}{\omega}\lied_{\vec{\xi}}\omega,\quad\csg{\phi}=\cs{\phi}+\frac{2}{\omega}\lied_{\vec{\xi}}\omega\spacef.
						\end{equation}
					\\
					
					As a matter of fact, the space-time KVF ($ \ct{H}{_{\alpha\beta}}=0 $) and CKVF ($ \ct{H}{_{\alpha\beta}}\propto \pt{g}{_{\alpha\beta}} $) only generate part of the asymptotic symmetries of \cref{def:asymptotic-symmetries} --if they also satisfy \eqref{eq:biconformal-m}.
				\subsection{Strongly equipped $ \scri $}\label{ssec:strong-structure} 

				We consider now the asymptotic symmetries of \cref{def:asymptotic-symmetries} for strongly equipped  $\scri$ of \cref{def:strong-structure}. Let us keep the notation that was used for denoting general ($ \ct{\xi}{^a} $), $ \ct{m}{^a} $-orthogonal ($ \ct{\eta}{^a} $) and $ \ct{m}{^a} $-tangent ($ \ct{\tau}{^a} $) symmetries, respectively. \\
				
				Then, for $ \ct{\xi}{^a}\defeq\beta\ct{m}{^a}+\ct{\chi}{^a} $:
					\begin{empheq}[
					left={\begin{cases}
						\lied_{\vec{\xi}}\ctcn{P}{_{ab}}= 2\psi\ctcn{P}{_{ab}}\spacef,\\
						\lied_{\vec{\xi}}\ct{m}{_{a}}= \phi\ct{m}{_{a}}\spacef,
						\end{cases}
							\iff\quad
					}\empheqlbrace]
					{align}
					&\lied_{\vec{m}}\beta-\ctcn{a}{^e}\ct{\chi}{_e}= \phi\spacef,\label{eq:c1c}\\
					&\beta\ctcn{a}{_b}+\cdcn{_b}\beta= 0 \spacef,\label{eq:c2c}\\
					&\lied_{\vec{m}}\ct{\chi}{^a}+\ct{\chi}{^e}\ctcn{a}{_e}\ct{m}{^a}= 0\spacef,\label{eq:c3c}\\
					&2\cdcn{_{(d}}\ct{\chi}{_{c)}}+\prn{\cscn{\kappa}\beta-2\psi}\ctcn{P}{_{cd}}= 0\spacef.\label{eq:c4c}
					\end{empheq}
				For $ \ct{\tau}{^a}\defeq\alpha\ct{m}{^a} $:
				\begin{empheq}[				
					left={
						\begin{cases}								
					\lied_{\vec{\tau}}\ctcn{P}{_{ab}}=			 	2\theta\ctcn{P}{_{ab}}\spacef,\label{eq:bitranslation-P}\\
						\lied_{\vec{\tau}}\ct{m}{_{a}}= \lambda\ct{m}{_{a}}\spacef,\label{eq:bitranslation-r}
						\end{cases}
						\iff\quad 
					}\empheqlbrace]
					{align}
					&\lied_{\vec{m}}\alpha= \lambda\spacef,\label{eq:c1t}\\
					&\alpha\ctcn{a}{_b}+\cdcn{_b}\alpha= 0\spacef,\label{eq:c2t}\\
					&\theta-\frac{1}{2}\alpha\cscn{\kappa}= 0 \spacef.\label{eq:c3t}					
					\end{empheq}
				Finally, for $ \ct{\eta}{^a} $ ($ \ct{\eta}{^e}\ct{m}{_e} =0$):
					\begin{empheq}[
						left={\begin{cases}
						\lied_{\vec{\eta}}\ctcn{P}{_{ab}}= 2\varphi\ctcn{P}{_{ab}}\spacef,\label{eq:biconformaltangent-P-strong}\\
						\lied_{\vec{\eta}}\ct{m}{_{a}}= \mu \ct{m}{_{a}}\spacef,\label{eq:biconformaltangent-r-strong}
						\end{cases}
						\iff\quad}\empheqlbrace
					]
					{align}
					&\ctcn{a}{^e}\ct{\eta}{_e}=-\mu\spacef,\label{eq:c1cs}\\
					&\lied_{\vec{m}}\ct{\eta}{^a}+\ct{\eta}{^e}\ctcn{a}{_e}\ct{m}{^a}= 0\spacef,\label{eq:c2cs}\\
					&\cdcn{_{(a}}\ct{\eta}{_{b)}}-\varphi\ctcn{P}{_{ab}}= 0\spacef.\label{eq:c3cs}
					\end{empheq}
			
				It can be shown that all the vector fields $ \ct{\tau}{^a} $ satisfying \cref{eq:c1t,eq:c2t,eq:c3t} form a subalgebra $ \ft $ which we call  `bitranslations'. Moreover, for any $ \ct{\xi}{^a}\in\fb $ and any $ \ct{\tau}{^a}\in\ft $
				\begin{equation}\label{eq:commutator-tau-xi}
				\lied_{\vec{\tau}}\ct{\xi}{^a}= \prn{\alpha\phi-\beta\lambda+\alpha\ct{\chi}{^e}\ctcn{a}{_e}}\ct{m}{^a}\spacef.
				\end{equation}
				Thus, the subalgebra $ \ft $ is a Lie ideal of $ \fb $. For any two $ \ct{\vec{\tau}}{_1},\ct{\vec{\tau}}{_2} \in \ft$,
				\begin{equation}
				\commute{\ct{\vec{\tau}}{_1}}{\ct{\vec{\tau}}{_2}}^a=\prn{\alpha_1 \lied_{\vec{m}}\alpha_2-\alpha_2 \lied_{\vec{m}}\alpha_1}\ct{m}{^a}\spacef,
				\end{equation}
				therefore, $ \ft $ is non-Abelian. Note that $ \ft $ has a subalgebra, $ \mathfrak{ct} $, of `conformal translations' defined by those elements of $ \ft $ for which $ \theta=\lambda $, and that this is Abelian\footnote{And an ideal of the Lie subalgebra of $ \fb $ consisting on CKVF ($ \phi=\psi $).}. Furthermore, given one element of $ \ft $, multiplying it by a function $ \nu $ such that $ \cdcn{_a}\nu= 0 $ produces a new element of $ \ft $; the subalgebra $ \ft $ is \emph{infinite dimensional} and by \cref{eq:cn-foliation-acceleration,eq:c2t} the general form of an element $ \ct{\tau}{^a}\in\ft $ is
					\begin{equation}\label{eq:general-translation-strong}
						\ct{\tau}{^a}=\nu\prn{v}F\ct{m}{^a},\quad\text{with}\quad\frac{1}{F}=\lied_{\vec{m}}v\spacef,
					\end{equation}
				where $ \nu $ is an arbitrary function of $ v $ and one has (using obvious notation) 
					\begin{equation}
						\commute{\ct{\vec{\tau}}{_1}}{\ct{\vec{\tau}}{_2}}^a=\prn{\nu_1 \lied_{\vec{m}}\nu_2-\nu_2 \lied_{\vec{m}}\nu_1}F\ct{m}{^a}\spacef.
					\end{equation}
				\\
				
				In the same way, it is easily proven that the vector fields $ \ct{\eta}{^a} $ form a subalgebra $ \fcs $ and are CKVF of the metric on each cut $ \Sc_v $. \\
				
				Importantly, wee see that any $ \ct{\xi}{^a} \in \fb$ is a composition of a $ \ct{\tau}{^a}\in\ft $, with  $ \lambda=\phi+\ctcn{a}{^e}\ct{\xi}{_e} $ and $ 2\theta= \cdcn{_d}\ct{\xi}{^d}+\cscn{\kappa}\beta-2\psi$ (from \cref{eq:c1c,eq:c2c,eq:c3c,eq:c1t,eq:c2t,eq:c3t}), and a $ \ct{\eta}{^a}\in\fcs  $, with $ \mu=\phi-\lied_{\vec{m}}\prn{\beta} $ and $ 2\varphi= 2\psi-\cscn{\kappa}\beta$ (from \cref{eq:c1c,eq:c2c,eq:c3c,eq:c1cs,eq:c2cs,eq:c3cs} ). Let us denote the groups associated to these algebras by $ \fB, \fT,\fCS $, respectively. Then, we have that $ \fT $ is a normal subgroup of $ \fB $ and that it makes sense to define the quotient group $ \fB/\fT$ whose Lie algebra we denote by $ \fb/\ft $. But the elements of $ \fb/\ft $ are precisely the elements of $ \fcs $: any symmetry $ \ct{\eta}{^a} $ modulo a bitranslation is in $ \fcs $. Furthermore, since these are the conformal transformations of $ \Sc_v $, if this has $ \mathbb{S}^2 $-topology, $ \fCS $ is isomorphic to the Lorentz Group $ \mathsf{SO}(1,3) $. Another easily verifiable property is that any $ \ct{\tau}{^a}\in \ft $ commutes with any $ \ct{\eta}{^a}\in\fcs $,
				\begin{equation}
				\commute{\vec{\tau}}{\vec{\eta}}^a= 0\spacef,
				\end{equation}
				as one simply has to set $ \beta=0 $, $ \ct{\chi}{^a}=\ct{\eta}{^a} $ and $ \phi=\mu=-\ct{\eta}{^e}\ct{a}{_e} $ in \cref{eq:commutator-tau-xi}.	Note that solutions $ \alpha $ to  \cref{eq:c2t} always exist because $ \ct{m}{^a} $ defines a foliation -- see \cref{eq:cn-foliation-acceleration}. Then, given $ \alpha $, one can take \cref{eq:c1t,eq:c3t} as definitions for $ \lambda$ and $ \theta $. Also, if we assume $ \mathbb{S}^2 $-topology for the cuts, there always exist (up to 6) conformal Killing vector fields satisfying \cref{eq:c3cs}.  \Cref{eq:c1cs} can be taken as the definition for $ \mu $ and \cref{eq:c2cs} is equivalent to
					\begin{equation}
						\ct{m}{^c}\lied_{\vec{\eta}}\ctcn{P}{_{c}^a}=0,
					\end{equation}
				which using $ \ctcn{\omega}{_{ab}}=0 $ can be expressed as
					\begin{equation}\label{eq:condition-strong-symm}
						\ct{\eta}{^e}\ctcn{\kappa}{_e^a}-\ctcn{P}{^a_d}\ct{m}{^e}\cds{_{e}}\ct{\eta}{^d}=0\quad
					\end{equation}
				and does not hold in general. Then, given \cref{def:strong-structure}, solutions $ \xi\in\fb $ to \cref{eq:c1c,eq:c2c,eq:c3c,eq:c4c} not necessarily exist\footnote{One has not only to study the solutions $ \ct{\eta}{^a} $ to \cref{eq:condition-strong-symm} but also the integrability conditions.}. In summary: 
				\begin{quotation}
					
				\emph{The asymptotic group of symmetries $ \fB $ that preserve the strong structure of \cref{def:strong-structure}  is the (semidirect) product of the (normal) subgroup of bitranslations $ \fT $ and 
				the subgroup of conformal transformations in two dimensions $ \fCS $
					\begin{equation}
					\fB=\fT	\ltimes \fCS\spacef.
					\end{equation}
					The subalgebra of bitranslations $ \ft $ is a non-Abelian Lie ideal and its elements commute with the ones of the algebra $ \fcs $ of the group of conformal transformations on $ \Sc_v $, $ \fCS$
				}
				\end{quotation}
			Let us conclude this section by briefly commenting on the units of $ \cs{\alpha} $. If eventually one wished to take the limit of the symmetries to $ \Lambda=0 $, assuming the limit exists and $ \alpha\evalat{\Lambda=0} $ is regular, one has to rescale any infinitesimal symmetry as --see  \cref{eq:limit-asymptotic-supermomentum}--
				\begin{equation}
					N\ct{\xi}{^a}=\alpha\ct{M}{^a}+N\ct{\chi}{^a},
				\end{equation}
			where $ N\ct{\xi}{^a} $ should be dimensionless to fit with the asymptotic symmetries of the $ \Lambda=0 $ case. Therefore, one has to assign $ \alpha $ the dimensions of length, $ \brkt{\alpha}=L $. Another way of seeing this is that by a conformal rescaling
				\begin{equation}
					\ctg{m}{^a}=\frac{1}{\omega}\ct{m}{^a}\spacef,
				\end{equation}
			and as any infinitesimal symmetry $ \ct{\xi}{^a} $ must be conformally invariant
				\begin{equation}
					\csg{\alpha}=\omega\alpha.
				\end{equation}
			Because $ \omega $ is dimensionless and lengths rescale with $ \omega $, we arrive at the same conclusion, i.e., $ \brkt{\alpha}=L $.
			\subsection{Relation between the tensor $ \rho $ and asymptotic translations}
			It is possible to relate the asymptotic symmetries to $ \ctcn{\rho}{_{ab}} $ and the vanishing of $ \ctcn{V}{_{ab}} $. For a general foliation, using \cref{eq:VAB-acceleration}, one has that $ \ctcn{V}{_{ab}}=0 $ if and only if
				\begin{equation}
				\ctcn{\rho}{_{AB}}=\ct{f}{_{AB}}+\cdcn{_A}\ctcn{a}{_B}-\ctcn{a}{_A}\ctcn{a}{_B}-\frac{1}{2}\mcn{_{AB}}\prn{\cdcn{_E}\ctcn{a}{^E}-\ctcn{a}{_E}\ctcn{a}{^E}-\cscn{K}}\spacef.
				\end{equation}
			Hence, using the function $ F $ of \cref{eq:cn-foliation-acceleration}
				\begin{lemma}
					Assume $ \scri $ is strictly equipped. Then
						\begin{equation}\label{eq:no-v-translations-rho}
							\ctcn{V}{_{AB}}=0 \iff \ctcn{\rho}{_{AB}}=\ct{f}{_{AB}}-\frac{1}{F}\cdcn{_{A}}\cdcn{_{B}}F+\frac{1}{2}\mcn{_{AB}}\brkt{\frac{1}{F}\cdcn{_{C}}\cdcn{^C}F+\cscn{K}}\spacef.
						\end{equation}
				\end{lemma}
		 	As an immediate consequence,
				\begin{corollary}\label{thm:col-rho-noV-alpha}
					If $ \scri $ is strongly equipped and $ \ct{\tau}{^a}=\alpha\ct{m}{^a} $ is a bitranslation, then
						\begin{equation}
							\ctcn{V}{_{AB}}=0 \iff \ctcn{\rho}{_{AB}}= -\frac{1}{\alpha}\cdcn{_{A}}\cdcn{_{B}}\alpha+\frac{1}{2}\mcn{_{AB}}\brkt{\frac{1}{\alpha}\cdcn{_{C}}\cdcn{^C}\alpha+\cscn{K}}\spacef.
						\end{equation}
				\end{corollary}
			The last result follows by noting that if $ \ct{\tau}{^a} \in \ft$ then $ \alpha $ satisfies \cref{eq:c2t} and that for umbilical cuts $ \ct{f}{_{AB}}=0 $ (see \cref{eq:fakenewsdef}). 
				\begin{remark}
				Notice that the previous equation is precisely \eqref{eq:Hess-gen}, providing a neat interpretation for the functions $\alpha$: if the leaves $ \Sc_{v} $ have topology $ \mathbb{S}^2 $ they correspond to the $ l=0,1 $ \emph{spherical harmonics}; in fact, they are exactly a linear combination of the $ l=0,1 $ \emph{spherical harmonics} in the round gauge with
$ 2\ct{\rho}{_{AB}}=\cs{K}\mc{_{AB}} $. In other words, if $ \scri $ is strongly equipped and $ \ct{\tau}{^a}=\alpha\ct{m}{^a} $ is a bitranslation, 
the function $F$ appearing in  \cref{eq:general-translation-strong} is,
 on every leaf, a solution of \eqref{eq:Hess-gen} if and only if $\ct{V}{_{AB}}\eqSv{v} 0$ there.

				\end{remark}
			In view of this remark, we are induced to distinguish a class of \emph{asymptotic translations},
				\begin{deff}[Asymptotic translations]\label{def:asymptotic-translation}
					Let $ \scri $ be strongly equipped. We say that a bitranslation $ \ct{\tau}{^a}=\alpha\ct{m}{^a}\in\ft $ is an asymptotic translation if and only $\alpha$ satisfies \cref{eq:Hess-gen}. In particular, if the leaves have topology $\mathbb{S}^2$, in a round gauge the restriction $ \csC{\alpha} $ of $ \alpha $ is a linear combination of the $l=0,1$ spherical harmonics.
				\end{deff}
			Observe that these results provide us with a notion of translations intrinsic to $ \scri $ which, as far as we know, have not been characterised before for $ \Lambda>0 $\footnote{For $ \Lambda=0 $, see \cite{Geroch1977}.}. Although we have required a strongly equipped $ \scri $ --that is, the existence of a foliation by umbilical cuts--  important examples have this structure, as the Kottler, Kerr-de Sitter and Robinson-Trautman metrics or the C-metric. \Cref{def:asymptotic-translation} is supported
			by the fact that the restriction of the four- dimensional group of translational KVF in de Sitter space-time are asymptotic translations in this sense. This is explicitly proven in subsection \ref{subsec:dS} of the next chapter, where de Sitter space-time is studied. For instance, for the asymptotic translation studied in subsection \ref{subsec:dS}, $F(\chi) \partial_\chi$, one easily gets $\alpha =F(\chi)/a$ and 
				\begin{equation}
					\cdcn{_{A}}\alpha=0\spacef,
				\end{equation}
			and the restriction $ \csC{\alpha} $ of $ \alpha $ to each cut is a constant 
			Moreover, for the cuts associated to each translation in $\scri$ of de Sitter, $ \ct{V}{_{AB}}=0 $.

			\begin{prop}\label{thm:ckvf-onscri}
				Let $ \pt{\xi}{^\alpha} $ be a CKVF of $ \prn{\cs{M},\ct{g}{_{\alpha\beta}}} $ with non-vanishing restriction to $ \scri $ and define 
				\begin{equation}
				\ct{\xi}{^a}\defeqs\ct{\omega}{_{\alpha}^a}\pt{\xi}{^\alpha}\quad
				\end{equation}
				and 
				\begin{equation}
				\ct{m}{^{a}}\defeq \frac{1}{\alpha}\ct{\xi}{^a}\quad\text{with}\quad\alpha\defeq \sqrt{\ct{\xi}{_{a}}\ct{\xi}{^a}}.
				\end{equation}
				Assume that $ \ct{\xi}{^a} $ is orthogonal to cuts with $ \mathbb{S}^2 $-topology. Then
					\begin{enumerate}
						\item $ \ct{\xi}{^{a}} $ is a CKVF of $ \prn{\scri,\ms{_{ab}}} $ and a BCKVF of $ \prn{\ct{m}{_{a}},\ctcn{P}{_{ab}}}$ that belongs to $ \ft $.\label{it:xi-bckvf}
						\item $ \ct{m}{^a} $ is shear-less ($ \ctcn{\Sigma}{_{ab}}=0 $).\label{it:m-shearless}
						\item The restriction to the the leaves $ \csC{\alpha} $ of the function $ \alpha $ is 
						a solution of \eqref{eq:Hess-gen} 
					(and thus proportional to a combination of the first four spherical harmonics in a round gauge) if and only if $ \ctcn{V}{_{ab}}=0 $ and if and only if $ \ctcn{C}{_{A}}\defeq\ctcn{E}{^a_{A}}\ct{m}{^b}\ct{C}{_{ab}}=0$ (equivalently $ \ctcnp{C}{_A}=-\ctcnm{C}{_{A}} $). \label{it:translation-z}
					\end{enumerate}
			\end{prop}
		\begin{proof}
			Point \ref{it:xi-bckvf} is trivial. Point \ref{it:m-shearless} follows by noting that bitranslations satisfy \cref{eq:bitranslation-alpha-theta}. Hence, $ \ctcn{\Sigma}{_{ab}}=0 $, which together with the assumption that $ \ct{m}{_{a}} $ is surface-orthogonal make $ \scri $ strongly equipped. Then, by \cref{thm:col-rho-noV-alpha} and its remark it follows that the restriction $ \csC{\alpha} $ of $ \alpha $ to the leaves is a solution of \eqref{eq:Hess-gen} if and only if $V_{ab}=0$. 
			Now, the fact that $ \ct{T}{_{AB}^C}=0 $ (which follows from $ \ctcn{\Sigma}{_{ab}}=0 $) \cref{eq:CApV-cut-foliation,eq:CAmV-cut-foliation}, together with the $ \mathbb{S}^2 $-topology of the cuts gives 
			\begin{equation}
			\ctcn{V}{_{ab}}=0\iff \ctcn{C}{_{A}}=0.
			\end{equation}
		\end{proof}

		\subsection{Conserved charges and Balance laws}\label{ssec:balance-law}
				Once we have several definitions of symmetries at hand, we can address the question of charges and conservation laws. With that purpose we are going to treat two types of charges and conserved currents associated with symmetries. The first class is defined using symmetric tensor fields and symmetries intrinsic to $ \scri $; the second, employs the rescaled Bel-Robinson tensor $ \ct{\D}{_{\alpha\beta\gamma}^\delta} $ together with conformal symmetries of $ \prn{\cs{M},\ct{g}{_{\alpha\beta}}} $ and/or asymptotic symmetries. We comment on why the first or second class currents presented below cannot give the right answer for a gravitational energy on $ \scri $. The use of these charges can be fruitful in other investigations though.
		
					\subsubsection{First class charges}
						Let $ \ct{t}{_{ab}} $ be any rank-two, symmetric tensor field on $ \scri $ and $ \ct{\eta}{^a} $  a CKVF of $ \prn{\scri,\ms{_{ab}}} $. Define the current
							\begin{equation}
								\ct{j}{^a}\defeq \ct{t}{^{ab}}\ct{\eta}{_b}.
							\end{equation}
						The divergence of this current reads
							\begin{equation}
								 \cds{_{a}} \ct{j}{^a}= \ct{\eta}{_{b}}\cds{_{a}}\ct{t}{^{ab}}+ \lambda\ms{_{ab}}\ct{t}{^{ab}}\spacef,
							\end{equation}
						where $ 3\lambda\defeq \cds{_{a}}\ct{\eta}{^a} $. If instead one uses a biconformal infinitesimal symmetry of \cref{def:asymptotic-symmetries} and defines the current
							\begin{equation}\label{eq:bc-first-charge}
								\ct{y}{^a}\defeq\ct{t}{^{ab}}\ct{\xi}{_{b}}\spacef,
							\end{equation}
						its divergence gives
							\begin{equation}
								\cds{_{a}}\ct{y}{^a}= \ct{\xi}{_{b}}\cds{_{a}}\ct{t}{^{ab}}+ \psi\ctcn{P}{_{ab}}\ct{t}{^{ab}}+\phi\ct{m}{_{a}}\ct{m}{_{b}}\ct{t}{^{ab}}.
							\end{equation}
						Particular cases of conserved $ \ct{j}{^a} $-currents include those constructed with any  TT-tensor, and for them the charge
							\begin{equation}
								\J\defeq\int_{\Sc} \ct{j}{^a}\ct{r}{_{a}}\csC{\epsilon}
							\end{equation}
						is conserved, where $ \Sc $ is \emph{any} cut with normal $ \ct{r}{_{a}} $ and volume form $ \csC{\epsilon} $. This follows from Stokes theorem, assuming a region $ \Delta $ bounded by two such cuts . An example of conserved $ \ct{y}{^{a}} $-currents is obtained using  TT-tensors satisfying $ \ct{m}{_{a}}\ct{m}{_{b}}\ct{t}{^{ab}}=0 $. For any such current, 
							\begin{equation}
								\Y\defeq\int_{\Sc} \ct{y}{^a}\ct{r}{_{a}}\csC{\epsilon}
							\end{equation}
						is a conserved charge. Observe that in this particular case $ \Y $ is \emph{trivial} for cuts orthogonal to $ \ct{m}{_{a}} $ --if they exist-- , that is, when $ \ct{m}{_{a}}=\ct{r}{_{a}} $ because of the requirement $ \ct{y}{_{d}}\ct{r}{^d}=0 $. Another case, is a $ \ct{y}{^a} $-current constructed with a TT-tensor and a BCKVF with $ \phi=0 $. Despite being obvious, it is necessary to remark that charges defined with $ \ct{y}{^a} $ or $ \ct{j}{^a} $ may be conserved even when the current itself is not -- it is enough that the integral over the region $ \Delta $ of the divergence of the current vanishes. 
						
						It is tempting to define charges using $ \ct{D}{_{ab}} $ or $ \ct{C}{_{ab}} $ ---or a linear combination thereof, or $\ctp{D}{_{ab}}$, $\ctm{D}{_{ab}}$, etcetera--- for the tensor $ \ct{t}{_{ab}} $, as it has been already proposed in the literature for $ \ct{D}{_{ab}} $ \cite{Ashtekar2014}. The balance law  associated to these charges that results from the application of Stokes theorem is not affected by the presence of gravitational radiation, and to illustrate this with our formalism consider the specific case of a strongly equipped $ \scri $ (\cref{def:strong-structure}). Let $ \ct{\xi}{^a} $ be a member of the algebra of biconformal transformations $ \fb $. We know that, in general, it will be composed by a member $ \ct{\tau}{^a} $ of the bitranslations $ \ft $ and an element $ \ct{\chi}{^a} $ of the CKVF $ \fcs $ of the projector $ \ctcn{P}{_{ab}} $ --see the end part of \cref{ssec:symmetries}. 
						For $ \ct{t}{_{ab}}=\ct{D}{_{ab}} $ in \cref{eq:bc-first-charge},
						\begin{equation}\label{eq:Dcharge}
						\Y=\int_{\Sc}\ct{y}{^a}\ct{m}{_{a}}\cscn{\epsilon}=\int_{\Sc}\cbrkt{\alpha\cscn{D}+\ct{\chi}{_{a}}\ctcn{D}{^a}}\cscn{\epsilon}\spacef.
						\end{equation}
						 Let $ \Delta $ be a region bounded by two cuts $ \Sc_{1,2} $ of the foliation given by $ \ct{m}{^a} $, then
							\begin{equation}\label{eq:div-general-1-class}
								\Y\evalat{\Sc_{2}}-\Y\evalat{\Sc_{1}}=\int_{\Delta}\cds{_{a}}\ct{y}{^a}\cs{\epsilon}=\int_{\Delta}\brkt{\prn{\ct{\xi}{_{a}}\cds{_{d}}\ct{D}{^{da}}}+\prn{\phi-\psi}\cscn{D}}\cs{\epsilon}\spacef,
							\end{equation}
						The divergence $ \cds{_{d}}\ct{D}{^{da}} $ is sourced by the matter fields (see \cref{eq:divD}), whereas the second term only contains Coulomb contributions and vanishes identically for conformal symmetries of $ \prn{\scri,\ms{_{ab}}} $, in particular for the asymptotic basic symmetries of definition \ref{def:basic-symmetries}. Observe that there is no gravitational radiation contribution in \eqref{eq:div-general-1-class},  even when gravitational waves \emph{can be arriving at}  $ \scri $ according to \cref{def:criterionGlobal}. The same formula holds interchanging $ \ct{D}{_{ab}} $ by $ \ct{C}{_{ab}} $, only that now the first term in the integrand vanishes identically due to  \cref{eq:divC}. And similar results can be found for linear combinations of $ \ct{D}{_{ab}} $ and $ \ct{C}{_{ab}} $, and for $\ctp{D}{_{ab}}$ or $\ctm{D}{_{ab}}$. This is surprising, as the charge \eqref{eq:Dcharge}, or the analogous ones using linear combinations of $ \ct{D}{_{ab}} $ and $ \ct{C}{_{ab}} $ or $\ctp{D}{_{ab}}$, $\ctm{D}{_{ab}}$, etcetera, include terms of type $\ct{\chi}{_{a}}\ctcn{D}{^a}$ which are associated to the radiative sector of the gravitational field. This opens the door for modifications of these currents associated to $ \ct{D}{_{ab}} $ and $ \ct{C}{_{ab}} $  by adding extra terms that may lead to a more satisfactory balance law. This is work in progress.
						
						Next, assume that \cref{def:noIncomingRadDefOpen} holds, and thus $ -\ct{m}{^a} $ points along the spatial projection of  the propagation direction of radiation, as discussed in \cref{ssec:lightlike-condition-directional,ssec:noincomingrad}. Then, charges defined on the `natural' cuts orthogonal to $ \ct{m}{^a} $ might be sensible to radiative contributions. But now $ \ctcn{D}{^a} $ is the divergence of the symmetric traceless tensor field $ \ctcn{V}{_{ab}} $, the first component of news, 
						\begin{equation}
						N\ctcn{D}{_b}= -\cdcn{_c}\ctcn{V}{_b^c}\spacef,
						\end{equation}
						which on any cut of the foliation is written by means of the intrinsic connection as
						\begin{equation}
						N\ctcn{D}{_B}= -\cdc{_C}\ctcn{V}{_B^C}\spacef.
						\end{equation}
						Then, for topological spheres, and in general for compact cuts, the term $\ct{\chi}{_{a}}\ctcn{D}{^a}$ integrates out  on using \eqref{eq:c4c} and the charge reads
						\begin{equation}
						\Y=\int_{\Sc}\ct{y}{^a}\ct{m}{_{a}}\cscn{\epsilon}=\int_{\Sc}\alpha\cscn{D}\cscn{\epsilon}\spacef,
						\end{equation}
						which only contains the Coulomb contribution $ \cscn{D} $. 	A very similar cancellation occurs if one uses a CKVF of $ \scri $ because the tangent part to an umbilical cut of the conformal symmetry is a CKVF of the metric on that cut too. Hence, neither these charges, nor their difference given by the general \cref{eq:div-general-1-class}, contain explicit radiative terms. Of course, the discussion of \cref{ssec:lightlike-condition-directional,ssec:noincomingrad} on the interpretation of the Coulomb and radiative  terms as such depends on the choice of $ \ct{m}{_{a}} $. Still, the fact that a general firs-class current $ \ct{y}{^a} $ is identically conserved in the absence of matter fields and for any conformal transformation shows that the associated charges $ \Y $ for any choice of cut are insensible to gravitational radiation. Indeed, for $ \scri=\mathbb{S}^3 $ or $ \scri=\mathbb{R}^3 $ the radius of the topological 2-spheres can be shrunk to $ 0 $, hence making these charges to vanish identically. This is not the case for $ \mathbb{R}\times\mathbb{S}^2 $ and thus one could consider the vanishing of these charges as a topological feature.
						
						Of course, the interest of having conserved charges is not only related to the existence of gravitational radiation, and in that sense the above charges may be very useful in different contexts. 
					\subsubsection{Second class charges}
						When dealing with fields other than gravity, the standard approach is to consider charges associated to the energy-momentum tensor of the field theory. However, as mentioned at the beginning and is well-known, there is not such a thing in General Relativity. Still, we can resort to using the rescaled Bel-Robinson tensor, which possesses many properties of an energy-momentum tensor, and thereby define a second class of charges. One has to be aware of the dimensionality of such charges and currents, since they are of tidal nature and do not carry, in general, units of energy-momentum. 
						
						Consider first a triplet of CKVF $ \prn{\ptrd{(i)}{\xi}{^\alpha},\ptrd{(j)}{\xi}{^\alpha},\ptrd{(k)}{\xi}{^\alpha}} $ of the space-time $ \prn{\cs{M},\ct{g}{_{\alpha\beta}}} $, which can contain repeated elements. 
						Assume that in  a neighbourhood of $ \scri $, $ \ct{T}{_{\alpha\beta}}=0 $, --note that this is a more restrictive condition that the one taken in the rest of the work, see \cref{it:energytensorassumption} on page \pageref{it:energytensorassumption}. Then, in that neighbourhood of $ \scri $
							\begin{equation}
								\cd{_{\mu}}\ct{\D}{^\mu_{\alpha\beta\gamma}}=0\spacef.
							\end{equation}
						It is easy to check that the currents \cite{Senovilla2000,Lazkoz2003}
							\begin{equation}
								\ct{\B}{^\alpha}\defeq\ptrd{(i)}{\xi}{^\mu}\ptrd{(j)}{\xi}{^\nu}\ptrd{(k)}{\xi}{^\rho}\ct{\D}{^\alpha_{\mu\nu\rho}}
							\end{equation}
						are divergence-free in that region of the space-time (including $ \scri $)
							\begin{equation}
								\cd{_{\mu}}\ct{\B}{^\mu}=0\spacef.
							\end{equation}
						Then, the quantity defined on any spacelike hypersurface $ \Sigma $ 
						orthogonal to some timelike $ \ct{t}{_{\alpha}} $
							\begin{equation}
								\cs{\B}_{\Sigma}\defeq\int_{\Sigma}\ct{t}{_{\mu}}\ct{\B}{^\mu}\cs{\epsilon}
							\end{equation}
						is conserved in a space-time region $ \Delta_{M} $ bounded by any two $ \Sigma_1 $ and $ \Sigma_2 $ 
						orthogonal to any two future-pointing timelike $ \ctrd{1,2}{t}{_{\alpha}} $ and with $ \Sigma_{2} $ to the future of $ \Sigma_{1} $,
							\begin{equation}
							0=\int_{\Delta_{M}}\cd{_{\mu}}\ct{\B}{^\mu}\eta=\cs{\B}_{\Sigma_{2}}-\cs{\B}_{\Sigma_{1}}\spacef.
							\end{equation}
						In particular, $ \Sigma $ can be chosen to be $ \scri $. 
						
						Suppose, first, that  $ \ptrd{(i)}{\xi}{^\alpha} $ are completely tangent to $ \scri $. Then
							\begin{equation}
								\cs{\B}_{\scri}=\int_{\scri} \ct{Q}{_{abc}}\ptrd{(i)}{\xi}{^a}\ptrd{(j)}{\xi}{^b}\ptrd{(k)}{\xi}{^c}\cs{\epsilon}\spacef,
							\end{equation}
						where $ \ct{Q}{_{abc}} $ is defined for the rescaled Bel-Robinson tensor on $ \scri $ as in \cref{eq:Qalfonso}, and one can write
							\begin{equation}
							\cs{\B}_{\scri}=\int_{\scri}\prn{\ctrd{(i)}{\xi}{^m}\ctrd{(j)}{\xi}{_m}\cts{\Pc}{_d}\ctrd{(k)}{\xi}{^d}-4\ctrd{(i)}{\xi}{^e}\ct{C}{_{ec}}\ctrd{(j)}{\xi}{^f}\ct{D}{_{fd}}\ct{\epsilon}{^{pcd}}\ctrd{(k)}{\xi}{_{p}}}\cs{\epsilon}\spacef.
							\end{equation}
						 Notice: if the condition in \cref{def:noIncomingRadDefOpen} holds on $ \scri $ and there is no radiation $ \cts{\P}{^a}=0 $, $ \cs{\B}_{\scri}=0 $. 
						 
						 Suppose, now, that $ \ptrd{(i)}{\xi}{^\alpha}\eqs\csrd{(i)}{\beta}\ct{n}{^\alpha}  $ for some non-vanishing functions $\csrd{(i)}{\beta}$, then
							\begin{equation}
							\cs{\B}_{\scri}=\int_{\scri} \csrd{(i)}{\beta}\csrd{(j)}{\beta}\csrd{(k)}{\beta}\cs{\W}\defeq \C_{\Sigma} \spacef,
							\end{equation}
						where $ \cs{\W} $ is the asymptotic canonical supermomenta density --see \cref{eq:canonical-se-decomposition}. Observe that $ \C_{\Sigma} $ vanishes if and only if $ \cs{\W}=0 $ at $ \scri $, i.e., $ \ct{d}{_{\alpha\beta\gamma\delta}}= 0 $ there. Because the charge is conserved in $ \Delta_{M} $, $ \C_{\Sigma}=0 $ for all $ \Sigma $ in $ \Delta_{M} $. In particular, de Sitter space-time has $ \C_{\Sigma}=0 $ everywhere, including at $\scri$, and any other space-time having $ \cs{\B}_{\scri}=0 $ is de Sitter space-time in the domain of dependence of $ \scri $.\\
							
						Now, let us focus on a strongly equipped $ \scri $ and consider a bitranslation $ \ct{\tau}{^a}=\alpha\ct{m}{^a} $. Let us define  the current
							\begin{equation}
								\ct{\R}{^a}\defeqs-\ct{\D}{^\alpha_{\mu\nu\rho}}\ct{\omega}{_\alpha^a}\ct{e}{^\mu_{b}}\ct{e}{^\nu_{c}}\ct{e}{^\rho_{d}}\ct{\tau}{^b}\ct{\tau}{^c}\ct{\tau}{^d}.
							\end{equation}
						Notice that according to the discussion at the end of \cref{ssec:strong-structure}, the dimensions of this quantity are
							\begin{equation}
								\brkt{\ct{\R}{^a}}=L^{-1}\spacef,
							\end{equation}
						despite of being constructed with a superenergy tensor --indeed, $ \ct{\R}{^a} $ has physical units of $ MT^{-2} $. Its integral over any cut gives a charge with units of energy. If $ \ct{m}{^a} $ is orthogonal to cuts $ \Sc_{i} $, the divergence of the current $ \ct{\R}{^a} $ integrated over the compact region $ \Delta $ bounded by $ \Sc_{1,2} $ gives a balance law
							\begin{equation}
								\int_{\Delta}\cds{_{a}}\ct{\R}{^a}\cs{\epsilon}=\cs{\R}_{\Sc_{2}}-\cs{\R}_{\Sc_{1}}\spacef,
							\end{equation}
						where 
							\begin{equation}
								\cs{\R}_{\Sc_{i}}\defeq\int_{\Sc_{i}}\ct{\R}{^a}\ct{m}{_{a}}\csC{\epsilon}=\int_{\Sc_{i}}\alpha^3\brkt{\frac{1}{4}\prn{\csp{\W}+\csm{\W}}-\prn{\csp{\Z}+\csm{\Z}}+\frac{3}{2}\cs{\V}}\csC{\epsilon}\spacef.
							\end{equation}
						The left-hand side follows by decomposing $ \sqrt{2}\ct{m}{_{\alpha}}=\ctp{k}{_{\alpha}}-\ctm{k}{_{\alpha}} $ and introducing the definitions of \cref{eq:Wpdef,eq:Wmdef,eq:Zpdef,eq:Zmdef,eq:Vdef}. This charge contains both radiative and Coulomb contributions. In order to compute the intrinsic divergence of $ \ct{\R}{^a} $, one has to know the Lie derivative of $ \ct{\D}{_{\alpha\beta\gamma\delta}} $  along $ \ct{n}{^\alpha} $ at $ \scri $. 
						

	\section{Examples}\label{sec:Examples}

		\subsection{de Sitter space-time}\label{subsec:dS}
		The paradigmatic space-time with $\Lambda>0$ is de Sitter, the 4-dimensional space-time with positive constant curvature. Its metric can be written (with $c=1$) as
		\be\label{eq:dSphys}
		\hat g_{{\rm dS}} =-dt^2 +a^2 \cosh^2(t/a) h^{\mathbb{S}^3}
		\ee
		where $a$ is a shorthand for $a=\sqrt{3/\lambda}$, $t\in(-\infty,\infty)$ and $h^{\mathbb{S}^3}$ is the round metric (unit constant curvature) on the standard 3-sphere $\mathbb{S}^3$. The conformal completion is easily achieved by means of the change
		$$
		t\rightarrow \tau , \hspace{1cm} \sin \tau = \frac{1}{\cosh(t/a)} 
		$$
		which brings the metric to the form 
		\be\label{eq:dSunphys}
		g_{{\rm dS}} =\frac{a^2}{\sin^2\tau} \left(-d\tau^2 +h^{\mathbb{S}^3} \right).
		\ee
		Thus, the Einstein space metric is the unphysical metric, $\Omega =  \sin \tau$, and $\scri^\pm$ are given, respectively, by $\tau =0,\pi$ corresponding to $|t|\rightarrow \infty$. In the unphysical space-time $\scri^\pm$ have toplogy $\mathbb{S}^3$ and inherit the round metric $h =a^2 h^{\mathbb{S}^3}$ os `radius' $a$.
		
		The KVF of \eqref{eq:dSphys} are
		$$
		\vec\xi = \frac{1}{a} \tanh (t/a) {\rm grad} X + X \partial_t + \vec\xi^{\mathbb{S}^3}
		$$
		where $\vec\xi^{\mathbb{S}^3}$ are the KVF of $\mathbb{S}^3$, $X$ is any function on $\mathbb{S}^3$ (and grad $X$ its gradient) satisfying
		\be\label{eq:DDX} 
		\cds{_i}\cds{_j} X =-\frac{X}{a^2} h_{ij} 
		\ee
		so that each $\cds{_i} X$ defines a CKVF of $(\scri,h_{ij})$. There are four (linearly independent) of these, corresponding to the four Cartesian coordinates of $\mathbb{R}^4$ into which one can embed canonically $\mathbb{S}^3$ with
		$$
		X_1^2 +X_2^2 +X_3^2 +X^2_4=1 .
		$$
		These four functions provide four KVFs of \eqref{eq:dSphys} which together with the six KVFs of $\mathbb{S}^3$ add up to the ten KVFs of de Sitter. All these KVFs extend to $\scri$ in the unphysical metric and they are tangent to $\scri$ there, as follows from their expression in terms of the coordinate $\tau$
		$$
		\vec\xi = \frac{1}{a} \cos\tau {\rm grad} X-\frac{1}{a} \sin\tau X \partial_\tau + \vec\xi^{\mathbb{S}^3}
		$$
		so that, at $\scri$ they become
		\be\label{eq:CKVdS}
		\vec\eta :=\vec\xi|_{\scri^\pm} = \pm {\rm grad} X + \vec\xi^{\mathbb{S}^3}.
		\ee
		These are the ten CKVF of $\scri$ and all of them are good {\em basic asymptotic infinitesimal symmetries} according to \cref{def:basic-symmetries} because in de Sitter $C_{ab}=D_{ab}=0$. The four ones given by $\eta_i =\cds{_i} X $ are the proper (non-KVF) ones. These are the 4 CKV used to define the de-Sitter 4-momentum and represent the {\em translations}.
		
		Using standard angular coordinates $\{\chi,\theta,\varphi\}$ on the 3-sphere , the four independent solutions of \eqref{eq:DDX} can be chosen as usual: $X_1 \sin\chi \sin\theta \sin\varphi$, $X_2 =\sin\chi\sin\theta \cos\varphi$, $X_3 = \sin\chi \cos\theta$ and $X_4 =\cos\chi$. Now, given the symmetry of the problem, any of the 4 translations is equivalent to any other and thus one can concentrate to study just one of them, say the `simplest' 
		$$
		\vec\eta_4 = {\rm grad} X_4= -\sin\chi \partial_\chi .
		$$
		This vector field is surface-orthogonal and shear-free, defining a foliation by {\em umbilical} cuts, each leaf $\chi=$constant being a round 2-sphere. To prove this it is enough to recall \eqref{eq:DDX}, which translates for the unit $m_i =a \cds{_i} \chi $ into 
		$$
		\cds{_i} m_j =- \frac{1}{a} \cot\chi (h_{ij} -m_i m_j) = - \frac{1}{a} \cot\chi  P_{ij} .
		$$
		Hence, the 4 translations define, themselves, umbilical foliations, and any of them serves to {\em strongly equip} $\scri$. The question arises, for any of these strong equipments, what are the corresponding asymptotic infinitesimal symmetries of definition \ref{def:asymptotic-symmetries}. To answer, consider the (strong) equipment defined by $m^a =a \eta_4^a$ as before (remember that all the strong equipments are equivalent). Then a straightforward calculation solving the equations \eqref{eq:biconformal-P} and \eqref{eq:biconformal-m} leads to the general solution given by
		$$
		F(\chi) \partial_\chi +\vec\xi^{\mathbb{S}^2} 
		$$
		where $F(\chi)$ is arbitrary and $\vec\xi^{\mathbb{S}^2}$ are the CKVF of the round $\mathbb{S}^2$. This algebra is infinite-dimensional.
		
		To end this subsection, observe that as the Weyl tensor of \eqref{eq:dSphys} vanishes, the rescaled Weyl tensor so does and there is no notion of (strong) orientation for $\scri$ in de Sitter, which is logical given the absolute symmetry of the space-time: any point is equivalent to any other point.

		\subsection{The Kerr-de Sitter and Kottler metrics}	
			Let us start with the conformal Kerr-de Sitter metric
			\begin{align}\label{metricKerrdS}
			\df s^2&=\frac{1}{r^2}\Bcbrkt{\prn{-\frac{\Delta_r}{\rho^2}+\frac{\Delta_\theta}{\rho^2}a^2 \sin^2\theta} \df{t}^2+\frac{\rho^2}{\Delta_r}\df{r}^2+\frac{1}{\Xi^2}\brkt{-\frac{\Delta_r}{\rho^2}a^2\sin^4\theta+\frac{\Delta_\theta}{\rho^2}(r^2+a^2)^2\sin^2\theta}\df{\phi}^2+\nonumber\\
			&+\frac{1}{\Xi}\brkt{\frac{\Delta_r}{\rho^2}a\sin^2\theta-\frac{\Delta_\theta}{\rho^2}a\sin^2\theta(r^2+a^2)}\prn{\df{\phi} \df{t}+\df{t} \df{\phi}}+ \frac{\rho^2}{\Delta_\theta}\df{\theta}^2\quad}.
			\end{align}	
			These are Boyer-Lindquist-type coordinates, with
			\begin{equation}
			t\in\mathbb{R},\quad r\in\mathbb{R},\quad \theta\in[0,\pi),\quad \phi\in[0,2\pi]
			\end{equation}
			and 
			\begin{equation}
			\Lambda>0,\quad a\in\mathbb{R}, \quad m\in\mathbb{R}\setminus 0\quad.
			\end{equation}
			The metric functions are defined as
			\begin{align}
			\rho^2&\defeq r^2+a^2 \cos^2\theta\quad,\\
			\Delta_r&\defeq (a^2+r^2)\prn{1-\frac{\Lambda}{3}r^2}-2mr\quad,\\
			\Delta_\theta & \defeq 1+\frac{\Lambda}{3}a^2\cos^2\theta\quad,\\
			\Xi&\defeq 1+\frac{\Lambda}{3}a^2\quad.
			\end{align}
			The particular case with $a=0$ gives the Kottler (sometimes called Schwarzschil-de Sitter) spherically symmetric conformal metric.
			
			Infinity is located at $ r\rightarrow \infty $, and we have chosen
			\begin{equation}
			\Omega\defeq \frac{A}{r}\quad,
			\end{equation}
			with $ A= $constant with dimensions $ \brkt{A}=L $, so that $ \brkt{\Omega}=1 $. From now on we set $ A=1 $. This choice of $ \Omega $ indeed belongs to the divergence-free family of conformal gauges \eqref{eq:divergence-freeGauge}. Hence, the normal to $ \scri $ is
				\begin{equation}
					\ct{N}{_{\alpha}}=-\frac{1}{r^2}\cd{_{\alpha}}r\quad.
				\end{equation}
			 Notice that $ \df{r}^2=r^4\df{\Omega}^2 $ and that
			\begin{align}
			N^2&\eqs\frac{\Lambda}{3}\quad,\\
			\frac{\rho^2}{\Delta_r}\Omega^2r^4&\eqs -\frac{1}{N^2}\quad,\\
			\frac{\rho^2}{\Delta_\theta}\Omega^2&\eqs \frac{1}{1+N^2a^2\cos^2\theta} \quad,\\
			\frac{\Delta_r}{\rho^2}\Omega^2&\eqs  -N^2\quad,\\				
			\frac{\Delta_\theta}{\rho^2}\Omega^2&\eqs 0\quad,\\
			\frac{\Delta_\theta}{\rho^2}\prn{r^2+a^2}^2\Omega^2&\eqs 1+N^2a^2\cos^2\theta\quad.				
				\end{align}
			Using these formulae, one can write the metric of $ \scri $ as
				\begin{equation}
			h=N^2\df{t}^2+\frac{1}{\Xi^2}\prn{1+N^2a^2}\sin^2\theta \df{\phi}^2-\frac{1}{\Xi}N^2a\sin^2\theta\prn{\df{\phi} \df{t} + \df{t}\df{\phi}}+\prn{1+N^2a^2\cos^2\theta}^{-1}\df{\theta}^2\quad.
				\end{equation}
			The electric and magnetic parts of the rescaled Weyl tensor at $ \scri $ read respectively
				\begin{align}
				\ct{C}{_{ab}}&= 0\quad,\\
				\ct{D}{_{ab}}&=- \frac{2}{3}\Lambda m \cds{_{a}}t\cds{_{b}}t+\frac{2}{\Xi}N^2a m \sin^2\theta \prn{\cds{_{a}}\phi\cds{_{b}}t+\cds{_{a}}t\cds{_{b}}\phi}+\nonumber\\
				&+m\prn{1+a^2N^2\cos^2\theta}^{-1}\cds{_{a}}\theta\cds{_{b}}\theta+\frac{1}{\Xi^2}m\sin^2\theta\prn{1+a^2\Lambda\cos^2\theta-\frac{2}{3}a^2\Lambda}\cds{_{a}}\phi\cds{_{b}}\phi\quad.\label{eq:electric-Kerr-metric}
				\end{align}
			The intrinsic Ricci tensor, scalar curvature and Schouten tensor have the following expressions:
				\begin{align}
					\cts{R}{_{ab}}&=2a^2N^4\cos^2\theta\cds{_{a}}t\cds{_{b}}t-\frac{2}{\Xi}aN^2\sin^2\theta\prn{1+3a^2N^2\cos^2\theta}\cds{_{(a}}t\cds{_{b)}}\phi\nonumber\\
					&+\frac{1}{\Xi^2}\brkt{1-\prn{1-3a^2N^2}\cos^2\theta-3a^2N^2\cos^4\theta}\prn{1+a^2N^2}\cds{_{a}}\phi\cds{_{b}}\phi\nonumber\\
					&+\frac{1+4a^2N^2\cos^2\theta-a^2N^2}{1+a^2N^2\cos^2\theta}\cds{_{a}}\theta\cds{_{b}}\theta\quad,\\
					\csS{R}&=2-2a^2N^2+10N^2a^2\cos^2\theta\quad,\\
					\cts{S}{_{ab}}&=\frac{1}{2}N^2\prn{a^2N^2-a^2N^2\cos^2\theta-1}\cds{_{a}}t\cds{_{b}}t-\frac{1}{\Xi}aN^2\sin^2{\theta}\prn{1+a^2N^2+a^2N^2\cos^2\theta}\cds{_{(a}}t\cds{_{b)}}\phi\nonumber\\
					&+\frac{1}{2\Xi^2}\prn{1+a^2N^2}\sin^2\theta\prn{1+a^2N^2+a^2N^2\cos^2\theta}\cds{_{a}}\phi\cds{_{b}}\phi\nonumber\\
					&+\frac{1-a^2N^2+3a^2N^2\cos^2\theta}{2\prn{1+a^2N^2\cos^2\theta}}\cds{_{a}}\theta\cds{_{b}}\theta\quad.\label{eq:schouten-kerr-dS}
				\end{align}
			There are two repeated PND $ \ctrd{1}{\ell}{_{\alpha}} $ and $ \ctrd{2}{\ell}{_{\alpha}} $ which read at $ \scri $
					\begin{align}
					\ctrd{1}{\ell}{_\alpha}&\eqs\frac{1}{\sqrt{2}}\prn{-\frac{1}{Nr^2}\cd{_\alpha}r-N\cd{_\alpha}t+\frac{1}{\Xi}aN\sin^2\theta\cd{_{\alpha}}\phi}\quad,\\
					\ctrd{2}{\ell}{_\alpha}&\eqs\frac{1}{\sqrt{2}}\prn{-\frac{1}{Nr^2}\cd{_\alpha}r+N\cd{_\alpha}t-\frac{1}{\Xi}aN\sin^2\theta\cd{_{\alpha}}\phi}\quad.
				\end{align}
			Accordingly, there are two different strong orientations (see \cref{def:orientationStrong} and \cref{rmk:number-strong-orientations}). We choose one of them by defining
				\begin{align}
					\ctm{k}{_\alpha}&\defeqs \ctrd{1}{\ell}{_\alpha}\quad,\label{eq:km-Kerr-metric-a}\\
					\ct{m}{_\alpha}&\defeqs \ct{n}{_\alpha}-\sqrt{2}\ctm{k}{_\alpha}=N\prn{\cd{_\alpha}t-\frac{1}{\Xi}a\sin^2\theta\cd{_{\alpha}}\phi}\quad,\\
					\ctp{k}{_\alpha}&\defeqs \frac{1}{\sqrt{2}}\prn{\ct{n}{_\alpha}+\ct{m}{_\alpha}}=\frac{1}{\sqrt{2}}\prn{-\frac{1}{Nr^2}\cd{_\alpha}r+N\cd{_\alpha}t-\frac{1}{\Xi}aN\sin^2\theta\cd{_{\alpha}}\phi}=\ctrd{2}{\ell}{_{\alpha}}\quad,\label{eq:kp-Kerr-metric-a}
				\end{align}		
			where $ N\ct{n}{_{\alpha}}\defeqs\ct{N}{_{\alpha}} $, such that  $ \ctm{k}{^\alpha}\ctp{k}{_\alpha}\eqs-1 $, $ \ct{m}{^\alpha}\ctm{k}{_\alpha}\eqs-1/\sqrt{2} $ and $ \ct{m}{^\alpha}\ctp{k}{_\alpha}\eqs1/\sqrt{2} $. Notice that both repeated PND are coplanar with the normal $ \ct{N}{_{\alpha}} $ which makes the two strong orientations equivalent from the viewpoint of $\scri$, in the sense that they define, {\em up to sign}, the same vector field $m^a$ there. The pullback to $ \scri $ of $ \ct{m}{_{\alpha}} $ is
				\begin{align}
				\ct{m}{_{a}}&\eqs N\prn{\cds{_a}t-\frac{1}{\Xi}a\sin^2\theta\cds{_{a}}\phi},\label{eq:m-Kerr-dS-1}\\
				\ct{m}{^a}&\eqs \frac{1}{N}\delta_t^a\quad,
				\end{align}
			$ \dpart{_{t}} $ being a KVF of $ \prn{\scri,\ms{_{ab}}} $. The non-vanishing intrinsic connection coefficients are
				\begin{align}
					\cts{\Gamma}{^\theta_{t\phi}}&=\frac{1}{\Xi}aN^2 \cos\theta\sin\theta\prn{1+a^2N^2\cos^2\theta}\quad,\\
					\cts{\Gamma}{^\theta_{\phi\phi}}&=-\frac{1}{\Xi^2}\prn{1+a^2N^2}\cos\theta\sin\theta\prn{1+a^2N^2\cos^2\theta},\\
					\cts{\Gamma}{^\theta_{\theta\theta}}&=\frac{a^2N^2\cos\theta\sin\theta}{1+a^2N^2\cos^2\theta}\quad,\\
					\cts{\Gamma}{^\phi_{\theta t}}&=-\frac{\Xi aN^2\cos\theta}{\sin\theta\prn{1+a^2N^2\cos^2\theta}}\quad,\\
					\cts{\Gamma}{^\phi_{\phi\theta}}&=\frac{\cos\theta}{\sin\theta}\quad,\\
					\cts{\Gamma}{^t_{\theta t}}&=-\frac{a^2N^2\cos\theta\sin\theta}{1+a^2N^2\cos^2\theta}\quad.
				\end{align}
				One does not need them to compute the kinematics of $ \ct{m}{_{a}} $ (see definitions in \cref{app:congruences}) though; noting the fact that $ \ct{m}{^a} $ is a KVF, $ \ctcn{\kappa}{_{ab}} $ vanishes\footnote{One can check that this is the case by doing the explicit calculation using the non-vanishing components of $ \cts{\Gamma}{^a_{cb}} $.},  whereas $ \ctcn{a}{_b} $ vanishes by symmetrising in \cref{eq:derivative-r-decomposition} and contracting once with $ \ct{m}{^a} $, and $ \ctcn{\omega}{_{ab}} $ does not involve the connection:
				\begin{align}
				\ctcn{a}{_{b}}&= 0\quad,\label{eq:acceleration-metric-Kerr}\\
				\ctcn{\kappa}{_{ab}}&=0\quad,\label{eq:kappa-metric-Kerr}\\
				\ctcn{\omega}{_{ab}}&=\frac{2}{\Xi}a\sin\theta\cos\theta\cds{_{[a}}\phi\cds{_{b]}}\theta\quad.\label{eq:vorticity-metric-Kerr}
				\end{align}		
			\Cref{eq:vorticity-metric-Kerr} implies that $ \ct{m}{_{a}} $ is not surface-orthogonal, that is, it does not give a foliation. 
			 The projector to $ \Scn $ (see \cref{app:congruences}) reads
				\begin{equation}
				\ctcn{P}{_{ab}}= \frac{1}{1+a^2N^2\cos^2\theta}\cds{_{a}}\theta\cds{_{b}}\theta+\frac{1}{\Xi^2}\prn{1+a^2N^2\cos^2\theta}\sin^2\theta \cds{_{a}}\phi\cds{_{b}}\phi\quad.
				\end{equation}
			The pair $ \prn{\ctcn{P}{_{ab}},\ct{m}{_{a}}} $ characterises the congruence of curves given by $ \ct{m}{^a} $ and the projected surface $ \Scn $; we say, according to \cref{def:additional-structure}, that $ \scri $ is equipped.\\
			
			All the quantities corresponding to the decomposition of $ \ct{D}{_{ab}} $ and $ \ct{C}{_{ab}} $ (see \cref{ssec:lightlike-projections} and \cref{eq:notation-symmetric-tensor-m,eq:notation-symmetric-tensorLatin-m}) vanish except for
					\begin{equation}
						\cs{D}=-\ctcn{D}{^M_M}= -2m \quad
					\end{equation}
					and thus we have
					\be\label{eq:DKerrdS}
					\ct{D}{_{ab}} = -m \left(3 m_a m_b - \ct{h}{_{ab}} \right).
					\ee
					
			\subsubsection{Asymptotic symmetries}	
					It is known that due to the $ \mathbb{R}\times\mathbb{S}^2 $ topology of $ \scri $ \cite{Mars2017} the group of CKVF of $ \prn{\scri,\ms{_{ab}}} $ is 4-dimensional \cite{Ashtekar2014}. This, however, misses the TT-tensor $D_{ab}$, which must be taken into account as essential part of the asymptotic structure. Taking the Lie derivative of \eqref{eq:DKerrdS} one easily finds that the definition \ref{def:basic-symmetries} requires the solutions to be actually KVF of $ \prn{\scri,\ms{_{ab}}}$. In this sense, the generators of the basic symmetries are given by  $\partial_t $ and $\partial_\phi $, which indeed are KVF of $ \prn{\scri,\ms{_{ab}}}$. Hence, this group is just 2-dimensional ---unless in the Kottler metric case, $a=0$, which is 4-dimensional. 
					
					We can also study the asymptotic symmetries of  (\cref{def:asymptotic-symmetries}). The algebra of biconformal transformations $ \fb $ is constituted by elements of the form
						\begin{equation}
							\ct{\xi}{^a}=\beta\ct{m}{^a}+\ct{\chi}{^a}\quad,
						\end{equation}
					where $ \beta $ and $ \ct{\chi}{^a} $ satisfy \cref{eq:biconformal-mbeta,eq:biconformal-Dbeta,eq:biconformal-m-chi,eq:biconformal-beta-psi-chi}, that is,
						\begin{equation}
							\lied_{\vec{m}}\ct{\chi}{^a}=0,\quad\cdcn{_{b}}\beta=-2\ctcn{\omega}{_{eb}}\ct{\chi}{^e},\quad 2\cdcn{_{(a}}\ct{\chi}{_{b)}}=2\psi\ctcn{P}{_{ab}}
						\end{equation}
					and one defines $ \phi\defeq \ct{m}{^e}\cds{_e}\beta $. On the one hand, from \cref{eq:bitranslation-ralpha,eq:bitranslation-Dalpha,eq:bitranslation-alpha-theta}, it follows that the elements of the subalgebra of bitranslations $ \ft $ have the form
					\begin{equation}
						\ct{\tau}{^a}=\alpha\ct{m}{^a}\quad\text{with}\quad\cdc{_{a}}\alpha=0\quad,
					\end{equation}
						and $ \lambda \defeq \ct{m}{^e}\cds{_{e}}\alpha$.  However, according to \cref{eq:commutator-Es}, $ \cdcn{_{[a}}\cdcn{_{b]}}\alpha=-\ctcn{\omega}{_{ab}}\ct{m}{^e}\cds{_{e}}\alpha $. Thus the only possibility is $ \alpha= $constant. In other words, there is just one element of $ \ft $ and this is the KVF $ \dpart{t} $. On the other hand, it is easily seen that the non-vanishing $ \ctcn{\omega}{_{ab}} $ spoils the existence of a subalgebra of conformal transformations of the projector $ \fcs $ --see comments below \cref{eq:conformalS-Deta}. Finally, a more detailed calculation shows that the remaining general biconformal symmetries $ \ct{\xi}{^a}\in\fb $ associated to the orientation given by \cref{eq:m-Kerr-dS-1} are of the form:
							\begin{equation}
								\ct{\xi}{^a}=\alpha\delta_t^a+b\delta_\phi^a\quad,
							\end{equation}
						where $ b $ is a constant. Therefore, the basic infinitesimal symmetries are precisely the biconformal infinitesimal symmetries of the pairs $ \prn{\ct{m}{_{a}},\ctcn{P}{_{ab}}} $ that define the strong orientation, unless in the particular Kottler metric with $a=0$, where the equipped symmetries constitute an infinite-dimensional algebra. This algebra will be given as a particular case of the different equipments that we are going to consider next.
						
				\subsubsection{Strong equipment
				}
				
					There are other interesting equipments on $ \scri $, that is, other choices for $ \ct{m}{_{a}} $.\footnote{This is a different $m^a$ from the previous section: we keep the notation but this should not lead to confusion} Specifically, the vector field
						\begin{align}
							\ct{m}{_{a}}&\defeq \frac{N}{\Xi}\prn{1+N^2a^2\cos^2\theta}^{\frac{1}{2}}\cds{_{a}}t\quad,\\
							\ct{m}{^{a}}&=\Xi\frac{1}{N\prn{1+N^2a^2\cos^2\theta}^{\frac{1}{2}}}\prn{\delta_t^a+aN^2\delta_\phi^a}\quad\label{eq:m2-kerr-dS}
						\end{align} 
					is worth attention. All its kinematic 
					vanish except the acceleration,
						\begin{equation}
							\ctcn{a}{_{b}}=a^2N^2\frac{\cos\theta\sin\theta}{1+a^2N^2\cos^2\theta}\cdcn{_{b}}\theta\quad.
						\end{equation}
					Hence, it is orthogonal to a foliation of umbilical cuts with metric
						\begin{equation}\label{eq:qabKerrdS}
							\mcn{_{AB}}=\frac{\sin^2\theta}{\Xi}\cdcn{_A}\phi\cdcn{_B}\phi+\frac{1}{1+a^2 N^2\cos^2\theta}\cdcn{_{A}}\theta\cdcn{_{B}}\theta\quad
						\end{equation}
					and Gaussian curvature
						\begin{equation}
							\cscn{K}=1+2a^2N^2\cos^2\theta\quad.
						\end{equation}
				 The projector to these cuts is written as
						\begin{equation}
							\ctcn{P}{_{ab}}=\frac{\sin^2\theta}{\Xi}\prn{aN^2\cds{_{a}}t-\cds{_{a}}\phi}\prn{aN^2\cds{_{b}}t-\cds{_{b}}\phi}+\frac{1}{1+a^2N^2\cos^2\theta}\cds{_{a}}\theta\cds{_{b}}\theta\quad.
						\end{equation}
				Observe that this provides a \emph{strongly} equipped $ \scri $ --\cref{def:strong-structure}. 
				
				The tensor $ \ctcn{\rho}{_{ab}} $ can be computed using that \eqref{eq:qabKerrdS} has axial symmetry and  the general expressions for axially-symmetric metrics of \cref{sssec:rho-axisymm}. This calculation also provides the conformal factor 
				\be\label{eq:omegaKerrdS}
				\omega = K\Xi \sqrt{\Delta_\theta}=K\Xi \prn{1+N^2a^2\cos^2\theta}^{\frac{1}{2}}
				\ee
				between \eqref{eq:qabKerrdS} and the round metric \eqref{eq:qround}.
				Concerning $ \ctcn{\rho}{_{AB}} $ we get
						\begin{align}
							\ctcn{\rho}{_{AB}}&=\frac{\sin^2\theta}{2\Xi}\prn{1+a^2N^2+a^2N^2\cos^2\theta}\cdcn{_{A}}\phi\cdcn{_{B}}\phi\nonumber\\
							&+\frac{1}{2\prn{1+a^2N^2\cos^2\theta}}\brkt{1+3a^2N^2\cos^2\theta-a^2N^2}\cdcn{_{A}}\theta\cdcn{_{B}}\theta.
						\end{align}
						From \eqref{eq:schouten-kerr-dS} and this expression we realize that $\ctcn{S}{_{AB}} =\ctcn{\rho}{_{AB}}$. Thus, noting that $ \ctcn{T}{_{AB}^C}=0 $ and that this implies $ \ctcn{U}{_{AB}}=\ctcn{S}{_{AB}} $, one can compute $ \ctcn{V}{_{AB}} $ using \cref{eq:schouten-kerr-dS}:
						\begin{equation}\label{eq:V=0KerrdS}
							\ctcn{V}{_{AB}}=\ctcn{U}{_{AB}}-\ctcn{\rho}{_{AB}}=\ctcn{S}{_{AB}}-\ctcn{\rho}{_{AB}}=0\quad.
						\end{equation}
					This is the expected result taking into account $ \ct{C}{_{ab}}=0 $ and \cref{eq:CA-V-foliation}. 
					
					The asymptotic symmetries of definition \ref{def:asymptotic-symmetries} for the new equipment are the biconformal symmetries acting on the new pairs $ \prn{\ct{m}{_{a}},\ctcn{P}{_{ab}}} $ and they take the form
						\begin{equation}
							\ct{\xi}{^a}=\alpha\ct{m}{^a}+\ct{\eta}{^a}\quad,
						\end{equation}	
					where the restriction to the cuts $ \ctcn{\eta}{^A} $ of $ \ct{\eta}{^a} $ are CKVF of $ \prn{\mcn{_{AB}},\Sc} $,  $ \ct{m}{^a} $ is given in \cref{eq:m2-kerr-dS} and 
						\begin{equation}
							\alpha=\nu(t)\prn{1+N^2a^2\cos^2\theta}^{\frac{1}{2}}
						\end{equation}
					Here, $v=t$ is the parameter defining the foliation as in \cref{eq:m-foliations} and $ \nu(t) $ is an arbitrary function of $ t $ which makes the dimension of the subalgebra $ \ft $ infinite. Indeed, this is the canonical form of a bitranslation, see \cref{eq:general-translation-strong}, with $ F=(N/\Xi)\prn{1+N^2a^2\cos^2\theta}^{\frac{1}{2}} $. Now, since the leaves of the foliation are topologically $ \mathbb{S}^2 $, $ \ctcn{\eta}{^A} $ are the CKVF of the sphere, that is,  the Lorentz Group $ \mathsf{SO}(1,3) $. This agrees with the general results of \cref{ssec:strong-structure}. As a last  remark concerning this strong equipment, observe that the restriction to the leaves $ \csC{\alpha} $ of $ \cs{\alpha} $ is, on using \eqref{eq:omegaKerrdS} 
					$$
					\csC{\alpha} ={\rm constant} \times \omega
					$$
					and therefore it corresponds to the `$ l=0 $'  solution of the equation \eqref{eq:Hess-gen}, in agreement with \eqref{eq:V=0KerrdS} and \cref{thm:col-rho-noV-alpha}. Hence the subgroup of bitranslations given by $ \ct{\tau}{^a}=\alpha\ct{m}{^a} $ acting on the strong equipment correspond to infinitesimal asymptotic translations of \cref{def:asymptotic-translation}. This structure is also the general solution for the Kottler metric with $a=0$, for which both equipments are actually the same.
					
				\subsubsection{Asymptotic supermomenta}
					We compute the asymptotic canonical super-Poynting vector field $ \cts{P}{^a} $ and  canonical superenergy density $ \cs{\W} $ with the following outcome:
						\begin{align}
							\cts{\Pc}{^a}&\eqs 0\quad,\\
							\cs{\W}&\eqs  6m^2\quad.
						\end{align}
						The vanishing of $ \cts{\Pc}{^a} $ indicates that the space-time contains no gravitational radiation at infinity. This agrees with the fact that the two repeated PND $\ctrd{1}{\ell}{^\alpha}$ and $ \ctrd{2}{\ell}{^\alpha} $ are coplanar with $ \ct{N}{_\alpha} $ --see \cref{rmk:criterion-commutator,rmk:equivalent-statement-criterion}-- and that $ \ms{_{ab}} $ is conformally flat.\\
						
							\begin{figure}[h!]
								\centering
								\includegraphics[scale=1]{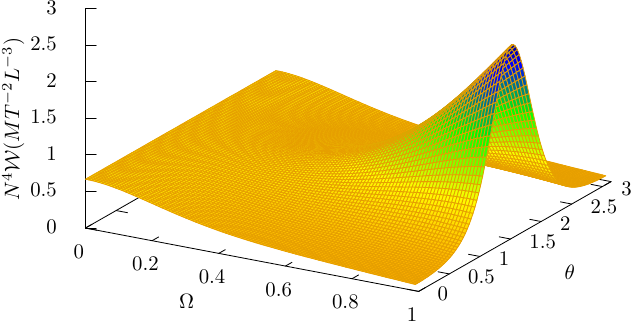}
								\caption[Asymptotic superenergy for the Kerr-dS metric]{Asymptotic superenergy for the Kerr-dS metric around $ \scri $ for $ a=m=\Lambda=1 $. The dependence on the angular parameter $ a $ fades away as approaching $ \scri $. The peak in the superenergy, then, occurs at the equator ($ \theta=\pi/2 $).}
								\label{fig:canonicalSuperKerrdS}
							\end{figure}
						There is an interesting feature of the canonical superenergy $ \cs{\W} $: it does not depend on $ a $ at $ \scri $, so that it has the same constant value as the one for $ a=0 $.\footnote{That is, the Kottler metric.} We can compute the superenergy density associated to $ \ct{N}{^\alpha} $ outside $ \scri $, its expression is
							\begin{equation}
							N^4\cs{\W}=\frac{6\Delta_r^2m^2\Omega^4}{\rho^4\prn{1+a^2\Omega^2\cos^2\theta}^3}\quad,
							\end{equation}
						and its Taylor expansion around $ \Omega=0 $ yields
							\begin{equation}
							N^4\cs{\W}\eqs\frac{2}{3}\Lambda^2m^2-\frac{2}{3}\Lambda m^2\brkt{6+3a^2\Lambda-5a^2\Lambda\sin^2\theta}\Omega^2+...\quad.
							\end{equation}
						Therefore, we see that $ a $ enters only at second leading order. This effect can be appreciated in \cref{fig:canonicalSuperKerrdS}.

		\subsection{The C-metric}	
			The existence of exact solutions of Einstein's Field Equations containing gravitational radiation at infinity when $ \Lambda=0 $ was demonstrated in \cite{Ashtekar-Dray81} by showing that the so called C-metric has a non-vanishing news tensor at $ \scri $. For $ \Lambda>0 $, the first proof of an exact solution having gravitational waves at infinity according to \cref{def:criterionGlobal} was presented in \cite{Fernandez-Alvarez_Senovilla20b} using precisely the C-metric but now with $ \Lambda>0 $. In the present work, we expand that analysis in several directions and in particular we suggest that two news tensors on $ \scri $ exist using the results of \cref{sec:news}.\\
			
			The C-metric with $ \Lambda>0 $ describes two accelerating black holes in a de Sitter background \cite{Podolsky2000}. As such, one expects the presence of gravitational radiation at $ \scri $. We will consider this metric in the form of a particular sub-case of the accelerating, charged, rotating Plebański-Demiański solution 
			\cite{Podolsky2009} --see also \cite{Griffiths-Podolsky2009}. The conformal metric, selecting a gauge according to \cref{eq:divergenc-freeEquation}, reads
				\begin{equation}
				\df{s^2}=\frac{\prn{\eta f(\eta)}^2}{S}\prn{-T\df{\tau^2}+\frac{1}{T}\df{q^2}+\frac{1}{S}\df{p^2}+S\df{\sigma^2}}\quad,
				\end{equation}
				where 
				\begin{align}
				T(q) &\defeq (q^2-a^2)(1+2 mq) -\Lambda/3 
				\quad,\\
				S(p) &\defeq (1-p^2)(1-2a mp)\quad,
				\end{align}
			 $ \eta $ is a conformal-gauge function and $ f(\eta) $ an arbitrary function regular and different from zero at $ \eta=0 $, both to be specified. The conformal boundary $ \scri $ is defined by $ q=-a p $, the conformal factor being
			 	\begin{equation}
			 	\Omega^2\defeq \frac{\prn{\eta f(\eta)}^2}{S}\prn{q + a p}^2
			 	\end{equation}
			 and the normal to $ \scri $
			 	\begin{equation}\label{eq:normal-C-metric}
			 	\ct{N}{_\alpha}\eqs-\frac{\eta f(\eta)}{\sqrt{S}}\prn{\cd{_\alpha}q+a\cd{_\alpha}p}.
			 	\end{equation}
			The gauge function $ \eta $ is a first integral of $ \ct{N}{^\alpha} $ and we choose it to be
			 	\begin{equation}
			 		\eta\defeq e^{a\prn{1-2a m}F(q)}\frac{\prn{1-p}^\frac{1}{2}}{\prn{1-2a mp}^{2a m/\prn{1+2a m}}\prn{1+p}^{\prn{1-2a m}/2\prn{1+2a m}}}\quad,
			 	\end{equation}
			 with
			 	\begin{equation}
			 		F(q)=-\int \frac{1}{T(q)}\df{q}\quad.
			 	\end{equation}
			 It is possible to set
			 	\begin{equation}
			 		\frac{\prn{\eta  f(\eta)}^2}{S}\eqs 1,
			 	\end{equation}	
			 and for that we take
			 	\begin{equation}
				 		f(\eta)\defeq e^{-a\prn{1-2a m}F\prn{-a P^{-1}(\eta^2)}}\prn{1+P^{-1}(\eta^2)}^{\frac{1}{\prn{1+2a m}}}\prn{1-2a m P^{-1}(\eta^2)}^{\frac{\prn{1+6a m}}{2\prn{1+2a m}}}
			 	\end{equation}
			 where $ P^{-1}\prn{\eta^2} $ is the inverse function of $ P(p) $ such that $ P(p)\eqs\eta^2 $.\\
			 	
			 There are four constant parameters, namely the acceleration $ a $, the mass $ m $ , $ \Lambda $ and $ C $. The metric may present two conical singularities at $ p= 1 $ and/or at $ p=-1 $. One can fix $ C $ to cure one of these singularities but never both of them at the same time. Additionally, this fixing defines the range of the coordinate $ \sigma\in \left[0,2\pi C\right) $ \cite{Griffiths-Podolsky2009}. Since  both singularities are not curable at the same time, one has to restrict the range of the coordinate $ p $ in order to exclude the persisting one. We fix $ C $ to
			 	\begin{equation}
			 		C=\frac{1}{\prn{1-2a m}}
			 	\end{equation}
			 so that $ p=1 $ defines a regular axis for the KVF $ \dpart{\sigma} $  and restricts the range $p\in (-1,1]$ in order to avoid the singular point at $ p=-1 $. This, together with the further condition 
			 	\begin{equation}
			 		2a m <1,
			 	\end{equation}
			 makes $ S\geq0 $, only vanishing at $ p=1 $ --thus preserving the signature of the metric.\\
			 
			 There are two KVF of $ \ct{g}{_{\alpha\beta}} $: $\partial_\tau$ and $\partial_\sigma$. The former has $\mathbb{R}$-orbits whereas the latter has cyclic orbits. Observe that $ \ct{g}{_{\tau\tau}}\eqs - T $ and $ \ct{g}{_{qq}}\eqs 1/T$ and that $T(q) <0$ at $ \scri $, hence the space-time is non-stationary around the conformal boundary, as one expects. Because we are interested in studying $ \scri $, we further restrict ourselves to $q \in (-a,a)$ --which keeps $ T(q)$ between two roots and negative. One more feature is that the roots of $ T(q) $ represent horizons which, by our previous remarks, do not meet $ \scri $. The Weyl tensor has two repeated principal null directions (which also become repeated PND of $ \ct{d}{_{\alpha\beta\gamma}^\delta} $ at $ \scri $) given by:
	 			\begin{align}
	 				\ctrd{1}{\ell}{_\alpha}&=\frac{1}{\sqrt{2}}\frac{N}{T}\prn{-T\cd{_\alpha}\tau+\cd{_\alpha}q}\quad,\\
	 				\ctrd{2}{\ell}{_\alpha}&=\frac{1}{\sqrt{2}}\frac{N}{T}\prn{T\cd{_\alpha}\tau+\cd{_\alpha}q}\quad.
	 			\end{align}
	 		Notice that we have chosen them such that $ \ctrd{1}{\ell}{^\alpha}\ctrd{2}{\ell}{_\alpha}\eqs-1 $, but this does not hold outside $\scri$. \\
			
			From now on we focus on $ \scri $. Note that $ T\eqs -a^2S-N^2 =: \csS{T}$. The metric there reads
				\begin{equation}
				h= (a^2S+N^2) \df{\tau^2}+\frac{N^2}{S(a^2S+N^2)}\df{p^2}+S\df{\sigma^2}\quad.
				\end{equation}
			This is positive definite and has a regular limit when $\Lambda \rightarrow 0$ leading to a degenerate metric 
			The intrinsic connection on $ \scri $ is
				\begin{align}
				\cts{\Gamma}{^\tau_{ab}}&\eqs -\frac{\alpha^2}{\bar{T}}\dpart{p}S \cds{_{(a}}\tau\cds{_{b)}}p\quad,\\
				\cts{\Gamma}{^p_{ab}}&\eqs \frac{3}{2\Lambda}S\bar{T}\alpha^2\dpart{p}S\cds{_a}\tau\cds{_b}\tau+\frac{6S\alpha^2+\Lambda}{6S\bar{T}}\dpart{p}S\cds{_a}p\cds{_b}p+\frac{3}{2\Lambda}\bar{T}S\dpart{p}S\cds{_a}\sigma\cds{_b}\sigma\quad.
				\end{align}
			and the intrinsic Ricci tensor of $ \scri $ reads
				\begin{align}
				\cts{R}{_{ab}}&=\frac{a^2}{2N^2}\csS{T}\brkt{S\dpart{p}^2S+\prn{\dpart{p}S}^2}\cds{_{a}}\tau\cds{_{b}}\tau+\frac{1}{S\csS{T}}\brkt{a^2S\dpart{p}^2S+\frac{1}{2}a^2\prn{\dpart{p}S}^2+\frac{1}{2}N^2\dpart{p}^2S}\cds{_{a}}p\cds{_{b}}p\nonumber\\
				&-\frac{S}{2N^2}\brkt{a^2S\dpart{_{p}}^2S+a^2\prn{\dpart{p}S}^2+N^2\dpart{p}^2S}\cds{_{a}}\sigma\cds{_{b}}\sigma\quad,\\
				\csS{R}&=-\frac{1}{2N^2}\brkt{4Sa^2\dpart{p}^2S+3a^2\prn{\dpart{p}S}^2+2N^2\dpart{p}^2S}\quad,
				\end{align}
			thus the intrinsic Schouten tensor follows
				\begin{align}\label{eq:schouten-C-metric}
				\cts{S}{_{ab}}&\eqs\frac{3\csS{T}}{8N^2}\brkt{\prn{\dpart{p}S}^2a^2-2N^2\dpart{p}^2S}\cds{_{a}}\tau\cds{_{b}}\tau+\frac{1}{8S\csS{T}}\brkt{4a^2S\dpart{p}^2S+a^2\prn{\dpart{p}S}^2+2N^2\dpart{p}^2S}\cds{_{a}}p\cds{_{b}}p\nonumber\\
				&-\frac{S}{8N^2}\brkt{\prn{\dpart{p}S}^2a^2+2N^2\dpart{p}^2S}\cds{_a}\sigma\cds{_{b}}\sigma\quad.
				\end{align}	
			One can also obtain the electric and magnetic parts of the rescaled Weyl tensor on $ \scri $ whose non-vanishing components are
			\begin{align}
				\ct{C}{_{ab}}&= \frac{6}{\Lambda}a m S(3Sa^2+\Lambda)\cds{_{(a}}\tau\cds{_{b)}}\sigma\quad,\label{eq:magneticC-metric}\\
				\ct{D}{_{ab}}&= -\frac{m}{\Lambda}\prn{9S^2a^4+5\Lambda Sa^2+\frac{2}{3}\Lambda^2}\cds{_a}\tau \cds{_b}\tau \nonumber\\
				&+ \frac{m\Lambda}{S\prn{\Lambda+3Sa^2}}\cds{_a}p\cds{_b}p+\frac{m}{\Lambda}S\prn{\Lambda+9Sa^2}\cds{_a}\sigma\cds{_b}\sigma\quad.\label{eq:electricC-metric}
			\end{align}
			Now we make a choice of \emph{strong orientation} --see \cref{def:orientationStrong}. For that, define 
				\begin{align}
					\ctm{k}{_\alpha}&\defeqs \ctrd{1}{\ell}{_\alpha}\quad,\label{eq:km-C-metric-a}\\
					\ct{m}{_\alpha}&\defeqs \ct{n}{_\alpha}-\sqrt{2}\ctm{k}{_\alpha}=\brkt{N\cd{_\alpha}\tau -\prn{\frac{N}{T}+\frac{1}{N}}\cd{_\alpha}q-\frac{a}{N}\cd{_\alpha}p}\quad,\\
					\ctp{k}{_\alpha}&\defeqs \frac{1}{\sqrt{2}}\prn{\ct{n}{_\alpha}+\ct{m}{_\alpha}}=\frac{1}{\sqrt{2}}\brkt{N\cd{_\alpha}\tau-\prn{\frac{N}{T}+\frac{2}{N}}\cd{_\alpha}q-\frac{2a}{N}\cd{_\alpha}p	}\quad,\label{eq:kp-C-metric-a}
				\end{align}	
			where, as usual, $ \ct{n}{_{\alpha}} $ is the unit version of the normal to $ \scri $, in this case given by \eqref{eq:normal-C-metric}. Notice that $ \ctm{k}{^\alpha}\ctp{k}{_\alpha}\eqs -1 $, $ \ct{m}{^\alpha}\ctm{k}{_\alpha}\eqs -1/\sqrt{2} $, $ \ct{m}{^\alpha}\ctp{k}{_\alpha}\eqs 1/\sqrt{2} $. This choice of null directions constitutes an example of the lightlike set up of \cref{ssec:lightlike-condition-directional} with $ \ct{m}{_{\alpha}} $ defining a strong orientation. The pullback of $ \ct{m}{_{\alpha}} $ to $ \scri $ is
				\begin{align}
					\ct{m}{_{a}}=N\prn{\cds{_{a}}\tau+\frac{a}{\csS{T}}\cds{_{a}}p}\quad,\label{eq:m-C-metric}\\
					\ct{m}{^a}=\prn{-\frac{N}{\csS{T}}\delta_\tau^a-\frac{a S}{N}\delta_p^a}\quad.
				\end{align}
		The projector to $ \Scn $ and its metric 
		read, respectively,
				\begin{align}
					\ctcn{P}{_{ab}}&= Sa^2\cds{_{a}}\tau\cds{_{b}}\tau +\frac{N^4}{\csS{T}^2 S}\cds{_{a}}p\cds{_{b}}p-\frac{N^2a}{\csS{T}}\prn{\cds{_a}\tau\cds{_b}p+\cds{_a}p\cds{_b}\tau}+S\cds{_{a}}\sigma\cds{_{b}}\sigma\quad,\label{eq:projector-C-metric}\\
					\ctcn{q}{}&=	\frac{1}{S}\df{p}^2+S\df{\sigma}^2\quad.\label{eq:cut-metric-C-metric}
				\end{align}
			Recall that \cref{eq:m-C-metric,eq:projector-C-metric} characterise the projected surface $ \Scn $ --see \cref{def:strong-structure}. After this, we can study the kinematics of $ \ct{m}{_{a}} $ namely the acceleration, vorticity and expansion tensor (see \cref{eq:acceleration,eq:vorticity,eq:expansion-tensor}):
				\begin{align}
				\ctcn{a}{_{b}}&= 0\quad,\label{eq:acceleration-metric-C}\\
				\ctcn{\kappa}{_{ab}}&= -\dpart{p}S\frac{a}{2N}\mcn{_{ab}}\quad,\label{eq:kappa-metric-C}\\
				\ctcn{\omega}{_{ab}}&=0\quad.\label{eq:vorticity-metric-C}
				\end{align}
			\Cref{eq:vorticity-metric-C} tells us that $ \ct{m}{_{a}} $ is surface-orthogonal, and thus defines a foliation by cuts; \cref{eq:kappa-metric-C} indicates that $ \ctcn{\Sigma}{_{ab}}=0 $, therefore the cuts are umbilical; \cref{eq:acceleration-metric-C} is consistent with $ \ct{m}{_{a}} $ defining a foliation and in addition shows that $ \ct{m}{^a} $ is geodesic. From \cref{eq:cn-foliation-acceleration} we deduce that the function $ 1/F=\ct{m}{^a}\cds{_{a}}v$ is constant on the cuts, $ \cdcn{_{A}}F=0 $, where $ v $ is the parameter selecting the leaves \eqref{eq:m-foliations}. From \cref{eq:m-C-metric} one deduces
				\begin{equation}
					v=\tau+a\int\frac{1}{\csS{T}}\df{p}\quad,\label{eq:v-C-metric}
				\end{equation}
			and $ F=N $. Therefore, with this choice of $ \ct{m}{_{a}} $ the C-metric possesses a \emph{strongly equipped} $ \scri $ --see \cref{def:strong-structure}. 
			
			On each cut one has the following non-vanishing connection symbols
				\begin{equation}
					\ctc{\Gamma}{^p_{pp}}\eqSv{v} -\frac{1}{2S}\dpart{p}S,\quad\ctc{\Gamma}{^\sigma_{p\sigma}}\eqSv{v} \frac{1}{2S}\dpart{p}S,\quad\ctc{\Gamma}{^p_{\sigma\sigma}}\eqSv{v} -\frac{1}{2}S\dpart{p}S.
				\end{equation}
			and the Gaussian curvature
				\begin{equation}
				\cscn{K}=-\frac{1}{2}\dpart{p}^2S=1-6a m p\quad.
				\end{equation}
			 The projections of \cref{eq:schouten-C-metric} to any cut give
				\begin{equation}
					\ctcn{S}{_{AB}}\eqSv{v}-\frac{1}{8N^2S}\brkt{\prn{\dpart{p}S}^2\alpha^2+2N^2\dpart{p}^2S}\cdcn{_{A}}p\cdcn{_{B}}p-\frac{S}{8N^2}\brkt{a^2\prn{\dpart{p}S}^2+2N^2\dpart{p}^2S}\cdcn{_{A}}\sigma\cdcn{_{B}}\sigma.
				\end{equation}
			Note that all these quantities defined on each cut of the foliation hold on $ \scri $, thus one can underline them as they belong to $ \Scn $ (see \cref{app:congruences}). Another important consequence of having $ \ctcn{\sigma}{_{AB}}=0=\ctcn{\omega}{_{AB}} $ is that from \cref{eq:TABC-congruences,eq:UABdef-congruences,eq:LAB-congruences,eq:SABC-congruences,eq:WABC-congruences}
				\begin{align}
					\ctcn{T}{_{ABC}}&=\ctcn{S}{_{ABC}}=\ctcn{W}{_{ABC}}=0\quad,\label{eq:T-C-metric}\\
					\ctcn{L}{_{AB}}&=\frac{1}{8}\cscn{\kappa}^2\mcn{_{AB}}\quad,\\
					\ctcn{U}{_{AB}}&=\ctcn{S}{_{AB}}+\ctcn{L}{_{AB}}= -\frac{1}{4}\dpart{p}^2S\prn{\frac{1}{S}\cdcn{_{A}}p\cdcn{_{B}}p+S\cdcn{_{A}}\sigma\cdcn{_{B}}\sigma}\quad.\label{eq:U-C-metric}
				\end{align}
			A quick check shows that $ \ctcn{U}{^C_C}=-\dpart{p}^2S/2=\cscn{K} $. We are interested in the lightlike projections of the rescaled Weyl tensor defined in \cref{ssec:lightlike-projections} because they are extensively used in the search of news and very useful for computing the asymptotic radiant superenergy. The non-vanishing ones, written in our notation for congruences \eqref{eq:notation-symmetric-tensor-m},\eqref{eq:notation-symmetric-tensorLatin-m}, are:
				\begin{align}
					\cs{D}&= -2m\quad,\\
					\ctcn{C}{_A}&=\ctcnp{C}{_A}\eqc \frac{3}{N}a mS\cdcn{_A}\sigma\quad,\label{eq:F-C-metric}\\
					\ctcn{D}{_A}&=\ctcnp{D}{_A} \eqc -\frac{3}{N}a m\cdcn{_A}p\quad,\label{eq:FA-C-metric}\\
					\cttcn{C}{_{AB}}&= \frac{1}{2}\ctcnp{C}{_{AB}}= -\frac{9}{N^2}Sa^2m \cdcn{_{(A}}p\cdcn{_{B)}}\sigma\quad,\label{eq:CAB-C-metric}\\
					\cttcn{D}{_{AB}}&=\frac{1}{2}\ctcnp{D}{_{AB}}= -\frac{3}{N^2}ma^2\cdcn{_A}p\cdcn{_B}p+\frac{3}{N^2}S^{2}a^2m\cdcn{_A}\sigma\cdcn{_B}\sigma\quad.\label{eq:DAB-C-metric}
				\end{align}
			
				\subsubsection{Asymptotic symmetries}
					The metric $ \ms{_{ab}} $ at $ \scri $ inherits as KVF $ \dpart{\tau} $ and $ \dpart{\sigma} $ which, in addition, leave invariant $ \ct{m}{_{a}} $ and thus they belong to the algebra of biconformal symmetries $ \fb $ --see \cref{ssec:symmetries}. Apart from those, we can study the asymptotic symmetries of  \cref{def:asymptotic-symmetries}. Because we are in the case in which $ \scri $ is strongly equipped (\cref{def:strong-structure}), we can use \cref{eq:general-translation-strong} and $ F=N $ to write the general form of the elements of the subalgebra of bitranslations $ \ft $:
						\begin{equation}
							\ct{\tau}{^a}=\nu (v) \ct{m}{^a}\quad,
						\end{equation}
					with $ \nu(v) $ an arbitrary function depending on $ v $ as defined in \eqref{eq:v-C-metric}. As we have shown in \cref{ssec:symmetries}, the general elements of the biconformal symmetries $ \fb $ preserving the strongly equipped $ \scri $ are the sum of an element of $ \ft $ and an element $ \ct{\eta}{^a} $ of the conformal transformations $ \fcs $ of the projector, which in this case is given by \eqref{eq:projector-C-metric}. On each cut $ \Sc_{v} $ we can project \cref{eq:c3cs} to give
						\begin{equation}\label{eq:conformalKVF-cuts-C-metric}
							\cdc{_{(A}}\ct{\eta}{_{B)}}\eqSv{v} 2\varphi\mc{_{AB}}\quad,
						\end{equation}
					where $ \ct{\eta}{_B} \eqSv{v}\ct{E}{^b_B}\ct{\eta}{_{b}} $ on each cut. Thus the restriction to each cut of $ \ct{\eta}{_{b}} $ gives locally the CKVF of the conformally flat $ \mc{_{AB}} $. To know the number of these CKVF we need to consider the global topology of the leaves. 
					Notice that the topology of the cuts is $ \mathbb{R}^2 $ ---since we have to remove the point $ p=-1 $ from a 2-sphere--- even though the distances remain finite. From \cref{app:conformal-miscellaneous} we then know that there is an infinite number of CKVF. In summary, the algebra of asymptotic symmetries for this strong equipment is multiply infinite, depending on an arbitrary function of $v$ and on an arbitrary entire function on the complex plane.

				\subsubsection{Asymptotic supermomenta}
\begin{figure}[t!]
						\centering
						\includegraphics[scale=1]{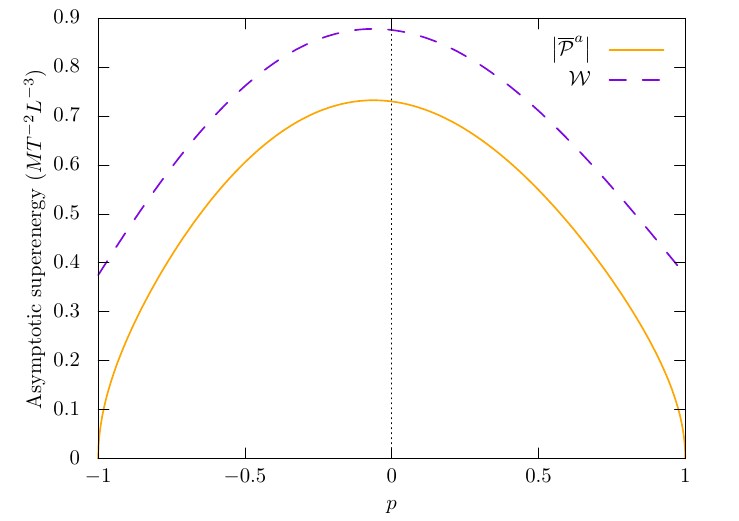}
						\caption[Canonical asymptotic superenergy for the C-metric with $ \Lambda>0 $.]{Canonical asymptotic superenergy $ \cs{\W} $ and super-Poynting vector $ \cts{\Pc}{^a} $ for the C-metric with $ \Lambda>0 $. The constant parameters have been set to $ \Lambda=1 $, $ a=1/4 $, $ m=1/4 $.}\label{fig:canonical-se-c-metric}
						\end{figure}

					The asymptotic canonical super-Poynting vector and superenergy are represented in \cref{fig:canonical-se-c-metric} and have the following expressions:
						\begin{equation}\label{eq:canonical-super-P-C-metric}
						\cts{\Pc}{^a}\eqs \sqrt{\frac{3}{\Lambda}}18a m^2 S  \prn{1+\frac{6}{\Lambda}Sa^2}\delta_{p}^a\quad,
						\end{equation}
						\begin{equation}\label{eq:canonical-superenergy-C-metric}
						\cs{\W}\eqs 6m^2  \prn{1+\frac{54}{\Lambda^2}S^2a^4+\frac{18}{\Lambda}Sa^2}\quad.
						\end{equation}
					Observe that the super-Poynting vector field does not vanish anywhere on $ \scri $. This fact, according to \cref{def:criterionGlobal}, indicates that \emph{there is gravitational radiation at} $ \scri $. This is the expected result. Note that the canonical asymptotic super-Poynting \eqref{eq:canonical-super-P-C-metric} vanishes if and only if the acceleration parameter $ a $ is zero (which implies the absence of radiation in that case). However, the canonical superenergy density \cref{eq:canonical-superenergy-C-metric} is different from zero even for $ a=0 $. Another feature characterising strong orientation is \cref{eq:flux-direction}, which can be easily verified for the present example contracting \cref{eq:canonical-super-P-C-metric} with $ \ct{m}{_{a}} $ given by \cref{eq:m-C-metric}
						\begin{equation}
							\ct{m}{_{a}}\cts{\Pc}{^a}\eqs -\frac{54a^2 m^2 S}{\Lambda\prn{a^2S+\Lambda/3}} \prn{1+\frac{6}{\Lambda}Sa^2}\leq 0\quad.
						\end{equation}
					\\
					
					Now we can take the limit to $ \Lambda=0 $ --see \cref{ssec:afs-limit}. For that one has to use the asymptotic supermomentum \eqref{eq:asymptotic-supermomentum}
						\begin{equation}
						\ct{p}{^\alpha} \eqs 2m^2\brkt{\prn{a^2S+\frac{\Lambda}{3}}\prn{\Lambda+9Sa^2}\delta^\alpha_q+a S\prn{2\Lambda+9Sa^2}\delta^\alpha_p}\quad. \label{eq:asymptotic-sumpermomentum-C-metric}
						\end{equation} 
					Then, we set $ \Lambda=0 $ in \cref{eq:asymptotic-sumpermomentum-C-metric} which by \cref{eq:limit-asymptotic-supermomentum} gives the asymptotic radiant supermomentum,
						\begin{equation}
							\ct{\Q}{^\alpha} \eqsflat 18m^2 S^{2}a^3\prn{a\delta^\alpha_q+\delta^\alpha_p}\quad.\label{eq:asymptotic-radiant-supermomentum-C-metric}
						\end{equation}
					The manifestly non-vanishing asymptotic radiant supermomentum for $ \Lambda=0 $ implies the presence of a non-vanishing news tensor \cite{Fernandez-Alvarez_Senovilla20} and, in consequence, that gravitational waves arrive at infinity.
				
				\subsubsection{Radiant quantities}
				We turn now to the study of the radiant asymptotic superenergy. Following \cref{ssec:lightlike-projections,ssec:radiant-superenergy}, we compute the quantities associated to the $ \ctpm{k}{^\alpha} $ of \cref{eq:km-C-metric-a,eq:kp-C-metric-a}. The procedure is straight-forward using \cref{eq:DAB-C-metric,eq:F-C-metric,eq:FA-C-metric} and recalling \cref{eq:Wpdef,eq:Wmdef,eq:Zpdef,eq:Zmdef,eq:Vdef}. The non-vanishing quantities are:
					\begin{align}
					\csp{\W}&= \frac{1296}{\Lambda}a^4m^2S^2\quad,\\
					\csp{\Z}&= \frac{108}{\Lambda}Sa^2m^2\quad,\\
					\ctp{\Q}{^\alpha}&=- \sqrt{\frac{3}{\Lambda}}\frac{36}{\sqrt{2}}a^2m^2\frac{S}{\csS{T}}\delta_\tau^\alpha+2\sqrt{2}\sqrt{\frac{3}{\Lambda}}\frac{a^2 m^2}{\Lambda}\prn{-54S\csS{T}-9\Lambda S}\delta_q^\alpha\nonumber\\
					&+\sqrt{2}108\sqrt{\frac{3}{\Lambda}}\frac{a^3m^2}{\Lambda}S^{2}\delta_p^\alpha\quad,\label{eq:radiantMomentumC-metric}\\
					\cs{\V}&= 4m^2\quad,
					\end{align}
				while $ \ctm{\Q}{^\alpha}=0 $ --hence $ \csm{\W}=\csm{\Z}=0 $. 
				\begin{figure}[ht!]\label{fig:radiant-se-c-metric}
					\centering
					\includegraphics[scale=1]{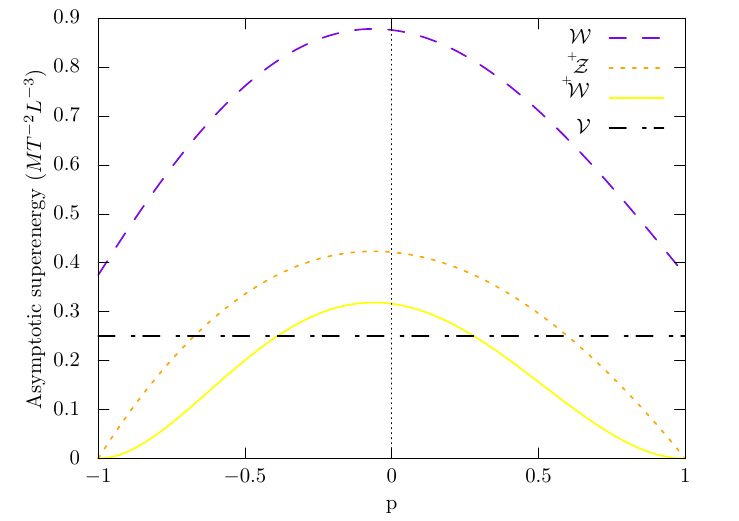}
					\caption[Asymptotic superenergy for the C-metric with $ \Lambda>0 $.]{Radiant and Coulomb components of the asymptotic superenergy on $ \scri $ together with the canonical supernergy density $ \cs{\W} $ for the C-metric with $ \Lambda>0 $. The constant parameters have been set to $ \Lambda=1 $, $ a=1/4 $, $ m=1/4 $.}
				\end{figure}
				Another useful check is to note that
					\begin{equation}
						4\cs{\W}-\csp{\W}-4\csp{\Z}-6\cs{\V}\eqs 0\quad,
					\end{equation}
				which shows that \cref{eq:superneregySum} is satisfied.
				Note that \cref{def:noIncomingRadDefOpen} is fulfilled too, i.e., there is no incoming radiation along $m^a$. 
				Then, \cref{thm:noincoming-np-non-compact} tells us that the first component of news exists.				
				\subsubsection{Radiant news}
				 	If we want to find a news tensor as proposed in \cref{ssec:news-congruences}, the first thing to notice is that due to \cref{eq:T-C-metric,eq:U-C-metric,eq:FA-C-metric} one has on each cut
			 			\begin{equation}\label{eq:divV-Cmetric}
					 		N\ctc{\epsilon}{_{BE}}\ctp{C}{^E}\eqSv{v} N\ctc{F}{_B}\eqSv{v}-\cdc{_E}\ct{V}{_B^E}\quad.
				 		\end{equation}	
				 We know that the solution of this equation gives the first component $ \ct{V}{_{AB}} $ of news see \cref{thm:onepiece-news}. To compute it, write \cref{eq:divV-Cmetric} explicitly in terms of the right-hand side of \cref{eq:FA-C-metric},
				 \begin{equation}
				 - 3a m \cdc{_A}p\eqc-S\dpart{p}\ct{V}{_{Ap}}-\frac{1}{S}\dpart{\sigma}\ct{V}{_{A\sigma}}+S\ctc{\Gamma}{^C_{pA}}\ct{V}{_{Cp}}+S\ctc{\Gamma}{^C_{pp}}\ct{V}{_{CA}}+\frac{1}{S}\ctc{\Gamma}{^C_{\sigma\sigma}}\ct{V}{_{CA}}+\frac{1}{S}\ctc{\Gamma}{^C_{A\sigma}}\ct{V}{_{C\sigma}
				 }.
				 \end{equation}
				 We have to set $ \ct{V}{_{AB}} $  traceless ($ \ct{V}{_{\sigma\sigma}}+S^2\ct{V}{_{pp}} \eqc 0$) and symmetric ($ \ct{V}{_{p\sigma}}\eqc\ct{V}{_{\sigma p}} $) and we further assume that $ \ct{V}{_{AB}} $ is axially symmetric due to the existence of the axial KVF $ \dpart{\sigma} $, that is
				  \begin{equation}
				  	 \dpart{\sigma}\ct{V}{_{AB}}\eqc 0. 
				  \end{equation}
				The solution reads
				 \begin{equation}
				 \ct{V}{_{p\sigma}}\eqc \frac{c_1}{S}\quad\text{with}\quad c_1=\text{constant}\quad,
				 \end{equation}
				 \begin{equation}
				 \ct{V}{_{pp}}\eqc \frac{H}{S^2},
				 \end{equation}
				 where 
				 \begin{equation}
				 H\defeqc \int 3a m S dp\eqc 3a m\prn{\frac{1}{2}a mp^4-\frac{1}{3}p^3-a m p^2+p}+c_2\quad\text{with}\quad c_2=\text{constant}.
				 \end{equation}
				 This function $ \cs{H} $ must be positive where $ S>0 $ and because we assume $ a>0 $, $m>0 $. Regularity at $ p=\mu $ with $ \mu=\pm1 $   requires
				 \begin{equation}
				 c_1\eqc 0,\quad c_2\eqc  \frac{3}{2}m^2a^2-\mu 2a m	\quad,					\end{equation} 
				 and cannot be achieved on both poles, $ p=-1,1 $, simultaneously. Because with our gauge fixing $ p\in(-1,1] $, we have to choose $ \mu=1 $. Then,
				 \begin{align}
				 \ctcn{V}{_{AB}}&= \frac{H}{S^2}\cdcn{_A}p\cdcn{_B}p - H \cdcn{_A}\sigma\cdcn{_B}\sigma\quad,\\
				 \ctcn{V}{_{AB}}\evalat{p=1}&=0.
				 \end{align}
				 It is possible now to deduce what $ \ctcn{\rho}{_{AB}} $ is:
				 \begin{align}
				 \ctcn{\rho}{_{AB}}&= \ct{U}{_{AB}}-\ct{V}{_{AB}}\nonumber\\
				 &=-\frac{3a^2m^2p^4-2a mp^3-6a^2m^2p^2+18a mp+3a^2m^2-4a m-2}{4(p-1)(p+1)(2a mp-1)}\cdcn{_A}p\cdcn{_B}p\nonumber\\
				 &+\frac{1}{4}(p-1)(p+1)(2a mp-1)(3a^2 m^2p^4-2a mp^3-6a^2m^2p^2-6a mp+3a^2 m^2-4a m+2)\cdcn{_{A}}\sigma\cdcn{_B}\sigma.
				 \end{align}
				The radiant news on each leaf of this strong equipment is simply given by $ \ctcnp{n}{_{AB}}=2\ctcn{V}{_{AB}} $. Observe that
				\begin{equation}
					\ctp{n}{_{AB}}\eqSv{v} 0\quad\forall v\iff \cts{\P}{^a}\eqs 0\iff\ctp{\Q}{^\alpha}=0\quad.
				\end{equation}
				
				\subsubsection{The other strong orientation}
					If we choose $ \ctm{k}{^\alpha} $ aligned with the other repeated PND, that is
						\begin{align}
							\ctm{k}{_\alpha}&\defeqs \ctrd{2}{\ell}{_\alpha}\quad,\label{eq:km-C-metric-b}\\
							\ct{m}{_\alpha}&\defeqs \ct{n}{_\alpha}-\sqrt{2}\ctm{k}{_\alpha}=\brkt{-N\cd{_\alpha}\tau -\prn{\frac{N}{\csS{T}}+\frac{1}{N}}\cd{_\alpha}q-\frac{a}{N}\cd{_\alpha}p}\quad,\label{eq:m-C-metric-b}\\
							\ctp{k}{_\alpha}&\defeqs \frac{1}{\sqrt{2}}\prn{\ct{n}{_\alpha}+\ct{m}{_\alpha}}=\frac{1}{\sqrt{2}}\brkt{-N\cd{_\alpha}\tau-\prn{\frac{N}{\csS{T}}+\frac{2}{N}}\cd{_\alpha}q-\frac{2a}{N}\cd{_\alpha}p	}\quad,\label{eq:kp-C-metric-b}
						\end{align}	
					neither the asymptotic super-Poynting nor the asymptotic superenergy change, as they do not depend on this choice. The radiant superquantities $ \csp{\Z} $ and $ \csp{\W} $ in general would be different, nevertheless for the new $ \ctp{k}{^\alpha} $ they have the same value as for the old $ \ctp{k}{^\alpha} $; one also finds $ \ctm{\Q}{^\alpha} =0$. There is a change in the direction of the radiant supermomentum $ \ctp{\Q}{^\alpha} $ though --compare with \cref{eq:radiantMomentumC-metric} --
						\begin{align}
							\ctp{\Q}{^\alpha}&\eqs \sqrt{\frac{3}{\Lambda}}\frac{36}{\sqrt{2}}a^2m^2\frac{S}{\csS{T}}\delta_\tau^\alpha+2\sqrt{2}\sqrt{\frac{3}{\Lambda}}\frac{a^2 m^2}{\Lambda}\prn{-54S\csS{T}-9\Lambda S}\delta_q^\alpha+\nonumber\\
							&+\sqrt{2}108\sqrt{\frac{3}{\Lambda}}\frac{a^3m^2}{\Lambda}S^{2}\delta_p^\alpha\quad.
						\end{align}
					An intuitive interpretation of this difference is that on the first case, with $ \ctm{k}{_\alpha}=\ctrd{1}{\ell}{_\alpha}$, $ -\ct{m}{_{a}} $ points along the spatial propagation direction of the gravitational radiation coming from one of the two black holes, while with $ \ctm{k}{_\alpha}=\ctrd{2}{\ell}{_\alpha}$, $ -\ct{m}{_{a}}$ gives the propagation direction of the radiation coming from the other one. Notice that in each case the no incoming radiation condition holds, a fact that is compatible with the existence of two different propagation directions: with $ \ctm{k}{_{\alpha}}=\ctrd{1}{\ell}{_{\alpha}} $, \cref{def:noIncomingRadDefOpen} tells that there is no radiation travelling along the spatial direction $ \ct{m}{_{\alpha}} $ of \cref{eq:m-C-metric}; with $ \ctm{k}{_{\alpha}}=\ctrd{2}{\ell}{_{\alpha}} $, \cref{def:noIncomingRadDefOpen} tells that there is no radiation travelling along the spatial direction $ \ct{m}{_{\alpha}} $ of \cref{eq:m-C-metric-b}.

		\subsection{Robinson-Trautman metrics }
				We explore now the Robinson-Trautman family of solutions 
				with a positive cosmological constant and admitting $ \scri $ --for details, see  \cite{Stephani2003,Griffiths-Podolsky2009} and also \cite{Griffiths2002}. We write the conformal metric as
				\begin{equation}
				\df{s}^2=P^2\prn{\df{u}\df{\ell}+\df{\ell}\df{u}-\prn{2\ell^2H}\df{u}^2+\frac{1}{P^2}\prn{\df{\zeta}\df{\bzeta}+\df{\bzeta}\df{\zeta}}}\quad,
				\end{equation}
					where $ u $ is a retarded time coordinate, $ \ell $ an inverse radius and $ \zeta,\bzeta $ a couple of complex stereographic coordinates. The gauge has been chosen such that
						\begin{equation}
							\cd{_{\mu}}\ct{N}{^\mu}\eqs 0.
						\end{equation}
				  The metric functions are defined as
				\begin{align}
				-2\ell^2H&\defeq\frac{\Lambda}{3}+2\ell\partial_u\ln P-\ell^2 K+2m\ell^3\quad,\\
				K&\defeq  2P^2 \partial_\zeta \partial_\bzeta \ln P\quad.
				\end{align}
				Depending on the matter content of the physical space-time, the function $ P=P(u,\zeta,\bzeta) $ and the function $ m(u) $ satisfy the so called Robinson-Trautman equation, which for $\Lambda$-vacuum is a fourth-order differential equation. Infinity is located at $ \Omega\defeq \ell =0 $, therefore the normal to $ \scri $ is 
					\begin{equation}
						\ct{N}{_{\alpha}}\defeqs\cd{_{\alpha}}\ell
					\end{equation}
				and the metric at $ \scri $ reads
				\begin{equation}\label{eq:hRob-Tra}
				h=N^2P^2\df{u}^2+\df{\zeta}\df{\bzeta}+\df{\bzeta}\df{\zeta}\quad,	
				\end{equation}
				where for simplicity we use the same letter $ u $ to denote the restriction to $ \scri $ of the retarded time.
				
				As these metrics are algebraically special, with multiple PND given at $ \scri $ by
				 	\begin{equation}
				 		\ctrd{1}{\ell}{^\alpha}=-\frac{N}{P\sqrt{2}}\delta_\ell^\alpha\quad.
				 	\end{equation} 
				a strong orientation (\cref{def:orientationStrong}) always exists, determined by setting
				 	\begin{align}
					 	\ctm{k}{^\alpha}&\defeqs \ctrd{1}{\ell}{^\alpha}\quad,\label{eq:km-RT-metric}\\
					 	\ct{m}{^\alpha}&\defeqs \ct{n}{^\alpha}-\sqrt{2}\ctm{k}{^\alpha}=\frac{1}{NP}\delta_u^\alpha\quad,\\
					 	\ctp{k}{^\alpha}&\defeqs \frac{1}{\sqrt{2}}\prn{\ct{n}{^\alpha}+\ct{m}{^\alpha}}=-\frac{N}{\sqrt{2}P}\delta_\ell^\alpha+\frac{\sqrt{2}}{NP}\delta_u^\alpha\quad,\label{eq:kp-RT-metric}
				 	\end{align}	
				 where as usual $ N\ct{n}{^{\alpha}}\defeq\ct{N}{^{\alpha}} $. The pullback of $ \ct{m}{_\alpha} $ to $ \scri $ is
				 	\begin{align}
				 		\ct{m}{^a}=\frac{1}{NP}\delta^a_u\quad,
				 		\ct{m}{_a}=NP\cds{_a}u\quad.
				 	\end{align}
				 Using the connection coefficients one can compute the kinematic quantities of $ \ct{m}{_{a}} $; they read
				 	\begin{equation}
				 		\ctcn{\omega}{_{ab}}=0,\quad\ctcn{\kappa}{_{ab}}=0,\quad\ctcn{a}{_{b}}=-\ctcn{D}{_{b}}\ln P\quad.
				 	\end{equation}
				 Therefore, $ \ct{m}{_{a}} $ is orthogonal to an umbilical foliation --see \cref{eq:vorticity,eq:expansion-tensor,eq:acceleration}. This was expected, as the conditions in \cref{thm:noSigma} are met --see \cref{rmk:asymp-gs-remarkPND,rmk:asymp-gs-remarkY}. The leaves, in general, contain singularities which  depend on the choice of solutions for the PDEs involved, and on the matter contents, etc. 
				 The projector is 
				 	\begin{equation}
				 	\ctcn{P}{_{ab}}= \cds{_{a}}{\zeta}\cds{_{b}}{\bzeta}+\cds{_{a}}{\bzeta}\cds{_{b}}{\zeta}\quad.
				 	\end{equation}
					and the metric on the leaves is simply locally flat. 
					
\subsubsection{Asymptotic symmetries}\label{sssec:symm-rt}
				For the given strong equipment, the asymptotic symmetries are the infinitesimal biconformal symmetries acting on the pair  $ \prn{\ct{m}{_a},\ctcn{P}{_{ab}}} $ and must take form
					\begin{equation}
						\ct{\xi}{^a}=\alpha\ct{m}{^a}+\ct{\chi}{^a}\quad,
					\end{equation}
				with
					\begin{equation}
						\alpha=\nu(u)P\quad,
					\end{equation}
				$ \nu(u) $ being an arbitrary function of the coordinate $ u $ and the restriction $ \ct{\chi}{^A} $  of $ \ct{\chi}{^a} $ to each cut a CKVF of the flat metric. This time, the dimension of the algebra of biconformal infinitesimal symmetries can be `doubly infinite' depending on the topology of the leaves. For instance, this will be the case for $ \mathbb{R}^2 $ and $ \mathbb{R}\times\mathbb{S}^1 $ topologies, which also warrants the existence of CKVFs with fixed points --see \cref{rmk:fixed-points}. In general, there are no CKVF that are infinitesimal basic symmetries, what is to be expected as there are no KVFs for Robinson-Trautman metrics 
				in the generic case.
				
				Observe that in this case the restriction of the function $\alpha$ to the leaves $u=u_0$ is not one of the solutions of \eqref{eq:Hess-gen}. This can be seen by noticing that the metric on the leaves is flat and its conformal factor with respect to the round metric is given, from \eqref{eq:q-round}, by $\omega = 1+\zeta\bar\zeta/2$ and comparing it with $P(u_0,\zeta,\bar\zeta$. This result was to be expected, as these space-times contain, in general, gravitational radiation --as we are going to prove, and therefore the component $V_{AB}$ of news will not vanish.

				So far, this applies to the general Robinson-Trautman metric with positive $ \Lambda $. 
				
				Now we concentrate on two particular simple but relevant examples, which are illustrative of several points we would like to remark.

				\subsubsection{Petrov type D solution for null electromagnetic field}
				This is a very atypical example, as it corresponds to the unique Einstein-Maxwell type D solution for a null electromagnetic field. This case is defined by 
				$$
				P=1,\hspace{1cm} K=0
				$$
				while $m(u)$ is determined by the (physical) {\em null} electromagnetic field ($\sqrt{2}\zeta :=(x+iy)$)
				\be
				\mathbf{F}=du \wedge (h_x(u)dx +h_y(u) dy) \label{eq:FinR-T}
				\ee
				via
				\be\label{eq:m(u)}
				m(u) =\frac{\varkappa}{2}  \int^u_{u_0} [h_x^2(w)+h_y^2(w)]dw + M 
				\ee
				where $M=m(u_0)$ is a constant. Here, $h_x(u)$ and $h_y(u)$ are arbitrary functions which determine the amplitud {\em and polarization} of the electromagnetic waves. See \cite{Senovilla2015} for details.
				
				Clearly the metric \eqref{eq:hRob-Tra} on $\scri$ is flat in this case, implying in particular $C_{ab}=0$. No gravitational radiation. And {\em no trace of the electromagnetic radiation} appears either in $(\scri,\ms{_{ab}})$. But this electromagnetic radiation surely arrives at $\scri$! How does it manifest itself in the conformal picture? One may think that this is encoded in the energy-momentum tensor. However, the physical energy-momentum tensor is
				\be
				\pt{T}{_{\mu\nu}}= \frac{2\ell^2}{N^2} (h_x^2(u)+h_y^2(u))k_\mu k_\nu \label{eq:TRob-Tra}
				\ee
				and thus $\ct{T}{_{\mu\nu}}= \pt{T}{_{\mu\nu}}/\ell \eqs 0$ also vanishes at $\scri$. Where is the electromagnetic radiation?
				
				The answer is simple: in $\ct{D}{_{ab}}$. And this is a very neat, as well as truly illustrative, example of our main message, that is, one must take $\ct{D}{_{ab}}$ into account and the whole picture, including radiation(s) (of any type), is encoded at infinity in the structure defined by $(\scri, \ms{_{ab}}, \ct{D}{_{ab}})$. 
				
				Observe that in the present situation
				\be
					\ct{D}{_{ab}} = -m(u) \left(3 m_a m_b - \ct{h}{_{ab}} \right)
				\ee
				and $m(u)$ carries the information of the electromagnetic radiation arriving at $\scri$, as follows from \eqref{eq:m(u)}. Notice further that, actually, the electromagnetic radiation is source for $\ct{D}{_{ab}}$, in agreement with the general equation \eqref{eq:divD}, which can be checked to hold now by simply using the derivative of \eqref{eq:m(u)} with respect to $u$ and taking \eqref{eq:TRob-Tra} into account.

			\subsubsection{Petrov type N Robinson-Trautman $\Lambda$-vacuum solutions}
				
				We now consider another different situation within the Robinson-Trautman family, the $\Lambda$-vacuum 
				 Petrov type-N case. The conditions that particularise the metric to that subfamily of space-times are:
				\begin{equation}
				m=0\quad,\quad K=K(u)\quad.
				\end{equation}
				For type N, the general solution for $ P $ is (see \cite{Griffiths2002}) 
					\begin{equation}\label{eq:general-P-RT-N}
					P=\frac{1}{\sqrt{\dpart{\bzeta}\bar{F}\dpart{\zeta}F}}\prn{1+\epsilon F\bar{F}},
					\end{equation}
				with $ \epsilon=-1,0,1 $ and $ F(u,\zeta) $ any function analytic on $ \zeta $.
				The non-vanishing components of the intrinsic connection in these coordinates are
				\begin{align}
						\cts{\Gamma}{^u_{au}}&=\dpart{a}\ln P\quad,\\
						\cts{\Gamma}{^\zeta_{uu}}&=-N^2P\dpart{\bzeta}P\quad,\\
						\cts{\Gamma}{^\bzeta_{uu}}&=-N^2P\dpart{\zeta}P\quad,
				\end{align}
				and the curvature and Schouten tensor are given by
				\begin{align}
					\cts{R}{_{ab}}&=-2N^2P\dpart{\zeta}\dpart{\bzeta}P \cds{_{a}}u\cds{_{b}}u-\frac{2}{P}\dpart{\zeta}\dpart{\bzeta}P\cds{_{(a}}\zeta\cds{_{b)}}\bzeta\nonumber\\
					&-\frac{1}{P}\dpart{\zeta}\dpart{\zeta}P \cds{_{a}}\zeta\cds{_{b}}\zeta-\frac{1}{P}\dpart{\bzeta}\dpart{\bzeta}P\cds{_{a}}\bzeta\cds{_{b}}\bzeta\quad,\\
					\csS{R}&=-\frac{4}{P}\dpart{\zeta}\dpart{\bzeta}P\quad,\\
					\cts{S}{_{ab}}&=-N^2P\dpart{\zeta}\dpart{\bzeta}P\cds{_{a}}u\cds{_{b}}u-\frac{1}{P}\dpart{\zeta}\dpart{\zeta}P\cds{_{a}}\zeta\cds{_{b}}\zeta-\frac{1}{P}\dpart{\bzeta}\dpart{\bzeta}P\cds{_{a}}\bzeta\cds{_{b}}\bzeta.\label{eq:schouten-RT-metric}
				\end{align}
				 and the electric and magnetic parts of the rescaled Weyl tensor at $ \scri $ read
	 				\begin{align}
				 	\ct{D}{_{ab}}&=\frac{1}{N^2 P^3}\dpart{\zeta}\prn{P^2\dpart{u}\dpart{\zeta}\ln P}\cds{_{a}}\zeta\cds{_{b}}\zeta+ \frac{1}{N^2 P^3}\dpart{\bzeta}\prn{P^2\dpart{u}\dpart{\bzeta}\ln P}\cds{_{a}}\bzeta\cds{_{b}}\bzeta\quad,\\
				 	\ct{C}{_{ab}}&= i\frac{1}{N^2 P^3}\dpart{\zeta}\prn{P^2\dpart{u}\dpart{\zeta}\ln P}\cds{_{a}}\zeta\cds{_{b}}\zeta-i\frac{1}{N^2 P^3}\dpart{\bzeta}\prn{P^2\dpart{u}\dpart{\bzeta}\ln P}\cds{_{a}}\bzeta\cds{_{b}}\bzeta\quad.
				 	\end{align}
				 In this situation $\ctrd{1}{\ell}{^\alpha}$ is the quadruple PND of the type-N Weyl tensor (and  hence of $ \ct{d}{_{\alpha\beta\gamma}^\delta} $).

				  Gravitational waves are expected at infinity. Indeed, this is the case according to \cref{def:criterionGlobal} because the asymptotic canonical super-Poynting vector field and superenergy read
				  	\begin{align}
				  		\cts{\Pc}{^a}&\eqs -\frac{4}{N^4P^7}\dpart{\zeta}\prn{P^2\dpart{u}\dpart{\zeta}\ln P}\dpart{\bzeta}\prn{P^2\dpart{u}\dpart{\bzeta}\ln P}\ct{m}{^a}\quad,\label{eq:sp-RT-N}\\
				  		\cs{\W}&\eqs \frac{4}{N^4P^6}\dpart{\zeta}\prn{P^2\dpart{u}\dpart{\zeta}\ln P}\dpart{\bzeta}\prn{P^2\dpart{u}\dpart{\bzeta}\ln P}\quad,\label{eq:se-RT-N}
				  	\end{align}
				  thus $ \cts{\Pc}{^a} $ is non-vanishing everywhere on $ \scri $, pointing along $ -\ct{m}{^a} $. There is just one radiant quantity different from $ 0 $, as corresponds to a $ \ct{d}{_{\alpha\beta\gamma}^\delta} $ of Petrov type-N when strong orientation is chosen --see \cref{fig:superenergy-flow}:
						\begin{equation}
							\csp{\W}\eqs \frac{16}{N^4P^6}\dpart{\zeta}\prn{P^2\dpart{u}\dpart{\zeta}\ln P}\dpart{\bzeta}\prn{P^2\dpart{u}\dpart{\bzeta}\ln P}\quad.
						\end{equation}
					From this expression and \cref{eq:se-RT-N}, clearly
						\begin{equation}
							4\cs{\W}=\csp{\W}\quad,
						\end{equation}
					fulfilling \cref{eq:superneregySum}, and by \cref{eq:sp-RT-N} 
						\begin{equation}
							\cts{P}{^a}\eqs -\cs{\W}\ct{m}{^a},
						\end{equation}
					which is fine with the general expression \eqref{eq:sPspecial} of $ \cts{\Pc}{^a} $ for algebraically special $ \ct{d}{_{\alpha\beta\gamma}^\delta} $.
				
					It is easy to see what the radiant news is in this case. Since the leaves defined by the strong equipment are umbilical, the tensor $ \ct{f}{_{AB}} $ of \cref{eq:fakenewsdef} vanishes. The same argument applies to $ \ctcn{T}{_{ABC}} $ of \cref{eq:TABC-congruences}. Furthermore, because $ \cscn{\kappa}=0 $, from \cref{eq:UABdef-congruences} one has
						\begin{equation}
							\ctcn{U}{_{AB}}=\ctcn{S}{_{AB}},
						\end{equation}
					or, by means of the decomposition of \cref{thm:rho-tensor-noncompact-foliation},
						\begin{equation}
							\ctcn{V}{_{AB}}+\ctcn{\rho}{_{AB}}=\ctcn{S}{_{AB}}.
						\end{equation}
					As it is pointed out in \cref{sssec:symm-rt}, if we assume  $ \mathbb{R}^2 $ or $ \mathbb{R}\times\mathbb{S}^1 $ topology for the cuts, the existence a CKVF with a fixed point on each cut is ensured. In that case, by \cref{thm:rho-tensor-noncompact-foliation}, we now that a flat metric on the cuts, as it is the case, implies $ \ctcn{\rho}{_{AB}}=0 $. Hence, the news tensor of \cref{thm:noincoming-np-non-compact}, given by
						\begin{equation}
							\ctcnp{n}{_{AB}}=2\ctcn{V}{_{AB}}
						\end{equation}
					is simply the tangent part of $ \cts{S}{_{ab}} $ \cref{eq:schouten-RT-metric} to the cuts, that is
						\begin{equation}
							\ctcnp{n}{_{AB}}=-\frac{2}{P}\dpart{\zeta}\dpart{\zeta}P\cdcn{_{A}}\zeta\cdcn{_{B}}\zeta-\frac{2}{P}\dpart{\bzeta}\dpart{\bzeta}P\cdcn{_{A}}\bzeta\cdcn{_{B}}\bzeta\quad.
						\end{equation}
					It is possible indeed to write the asymptotic canonical supermomenta and super-Poynting	in terms of $ \ctcnp{n}{_{AB}} $,
						\begin{align}
							\cts{\Pc}{^a}&=-\frac{1}{N^4P^6}\brkt{-\dpart{u}P+P\dpart{u}}\prn{\ctcnp{n}{_{\zeta\zeta}}}\brkt{-\dpart{u}P+P\dpart{u}}\prn{\ctcnp{n}{_{\bzeta\bzeta}}}\ct{m}{^a}.\\
							\cs{\W}&=\frac{1}{N^4P^6}\brkt{-\dpart{u}P+P\dpart{u}}\prn{\ctcnp{n}{_{\zeta\zeta}}}\brkt{-\dpart{u}P+P\dpart{u}}\prn{\ctcnp{n}{_{\bzeta\bzeta}}}.
						\end{align}
					From these expressions we find
						\begin{equation}
							\ctcnp{n}{_{AB}}=0 \implies\cts{\Pc}{^a}=0,\quad\cs{\W}=0.
						\end{equation}

	\section{Discussion}
			
Let us remark some of the results presented in this work:
	\begin{itemize}
		\item Based on the rescaled Bel-Robinson tensor $ \ct{\D}{_{\alpha\beta\gamma\delta}} $ at infinity, the asymptotic supermomentum determines the presence of gravitational radiation escaping from --or entering into-- the space-time. At the same time, it provides a direct connection between the existence of gravitational radiation and the algebraic classification of the rescaled Weyl tensor $ \ct{d}{_{\alpha\beta\gamma}^\delta} $ at $ \scri $. 
		\item The radiation criteria thus defined have a neat correspondence in the two considered scenarios, $ \Lambda>0 $ and $ \Lambda=0 $, and share the same geometric and algebraic meaning: there is no gravitational radiation on $\scri$ if and only if $N^\alpha|_\scri$ is a principal vector of $ \ct{d}{_{\alpha\beta\gamma}^\delta} $ in the sense of Pirani. In fact, it is possible to take the limit from $ \Lambda>0 $ to $ \Lambda=0 $ explicitly. This feature exceeds the capability of traditional methods.
		\item A general method has been presented for computing news-like tensors at $ \scri $ and we have identified a first component $ \ct{V}{_{AB}} $ for all of them and which has an origin similar to that of the news tensor field in $ \Lambda=0 $. Especifically, its existence relies on a tensor field $ \ct{\rho}{_{AB}} $ that has been found for general Riemannian two-dimensional manifolds. The tensor $ \ct{V}{_{AB}} $ is determined at every cut $ \prn{\Sc,\mc{_{AB}}} $ in  $ \prn{\scri,\ms{_{ab}}} $. A particularly outstanding class of such news have been identified --the radiant news--, and the necessary conditions for their existence have been found. They include a second component, called $ \ctpm{X}{_{AB}} $. Although these have the algebraic properties of $ \ct{V}{_{AB}} $, their origin is different as they are determined by $ \ct{D}{_{ab}} $ and the extrinsic curvature of the cuts where they are defined.
		\item The notion of (strictly, strongly) equipped $\scri$ has been proposed by selecting a particularly distinguished vector field on $\scri$. Thereby, a new set of infinitesimal symmetries, respecting the equipment, has been defined leading to Lie algebras which may have infinite dimension in some situations.
	\end{itemize}
Also, there are open lines that require further investigation:
	\begin{itemize}
			\item A more general class of news-like tensor in space-times with $ \Lambda>0 $ can be sought by means of the general method here presented. Also, a refined study of the radiant news tensor and their connection with the radiation condition is possible. Particularly, a transport equation for $ \ct{\rho}{_{ab}} $ along $ \ct{m}{^a} $ would shed light on these two issues.
			\item The definition of an energy-momentum at $ \scri $ with $ \Lambda>0 $ is still an open problem. The exploration of conserved charges and the study of the $ \Lambda=0 $ scenario suggest that a definition of momentum associated to the symmetries of an equipped $ \scri $ is plausible.
			\item The application of the tidal approach to the $ \Lambda<0 $ scenario has not been considered in this and the companion paper and should be addressed. If the outcome is successful, one could then talk of a \emph{universal} radiation condition at infinity.
	\end{itemize}

	\subsection*{Acknowledgments}
		Work supported under Grants No. FIS2017-85076-P (Spanish MINECO/AEI/FEDER, EU) and No. IT956-16 (Basque Government). Some of the calculations presented in \cref{sec:Examples} were performed using the computer algebra system Maxima --distributed under GNU GPL license. The authors would like to thank Robert Beig for pointing out reference \cite{Besse1987}, and also Raül Vera  for a careful reading of the two manuscripts and correction of typographical or minor errors.
		\appendix
		\renewcommand{\thesubsection}{\arabic{subsection}}
			
	\section{Geometry of spatial hypersurfaces, cuts and congruences}\label{app:spatial-hypersurfaces}
				We introduce some geometric tools for a general 3-dimensional, spacelike hypersurface $ \I $ embedded in a 4-dimensional space-time $\prn{M,\ct{g}{_{\alpha\beta}}} $. We will also consider the geometrical objects associated to a single cut $ \Sc $  on $ \I $ and to a general congruence $ \C $ given by a vector field $ \ct{r}{^a} $ on $ \I $.

		\subsection{Induced connection}\label{app:spatial-hypersurface}
			Consider a general spacelike hypersurface $ \I $ embedded in a 4-dimensional space-time $ \prn{M,\ct{g}{_{\alpha\beta}}} $. Let $ \ct{n}{_\alpha} $ be the timelike normal one-form at each point of $ \mathcal{I} $ normalised to $ \ct{n}{_\mu}\ct{n}{_\nu} \ct{g}{^{\mu\nu}} = -1 $.  Also, at each point, consider a set of linearly independent tangent vector fields $ \{\ct{\vec{e}}{_a}\} $, $ a=1,2,3 $. By definition, $ \ct{n}{_\mu} \ct{e}{^\mu_a} =0$ and $ \{\ct{\vec{e}}{_a}\} $ constitutes a basis for $ \fX_{\I} $, the set of vector fields of $ \I $. Use the inverse space-time metric to define the normal vector $ \ct{n}{^\alpha}\defeq \ct{g}{^{\alpha\mu}}\ct{n}{_\mu} $. This field completes a basis, $ \{\vec{n},\vec{\ct{e}{_a}}\} $, for the set of vector fields of $ \cs{M} $, $ \fX_{\cs{M}}$, at $ \I $. Analogously, consider a set of linearly independent one-forms orthogonal to $ \vec{n} $, $ \{\cts{\form{\omega}}{^a}\} $. They constitute a basis for the set of one-forms of $ \I $, $ \Lambda_\I$, and  $ \{-\form{n},\cts{\form{\omega}}{^a}\} $, for the set of one-forms of $ M $, $ \Lambda_\cs{M}$, at $ \I $. \\
			
			The hypersurface $ \I $ is endowed with an intrinsic Riemannian metric $ \ms{_{ab}} $, given by the pullback of the space-time metric to $ \mathcal{I} $ --- the first fundamental form of $ \I $:
				\begin{equation}
				\ms{_{ab}}\eqI \ct{e}{^\mu_a}\ct{e}{^\nu_b}\ct{g}{_{\mu\nu}}\quad .
				\end{equation}
			It is non-degenerate and its inverse is uniquely defined by
				\begin{equation}
				\ms{^{ab}} = \ct{\omega}{_\mu^a}\ct{\omega}{_\nu^b}\ct{g}{^{\mu\nu}}\quad .
				\end{equation}
			The second fundamental form of $ \I $ is defined by
				\begin{equation}\label{eq:secondFundamentalFormI}
					\ct{\kappa}{_{ab}} = \ct{e}{^\mu_a}\ct{e}{^\nu_b} \cd{_\mu}\ct{n}{_\nu}\quad.
				\end{equation}
			Any space-time vector field $ \ct{v}{^\alpha} $ can be decomposed into parts tangent and normal to $ \I $,
				\begin{equation}\label{eq:vector-parts-I}
					\ct{v}{^\alpha}\eqI-\ct{n}{^\alpha}\ct{n}{_\mu}\ct{v}{^\mu}+\cts{v}{^\alpha},\quad\cts{v}{^\mu}\ct{n}{_\mu}=0,\quad\cts{v}{^\alpha}=\ct{e}{^\alpha_a}\cts{v}{^a}\quad,
				\end{equation}
			with $ \vec{v}\in\fX_\I $. This decomposition and notation can be generalised to any tensor (field). The tangent part $ \cts{v}{^\alpha} $ can be obtained by the action of the projector
				\begin{equation}
				\ct{P}{^\alpha_\beta}\defeq \ct{e}{^\alpha_p}\ct{\omega}{_\beta^p},\quad \ct{P}{^\alpha_\beta} \ct{n}{_\alpha}= 0,\quad\cts{v}{^\alpha}=\ct{P}{^\alpha_\mu}\ct{v}{^\mu}\quad.
				\end{equation}
			Its covariant version reads
			\begin{equation}
			\ct{P}{_{\alpha\beta}} \eqI \ct{P}{^\mu_\beta}\ct{P}{^\nu_\beta}\ct{g}{_{\mu\nu}} \eqI \ct{g}{_{\alpha\beta}} + \ct{n}{_{\alpha}} \ct{n}{_{\beta}}\eqI\ct{P}{_{\beta\alpha}}\quad .
			\end{equation}
			The intrinsic volume form of $ \prn{\I,\ms{_{ab}}} $ is determined by
				\begin{align}
				-\ct{n}{_{\alpha}}\ct{\epsilon}{_{abc}} &= \ct{\eta}{_{\alpha\mu\nu\rho}}\ct{e}{^\mu_a}\ct{e}{^\nu_b}\ct{e}{^\rho_c}\quad,\label{eq:volume-I}\\
				-\ct{n}{^{\alpha}}\ct{\epsilon}{^{abc}} &= \ct{\eta}{^{\alpha\mu\nu\rho}}\ct{\omega}{_\mu^a}\ct{\omega}{_\nu^b}\ct{\omega}{_\rho^c}\quad,
				\end{align}
			such that $ \ct{\epsilon}{^{abc}}\ct{\epsilon}{_{abc}}= 6 $. This also fixes the orientation\footnote{According to the orientation for the unphysical space-time, $ \ct{\eta}{_{0123}}=1 $. This coincides with the one we chose for the physical space-time ---see the conventions in the companion paper \cite{Fernandez-Alvarez_Senovilla-afs}.} to $ \cts{\epsilon}{_{123}}=1 $, and $ \cts{\epsilon}{_{abc}} $ is the canonical volume element defined by $ \ms{_{ab}} $.\\
			
			Given the space-time connection, one can define an intrinsic covariant derivative on $ \mathcal{I} $ as
				\begin{equation}
					\ct{v}{^m}\cds{_m}\ct{u}{^a}\defeqI\ct{\omega}{_\mu^a}\ct{v}{^\nu}\cd{_\nu}\ct{u}{^\mu}, \text{ for } \ct{u}{^\alpha}\ct{n}{_\alpha}= \ct{v}{^\alpha}\ct{n}{_\alpha}=0,\quad\ct{u}{^\alpha}=\ct{e}{^\alpha_a}\ct{u}{^a},\quad \ct{v}{^\alpha}=\ct{e}{^\alpha_a}\ct{v}{^a}\quad,
				\end{equation}
			and extend this operator to act on any field on $ \mathcal{I} $. For any tensor field $ \ct{T}{^{\alpha_1 ...\alpha_r}_{\beta_1...\beta_q}} $ defined \emph{at least} on $ \mathcal{I} $, one has
				\begin{align}\label{eq:intrinsicDevI}
				&\ct{\omega}{_{\mu_{1}}^{a_1}}...\ct{\omega}{_{\mu_r}^{a_r}}\ct{e}{^{\nu_1}_{b_q}}...\ct{e}{^{\nu_q}_{b_q}}\ct{e}{^\rho_c}\cd{_\rho}\ct{T}{^{\mu_1...\mu_r}_{\nu_1...\nu_q}}=\cds{_c}\ct{\overline{T}}{^{a_1...a_r}_{b_1...b_q}}\nonumber\\
				&-\sum_{i=1}^{r}\ct{T}{^{a_1...a_{i-1}\sigma a_{i+1}...a_r}_{b_1...b_q}}\ct{n}{_\sigma}\ct{\kappa}{^{a_i}_c} -\sum_{i=1}^{q}\ct{T}{^{a_1...a_r}_{b_1...b_{i-1}\sigma b_{i+1}...b_q}}\ct{\kappa}{_{cb_i}}\ct{n}{^{\sigma}}\nonumber\quad
				\end{align}
			where $ \ct{\overline{T}}{^{a_1...a_r}_{b_1...b_q}}\defeqI  \ct{\omega}{_{\mu_{1}}^{a_1}}...\ct{\omega}{_{\mu_r}^{a_r}}\ct{e}{^{\nu_1}_{b_q}}...\ct{e}{^{\nu_q}_{b_q}}\ct{T}{^{\mu_1...\mu_r}_{\nu_1...\nu_q}}$.
			The new derivative operator is torsion-less, metric and volume preserving ---the underlying connection is the Levi-Civita connection associated to $ \ms{_{ab}} $,
				\begin{align}
				\cds{_a}\ms{_{bc}} &= 0,\\
				\cds{_a} \ct{\epsilon}{_{bcd}}&= 0\quad.
				\end{align}
			The intrinsic curvature is defined by means of $ \cds{_a} $ as
				\begin{equation}
				\prn{	\cds{_a}\cds{_b}-\cds{_b}\cds{_a}}\ct{v}{_c} = \cts{R}{_{abc}^m} \ct{v}{_m},\quad \form{v}\in\Lambda_\I
				\end{equation}
			and the intrinsic Ricci tensor and scalar curvature by $ \cts{R}{_{ab}}\defeq \cts{R}{_{amb}^m}$, $  \csS{R}\defeq\cts{R}{^m_m}$. The relation with the space-time curvature is given by the Gauss equation and its traces:
				\begin{align} 
				\cts{R}{_{abc}^d} &=\ct{e}{^\alpha_a}\ct{e}{^\beta_b}\ct{e}{^\gamma_c}\ct{R}{_{\alpha\beta\gamma}^\delta}\ct{\omega}{_\delta^d}+\ct{\kappa}{_{bc}}\ct{\kappa}{_a^d}-\ct{\kappa}{_{ac}}\ct{\kappa}{_b^d}\quad,\label{eq:gaussrelIi}\\
				\cts{R}{_{ac}} &= \ct{e}{^\alpha_a}\ct{e}{^\gamma_c}\ct{R}{_{\alpha\gamma}}+\ct{n}{^\beta}\ct{n}{_\delta}\ct{e}{^\alpha_a}\ct{e}{^\gamma_c}\ct{R}{_{\alpha\beta\gamma}^\delta}+\ct{\kappa}{_{cd}}\ct{\kappa}{_a^d}-\ct{\kappa}{_{ac}}\ct{\kappa}{}\quad,\label{eq:gaussrelIii}\\
				\cts{R}{}&= \ct{R}{}+2\ct{n}{^\alpha}\ct{n}{^\gamma}\ct{R}{_{\alpha\gamma}}+\ct{\kappa}{_{cd}}\ct{\kappa}{^{cd}}-\ct{\kappa}{^2}\quad,\label{eq:gaussrelIiii}
				\end{align}
			with $ \cs{\kappa}\defeq\ct{\kappa}{^c_c} $, and the space-time curvature and the second fundamental form are related by the Codazzi equation:
				\begin{equation}
				\ct{e}{^\alpha_a}\ct{e}{^\beta_b}\ct{e}{^\gamma_c}\ct{R}{_{\alpha\beta\gamma}^\delta}\ct{n}{_{\delta}}= 2\cds{_{[a}}\ct{\kappa}{_{b]c}}\quad.
				\end{equation}
		\subsection{Cuts}\label{ssec:foliationAndCuts}
			Let $ \Sc $ be any two-dimensional submanifold  embedded in $ \mathcal{I} $
			and assume that it has $ \mathbb{S}^2 $-topology. Generically, we will refer to these kind of surfaces as `cuts'. Let $ \ct{r}{_a} $ be the (spacelike) normal one-form to the cut within $ \I $ ---$ \ct{n}{_\alpha} $ is orthogonal to the cut too, of course. In a similar fashion as we have done above, we introduce a couple of linearly independent vector fields $ \{\ct{E}{^a_A}\} $, $ A=2,3 $, orthogonal to $ \ct{r}{_a} $ and tangent to $ \I $, such that they constitute a basis for the set  $ \fX_\Sc $ of vector fields of $ \Sc $. Also, rise an index to the normal one-form using $ \ms{^{ab}} $ and define a dual basis $ \{\ct{W}{_a^A}\} $ orthogonal to $ \ct{r}{^a} $. These sets of vector fields, being completely tangent to $ \I $, can be written as space-time fields: $ \ct{r}{^\alpha}\defeqI \ct{e}{^\alpha_a}\ct{r}{^a} $, $ \ct{E}{^\alpha_A}\defeqc \ct{E}{^a_A}\ct{e}{^\alpha_a} $ and $ \ct{W}{_\alpha^A}\defeqc \ct{W}{_a^A}\ct{\omega}{_\alpha^a} $. The triads $ \{\vec{r},\vec{\ct{E}{_A}}\} $ and $ \{\form{r},\ct{\form{W}}{^A}\} $ constitute a basis for $ \fX_\I $ and $ \Lambda_\I$ at $ \Sc $, respectively. Pushforwards/pullbacks of intrinsic objects to $ \Sc $ can be written in terms of $ \ct{W}{_\alpha^A} $ and $ \ct{E}{^\alpha_A} $. \\
			
			The intrinsic metric of $ \Sc $ is given by the pullback of the metric of $ \I $ ---the first fundamental form of $ \Sc $,
				\begin{equation}
				\mc{_{AB}}\defeqc \ct{E}{^a_A}\ct{E}{^b_B}\ms{_{ab}}\quad ,
				\end{equation} 
			which concides  with the pullback of the metric of $ M $ with $ \ct{E}{^\alpha_A} $,
				\begin{equation}
				\mc{_{AB}}\eqc \ct{E}{^\alpha_A}\ct{E}{^\beta_B}\ct{g}{_{\alpha\beta}}\quad .
				\end{equation}
			The second fundamental form of $ \Sc $ in $ \I $ is defined as
				\begin{equation}
				\ctc{\kappa}{_{AB}}\defeqc \ct{E}{^a_A}\ct{E}{^b_B}\cds{_a}\ct{r}{_b}\quad ,
				\end{equation}
			and the projector to the cut as
				\begin{equation}\label{eq:projectorcut}
				\ctc{P}{^a_b}\defeqc \ct{E}{^a_A}\ct{W}{_b^A}\eqc \ct{\delta}{^a_b}-\ct{r}{^a}\ct{r}{_b}\quad .
				\end{equation}
			Its covariant version is symmetric
				\begin{equation}
				\ctc{P}{_{ab}}\eqc  \ms{_{ab}}-\ct{r}{_a}\ct{r}{_b}\quad 
				\end{equation}
			and
				\begin{equation}
				\ctc{P}{_{\alpha\beta}}\defeqc \ct{\omega}{_\alpha^a}\ct{\omega}{_\beta^b}\ctc{P}{_{ab}}\eqc\ct{g}{_{\alpha\beta}}+\ct{n}{_\alpha}\ct{n}{_\beta}-\ct{r}{_\alpha}\ct{r}{_\beta}\quad .
				\end{equation}
			Any $ \vec{v} \in \fX_\I$ can be split into a normal and tangent part to $ \I $ as before (see \cref{eq:vector-parts-I}). Now, in addition to that, the tangent part to $ \I $ is decomposed into its tangent and normal parts to $ \Sc $:
				\begin{equation}
					\ct{v}{^\alpha}=-\ct{n}{_\mu}\ct{v}{^\mu}\ct{n}{^\alpha}+\cts{v}{^\alpha}=-\ct{n}{_\mu}\ct{v}{^\mu}\ct{n}{^\alpha}+\ct{r}{_\mu}\ct{v}{^\mu}\ct{r}{^\alpha}+\ctc{v}{^\alpha},\quad\text{with}\quad\ct{r}{_\mu}\ctc{v}{^\mu}=0=\ct{n}{_\mu}\ctc{v}{^\mu}\quad,
				\end{equation}
			where, $\ctc{P}{^\alpha_\mu}\ct{v}{^\mu}\eqc \ctc{v}{^\alpha}=\ctc{v}{^A}\ct{E}{^\alpha_A} $, with $ \vec{\csC{v}} \in \fX_\Sc$. \\
			
			Also, the intrinsic volume two-form of $ \prn{\Sc,\mc{_{AB}}} $ is determined by
				\begin{align}
				\ct{r}{_{a}}\ctc{\epsilon}{_{AB}} &\eqc \ct{\epsilon}{_{amn}}\ct{E}{^m_A}\ct{E}{^n_B}\quad,\\
				\ct{r}{^{a}}\ctc{\epsilon}{^{AB}} &\eqc \ct{\epsilon}{^{amn}}\ct{W}{_m^A}\ct{W}{_n^B}\quad,
				\end{align}
			such that $ \ctc{\epsilon}{^{AB}}\ctc{\epsilon}{_{AB}}= 2 $ and fixing the orientation to $ \ctc{\epsilon}{_{23}}=1 $. Notice that using \cref{eq:volume-I} one can write the space-time version of this two-form as $  \ctc{\epsilon}{_{\alpha\beta}}\eqc\ctc{P}{^\sigma_\alpha}\ctc{P}{^\rho_\beta} \ct{\eta}{_{\mu\nu\sigma\rho}}\ct{n}{^\mu}\ct{r}{^\nu} $, $ \ctc{\epsilon}{^{\alpha\beta}}\eqc\ctc{P}{^\alpha_\sigma}\ctc{P}{^\beta_\rho} \ct{\eta}{^{\mu\nu\sigma\rho}}\ct{n}{_\mu}\ct{r}{_\nu} $.\\
			
				An intrinsic connection on the cut can be defined as
				\begin{equation}
				\ct{V}{^M}\cdc{_M}\ct{U}{^A}\defeqc\ct{W}{_m^A}\ct{V}{^n}\cds{_n}\ct{U}{^m},\quad\text{where}\quad \ct{U}{^a}\eqc\ct{E}{^a_A}\ct{U}{^A},\quad \ct{V}{^a}\eqc\ct{W}{^a_A}\ct{V}{^A}\quad.
				\end{equation}
			Or, equivalently, by 
				\begin{equation}
				\ct{V}{^M}\cdc{_M}\ct{U}{^A}\defeqc\ct{W}{_\mu^A}\ct{V}{^\nu}\cd{_\nu}\ct{U}{^\mu},\quad\text{where} \quad\ct{U}{^\alpha}=\ct{E}{^\alpha_A}\ct{U}{^A},\quad \ct{V}{^\alpha}=\ct{W}{^\alpha_A}\ct{V}{^A}\quad.
				\end{equation}
			The intrinsic covariant derivative of a tensor field $ \ct{T}{^{a_1 ...a_r}_{b_1...b_q}} $ defined \emph{at least} on $ \Sc $ is written as
				\begin{align}\label{eq:intrinsicdevCut}
				&\ct{W}{_{m_{1}}^{A_1}}...\ct{W}{_{m_r}^{A_r}}\ct{E}{^{n_1}_{B_q}}...\ct{E}{^{n_q}_{B_q}}\ct{E}{^r_C}\cds{_r}\ct{T}{^{m_1...m_r}_{n_1...n_q}}=\cdc{_C}\ctc{T}{^{A_1...A_r}_{B_1...B_q}}\nonumber\\&
				+\sum_{i=1}^{r}\ct{T}{^{A_1...A_{i-1}s A_{i+1}...A_r}_{B_1...B_q}}\ct{r}{_s}\ctc{\kappa}{^{A_i}_C} 
				+\sum_{i=1}^{q}\ct{T}{^{A_1...A_r}_{B_1...B_{i-1}s B_{i+1}...B_q}}\ctc{\kappa}{_{CB_i}}\ct{r}{^{s}}\nonumber\quad .
				\end{align}
			 Again, the underlying connection is the Levi-Civita connection associated to $ \mc{_{AB}} $:
				\begin{align}
				\cdc{_A}\mc{_{BC}} &= 0\quad,\\
				\cdc{_A} \ctc{\epsilon}{_{BC}}&= 0\quad.
				\end{align}
			The Gauss equation and its traces read
			\begin{align}
			\ctc{R}{_{ABC}^D} &=\ct{E}{^a_A}\ct{E}{^b_B}\ct{E}{^c_C}\cts{R}{_{abc}^d}\ct{W}{_d^D}-\ctc{\kappa}{_{BC}}\ctc{\kappa}{_A^D}+\ctc{\kappa}{_{AC}}\ctc{\kappa}{_B^D}\quad,\label{eq:gaussrelCi}\\
			\ctc{R}{_{AC}} &= \ct{E}{^a_A}\ct{E}{^c_C}\cts{R}{_{ac}}+\ct{r}{^b}\ct{r}{_d}\ct{E}{^a_A}\ct{E}{^c_C}\cts{R}{_{abc}^d}-\ctc{\kappa}{_{CD}}\ctc{\kappa}{_A^D}+\ctc{\kappa}{_{AC}}\ctc{\kappa}{}\quad\label{eq:gaussrelCii},\\
			\csC{R}&= \cts{R}{}+2\ct{r}{^a}\ct{r}{^c}\cts{R}{_{ac}}-\ctc{\kappa}{_{CD}}\ctc{\kappa}{^{CD}}+\ctc{\kappa}{^2}\quad\label{eq:gaussrelCiii},
			\end{align}
			and the Codazzi equation,
			\begin{equation}\label{eq:cod-cuts}
			\ct{E}{^a_A}\ct{E}{^b_B}\ct{E}{^c_C}\cts{R}{_{abc}^d}\ct{r}{_{d}}= 2\cdc{_{[A}}\ctc{\kappa}{_{B]C}}\quad.
			\end{equation}
			\subsection{Congruences}\label{app:congruences}	
			Assume $ \I $, or at least an open connected portion\footnote{In which case, the results below apply only to that region.} $ \Delta\in\I $ with the same topology as $ \I $, and let $ \C $ be a congruence of curves there  locally defined by
				\begin{equation}
					\ct{x}{^a}=\ct{X}{^a}\prn{v,\zeta^A},
				\end{equation}
			where $ \ct{X}{^a} $ are invertible functions such that
				\begin{equation}
					v=V(x^a),\quad \zeta^A=Z^A(x^a)\quad.
 				\end{equation}
			Each curve of $ \C $ is marked by constant values of $ \zeta^A $ and parametrised by $ v $.
				\begin{figure}[h]
				\includegraphics[scale=0.5]{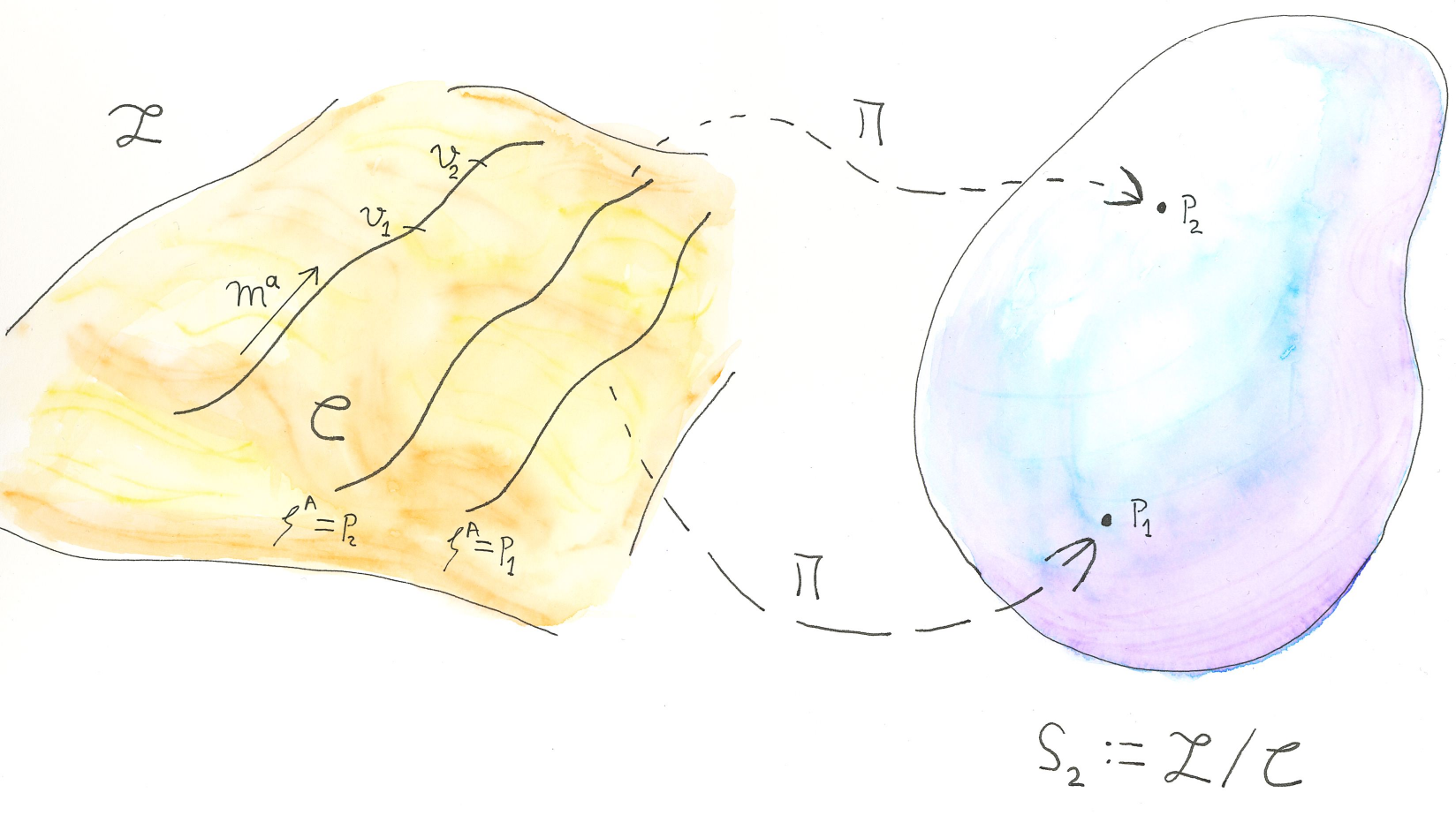}
				\caption[Canonical projection]{The space-like hypersurface $ \I $ equipped with a congruence $ \C $ of curves. The canonical projection $ \Pi $ maps each curve to a point on the projected `surface' $ \Scn $.}
				\end{figure}
			The unit vector field $ \ct{m}{^a} $ tangent to the curves can be written in the local basis $ \prn{\dpart{}/\dpart{}x^a} $,
				\begin{equation}\label{eq:congruence-vector}
					\ct{m}{^a}=\prn{\ms{_{cd}}\frac{\partial X^c}{\partial v}\frac{\partial X^d}{\partial v}}^{-\frac{1}{2}}\frac{\partial X^a}{\partial v},\quad\ct{m}{^a}\ct{m}{_a}=1\quad.
				\end{equation}

			It is easily checked that $ \ct{m}{^a}\cds{_a}V\neq0 $ and $ \ct{m}{^a}\cds{_a}Z^A=0 $. Notice that there is the following freedom in reparametrising and changing the markers of the curves:
				\begin{align}
					v \rightarrow v'\prn{v,\zeta^A},\quad\frac{\partial v'}{\partial v}\neq 0\quad,\\
					\zeta^A \rightarrow \zeta'^A\prn{\zeta^A},\quad\abs{\frac{\partial\zeta^A}{\partial\zeta^B}}\neq 0\quad.\label{eq:coordinate-change-projectiveS}
				\end{align}
			The quotient $ \Scn\defeq\I/\C $ is called the projected `surface'. It is a two-dimensional differential manifold although, in general, it is not Riemannian because it is not endowed with a natural metric as it will become clear later on. One can define a canonical projection $ \Pi $ that maps all points on a curve of $ \C $ to the same point on $ \Scn $. In this sense, each point on $ \Scn $ represents a curve of $ \C $ and $ \zeta^A $ are local coordinates on $ \Scn $ --indeed \cref{eq:coordinate-change-projectiveS} can be regarded as a local change of coordinates on $ \Scn $. The one-forms 	
				\begin{equation}\label{eq:Wcongruence}
					\ctcn{W}{_a^A}\prn{x}\defeq \prn{\Pi^*\prn{\df{\zeta^A}}}_a=\frac{\partial Z^A\prn{x}}{\partial x^a},\quad \ct{m}{^a}\ctcn{W}{_a^A}=0,\quad\lied_{\vec{m}} \ctcn{\form{W}}{^A}=0\quad,
				\end{equation}
			allow us to write the pullback $ \Pi^* $ to $ \I $ of any covariant tensor field $ \ct{T}{_{A_1...A_p}} $ on $ \Scn $ as
				\begin{equation}
					\ctcn{T}{_{a_1...a_p}}\prn{x}\defeq\brkt{\Pi^*T\prn{\zeta}}_{a_1...a_p}=\ct{T}{_{A_1...A_p}}\prn{Z\prn{x}}\ctcn{W}{_{a_1}^{A_1}}\prn{x}...\ctcn{W}{_{a_p}^{A_p}}\prn{x}\quad.
				\end{equation}
			The objects $ \ctcn{T}{_{a_1...a_p}} $ are covariant tensor fields on $ \I $ with no dependence on $ v $ and fully orthogonal to $ \ct{m}{^a} $. Thus there exists an isomorphism between covariant tensor fields on $ \Sc $ and covariant tensor fields on $ \I $ that have vanishing Lie derivative along $ \ct{m}{^a} $ and are orthogonal to $ \ct{m}{^a} $.\\
			
				Also, one can take the push-forward $ \Pi' $ of any contravariant tensor field $ \ct{T}{^{a_1...a_p}}$ at a point $ q\in \I $ to a point $ \Pi(q) $ on $ \Scn $,
				\begin{equation}
					\ctcn{T}{^{A_1...A_p}}\prn{\zeta}\evalat{\Pi(q)}\defeq\brkt{\Pi'T\prn{x}}^{A_1...A_p}=\brkt{\ct{T}{^{a_1...a_p}}\prn{x}\ctcn{W}{_{a_1}^{A_1}}\prn{x}...\ctcn{W}{_{a_p}^{A_p}}\prn{x}}\evalat{q}\quad.
				\end{equation}
			Because $ \ct{T}{^{a_1...a_p}} $ is defined everywhere on $ \I $ and $ \Pi' $ acts pointwise, the quantities $ \ctcn{T}{^{A_1...A_p}} $ are well defined at each point on $ \I $ and, thus, they can be considered as a set of \emph{scalar fields} on $ \I $. However, even though they change tensorially under the transformations \eqref{eq:coordinate-change-projectiveS}, they do not constitute tensor fields on $ \Sc_{2} $, in the sense that $ \ct{T}{^{a_1...a_p}} $ can give rise to different tensor fields on $ \Scn $ due to the dependence of $ \ctcn{T}{^{A_1...A_p}} $ on $ v $. Furthermore, as $ \ct{T}{^{a_1...a_p}} $ may contain transversal components along $ \ct{m}{^a} $, multiple tensor fields on $ \I $ can project to the same family of scalars $ \ctcn{T}{^{A_1...A_p}}  $. In any case, there exists an isomorphism between contravariant tensor fields on $ \I $ completely orthogonal to $ \ct{m}{^a} $ and with vanishing Lie derivative along $ \ct{m}{^a} $ and contravariant tensor fields on $ \Scn $.\\
			
			We can define a couple of linearly independent vector fields on $ \I $, $ \prn{\ctcn{E}{^a_A}} $, satisfying 
				\begin{equation}
					\ct{m}{_a}\ctcn{E}{^a_A}=0,\quad\ctcn{E}{^a_A}\ctcn{W}{_a^B}=\delta^B_A\quad.
				\end{equation}	
			Then, $ \prn{\ct{m}{^a},\ctcn{E}{^a_A}} $, $ \prn{\ct{m}{_a},\ctcn{W}{_a^A}} $ constitute a pair of dual bases. On the one hand, it is possible to lift contravariant tensor fields on $ \Scn $ to contravariant tensor fields on $ \scri $ by 
				\begin{equation}
					\ctcn{T}{^{a_1...a_p}}\prn{x}\defeq\ct{T}{^{A_1...A_p}}\prn{Z\prn{x}}\ctcn{E}{^{a_1}_{A_1}}\prn{x}...\ctcn{E}{^{a_p}_{A_p}}\prn{x}\quad
				\end{equation}
			which are orthogonal to $ \ct{m}{_a} $ and have, in general, non-vanishing Lie derivative along $ \ct{m}{^a} $. On the other hand, given a covariant tensor field on $ \scri $, one can construct pointwise a set of scalar fields on $ \scri $ as
				\begin{equation}
					\ctcn{T}{_{A_1...A_p}}\prn{x}\evalat{\Pi(q)}\defeq\brkt{\ct{T}{_{a_1...a_p}}\prn{x}\ctcn{E}{^{a_1}_{A_1}}\prn{x}...\ctcn{E}{^{a_p}_{A_p}}\prn{x}}\evalat{q}\quad.
				\end{equation}
			The projector orthogonal to $ \ct{m}{^a} $ is defined as
				\begin{equation}\label{eq:projector-foliation}
					\ctcn{P}{^a_b}\defeq \ctcn{E}{^a_C}\ctcn{W}{_b^C},\quad\ctcn{P}{^c_b}\ct{m}{_c}=0=\ctcn{P}{^a_c}\ct{m}{^c},\quad\ctcn{P}{^c_c}=2\quad,
				\end{equation}
			and in terms of $ \ct{m}{_a} $ its covariant version reads
				\begin{equation}\label{eq:projector-congruences}
					\ctcn{P}{_{ab}}=\ms{_{ab}}-\ct{m}{_a}\ct{m}{_b}\quad.
				\end{equation}
			This object gives a scalar product on $ \I $ of vectors orthogonal to $ \ct{m}{^a} $. It is possible to introduce a \emph{family} of inverse metric tensor fields on $ \Scn $ as
				\begin{equation}
				\mcn{^{AB}}\defeq \ctcn{W}{_a^A}\ctcn{W}{_b^B}\ms{^{ab}}\quad,
				\end{equation}	
			while the covariant version is given by the condition $ \mcn{_{AC}}\mcn{^{BC}}=\delta^B_A $.
			Alternatively, using $ \ctcn{E}{^a_A} $
				\begin{equation}
					\mcn{_{AB}}=\ctcn{E}{^a_A}\ctcn{E}{^b_B}\ms{_{ab}}\quad.
				\end{equation}
			Although one can use $ \mcn{^{AB}} $ and $ \mcn{_{AB}} $ to rise and lower indices on $ \Scn $, it \emph{is not a metric tensor}, since it depends on $ v $. More appropriately, it represents a one-parameter family of metric tensors. One can induce a `volume two-form' in a simple way,
				\begin{align}
					\ct{m}{_{a}}\ctcn{\epsilon}{_{AB}} &\eqc \ct{\epsilon}{_{amn}}\ctcn{E}{^m_A}\ctcn{E}{^n_B}\quad,\\
					\ct{m}{^{a}}\ctcn{\epsilon}{^{AB}} &\eqc \ct{\epsilon}{^{amn}}\ctcn{W}{_m^A}\ctcn{W}{_n^B}\quad,
				\end{align}
			which is completely antisymmetric and satisfies $ \ctcn{\epsilon}{^{AB}}\ctcn{\epsilon}{_{AB}}=2 $. However, this object depends on $ v $ too in general and, therefore, it constitutes a one-parameter family of volume forms. Observe that we have fixed the orientation to $ \ctc{\epsilon}{_{23}}=1 $.
			\\
			
			The covariant derivative on $ \I $ of $ \ct{m}{_a} $ is decomposed as	
				\begin{equation}\label{eq:derivative-r-decomposition}
				\cds{_a}\ct{m}{_b}= \ct{m}{_a}\ctcn{a}{_b}+\ctcn{\kappa}{_{ab}}+\ctcn{\omega}{_{ab}}\quad,	
				\end{equation}
			where 
				\begin{align}
				\ctcn{a}{_b}&\defeq \ct{m}{^c}\cds{_c}\ct{m}{_b}\quad\text{is the acceleration,}\label{eq:acceleration}\\
				\ctcn{\kappa}{_{ab}}&\defeq \ctcn{P}{^c_a}\ctcn{P}{^d_b}\cds{_{(c}}\ct{m}{_{d)}}\quad\text{is the expansion tensor,}\label{eq:expansion-tensor}\\
				\ctcn{\omega}{_{ab}}&\defeq \ctcn{P}{^c_a}\ctcn{P}{^d_b}\cds{_{[c}}\ct{m}{_{d]}}\quad\text{is the vorticity,}\label{eq:vorticity}
				\end{align}
			and the shear of $ \ct{m}{_a}  $ is defined as the traceless part of $ \ctcn{\kappa}{_{ab}} $,
				\begin{equation}\label{eq:kappa-congruence}
				\ctcn{\Sigma}{_{ab}}\defeq \ctcn{\kappa}{_{ab}}-\frac{1}{2}\ctcn{P}{_{ab}}\cscn{\kappa},\quad\cscn{\kappa}\defeq \ctcn{P}{^{cd}}\ctcn{\kappa}{_{cd}}\quad.
				\end{equation}
			It is easy to show that
				\begin{align}
					\lied_{\vec{m}}\ctcn{E}{^a_A}&=-\ct{m}{^a}\ctcn{E}{^c_A}\ctcn{a}{_c}\quad,\label{eq:liedEcongruence}\\
					\lied_{\vec{m}}\ctcn{P}{^a_b}&=-\ctcn{a}{_b}\ct{m}{^a}\quad,\label{eq:liedPcongruence}\\
					\lied_{\vec{m}}\ctcn{P}{_{ab}}&=2\ctcn{\kappa}{_{ab}}\quad.
				\end{align}	
			Also, defining
				\begin{equation}
					\ctcn{\epsilon}{_{ab}}\defeq\ct{m}{^e}\ctcn{P}{^c_a}\ctcn{P}{^d_b}\ct{\epsilon}{_{ecd}}=\ctcn{W}{_a^A}\ctcn{W}{_b^B}\ctcn{\epsilon}{_{AB}}\quad,
				\end{equation}
			and using \cref{eq:liedPcongruence} one derives
				\begin{equation}\label{eq:liedvolformcn}
					\lied_{\vec{m}}\ctcn{\epsilon}{_{ab}}=\cscn{\kappa}\ctcn{\epsilon}{_{ab}}\quad.
				\end{equation}
			Incidentally,
				\begin{equation}\label{eq:liedvolformcnsc}
				\lied_{\vec{m}}\ct{\epsilon}{_{abc}}=\cscn{\kappa}\ct{\epsilon}{_{abc}}\quad.
				\end{equation}
			As all the kinematic tensors are orthogonal to $ \ct{m}{^a} $, they are univocally determined by the one-parameter family of scalar fields on $ \I $ --which can be seen as objects on $ \Scn $--
				\begin{align}
					\ctcn{a}{_A}&\defeq\ctcn{E}{^a_A}\ctcn{a}{_a} \quad,\\
					\ctcn{\kappa}{_{AB}}&\defeq\ctcn{E}{^a_A}\ctcn{E}{^b_B}\ctcn{\kappa}{_{ab}}\quad,\\
					\ctcn{\Sigma}{_{AB}}&\defeq\ctcn{E}{^a_A}\ctcn{E}{^b_B}\ctcn{\Sigma}{_{ab}}\quad,\\
					\ctcn{\omega}{_{AB}}&\defeq\ctcn{E}{^a_A}\ctcn{E}{^b_B}\ctcn{\omega}{_{ab}}\quad.
				\end{align}
			The scalar fields $ \ct{T}{^{A_1...A_q}_{B_1...B_p}} $  associated to an arbitrary tensor field $ \ct{T}{^{a_1...a_q}_{b_1...b_p}} $ on $ \I $ can be differentiated along $ \ct{m}{^a} $: 
				\begin{align}
					\lied_{\vec{m}}\ct{T}{^{A_1...A_q}_{B_1...B_p}}&=\ct{m}{^j}\cds{_j}\ct{T}{^{A_1...A_q}_{B_1...B_p}}	\nonumber\\
					&=\ctcn{E}{^{b_1}_{B_1}}...\ctcn{E}{^{b_p}_{B_p}}\ctcn{W}{_{a_1}^{A_1}}...\ctcn{W}{_{a_q}^{A_q}}\lied_{\vec{m}}\prn{\ct{T}{^{a_1...a_q}_{b_1...b_p}}}\nonumber\\
					&-\sum_{i=1}^{p}\ct{T}{^{A_1...A_q}_{B_1...B_{i-1}\sigma B_{i+1}...B_p}}\ct{m}{^\sigma}\ctcn{a}{_{B_{i}}}\quad,\label{eq:liedScalarscongruence}
				\end{align}	
			where we have used the Leibniz property of the Lie derivative together with \cref{eq:liedEcongruence,eq:Wcongruence}.
			Then, the Lie derivative along $ \ct{m}{^a} $ of the one-parameter family of `metrics' $ \mcn{_{AB}} $ for fixed $ A, B $ can be computed to give
				\begin{equation}\label{eq:der-metric-congruences}
					\lied_{\vec{m}}\mcn{_{AB}}=\ctcn{\kappa}{_{AB}}.
				\end{equation}
			Now, let $ U $ be a function such that
				\begin{equation}
				\ct{m}{^c}\ct{s}{_c}=1,\quad\ct{s}{_a}\defeq \prn{\df{U}}_a\quad
				\end{equation}
			and expand $ \ct{s}{_a} $ in the $ \prn{\ct{m}{_a},\ctcn{W}{_a^A}} $ basis,
				\begin{equation}
					\ct{s}{_a}=\ct{m}{_a}+\ct{M}{_A}\ctcn{W}{_a^A}\quad.
				\end{equation}
			Taking the Lie derivative along $ \ct{m}{^a} $ of the functions $ \ctcn{M}{_A} $ for fixed $ A $ one finds
				\begin{equation}
					\ctcn{a}{_A}=-\lied_{\vec{m}}\ct{M}{_A}.
				\end{equation}
			Instead, if one takes de exterior derivative of $ \ct{s}{_a} $ --which vanishes by definition-- one gets the relation
				\begin{equation}\label{eq:commutator-Es}
					\commute{\ctcn{\vec{E}}{_A}}{\ctcn{\vec{E}}{_B}}^a=-2\ctcn{\omega}{_{AB}}\ct{m}{^a},
				\end{equation}
			which allows us to derive
				\begin{equation}
					\lied_{\ctcn{\vec{E}}{_A}}\ctcn{\form{W}}{^B}=0.
				\end{equation}
			Also,
				\begin{equation}
					\lied_{\ctcn{\vec{E}}{_A}}\ct{m}{_a}=-\ctcn{a}{_A}\ct{m}{_a}+2\ctcn{\omega}{_{AC}}\ctcn{W}{_a^C}.
				\end{equation}
			So far we have not introduced a connection on $ \Scn $, nor a covariant derivative. Note that in the basis $ \prn{\ct{m}{^a},\ctcn{E}{^a_A}} $ one has
				\begin{equation}
					\ctcn{E}{^c_A}\cds{_c}\ctcn{E}{^a_B}=-\prn{\ctcn{\kappa}{_{AB}}+\ctcn{\omega}{_{AB}}}\ct{m}{^a}+\ctcn{\gamma}{^C_{AB}}\ctcn{E}{^a_C}\quad,
				\end{equation}	
			where $ \ctcn{\gamma}{^C_{AB}} $ are functions such that $ \ctcn{\gamma}{^C_{AB}}=\ctcn{\gamma}{^C_{BA}} $, as one can check computing the commutator and using \cref{eq:commutator-Es}. Taking this into account it follows that
				\begin{equation}\label{eq:aux22}
					\ct{m}{^c}\cds{_c}\ctcn{E}{^a_A}=-\ctcn{a}{_A}\ct{m}{^a}+\prn{\ctcn{\kappa}{_{A}^C}+\ctcn{\omega}{_{A}^C}}\ctcn{E}{^a_C}\quad.
				\end{equation}
			 In addition, it can be shown that
				\begin{equation}
					\ctcn{E}{^c_A}\cds{_c}\ctcn{W}{_a^B}=-\prn{\ctcn{\kappa}{_{A}^B}+\ctcn{\omega}{_{A}^B}}\ct{m}{_a}-\ctcn{\gamma}{^B_{AC}}\ctcn{W}{_a^C}\quad.
				\end{equation}
			Contracting \cref{eq:liedvolformcn} with $ \ctcn{E}{^a_A}\ctcn{E}{^b_B} $ and using \cref{eq:liedScalarscongruence} one derives
				\begin{equation}\label{eq:der-volume-congruences}
					\lied_{\vec{m}}\ctcn{\epsilon}{_{AB}}=\cscn{\kappa}\ctcn{\epsilon}{_{AB}}.
				\end{equation}
				
			Under the change in \cref{eq:coordinate-change-projectiveS} , $ \ctcn{\gamma}{^C_{AB}} $ behaves like a connection. However, due to the dependence on $ v $, it \emph{is not a connection}, but a one-parameter family of such objects. Nevertheless, we can define a `covariant derivative' operator by
				\begin{equation}
					\cdcn{_A}\ct{v}{^B}\defeq \ctcn{E}{^a_A}\partial{_a}\ct{v}{^B}+\ctcn{\gamma}{^B_{AC}}\ct{v}{^C}, \text{ with } \ct{v}{^A}=\ct{v}{^a}\ctcn{W}{_a^A},\quad \ct{v}{^a}\ct{m}{_a}=0.
				\end{equation}
			For the same reasons stated above, this is not a tensor field on $ \Scn $. The definition can be extended to arbitrary-rank contravariant and covariant tensor objects. Its relation to the covariant derivative on $ \I $ acting on a tensor field $ \ct{T}{^{a_1 ...a_r}_{b_1...b_q}} $ is written as
				\begin{align}\label{eq:intrinsicdevS2}
				&\ctcn{W}{_{m_{1}}^{A_1}}...\ctcn{W}{_{m_r}^{A_r}}\ctcn{E}{^{n_1}_{B_q}}...\ctcn{E}{^{n_q}_{B_q}}\ctcn{E}{^r_C}\cds{_r}\ct{T}{^{m_1...m_r}_{n_1...n_q}}=\cdcn{_C}\ctcn{T}{^{A_1...A_r}_{B_1...B_q}}\nonumber\\
				&+\sum_{i=1}^{r}\ct{T}{^{A_1...A_{i-1}s A_{i+1}...A_r}_{B_1...B_q}}\ct{m}{_s}\prn{\ctcn{\kappa}{_C^{A_i}}+\ctcn{\omega}{_C^{A_i}}}+\sum_{i=1}^{q}\ct{T}{^{A_1...A_r}_{B_1...B_{i-1}s B_{i+1}...B_q}}\prn{\ctcn{\kappa}{_{CB_i}}+\ctcn{\omega}{_{CB_i}}}\ct{m}{^{s}}.
				\end{align}
			Then, for $ \ct{T}{^{m_1...m_r}_{n_1...n_q}} $ completely orthogonal to $ \ct{m}{^a} $ and $ \ct{m}{_a} $ one has
				\begin{equation}
				\ctcn{W}{_{m_{1}}^{A_1}}...\ctcn{W}{_{m_r}^{A_r}}\ctcn{E}{^{n_1}_{B_q}}...\ctcn{E}{^{n_q}_{B_q}}\ctcn{E}{^r_C}\cds{_r}\ct{T}{^{m_1...m_r}_{n_1...n_q}}=\cdcn{_C}\ctcn{T}{^{A_1...A_r}_{B_1...B_q}}\quad.
				\end{equation}
			This `covariant derivative' is `metric' and `volume-preserving' in the sense that
				\begin{align}
					\cdcn{_A}\ctcn{\epsilon}{_{BC}}&=0\quad,\label{eq:derCongruence-metric}\\
					\cdcn{_A}\mcn{_{BC}}&=0\quad,
				\end{align}
			and a typical calculation leads to an expression in terms of $ \mcn{_{AB}} $
			   \begin{equation}
			    \ctcn{\gamma}{^C_{AB}}=\frac{1}{2}\mcn{^{CD}}\prn{\ctcn{E}{^a_A}\dpart{a}\mcn{_{BD}}+\ctcn{E}{^b_B}\dpart{b}\mcn{_{AD}}-\ctcn{E}{^d_D}\dpart{d}\mcn{_{AB}}}\quad.
			   \end{equation}
			 Define a one-parameter family of tensor fields on $ \Scn $ by
			 	\begin{equation}
			 	\ctcn{R}{_{BAC}^D}\defeq \ctcn{E}{^a_A}\dpart{a}\ctcn{\gamma}{^D_{BC}}-\ctcn{E}{^b_B}\dpart{b}\ctcn{\gamma}{^D_{AC}}+\ctcn{\gamma}{^D_{AE}}\ctcn{\gamma}{^E_{BC}}-\ctcn{\gamma}{^D_{BE}}\ctcn{\gamma}{^E_{AC}},
			 	\end{equation}
			 which by construction has the symmetries
			 	\begin{equation}
			 		\ctcn{R}{_{ABC}^D}=-\ctcn{R}{_{BAC}^D},\quad\ctcn{R}{_{ABC}^D}+\ctcn{R}{_{BCA}^D}+\ctcn{R}{_{CAB}^D}=0\quad.
			 	\end{equation}
			 	 A direct calculation gives
			 	\begin{equation}
			 		\prn{\cdcn{_A}\cdcn{_B}-\cdcn{_B}\cdcn{_A}}\ct{V}{^D}=-\ctcn{R}{_{ABC}^D}\ct{V}{^C}-2\ctcn{\omega}{_{AB}}\lied_{\vec{m}}\ct{V}{^D}\quad,\label{eq:curvatureCongruences}
			 	\end{equation}
			 where $ \lied_{\vec{m}}\ct{V}{^C} $ is computed according to \cref{eq:liedScalarscongruence}. One can define a covariant version,
			 	\begin{equation}
			 		\ctcn{R}{_{ABCD}}\defeq \mcn{_{ED}}\ctcn{R}{_{ABC}^E}.
			 	\end{equation}
			 Note that this object does not have the antisymmetry property in the second pair of indices:
			 	\begin{equation}
			 		\ctcn{R}{_{AB(CD)}}=2\ctcn{\omega}{_{AB}}\ctcn{\kappa}{_{CD}}\quad,
			 	\end{equation}
			 where we have used \cref{eq:derCongruence-metric,eq:curvatureCongruences}. Hence
			 	\begin{equation}\label{eq:curvature-congruences-symmetric-antisymmetric}
			 		\ctcn{R}{_{ABCD}}=2\ctcn{\omega}{_{AB}}\ctcn{\kappa}{_{CD}}+	\ctcn{R}{_{AB[CD]}}\quad.
			 	\end{equation}
			The second term, since we are in 2 dimensions, can be identically written as
				\begin{equation}\label{eq:curvature-congruences}
					\ctcn{R}{_{AB[CD]}}=\cscn{K}\prn{\mcn{_{AC}}\mcn{_{BD}}-\mcn{_{AD}}\mcn{_{BC}}}\quad,
				\end{equation} 
			for some scalar function $ \cscn{K} $. The relation between the curvature tensor on $ \I $ and this one-parameter family of `curvature tensors' on $ \Scn $ can be determined by typical calculations. The result is a Gauss-like relation,
				\begin{align}
					\ctcn{R}{_{ABC}^D} =&\ct{E}{^a_A}\ct{E}{^b_B}\ct{E}{^c_C}\cts{R}{_{abc}^d}\ct{W}{_d^D}+2\ctcn{\omega}{_{AB}}\prn{\ctcn{\kappa}{_C^D}+\ctcn{\omega}{_C^D}}-\prn{\ctcn{\kappa}{_{BC}}+\ctcn{\omega}{_{BC}}}\prn{\ctcn{\kappa}{_A^D}+\ctcn{\omega}{_A^D}}\nonumber\\
					&+\prn{\ctcn{\kappa}{_{AC}}+\ctcn{\omega}{_{AC}}}\prn{\ctcn{\kappa}{_B^D}+\ctcn{\omega}{_B^D}}\quad,\label{eq:gaussrelCni}
				\end{align}
			which lowering an index can be written also as
				\begin{align}
				\ctcn{R}{_{AB[CD]}} =&\ct{E}{^a_A}\ct{E}{^b_B}\ct{E}{^c_C}\cts{R}{_{abcd}}\ct{E}{^d_D}+2\ctcn{\omega}{_{AB}}\ctcn{\omega}{_{CD}}-\prn{\ctcn{\kappa}{_{BC}}+\ctcn{\omega}{_{BC}}}\prn{\ctcn{\kappa}{_{AD}}+\ctcn{\omega}{_{AD}}}\nonumber\\
				&+\prn{\ctcn{\kappa}{_{AC}}+\ctcn{\omega}{_{AC}}}\prn{\ctcn{\kappa}{_{BD}}+\ctcn{\omega}{_{BD}}}\quad,\label{eq:gaussrelCnii}
				\end{align}
			and a Codazzi-like equation
				\begin{equation}
					\ctcn{E}{^a_A}\ctcn{E}{^b_B}\ctcn{E}{^c_C}\cts{R}{_{abc}^d}\ct{m}{_{d}}= 2\cdcn{_{[A}}\prn{\ctcn{\kappa}{_{B]C}}+\ctcn{\omega}{_{B]C}}}+2\ctcn{\omega}{_{AB}}\ctcn{a}{_C}\label{eq:codazzi-congruences}\quad.
				\end{equation}
			Now we are going to give an expression for the intrinsic Schouten tensor on $ \I $. \Cref{eq:intrinsicriemannScriSchoutenRelation} is valid in general for dimension 3, i.e., valid for $ \I $,
				\begin{equation}
				\cts{R}{_{abcd}}= 2\ms{_{a[c}}\cts{S}{_{d]b}}-2\ms{_{b[c}}\cts{S}{_{d]a}}\quad.
				\end{equation}
			Using this expression in \cref{eq:codazzi-congruences} one arrives at
				\begin{equation}
				2\mcn{_{C[A}}\ctcn{S}{_{B]}}=2\cdcn{_{[A}}\prn{\ctcn{\kappa}{_{B]C}}+\ctcn{\omega}{_{B]C}}}+2\ctcn{\omega}{_{AB}}\ctcn{a}{_C}\quad,
				\end{equation}
			which is equivalent to its trace,
				\begin{equation}
				\ctcn{S}{_{B}}=\cdc{_C}\prn{\ctcn{\kappa}{_B^C}+\ctcn{\omega}{_B^C}}-\cdcn{_B}\cscn{\kappa}+2\ctcn{\omega}{_{CB}}\ctcn{a}{^C}\quad.\label{eq:SAkappa-congruence-app}
				\end{equation}
			Notice also that using the same relations and contracting with $ \mcn{^{AC}}\mcn{^{BD}} $ in \cref{eq:curvature-congruences} one gets
				\begin{equation}
					\ctcn{S}{^E_E}= \cscn{K}+\frac{1}{2}\cscn{\Sigma}^2-\frac{1}{4}\cscn{\kappa}^2-\frac{3}{2}\cscn{\omega}^2\label{eq:traceSchouten-Congruence-app}\quad.
				\end{equation}
			A direct calculation together with \cref{eq:codazzi-congruences,eq:derivative-r-decomposition} leads to
				\begin{equation}\label{eq:der-connection-congruences}
					\lied_{\vec{m}}\ctcn{\gamma}{^C_{AB}}=2\cdcn{_{(A}}\ctcn{\kappa}{_{B)}^C}-\mcn{^{EC}}\cdcn{_E}\ctcn{\kappa}{_{AB}}+\ctcn{a}{^C}\ctcn{\kappa}{_{AB}}-2\ctcn{a}{_{(A}}\ctcn{\kappa}{_{B)}^C}\quad.
				\end{equation}
			This last equation provides a condition for the vanishing of $ \ct{m}{^e}\dpart{e}\ctcn{\gamma}{^C_{AB}} $ which appears below in \cref{eq:cn-goodGamma}. In general $ \Scn $ is  endowed with a one-parameter family of geometrical objects; only in the cases in which these quantities have vanishing derivative along $ \ct{m}{^a} $ --i.e., when they do not depend on $ v $-- they are a true metric, connection or volume form, respectively. Summarising, from \cref{eq:der-metric-congruences,eq:der-volume-congruences,eq:der-connection-congruences},
				\begin{align}
					\lied_{\vec{m}}\mcn{_{AB}}=0&\iff\ctcn{\kappa}{_{AB}}=0\quad,\\
					\lied_{\vec{m}}\ctcn{\epsilon}{_{AB}}=0&\iff \cscn{\kappa}=0\quad,\\
					\lied_{\vec{m}}\ctcn{\gamma}{^C_{AB}}=0&\iff \cdcn{_C}\ctcn{\kappa}{_{AB}}=\ctcn{a}{_C}\ctcn{\kappa}{_{AB}}\quad.\label{eq:cn-goodGamma}
				\end{align}
		Notice that $ \cscn{\kappa}=0 $ is not a  conformally-invariant equation; one can always achieve this condition by a conformal transformation of $ \ms{_{ab}} $\footnote{A conformal transformation  $ \ms{_{ab}}\rightarrow \Psi^2 \ms{_{ab}} $ implies $ \ct{m}{_a}\rightarrow\Psi\ct{m}{_a} $ according to \cref{eq:congruence-vector}, as well as  $ \mcn{_{AB}}\rightarrow \cscn{\Psi}^2 \mcn{_{AB}} $, with $ \cscn{\Psi}\defeq \Pi^*(\Psi) $.}. Observe that, for $ \cscn{\kappa}=0 $, $ \ctcn{\kappa}{_{AB}}=0 $ if and only if $ \ct{m}{^a} $ is 
			\emph{shear-free}\footnote{Note that the shear-free property is a conformally-invariant property.}, i.e., $ \ctcn{\Sigma}{_{AB}}=0 $ --see \cref{eq:kappa-congruence}. Additionally,  in that case, the condition on 
			the right-hand side of \cref{eq:cn-goodGamma} is trivially satisfied. Hence, for umbilical $ \ct{m}{^a} $ there is a conformal class of metrics	
			 $\cbrkt{ \ms{_{ab}} }$ for which $ \lied_{\vec{m}}\mcn{_{AB}}=\lied_{\vec{m}}\ctcn{\epsilon}{_{AB}}=\lied_{\vec{m}}\ctcn{\gamma}{^C_{AB}}=0 $.\\
			 
			Finally, there is a particular case of interest:
				\begin{equation}
					\ctcn{\omega}{_{AB}}=0 \iff \ctcn{m}{^a}\text{ orthogonal to cuts}.
				\end{equation}
			This is the case of a foliation, in which each leaf is a surface (a cut). Under this condition, the normal form can always be written as
				\begin{equation}\label{eq:m-foliations}
					\ct{m}{_a}=F\cds{_a}v\quad
				\end{equation}
			for some scalar function $ F $ such that
				\begin{equation}
					\frac{1}{F}=\lied_{_{\vec{m}}}v.
				\end{equation}
			 The calculation of the acceleration produces
				\begin{equation}\label{eq:cn-foliation-acceleration}
					\ctcn{a}{_b}=-\ctcn{P}{^c_b}\cds{_c}\ln F\quad.
				\end{equation}
			Let us point out that the geometrical objects induced by $ \ct{m}{^a} $ still depend on $ v $ and coincide on each leaf ($ v= $constant) with the intrinsic geometric quantities of the cuts, but only there. In general they are fields on $ \I $ associated to the particular family of curves. \\
				
			To end up with this appendix, let us mention that a very similar construction for congruences as the one above can be developed using the so called Cattaneo operator in substitution of the derivative operator $ \cdcn{_A} $,
				\begin{equation}\label{eq:cattaneo-operator}
					\cdcn{_c}\ctcn{T}{^{a_1...a_r}_{b_1...b_q}}\defeq\ctcn{P}{_{m_{1}}^{a_1}}...\ctcn{P}{_{m_r}^{a_r}}\ctcn{P}{^{n_1}_{b_q}}...\ctcn{P}{^{n_q}_{b_q}}\ctcn{P}{^r_c}\cds{_r}\ct{T}{^{m_1...m_r}_{n_1...n_q}}\quad,
				\end{equation}
			which is defined for arbitrary tensor fields $ \ct{T}{^{a_1...a_r}_{b_1...b_q}} $ on $ \I $.
	\section{Properties of the lightlike projections of a Weyl candidate tensor}\label{app:lightlike-projections}
		The following is a list of properties of the quantities defined in \cref{ssec:lightlike-projections} :
		\begin{properties}
			\item $ \ctp{k}{^\mu}\ctrmp{C}{_{\mu\nu}}=-\ctm{k}{^\mu}\ctp{C}{_{\mu\nu}} $, $ \ctp{k}{^\mu}\ctrmp{D}{_{\mu\nu}}=-\ctm{k}{^\mu}\ctp{D}{_{\mu\nu}} $.\label{it:nullDecompProp1}
			\item  $ \cs{C} =-\ctc{C}{^E_E}= \ctru{*}{C}{^{EF}_{EF}}$,  $ \cs{D} = -\ctc{D}{^E_E}=-\ct{C}{^{EF}_{EF}}$.
			\item $ \ctcmp{D}{_{AB}} = -\frac{1}{2}\cs{D}\mc{_{AB}}-\frac{1}{2}\cs{C}\ctc{\epsilon}{_{AB}}$.\label{it:nullDecompProp17}
			\item $ \ctcmp{C}{_{AB}} = \frac{1}{2}\cs{C}\mc{_{AB}}-\frac{1}{2}\cs{D}\ctc{\epsilon}{_{AB}}$.\label{it:nullDecompProp17i}
			\item $ \ctcpm{D}{_{AB}} = -\frac{1}{2}\cs{D}\mc{_{AB}}+\frac{1}{2}\cs{C}\ctc{\epsilon}{_{AB}}$.\label{it:nullDecompProp17ii}
			\item $ \ctcpm{C}{_{AB}} = \frac{1}{2}\cs{C}\mc{_{AB}}+\frac{1}{2}\cs{D}\ctc{\epsilon}{_{AB}}$.\label{it:nullDecompProp17iii}
			\item $ \ctp{\mathring{C}}{^B_{B}}=0 $, $ \ctp{\mathring{D}}{^B_{B}}=0 $.\label{it:nullDecompProp3}
			\item $ \ctp{\mathring{C}}{^{AB}} \ctrmp{\mathring{C}}{_{AB}}=0=\ctp{\mathring{C}}{^{AB}} \ctrmp{\mathring{D}}{_{AB}} $, $ \ctp{\mathring{D}}{^{AB}} \ctrmp{\mathring{D}}{_{AB}}=0=\ctp{\mathring{D}}{^{AB}} \ctrmp{\mathring{C}}{_{AB}}$. \label{it:nullDecompProp4}
			\item $ \ctm{k}{^\mu}\ctp{C}{_{A\mu}}=-\sqrt{2}\ctcp{C}{_A} $, $ \ctp{k}{^\mu}\ctm{C}{_{A\mu}}=\sqrt{2}\ctcm{C}{_A} $.\label{it:nullDecompProp5}
			\item $ \ctm{k}{^\mu}\ctp{D}{_{A\mu}}=-\sqrt{2}\ctcp{D}{_A} $, $ \ctp{k}{^\mu}\ctm{D}{_{A\mu}}=\sqrt{2}\ctcm{D}{_A} $.
			\item $ \ctc{C}{_A}= \ctcp{C}{_A}+\ctcm{C}{_A} $ , $ \ctc{D}{_A}= \ctcp{D}{_A}+\ctcm{D}{_A}  $.
			\item $ \ctcp{D}{_A}\ctc{\epsilon}{^{AB}}\ctcm{D}{_B}=\frac{1}{2}\ctc{D}{_A}\ctc{C}{^A} $ .
			\item $ 4\ctcp{D}{_A}\ctcm{D}{^A}=\ctc{D}{_A}\ctc{D}{^A}-\ctc{C}{_A}\ctc{C}{^A} $ . 
			\item $ \ctc{D}{_A}\ctc{\epsilon}{^{AB}}\ctc{C}{_B}=\ctp{D}{_A}\ctp{D}{^A}- \ctm{D}{_A}\ctm{D}{^A} $.
			\item $\ctcp{C}{_{AB}}=\ctt{C}{_{AB}}-\ctc{D}{^T_{(B}}\ctc{\epsilon}{_{A)T}} $.\label{it:nullDecompProp11}
			\item $\ctcp{D}{_{AB}}=\ctt{D}{_{AB}}+\ctc{C}{^T_{(B}}\ctc{\epsilon}{_{A)T}} $.\label{it:nullDecompProp11F}
			\item $\ctcm{C}{_{AB}}=\ctt{C}{_{AB}}+\ctt{D}{^T_{(B}}\ctc{\epsilon}{_{A)T}} $.\label{it:nullDecompProp12}
			\item $\ctcm{D}{_{AB}}=\ctt{D}{_{AB}}-\ctt{C}{^T_{(B}}\ctc{\epsilon}{_{A)T}} $.\label{it:nullDecompProp12F}
			\item $ \ctt{C}{_{AB}} =\frac{1}{2}\left(\ctcp{C}{_{AB}}+\ctcm{C}{_{AB}}\right) $.\label{it:nullDecompProp13}
			\item $ \ctt{D}{_{AB}} =\frac{1}{2}\left(\ctcp{D}{_{AB}}+\ctcm{D}{_{AB}}\right) $.\label{it:nullDecompProp13F}
			\item  $ \ctcp{C}{_A}=\frac{1}{2}\left(\ctc{C}{_A}-\ctc{\epsilon}{_A^E}\ctc{D}{_E}\right) $.\label{it:nullDecompProp14}
			\item $ \ctcm{C}{_A}=\frac{1}{2}\left(\ctc{C}{_A}+\ctc{\epsilon}{_A^E}\ctc{D}{_E}\right) $.\label{it:nullDecompProp15}
			\item  $ \ctcp{D}{_A}=\frac{1}{2}\left(\ctc{D}{_A}+\ctc{\epsilon}{_A^E}\ctc{C}{_E}\right) $.\label{it:nullDecompProp14-D}
			\item $ \ctcm{D}{_A}=\frac{1}{2}\left(\ctc{D}{_A}-\ctc{\epsilon}{_A^E}\ctc{C}{_E}\right) $.\label{it:nullDecompProp15-D}
			\item $ \ctcp{D}{_A}+\ctcm{D}{_A}=\ctc{\epsilon}{_A^B}\prn{\ctcp{C}{_B}-\ctcm{C}{_B}}$.\label{it:nullDecompProp16}
			\item $ \ct{r}{^a}\ctrmp{D}{_{aA}}= \ctcp{D}{_A} $.
			\item $ \ct{r}{^a}\ctrpm{D}{_{aA}}= \ctcm{D}{_A} $.
			\item$ 2\ct{r}{^a}\ct{r}{^b}\ctrpm{D}{_{ab}}=2 \ct{r}{^a}\ct{r}{^b}\ctrmp{D}{_{ab}}= \ctpm{k}{^\mu}\ctpm{k}{^\nu}\ctmp{D}{_{\mu\nu}}= \ct{r}{^a}\ct{r}{^b}\ct{D}{_{ab}} $.\label{it:nullDecompProp18}
			\item $ \mp\ctmp{k}{^\mu}\ct{E}{^\alpha_A}\ctpm{D}{_{\mu\alpha}}= \sqrt{2}\ctcupm{D}{_{A}}. $\label{it:FApmDef}
			\item $ \mp \ctmp{k}{^\mu}\ct{E}{^\alpha_A}\ctpm{C}{_{\mu\alpha}}= \sqrt{2}\ctcupm{C}{_{A}}. $\label{it:CApmDef}
			\item $ \ctc{\epsilon}{^{AB}}\ctc{D}{_B}= \ct{r}{_\mu}\ct{r}{_\rho}\ct{u}{_\sigma}\ct{W}{_\gamma^A}\ctru{*}{C}{^{\sigma\rho\gamma\mu}}. $\label{it:alternativeRelationF}
			\item $ \ctc{\epsilon}{^{AB}}\ctc{C}{_B}=- \ct{r}{_\mu}\ct{r}{_\rho}\ct{u}{_\sigma}\ct{W}{_\gamma^A}\ct{C}{^{\sigma\rho\gamma\mu}} $.\label{it:alternativeRelationC}
			\item $ \ctc{\epsilon}{^A_B}\ctcp{C}{_{A}}=-\ctcp{D}{_{B}} $.
			\item $ \ctc{\epsilon}{^A_B}\ctcm{C}{_{A}}=\ctcm{D}{_{B}} $.
			\item  $ \ctcp{D}{^A}\ctcp{C}{_{A}}=0=\ctcm{D}{^A}\ctcm{C}{_{A}} $.\label{it:nullDecompProp-bases}
			\item $ \ctcp{C}{^A}\ctcm{D}{_{A}}=\ctcm{C}{^A}\ctcp{D}{_{A}} $.
		\end{properties}
	\section{Bianchi identities}\label{app:bianchi-id}
		Assume that a general congruence of curves with tangent vector field $ \ct{m}{^a} $ exists, and define $ \ct{m}{^\alpha}\defeq \ct{e}{^\alpha_a}\ct{m}{^a} $ at $ \scri $ and on a neighbourhood ---this allows us to take its derivative along $ \ct{n}{^\alpha} $, though no particular extension of $ \ct{m}{^\alpha} $ is required. The plan is to write de components of \cref{eq:cefesScriDerWeyl} in terms of the lightlike projections of the rescaled Weyl tensor, i.e., the quantities appearing in \cref{ssec:lightlike-projections}. Objects that carry an over-ring will be substituted by objects carrying an underbar, for the same reason explained in \cref{ssec:lightlike-condition-directional}. Also, quantities originally defined with indices $ A, B, C, $ etc will be written fully/partially with space-time indices $ \alpha, \beta, \gamma, $ etc indicating that they have been contracted with $ \ct{W}{_\alpha^A} $ and $ \ct{E}{^\alpha_A} $ conveniently. The same mixed notation can appear in space-time tensors that have been contracted in some of their indices. As a couple of examples, one may find $ \ctcn{C}{_A} $ in substitution of $ \ctc{C}{_A} $ and $ \ctcn{C}{_\alpha} $ in correspondence to $ \ctcn{C}{_A}\ct{W}{_\alpha^A} $, or $ \ct{y}{_{\alpha AB}} $ corresponding to $ \ctcn{W}{^\mu_A}\ctcn{W}{^\nu_B}\ct{y}{_{\alpha\mu\nu}} $.\\
		
		Then, recast \cref{eq:cefesScriDerWeyl} as
		\begin{align}
		-\ct{y}{_{\alpha\beta\gamma}}&= \ct{g}{^{\mu\tau}}\cd{_\mu}\ct{d}{_{\alpha\beta\gamma\tau}}= \prn{-\ctp{k}{^\mu}\ctm{k}{^\tau}-\ctm{k}{^\mu}\ctp{k}{^\tau}+\ctcn{P}{^{\mu\tau}}}\cd{_\mu}\ct{d}{_{\alpha\beta\gamma\tau}}\nonumber\\
		&= -\ctp{k}{^\mu}\cd{_\mu}\prn{\ct{d}{_{\alpha\beta\gamma\tau}}\ctm{k}{^\tau}}-\ct{d}{_{\alpha\beta\gamma\tau}}\ctm{k}{^\tau}\ctp{k}{_\lambda}\ctp{k}{^\mu}\cd{_\mu}\ctm{k}{^\lambda}+\ct{d}{_{\alpha\beta\gamma\tau}}\ctcn{P}{^\tau_\lambda}\ctp{k}{^\mu}\cd{_\mu}\ctm{k}{^\lambda}\nonumber\\
		&-\ctm{k}{^\mu}\cd{_\mu}\prn{\ct{d}{_{\alpha\beta\gamma\tau}}\ctp{k}{^\tau}}-\ct{d}{_{\alpha\beta\gamma\tau}}\ctp{k}{^\tau}\ctm{k}{_\lambda}\ctm{k}{^\mu}\cd{_\mu}\ctp{k}{^\lambda}+\ct{d}{_{\alpha\beta\gamma\tau}}\ctcn{P}{^\tau_\lambda}\ctm{k}{^\mu}\cd{_\mu}\ctp{k}{^\lambda}\nonumber\\
		&+\ctcn{P}{^{\mu\tau}}\cd{_\mu}\prn{\ct{d}{_{\alpha\beta\gamma\tau}}}\quad
		\end{align}
		here $ \ctpm{k}{^\alpha} $ are defined as in \cref{eq:n-kp-km}, noting that this time they are extended outside $ \scri $. Next step is to contract this equation with $ \ctpm{k}{^\alpha} $ and the basis spanning the space of vectors orthogonal to $ \ct{m}{_a} $, $\cbrkt{ \ctcn{E}{^\alpha_A}} $ ---uppercase, Latin indices denote projections with this basis. This process is a straight-forward calculation. It is very long, though, and we just write down here the final outcome evaluated at $ \scri $:
		\begin{align}
		-\ctm{k}{^\alpha}\ctp{k}{^\beta}\ctm{k}{^\gamma}\ct{y}{_{\alpha\beta\gamma}}&\eqs -\ctm{k}{^\mu}\cd{_\mu}\cs{D}-2\sqrt{2}\ctcnp{D}{_\gamma}\ctm{k}{^\mu}\cd{_\mu}\ctm{k}{^\gamma}\nonumber\\
		&+2\sqrt{2}\ctcnm{D}{_\beta}\ctm{k}{^\mu}\cd{_\mu}\ctp{k}{^\beta}+\sqrt{2}\ctcn{P}{^{\mu\tau}}\cd{_\mu}\ctcnm{D}{_\tau}-\ctcn{P}{^{\mu\tau}}\cs{D}\cd{_\mu}\ctm{k}{_\tau}\nonumber\\
		&-\ctcnm{D}{_{\beta\tau}}\ctcn{P}{^{\mu\tau}}\cd{_\mu}\ctp{k}{^\beta}+\sqrt{2}\ctcnm{D}{^\mu}\ctm{k}{^\beta}\cd{_\mu}\ctp{k}{_\beta}\nonumber\\
		&+\ctcnmp{D}{_{\tau\alpha}}\ctcn{P}{^{\tau\mu}}\cd{_\mu}\ctm{k}{^\alpha}-2\ctcnmp{D}{_{[\gamma\tau]}}\ctcn{P}{^{\mu\tau}}\cd{_\mu}\ctm{k}{^\gamma}\quad\label{eq:idbianchi1},
		\end{align}
		\begin{align}
		-\ctm{k}{^\alpha}\ctp{k}{^\beta}\ctp{k}{^\gamma}\ct{y}{_{\alpha\beta\gamma}}&\eqs -2\sqrt{2}\ctcnm{D}{_\gamma}\ctp{k}{^\mu}\cd{_\mu}\ctp{k}{^\gamma}+2\sqrt{2}\ctcnp{D}{_\alpha}\ctp{k}{^\mu}\cd{_\mu}\ctm{k}{^\alpha}+\sqrt{2}\ctcn{P}{^{\mu\tau}}\cd{_\mu}\ctcnp{D}{_\tau}\nonumber\\
		&\cs{D}\ctcn{P}{^{\mu\tau}}\cd{_\mu}\ctp{k}{_\tau}+\sqrt{2}\ctcnp{D}{^\mu}\ctp{k}{^\alpha}\cd{_\mu}\ctm{k}{_\alpha}+\ctcnp{D}{_{\alpha\tau}}\ctcn{P}{^{\mu\tau}}\cd{_\mu}\ctm{k}{^\alpha}\nonumber\\
		&-\ctcnmp{D}{_{\beta\tau}}\ctcn{P}{^{\mu\tau}}\cd{_\mu}\ctp{k}{^\beta}-\ctcnmp{D}{_{[\gamma\tau]}}\ctcn{P}{^{\mu\tau}}\cd{_\mu}\ctp{k}{^\gamma}+\ctp{k}{^\mu}\cd{_\mu}\cs{D}\quad\label{eq:idbianchi2},
		\end{align}
		\begin{align}
		-\ctm{k}{^\beta}\ctp{k}{^\gamma}\ctcn{y}{_{A\beta\gamma}}&\eqs-\sqrt{2}\ctcn{E}{^\omega_A}\ctp{k}{^\mu}\cd{_\mu}\ctcnm{D}{_\omega}+\cs{D}\ctp{k}{^\mu}\ctcn{E}{^\omega_A}\cd{_\mu}\ctm{k}{_\omega}+\ctcnm{D}{_{A\gamma}}\ctp{k}{^\mu}\cd{_\mu}\ctp{k}{^\gamma}\nonumber\\
		&-\sqrt{2}\ctcnm{D}{_A}\ctp{k}{^\lambda}\ctp{k}{^\mu}\cd{_\mu}\ctm{k}{_\lambda}+ 2\ctcnmp{D}{_{[\beta A]}}\ctp{k}{^\mu}\cd{_\mu}\ctm{k}{^\beta}-\ctcnmp{D}{_{A\lambda}}\ctp{k}{^\mu}\cd{_\mu}\ctm{k}{^\lambda}\nonumber\\
		&-\ctcn{P}{^{\mu\tau}}\ctcn{E}{^\omega_A}\cd{_\mu}\ctcnmp{D}{_{\omega\tau}}-\sqrt{2}\ctcnm{D}{_A}\ctcn{P}{^{\mu\tau}}\cd{_\mu}\ctp{k}{_\tau}+\sqrt{2}\ctcnp{D}{_\tau}\ctcn{P}{^{\mu\tau}}\ctcn{E}{^\omega_A}\cd{_\mu}\ctm{k}{_\omega}\nonumber\\
		&-\ctcnm{t}{_{\gamma\tau A}}\ctcn{P}{^{\mu\tau}}\cd{_\mu}\ctp{k}{^\gamma}+\ctcnp{t}{_{A\beta\tau}}\ctcn{P}{^{\mu\tau}}\cd{_\mu}\ctm{k}{^\beta}\quad,\label{eq:idbianchi3}
		\end{align}
		\begin{align}
		-\ctp{k}{^\beta}\ctm{k}{^\gamma}\ctcn{y}{_{A\beta\gamma}}&\eqs\sqrt{2}\ctcn{E}{^\omega_A}\ctm{k}{^\mu}\cd{_\mu}\ctcnp{D}{_\omega}+\cs{D}\ctm{k}{^\mu}\ctcn{E}{^\omega_A}\cd{_\mu}\ctp{k}{_\omega}+\ctcnp{D}{_{A\gamma}}\ctm{k}{^\mu}\cd{_\mu}\ctm{k}{^\gamma}\nonumber\\
		&+\sqrt{2}\ctcnp{D}{_A}\ctm{k}{^\lambda}\ctm{k}{^\mu}\cd{_\mu}\ctp{k}{_\lambda}+ 2\ctcnmp{D}{_{[A\beta]}}\ctm{k}{^\mu}\cd{_\mu}\ctp{k}{^\beta}-\ctcnmp{D}{_{\lambda A}}\ctm{k}{^\mu}\cd{_\mu}\ctp{k}{^\lambda}\nonumber\\
		&-\ctcn{P}{^{\mu\tau}}\ctcn{E}{^\omega_A}\cd{_\mu}\ctcnmp{D}{_{\tau\omega}}+\sqrt{2}\ctcnp{D}{_A}\ctcn{P}{^{\mu\tau}}\cd{_\mu}\ctm{k}{_\tau}-\sqrt{2}\ctcnm{D}{_\tau}\ctcn{P}{^{\mu\tau}}\ctcn{E}{^\omega_A}\cd{_\mu}\ctp{k}{_\omega}\nonumber\\
		&-\ctcnp{t}{_{\gamma\tau A}}\ctcn{P}{^{\mu\tau}}\cd{_\mu}\ctm{k}{^\gamma}+\ctcnm{t}{_{A\beta\tau}}\ctcn{P}{^{\mu\tau}}\cd{_\mu}\ctp{k}{^\beta}\quad,
		\end{align}
		\begin{align}
		-\ctp{k}{^\beta}\ctp{k}{^\gamma}\ctcn{y}{_{A\beta\gamma}}&\eqs-\sqrt{2}\ctp{k}{^\mu}\ctcn{E}{^\omega_A}\cd{_\mu}\ctcnp{D}{_\omega}+\sqrt{2}\ctp{D}{_A}\ctp{k}{^\mu}\ctp{k}{^\lambda}\cd{_\mu}\ctm{k}{_\lambda}+\ctcnmp{D}{_{\gamma A}}\ctp{k}{^\mu}\cd{_\mu}\ctp{k}{^\gamma}\nonumber\\
		&+2\ctcnmp{D}{_{[\beta A]}}\ctp{k}{^\mu}\cd{_\mu}\ctp{k}{^\beta}-\cs{D}\ctp{k}{^\mu}\ctcn{E}{^\omega_A}\cd{_\mu}\ctp{k}{_\omega}-\ctcnp{D}{_{A\lambda}}\ctp{k}{^\mu}\cd{_\mu}\ctm{k}{^\lambda}\nonumber\\
		&-\ctcn{P}{^{\mu\tau}}\ctcn{E}{^\omega_A}\cd{_\mu}\ctcnp{D}{_{\omega\tau}}-\sqrt{2}\ctcnp{D}{_A}\ctcn{P}{^{\mu\tau}}\cd{_\mu}\ctp{k}{_\tau}-\sqrt{2}\ctcnp{D}{_\tau}\ctcn{P}{^{\tau\mu}}\ctcn{E}{^\omega_A}\cd{_\mu}\ctp{k}{_\omega}\nonumber\\
		&-\ctcnp{t}{_{\lambda\tau A}}\ctcn{P}{^{\mu\tau}}\cd{_\mu}\ctp{k}{^\lambda}-\ctcnp{t}{_{\lambda A\tau}}\ctcn{P}{^{\tau\mu}}\cd{_\mu}\ctp{k}{^\lambda}\quad,
		\end{align}
		\begin{align}
		-\ctm{k}{^\beta}\ctm{k}{^\gamma}\ctcn{y}{_{A\beta\gamma}}&\eqs\sqrt{2}\ctm{k}{^\mu}\ctcn{E}{^\omega_A}\cd{_\mu}\ctcnm{D}{_\omega}-\sqrt{2}\ctm{D}{_A}\ctm{k}{^\mu}\ctm{k}{^\lambda}\cd{_\mu}\ctp{k}{_\lambda}+\ctcnmp{D}{_{A\gamma}}\ctm{k}{^\mu}\cd{_\mu}\ctm{k}{^\gamma}\nonumber\\
		&+2\ctcnmp{D}{_{[A\beta]}}\ctm{k}{^\mu}\cd{_\mu}\ctm{k}{^\beta}-\cs{D}\ctm{k}{^\mu}\ctcn{E}{^\omega_A}\cd{_\mu}\ctm{k}{_\omega}-\ctcnm{D}{_{A\lambda}}\ctm{k}{^\mu}\cd{_\mu}\ctp{k}{^\lambda}\nonumber\\
		&-\ctcn{P}{^{\mu\tau}}\ctcn{E}{^\omega_A}\cd{_\mu}\ctcnm{D}{_{\omega\tau}}+\sqrt{2}\ctcnm{D}{_A}\ctcn{P}{^{\mu\tau}}\cd{_\mu}\ctm{k}{_\tau}+\sqrt{2}\ctcnm{D}{_\tau}\ctcn{P}{^{\tau\mu}}\ctcn{E}{^\omega_A}\cd{_\mu}\ctm{k}{_\omega}\nonumber\\
		&-\ctcnm{t}{_{\lambda\tau A}}\ctcn{P}{^{\mu\tau}}\cd{_\mu}\ctm{k}{^\lambda}-\ctcnm{t}{_{\lambda A\tau}}\ctcn{P}{^{\tau\mu}}\cd{_\mu}\ctm{k}{^\lambda}\quad,\label{eq:idbianchi6}
		\end{align}
		\begin{align}
			-\ctm{k}{^\beta}\ctcn{y}{_{A\beta C}}&\eqs -\ctcn{E}{^\omega_A}\ctcn{E}{^\sigma_C}\ctp{k}{^\mu}\cd{_\mu}\ctm{D}{_{\omega\sigma}}+\sqrt{2}\ctcnm{D}{_A}\ctcn{E}{^\sigma_C}\ctp{k}{^\mu}\cd{_\mu}\ctm{k}{_\sigma}+\sqrt{2}\ctcnm{D}{_C}\ctcn{E}{^\omega_A}\ctp{k}{^\mu}\cd{_\mu}\ctm{k}{_\omega}\nonumber\\
			&-2\ctcnm{D}{_{AC}}\ctp{k}{^\lambda}\ctp{k}{^\mu}\cd{_\mu}\ctm{k}{_\lambda}+\ctcnm{t}{_{A\lambda C}}\ctp{k}{^\mu}\cd{_\mu}\ctm{k}{^\lambda}+\ctcnm{t}{_{C\lambda A}}\ctp{k}{^\mu}\cd{_\mu}\ctm{k}{^\lambda}\nonumber\\
			&-\ctcn{E}{^\omega_A}\ctcn{E}{^\sigma_C}\ctm{k}{^\mu}\cd{_\mu}\ctcnmp{D}{_{\omega\sigma}}-\sqrt{2}\ctcnm{D}{_A}\ctcn{E}{^\sigma_C}\ctm{k}{^\mu}\cd{_\mu}\ctp{k}{_\sigma}+\sqrt{2}\ctcnp{D}{_C}\ctcn{E}{^\omega_A}\ctm{k}{^\mu}\cd{_\mu}\ctm{k}{_\omega}\nonumber\\
			&+\ctcnp{t}{_{A\lambda C}}\ctm{k}{^\mu}\cd{_\mu}\ctm{k}{^\lambda}+\ctcnm{t}{_{C\lambda A}}\ctm{k}{^\mu}\cd{_\mu}\ctp{k}{^\lambda}+\ctcn{E}{^\omega_A}\ctcn{E}{^\sigma_C}\ctcn{P}{^{\mu\tau}}\cd{_\mu}\ctcnm{t}{_{\sigma\tau\omega}}\nonumber\\
			&-\ctcnm{D}{_{AC}}\ctcn{P}{^{\tau\mu}}\cd{_\mu}\ctp{k}{_\tau}-\ctcnmp{D}{_{AC}}\ctcn{P}{^{\mu\tau}}\cd{_\mu}\ctm{k}{_\tau}+\ctcnm{D}{_{A\tau}}\ctcn{P}{^{\tau\mu}}\ctcn{E}{^\sigma_C}\cd{_\mu}\ctp{k}{_\sigma}\nonumber\\
			&+\ctcnmp{D}{_{A\tau}}\ctcn{P}{^{\mu\tau}}\ctcn{E}{^\sigma_C}\cd{_\mu}\ctm{k}{_\sigma}-2\ctcnmp{D}{_{[\tau C]}}\ctcn{P}{^{\mu\tau}}\ctcn{E}{^\omega_A}\cd{_\mu}\ctm{k}{_\omega}\nonumber\\
			&+\ctcnm{t}{_{C\tau A}}\ctp{k}{^\lambda}\ctcn{P}{^{\mu\tau}}\cd{_\mu}\ctm{k}{_\lambda}- \cs{D}\prn{ \ctcn{E}{^\lambda_C}\ctcn{E}{^\mu_A}-\mcn{_{AC}}\ctcn{P}{^{\lambda\mu}}}\cd{_\mu}\ctm{k}{_\lambda}\quad,\label{eq:idbianchi7}
		\end{align}
		\begin{align}
			-\ctp{k}{^\beta}\ctcn{y}{_{A\beta C}}&\eqs -\ctcn{E}{^\omega_A}\ctcn{E}{^\sigma_C}\ctm{k}{^\mu}\cd{_\mu}\ctcnp{D}{_{\omega\sigma}}-\sqrt{2}\ctcnp{D}{_A}\ctcn{E}{^\sigma_C}\ctm{k}{^\mu}\cd{_\mu}\ctp{k}{_\sigma}-\sqrt{2}\ctcnp{D}{_C}\ctcn{E}{^\omega_A}\ctm{k}{^\mu}\cd{_\mu}\ctp{k}{_\omega}\nonumber\\
			&-2\ctcnp{D}{_{AC}}\ctm{k}{^\lambda}\ctm{k}{^\mu}\cd{_\mu}\ctp{k}{_\lambda}+\ctcnp{t}{_{A\lambda C}}\ctm{k}{^\mu}\cd{_\mu}\ctp{k}{^\lambda}+\ctcnp{t}{_{C\lambda A}}\ctm{k}{^\mu}\cd{_\mu}\ctp{k}{^\lambda}\nonumber\\
			&-\ctcn{E}{^\omega_A}\ctcn{E}{^\sigma_C}\ctp{k}{^\mu}\cd{_\mu}\ctcnmp{D}{_{\sigma\omega}}+\sqrt{2}\ctcnp{D}{_A}\ctcn{E}{^\sigma_C}\ctp{k}{^\mu}\cd{_\mu}\ctm{k}{_\sigma}-\sqrt{2}\ctcnm{D}{_C}\ctcn{E}{^\omega_A}\ctp{k}{^\mu}\cd{_\mu}\ctp{k}{_\omega}\nonumber\\
			&+\ctcnm{t}{_{A\lambda C}}\ctp{k}{^\mu}\cd{_\mu}\ctp{k}{^\lambda}+\ctcnp{t}{_{C\lambda A}}\ctp{k}{^\mu}\cd{_\mu}\ctm{k}{^\lambda}+\ctcn{E}{^\omega_A}\ctcn{E}{^\sigma_C}\ctcn{P}{^{\mu\tau}}\cd{_\mu}\ctcnp{t}{_{\sigma\tau\omega}}\nonumber\\
			&-\ctcnp{D}{_{AC}}\ctcn{P}{^{\tau\mu}}\cd{_\mu}\ctm{k}{_\tau}-\ctcnmp{D}{_{CA}}\ctcn{P}{^{\mu\tau}}\cd{_\mu}\ctp{k}{_\tau}+\ctcnp{D}{_{A\tau}}\ctcn{P}{^{\tau\mu}}\ctcn{E}{^\sigma_C}\cd{_\mu}\ctm{k}{_\sigma}\nonumber\\
			&+\ctcnmp{D}{_{\tau A}}\ctcn{P}{^{\mu\tau}}\ctcn{E}{^\sigma_C}\cd{_\mu}\ctp{k}{_\sigma}-2\ctcnmp{D}{_{[C\tau]}}\ctcn{P}{^{\mu\tau}}\ctcn{E}{^\omega_A}\cd{_\mu}\ctp{k}{_\omega}\nonumber\\
			&+\ctcnp{t}{_{C\tau A}}\ctm{k}{^\lambda}\ctcn{P}{^{\mu\tau}}\cd{_\mu}\ctp{k}{_\lambda}-\cs{D}\prn{ \ctcn{E}{^\lambda_C}\ctcn{E}{^\mu_A}-\mcn{_{AC}}\ctcn{P}{^{\lambda\mu}}}\cd{_\mu}\ctp{k}{_\lambda}\quad.\label{eq:idbianchi8}
		\end{align}
		The number of independent components of the (rescaled) Cotton tensor $ \ct{y}{_{\alpha\beta\gamma}} $ in four dimensions is $ 16 $ (see \cite{Garcia2004}). Here we have written a total of $ 18 $ of which two can be expanded in terms of other ones: it is possible to write the $ (2,-,2) $  component of $ \ct{y}{_{\alpha\beta\gamma}} $ in terms of the $ (3,-,3) $ and $ (+,-,-) $, using $ \ct{y}{^\mu_{\alpha\mu}}=0 $ and $ \ct{y}{_{\alpha\beta\gamma}}=-\ct{y}{_{\beta\alpha\gamma}} $; the same for $ (2,+,2) $ in terms of $ (3,+,3) $ and $ (-,+,+) $.\\
	
	\section{NP formulation}\label{app:NPformulation}	
		Writing the lightlike projections of a Weyl candidate tensor (\cref{ssec:lightlike-projections}) and the superenergy quantities (\cref{ssec:basic-superenergy,ssec:radiant-superenergy}) in terms of the Weyl candidate scalars allows to have a first-glance interpretation of their significance. Not only that but having their components at hand can be helpful to check calculations. Consider a lightlike vierbein $ \prn{\ctm{k}{^\alpha},\ctp{k}{^\alpha},\ct{m}{^\alpha},\ct{\bar{m}}{^\alpha}} $ such that $ \ctp{k}{^\alpha}\ctm{k}{_\alpha}=-1 $, $ \ct{m}{^\alpha}\ct{\bar{m}}{_\alpha}=1 $, $ \ctpm{k}{^\alpha}\ct{m}{_\alpha}=0=\ct{m}{^\alpha}\ct{m}{_\alpha} $, with orientation fixed to $ \ct{\eta}{_{\hat{0}\hat{1}\hat{2}\hat{3}}}=i $. We use the following definitions for the Weyl candidate scalars:
		\begin{align}
		&\cs{\phi}{_0}\defeq \ct{C}{_{\hat{0}\hat{2}\hat{0}\hat{2}}} & & \cs{\phi}{_1}\defeq \ct{C}{_{\hat{0}\hat{1}\hat{0}\hat{2}}} & & \cs{\phi}{_2}=\frac{1}{2}\left(\ct{C}{_{\hat{0}\hat{1}\hat{0}\hat{1}}}-\ct{C}{_{\hat{0}\hat{1}\hat{2}\hat{3}}} \right)&\nonumber\\		  
		&\cs{\phi}{_3}\defeq -\ct{C}{_{\hat{0}\hat{1}\hat{1}\hat{3}}} && \cs{\phi}{_4}\defeq \ct{C}{_{\hat{1}\hat{3}\hat{1}\hat{3}}}&
		\end{align}
		Be aware that all the formulae below hold only with these definitions and choice of orientation. In this subsection, hatted Greek and Latin characters $ \hat{\alpha} $, $ \hat{A} $ represent basis indices.\\
		
		The lightlike `magnetic' and `electric' parts associated to $ \ctp{k}{^\alpha} $ and $ \ctm{k}{^\alpha} $ respectively.
		\begin{equation}
		\ctp{C}{^{\hat{\alpha}\hat{\beta}}}= i
		\begin{pmatrix}
		0 & 0 & 0 & 0 \\ 
		0 & -i2\Im(\ct{\phi}{_2}) & -\ct{\phi}{_3}  & \ct{\bar{\phi}}{_3}  \\ 
		0 &  -\ct{\phi}{_3}  & \ct{\phi}{_4}   &  0 \\ 
		0 & \ct{\bar{\phi}}{_3} &  0 &-\ct{\bar{\phi}}{_4} 
		\end{pmatrix} \quad ,
		\ctp{D}{^{\hat{\alpha}\hat{\beta}}}= 
		\begin{pmatrix}
		0 & 0 & 0 & 0 \\ 
		0 & 2\Re(\ct{\phi}{_2}) & - \ct{\phi}{_3} & - \ct{\bar{\phi}}{_3} \\ 
		0 &  - \ct{\phi}{_3} & \ct{\phi}{_4}   &  0 \\ 
		0 & - \ct{\bar{\phi}}{_3}  &  0 &\ct{\bar{\phi}}{_4} 
		\end{pmatrix} \quad .
		\end{equation}
		\begin{equation}
		\ctm{C}{^{\hat{\alpha}\hat{\beta}}}= i
		\begin{pmatrix}
		-i2\Im(\ct{\phi}{_2}) & 0 & \ct{\bar{\phi}}{_1} & -\ct{\phi}{_1} \\ 
		0 & 0 & 0  & 0\\ 
		\ct{\bar{\phi}}{_1}  & 0 & -\ct{\bar{\phi}}{_0} &  0 \\ 
		-\ct{\phi}{_1} & 0 &  0 & \ct{\phi}{_0}
		\end{pmatrix} \quad ,
		\ctm{D}{^{\hat{\alpha}\hat{\beta}}}= 
		\begin{pmatrix}
		2\Re(\ct{\phi}{_2}) & 0 & -\ct{\bar{\phi}}{_1} & -\ct{\phi}{_1} \\ 
		0 & 0 & 0  & 0\\ 
		-\ct{\bar{\phi}}{_1}  & 0 & \ct{\bar{\phi}}{_0} &  0 \\ 
		-\ct{\phi}{_1} & 0 &  0 & \ct{\phi}{_0}
		\end{pmatrix} \quad .
		\end{equation}
		The two dimensional components,
		\begin{equation}
		\ctcp{C}{^{\hat{A}\hat{B}}}= i
		\begin{pmatrix}s
		\ct{\phi}{_4} & 0 \\ 
		0  & -\ct{\bar{\phi}}{_4}
		\end{pmatrix} \quad ,
		\ctcp{D}{^{\hat{A}\hat{B}}}= 
		\begin{pmatrix}
		\ct{\phi}{_4} & 0 \\ 
		0  & \ct{\bar{\phi}}{_4}
		\end{pmatrix} \quad . 
		\end{equation}
		\begin{equation}
		\ctcm{C}{^{\hat{A}\hat{B}}}= i
		\begin{pmatrix}
		-\ct{\bar{\phi}}{_0} & 0 \\ 
		0  & \ct{\phi}{_0}
		\end{pmatrix} \quad ,
		\ctcm{D}{^{\hat{A}\hat{B}}}= 
		\begin{pmatrix}
		\ct{\bar{\phi}}{_0} & 0 \\ 
		0  & \ct{\phi}{_0}
		\end{pmatrix} \quad . 
		\end{equation}
		The traceless, two dimensional magnetic and electric parts,
		\begin{equation}
		\ctt{C}{^{\hat{A}\hat{B}}}= \frac{i}{2}
		\begin{pmatrix}
		\ct{\phi}{_4}-\ct{\bar{\phi}}{_0} & 0 \\ 
		0  & -(\ct{\bar{\phi}}{_4}-\ct{\phi}{_0})
		\end{pmatrix} \quad ,
		\ctt{D}{^{\hat{A}\hat{B}}}= \frac{1}{2}
		\begin{pmatrix}
		\ct{\ct{\phi}{_4}+\bar{\phi}}{_0} & 0 \\ 
		0  & \ct{\bar{\phi}}{_4}+\ct{\phi}{_0}
		\end{pmatrix} \quad. 
		\end{equation}
		The two dimensional vectors,
		\begin{equation}
		\ctcp{C}{^{\hat{A}}}= \frac{i}{\sqrt{2}}
		\begin{pmatrix}
		-\ct{\phi}{_3} \\ 
		\ct{\bar{\phi}}{_3}
		\end{pmatrix} \quad ,
		\ctcp{D}{^{\hat{A}}}= -\frac{1}{\sqrt{2}}
		\begin{pmatrix}
		\ct{\phi}{_3}\\ 
		\ct{\bar{\phi}}{_3}
		\end{pmatrix} \quad . 
		\end{equation}	
		\begin{equation}
		\ctcm{C}{^{\hat{A}}}= \frac{i}{\sqrt{2}}
		\begin{pmatrix}
		-\ct{\bar{\phi}}{_1} \\ 
		\ct{\phi}{_1}
		\end{pmatrix} \quad ,
		\ctcm{D}{^{\hat{A}}}= \frac{1}{\sqrt{2}}
		\begin{pmatrix}
		\ct{\bar{\phi}}{_1} \\ 
		\ct{\phi}{_1}
		\end{pmatrix} \quad . 
		\end{equation}	
		\begin{equation}
		\ctc{C}{^{\hat{A}}}= \frac{i}{\sqrt{2}}
		\begin{pmatrix}
		-(\ct{\phi}{_3}+\ct{\bar{\phi}}{_1}) \\ 
		\ct{\bar{\phi}}{_3}+\ct{\phi}{_1}
		\end{pmatrix} \quad ,
		\ctc{D}{^{\hat{A}}}= \frac{1}{\sqrt{2}}
		\begin{pmatrix}
		-\ct{\phi}{_3}+\ct{\bar{\phi}}{_1} \\ 
		-\ct{\bar{\phi}}{_3}+\ct{\phi}{_1}
		\end{pmatrix} \quad . 
		\end{equation}	
		And the traces,
		\begin{equation}\label{eq:np-coulomb}
		\cs{C}=-\ctc{C}{^{\hat{E}}_{\hat{E}}}=2\Im(\ct{\phi}{_2})\quad,\quad\cs{D}=-\ctc{D}{^{\hat{E}}_{\hat{E}}}=2\Re(\ct{\phi}{_2})\quad.
		\end{equation}
		For a general unit timelike vector field $ \ct{v}{^\alpha} $,
		\begin{equation}
		\ct{v}{^\alpha}\defeq \left( a \ctm{k}{^\alpha} + b \ctp{k}{^\alpha} + c \ct{m}{^\alpha} + \bar{c} \ct{\overline{m}}{^\alpha} \right), \quad \ct{v}{^\alpha}\ct{v}{_\alpha}=-1, \quad \forall a, b,c, \bar{c} \quad / \quad ab -c\bar{c}=\frac{1}{2}\quad,
		\end{equation} 

		it is a matter of direct calculation to get the expression for the super-Poynting vector
\begin{align}\label{eq:superPoyntingNP}
		\cts{\P}{^a}&=-4\ct{\omega}{_\alpha^a}\Biggl\{ 
		\ct{\phi}{_0}\ct{\overline{\phi}}{_0}\Bigl[-a^3\ctp{k}{^\alpha}\Bigr] +\ct{\phi}{_0}\ct{\overline{\phi}}{_1}\Bigl[-3a^2c\ctp{k}{^\alpha}+a^3\ct{\overline{m}}{^{\alpha}}\Bigr] + \text{c.c.} + \ct{\phi}{_0}\ct{\overline{\phi}}{_2}\Bigl[ -3ac^2\ctp{k}{^\alpha}+3a^2c\ct{\overline{m}}{^\alpha}\Bigr]  \nonumber\\ &+ \text{c.c.} + \ct{\phi}{_0}\ct{\overline{\phi}}{_3}\Bigl[-c^3\ctp{k}{^{\alpha}} +3ac^2\ct{\overline{m}}{^\alpha}\Bigr] + \text{c.c.} + \ct{\phi}{_0}\ct{\overline{\phi}}{_4}\Bigl[c^3\ct{\overline{m}}{^\alpha}\Bigr] + \text{c.c.} +\ct{\phi}{_1}\ct{\overline{\phi}}{_1}\Bigl[-a^3\ctm{k}{^{\alpha}} \\\nonumber &+(-3a^2b-6ac\overline{c})\ctp{k}{^\alpha}+3a^2c\ct{m}{^\alpha}+3a^2\bar{c} \ct{\overline{m}}{^\alpha}\Bigr] +
		\ct{\phi}{_1}\ct{\overline{\phi}}{_2}\Bigl[ -3a^2c\ctm{k}{^\alpha}+(-6abc-3c^2\bar{c})\ctp{k}{^\alpha}\\\nonumber &+3ac^2\ct{m}{^\alpha}+(6ac\bar{c}+3a^2b)\ct{\overline{m}}{^\alpha}\Bigr] + \text{c.c.} + \ct{\phi}{_1}\ct{\overline{\phi}}{_3}\Bigl[-3ac^2\ctm{k}{^\alpha} -3bc^2\ctp{k}{^\alpha}+c^3\ct{m}{^{(\alpha}}+(6abc \\\nonumber &+ 3c^2\bar{c})\ct{\overline{m}}{^\alpha}\Bigr] + \text{c.c.} +	\ct{\phi}{_1}\ct{\overline{\phi}}{_4}\Bigl[-c^3\ctm{k}{^\alpha} +3bc^2\ct{\overline{m}}{^\alpha}\Bigr] + \text{c.c.} +
		\ct{\phi}{_2}\ct{\overline{\phi}}{_2}\Bigl[(-3a^2b-6ac\bar{c})\ctm{k}{^\alpha}\\\nonumber &+(-3ab^2-6bc\bar{c})\ctp{k}{^\alpha}
		+(6abc+3c^2\bar{c})\ct{m}{^\alpha}+(6ab\bar{c}+3c\bar{c}^2)\ct{\overline{m}}{^\alpha}\Bigr] +				\ct{\phi}{_2}\ct{\overline{\phi}}{_3}\Bigl[(-6abc-3c^2\bar{c})\ctm{k}{^\alpha}\\\nonumber
		&-3b^2c\ctp{k}{^\alpha}+3bc^2\ct{m}{^\alpha}+(3ab^2+6bc\bar{c})\ct{\overline{m}}{^\alpha}\Bigr] + \text{c.c.} + 
		\ct{\phi}{_2}\ct{\overline{\phi}}{_4}\Bigl[-3bc^2\ctm{k}{^\alpha}+3b^2c\ct{\overline{m}}{^\alpha} \Bigr] + \text{c.c.} \\\nonumber 
		&+	\ct{\phi}{_3}\ct{\overline{\phi}}{_3}\Bigl[(-3ab^2-6bc\bar{c})\ctm{k}{^\alpha}-b^3\ctp{k}{^\alpha}+3b^2c \ct{m}{^\alpha} + 3b^2\bar{c}\ct{\overline{m}}{^\alpha}\Bigr] +
		\ct{\phi}{_3}\ct{\overline{\phi}}{_4}\Bigl[ -3b^2c\ctm{k}{^\alpha}+b^3\ct{\overline{m}}{^\alpha}\Bigr] + \text{c.c.} \\\nonumber &+  
		\ct{\phi}{_4}\ct{\overline{\phi}}{_4}\Bigl[-b^3\ctm{k}{^\alpha}\Bigr]
		\Biggr\}\spacef,
		\end{align}
		and the superenergy density					
		\begin{align}
		\cs{\W} &=4\Biggl\{ 
		\ct{\phi}{_0}\ct{\overline{\phi}}{_0}\Bigl[a^4\Bigr] +\ct{\phi}{_0}\ct{\overline{\phi}}{_1}\Bigl[4a^3c\Bigr] + \text{c.c.} + \ct{\phi}{_0}\ct{\overline{\phi}}{_2}\Bigl[6a^2c^2\Bigr] + \text{c.c.} + \ct{\phi}{_0}\ct{\overline{\phi}}{_3}\Bigl[4ac^3\Bigr] + \text{c.c.} \\\nonumber &+ \ct{\phi}{_0}\ct{\overline{\phi}}{_4}\Bigl[ c^4\Bigr] + \text{c.c.} +\ct{\phi}{_1}\ct{\overline{\phi}}{_1}\Bigl[4a^3b+12a^2c\bar{c}\Bigr] +
		\ct{\phi}{_1}\ct{\overline{\phi}}{_2}\Bigl[ 12a^2bc+12ac^2\bar{c} \Bigr] + \text{c.c.} \\\nonumber &+ \ct{\phi}{_1}\ct{\overline{\phi}}{_3}\Bigl[12abc^2+4c^3\bar{c}\Bigr] + \text{c.c.} +	\ct{\phi}{_1}\ct{\overline{\phi}}{_4}\Bigl[4bc^3\Bigr] + \text{c.c.} +
		\ct{\phi}{_2}\ct{\overline{\phi}}{_2}\Bigl[6a^2b^2+24abc\bar{c}+ 6c^2\bar{c}^2\Bigr] 	\\\nonumber &+			\ct{\phi}{_2}\ct{\overline{\phi}}{_3}\Bigl[12ab^2c + 12bc^2\bar{c}\Bigr] + \text{c.c.} + 	\ct{\phi}{_2}\ct{\overline{\phi}}{_4}\Bigl[6b^2c^2 \Bigr] + \text{c.c.} + \ct{\phi}{_3}\ct{\overline{\phi}}{_3}\Bigl[4ab^3+12b^2c\bar{c}\Bigr] \\\nonumber &+ 
		\ct{\phi}{_3}\ct{\overline{\phi}}{_4}\Bigl[4b^3c\Bigr] + \text{c.c.} +	\ct{\phi}{_4}\ct{\overline{\phi}}{_4}\Bigl[b^4\Bigr]
		\Biggr\}.
		\end{align}
		The radiant and Coulomb superenergy densities read
		\begin{align}
		\csp{\Z}&= 4\ct{\phi}{_3}\ct{\bar{\phi}}{_3}\quad\label{eq:ZpNP},\\
		\csm{\Z}&= 4\ct{\phi}{_1}\ct{\bar{\phi}}{_1}\quad\label{eq:ZmNP},\\
		\csp{\W}&= 4\ct{\phi}{_4}\ct{\bar{\phi}}{_4}\quad\label{eq:WpNP},\\
		\csm{\W}&= 4\ct{\phi}{_0}\ct{\bar{\phi}}{_0}\quad\label{eq:WmNP},\\
		\ctp{\Q}{^A}&= - 4\prn{\ct{\phi}{_4}\ct{\bar{\phi}}{_3}\ct{m}{^A}+\ct{\phi}{_3}\ct{\bar{\phi}}{_4}\ct{\bar{m}}{^A}}\quad\label{eq:QApNP},\\
		\ctm{\Q}{^A}&= - 4\prn{\ct{\phi}{_1}\ct{\bar{\phi}}{_0}\ct{m}{^A}+\ct{\phi}{_0}\ct{\bar{\phi}}{_1}\ct{\bar{m}}{^A}}\quad\label{eq:QAmNP}\quad,\\
		\cs{\V} &= 4\ct{\phi}{_2}\ct{\bar{\phi}}{_2}\quad\label{eq:VNP}\quad.
		\end{align}
		From here and \cref{eq:Qpdef,eq:Qmdef} it is easy to write the radiant supermomenta,
		\begin{align}
		\ctp{\Q}{^\alpha}&= 4\prn{\ct{\phi}{_4}\ct{\bar{\phi}}{_4}\ctm{k}{^\alpha}+\ct{\phi}{_3}\ct{\bar{\phi}}{_3}\ctp{k}{^\alpha}-\ct{\phi}{_4}\ct{\bar{\phi}}{_3}\ct{m}{^\alpha}-\ct{\bar{\phi}}{_4}\ct{\phi}{_3}\ct{\bar{m}}{^\alpha}}\quad,\label{eq:QpNP}\\
		\ctm{\Q}{^\alpha}&= 4\prn{\ct{\phi}{_0}\ct{\bar{\phi}}{_0}\ctp{k}{^\alpha}+\ct{\phi}{_1}\ct{\bar{\phi}}{_1}\ctm{k}{^\alpha}-\ct{\bar{\phi}}{_0}\ct{\phi}{_1}\ct{m}{^\alpha}-\ct{\phi}{_0}\ct{\bar{\phi}}{_1}\ct{\bar{m}}{^\alpha}}\quad.\label{eq:QmNP}
		\end{align} 
		Finally, the vector defined in \cref{eq:dA} has the expression
		\begin{equation}
		\ct{d}{^A}=\sqrt{2}\prn{\ct{\phi}{_1}\ct{\bar{\phi}}{_2}+\ct{\bar{\phi}}{_3}\ct{\phi}{_2}}\ct{\bar{m}}{^A}+\sqrt{2}\prn{\ct{\bar{\phi}}{_1}\ct{\phi}{_2}+\ct{\phi}{_3}\ct{\bar{\phi}}{_2}}\ct{m}{^A}\quad.
		\end{equation} 

	\section{Gauge transformations}\label{app:gauge-transformations}
			We present a collection of formulae giving the gauge behaviour of fields on $ \scri $ and also of those associated with single cuts $ \Sc $ and with the projected space $ \Scn $ associated to a general congruence. Recall that the gauge changes are residual transformations of the conformal factor,
				\begin{equation}
				\cs{\Omega}\rightarrow\csg{\Omega}=\cs{\omega}\cs{\Omega}\quad,
				\end{equation}
			with $ \cs{\omega} $ a positive definite function such that $ \ct{N}{^\mu}\cd{_\mu}\omega\eqs 0 $ ---according to our partial gauge fixing.
			\subsection{Metric, connection, volume form and curvature}
			Quantities of $ \prn{\ct{g}{_{\alpha\beta}},\cs{M}} $:
				\begin{align}
				\ctg{g}{_{\alpha\beta}}&=\omega^2\ct{g}{_{\alpha\beta}}\quad ,\\
				\ctg{\eta}{_{\alpha\beta\gamma\delta}}&=\omega^4\ct{\eta}{_{\alpha\beta\gamma\delta}}\quad,\\
				\ctg{\Gamma}{^\alpha_{\beta\gamma}} &= \ct{\Gamma}{^\alpha_{\beta\gamma}} +\ct{C}{^\alpha_{\beta\gamma}}\quad ,\ct{C}{^\alpha_{\beta\gamma}}=\frac{1}{\omega}\ct{g}{^{\alpha\tau}}\prn{2\ct{g}{_{
							\tau(\beta}}\ct{\omega}{_{\alpha)}}-\ct{g}{_{\alpha\beta}}\ct{\omega}{_\tau}}\quad\\
				\ctg{R}{_{\alpha\beta}}&=\ct{R}{_{\alpha\beta}}- 2\frac{1}{\omega}\cd{_\alpha}\ct{\omega}{_{\beta}}-\frac{1}{\omega^2}\ct{g}{_{\alpha\beta}}\ct{\omega}{_{\mu}}\ct{\omega}{^\mu}-\frac{1}{\omega}\ct{g}{_{\alpha\beta}}\cd{_{\mu}}\ct{\omega}{^\mu}+4\frac{1}{\omega^2}\ct{\omega}{_\alpha}\ct{\omega}{_\beta}\quad ,\label{eq:gaugeRicciTensorApp}\\
				\csg{R}&=\frac{1}{\omega^2}\cs{R}-6\frac{1}{\omega^3}\cd{_\mu}\ct{\omega}{^\mu} \quad ,\label{eq:gaugeRicciScalarApp}\\
				\ctg{N}{_{\alpha}}&=\omega\ct{N}{_{\alpha}}+\Omega\ct{\omega}{_\alpha}\quad ,\\
				\ctg{N}{}&=\frac{1}{\omega}\prn{\omega^2N^2-\Omega^2\ct{\omega}{_\mu}\ct{\omega}{^\mu}-2\Omega\ct{N}{_\mu}\ct{\omega}{^\mu}}^{1/2}\quad .\label{eq:gauge-NormN}
				\end{align}
	Quantities of $ \prn{\scri,\ms{_{ab}}} $ :
		\begin{align}
		\msg{_{ab}}&\eqs\omega^2\ms{_{ab}}\spacef,\label{eq:conformal-rescaling-scri}\\
		\ctg{\epsilon}{_{abc}}&\eqs\omega^3\ct{\epsilon}{_{abc}}\spacef,\\
		\ctsg{\Gamma}{^a_{bc}} &\eqs \cts{\Gamma}{^a_{bc}} +\cts{C}{^a_{bc}}\spacef,\cts{C}{^a_{bc}}=\frac{1}{\omega}\ms{^{at}}\prn{2\ms{_{
					t(b}}\cts{\omega}{_{c)}}-\ms{_{cb}}\cts{\omega}{_t}}\quad\\
		\ctsg{R}{_{ab}}&\eqs\cts{R}{_{ab}}- \frac{1}{\omega}\cds{_a}\cts{\omega}{_{b}}-\frac{1}{\omega^2}\ms{_{ab}}\cts{\omega}{_{m}}\cts{\omega}{^m}-\frac{1}{\omega}\ms{_{ab}}\cds{_{m}}\cts{\omega}{^m}+2\frac{1}{\omega^2}\cts{\omega}{_a}\cts{\omega}{_b}\spacef,\label{eq:gaugeRicciTensorScriApp}\\
		\ctsg{R}{}&\eqs\frac{1}{\omega^2}\cts{R}{}-4\frac{1}{\omega^3}\cds{_m}\cts{\omega}{^m} -2\frac{1}{\omega^4}\cts{\omega}{_{m}}\cts{\omega}{^m}\spacef,\label{eq:gaugeRicciScalarScriApp}\\
		\ctsg{S}{_{ab}}&\eqs \cts{S}{_{ab}}+2\frac{1}{\omega^2}\cts{\omega}{_a}\cts{\omega}{_b}-\frac{1}{\omega}\cds{_a}\cts{\omega}{_b}-\frac{1}{2\omega^2}\cts{\omega}{_s}\cts{\omega}{^s}\ms{_{ab}}\spacef.
		\end{align}
	Quantities associated to a cut $ \prn{\Sc,\mc{_{AB}}} $:
		\begin{align}
		\mcg{_{AB}}&\eqc\omega^2\mc{_{AB}}\spacef,\label{eq:gauge-mc}\\
		\ctcg{\epsilon}{_{AB}}&\eqc \omega^2\ctc{\epsilon}{_{AB}}\spacef,\\
		\ctcg{\Gamma}{^A_{BC}} &\eqc \ctc{\Gamma}{^A_{BC}} +\ctc{C}{^A_{BC}}\spacef,\ctc{C}{^A_{BC}}\eqc\frac{1}{\omega}\mc{^{AT}}\prn{2\mc{_{T(B}}\ctc{\omega}{_{A)}}-\mc{_{AB}}\ctc{\omega}{_T}}\quad\\
		\ctcg{R}{_{AB}}&\eqc\ctc{R}{_{AB}}+\frac{1}{\omega^2}\mc{_{AB}}\ctc{\omega}{_{M}}\ctc{\omega}{^M}-\frac{1}{\omega}\mc{_{AB}}\cdc{_{M}}\ctc{\omega}{^M}\spacef,\label{eq:gaugeRicciTensorCutApp}\\
		\ctcg{R}{}&\eqc\frac{1}{\omega^2}\ctc{R}{}+2\frac{1}{\omega^4}\ctc{\omega}{_{M}}\ctc{\omega}{^M}-2\frac{1}{\omega^3}\cdc{_M}\ctc{\omega}{^M} \spacef,\label{eq:gaugeRicciScalarCutApp}\\
		\ctcg{S}{_{AB}}&\eqc \ctc{S}{_{AB}} +2\frac{1}{\omega^2}\ctc{\omega}{_A}\ctc{\omega}{_B}-\frac{1}{\omega}\cdc{_A}\ctc{\omega}{_B}-\frac{1}{2\omega^2}\ctc{\omega}{_P}\ctc{\omega}{^P}\mc{_{AB}}-\ct{r}{^e}\cts{\omega}{_e}\prn{\frac{1}{\omega}\ctc{\kappa}{_{AB}}+\frac{1}{2\omega^2}\ct{r}{^d}\cts{\omega}{_d}\mc{_{AB}}}\spacef,\label{eq:gaugeSAB}\\
		\ctc{S}{_A}&	\eqc \frac{1}{\omega}\brkt{\ctc{S}{_A}-\frac{1}{\omega}\ctc{D}{_A}\prn{\ct{r}{^e}\cts{\omega}{_e}}+2\frac{1}{\omega^2}\ctc{\omega}{_A}\ct{r}{^e}\cts{\omega}{_e}+\frac{1}{\omega}\ctc{\omega}{_E}\ctc{\kappa}{_A^E}}\spacef,\\
		\csSg{S}&\eqc\frac{1}{\omega^2}\brkt{\csS{S}-\frac{1}{\omega}\ct{r}{^a}\ct{r}{^b}\cds{_a}\cts{\omega}{_b}+2\frac{1}{\omega^2}\prn{\ct{r}{^e}\cts{\omega}{_e}}^2-\frac{1}{2\omega^2}\cts{\omega}{_e}\cts{\omega}{^e}}\spacef.
		\end{align}
	Quantities associated to $ \Scn $:
		\begin{align}
		\mcng{_{AB}}&\eqcn\omega^2\mcn{_{AB}}\quad \label{eq:gauge-mcn},\\
		\ctcng{\epsilon}{_{AB}}&\eqcn\omega^2\ctcng{\epsilon}{_{AB}}\spacef,\\
		\ctcng{\Gamma}{^A_{BC}} &\eqcn \ctcn{\Gamma}{^A_{BC}} +\ctcn{C}{^A_{BC}}\spacef,\ctcn{C}{^A_{BC}}\eqcn\frac{1}{\omega}\mcn{^{AT}}\prn{2\mcn{_{T(B}}\ctcn{\omega}{_{A)}}-\mcn{_{AB}}\ctcn{\omega}{_T}}\quad\\
		\ctcng{S}{_{AB}}&\eqcn \ctcn{S}{_{AB}} +2\frac{1}{\omega^2}\ctcn{\omega}{_A}\ctcn{\omega}{_B}-\frac{1}{\omega}\cdcn{_{(A}}\ctcn{\omega}{_{B)}}-\frac{1}{2\omega^2}\ctcn{\omega}{_P}\ctcn{\omega}{^P}\mcn{_{AB}}-\ct{r}{^e}\cts{\omega}{_e}\prn{\frac{1}{\omega}\ctcn{\kappa}{_{AB}}+\frac{1}{2\omega^2}\ct{r}{^d}\cts{\omega}{_d}\mcn{_{AB}}}\spacef,\label{eq:gaugeSAB-congruence}\\
		\ctcn{S}{_A}&	\eqcn \frac{1}{\omega}\brkt{\ctcn{S}{_A}-\frac{1}{\omega}\ctcn{D}{_A}\prn{\ct{r}{^e}\cts{\omega}{_e}}+\frac{2}{\omega^2}\ctcn{\omega}{_A}\ct{r}{^e}\cts{\omega}{_e}+\frac{1}{\omega}\ctcn{\omega}{_E}\prn{\ctcn{\kappa}{_A^E}+\ctcn{\omega}{_A^E}}}\spacef,\\
		\csSg{S}&\eqs\frac{1}{\omega^2}\brkt{\csS{S}-\frac{1}{\omega}\ct{r}{^a}\ct{r}{^b}\cds{_a}\cts{\omega}{_b}+\frac{2}{\omega^2}\prn{\ct{r}{^e}\cts{\omega}{_e}}^2-\frac{1}{2\omega^2}\cts{\omega}{_e}\cts{\omega}{^e}}\spacef.
		\end{align}
			\subsection{Extrinsic geometry and kinematic quantities}
				For  a cut $ \Sc $:
				\begin{align}
				\ctg{r}{^a}	&\eqc \frac{1}{\omega}\ct{r}{^a}\quad,\\
				\ctcg{\kappa}{_{AB}} &\eqc \omega \ctc{\kappa}{_{AB}}+\mc{_{AB}}\ct{r}{^e}\cts{\omega}{_e}\quad,\\		\ctcg{\Sigma}{_{AB}} &\eqc \omega \ctc{\Sigma}{_{AB}}\quad,\\
				\csCg{\kappa}&\eqc \frac{1}{\omega}\csC{\kappa}+2\frac{1}{\omega^2}\ct{r}{^e}\cts{\omega}{_e}\quad.
				\end{align}
			For $ \Scn $:
				\begin{align}
				\ctg{m}{^a}	&\eqs \frac{1}{\omega}\ct{m}{^a}\quad\label{eq:gauge-m-vector},\\
				\ctcng{\kappa}{_{AB}} &\eqcn \omega \ctcn{\kappa}{_{AB}}+\mc{_{AB}}\ct{m}{^e}\cts{\omega}{_e}\quad,\\		\ctcng{\Sigma}{_{AB}} &\eqcn \omega \ctcn{\Sigma}{_{AB}}\quad,\\
				\csCg{\kappa}&\eqcn \frac{1}{\omega}\csC{\kappa}+2\frac{1}{\omega^2}\ct{m}{^e}\cts{\omega}{_e}\quad,\\
				\ctcng{a}{_A}&\eqcn \ctcn{a}{_A}-\frac{1}{\omega}\cdcn{_A}\omega\quad,\\
				\ctcng{\omega}{_{AB}}&\eqcn \omega\ctcn{\omega}{_{AB}}\quad.
				\end{align}
			\subsection{(rescaled) Weyl decomposition}
			Let $ \ct{r}{^a} $ be the vector field giving a congruence on $ \scri $, changing as
				\begin{equation}
					\ctg{m}{^a}\eqs\frac{1}{\omega}\ct{r}{^a}.
				\end{equation}
			The parts of the rescaled Weyl tensor in the decomposition with respect to this vector field on $ \scri $ transform as:
				\begin{align}
				\ctg{D}{_{ab}}&\eqs \frac{1}{\omega} \ct{D}{_{ab}}\quad,	&\ctg{C}{_{ab}}&\eqs \frac{1}{\omega} \ct{C}{_{ab}}\quad,\\
				\ctcng{D}{_{AB}}&\eqs\frac{1}{\omega}\ctcn{D}{_{AB}}\quad,&\ctcng{C}{_{AB}}&\eqs\frac{1}{\omega}\ctcn{C}{_{AB}} \quad,\\
				\ctcng{D}{_{A}}&\eqs\frac{1}{\omega^2}\ctcn{D}{_{A}}\quad,&\ctcng{C}{_{A}}&\eqs\frac{1}{\omega^2}\ctcn{C}{_{A}} \quad,\\
				\csg{D}&\eqs\frac{1}{\omega^3}\cs{D}\quad,&\csg{C}&\eqs\frac{1}{\omega^3}\cs{C} \quad,\\
				\ctpm{\tilde{D}}{_{\alpha\beta}}&\eqs\frac{1}{\omega}\ctpm{D}{_{\alpha\beta}}\quad,&\ctpm{\tilde{C}}{_{\alpha\beta}}&\eqs\frac{1}{\omega}\ctpm{C}{_{\alpha\beta}} \quad,\\
				\ctcnupm{\tilde{D}}{_{AB}}&\eqs\frac{1}{\omega}\ctcnupm{D}{_{AB}}\quad,&\ctcnupm{\tilde{C}}{_{AB}}&\eqs\frac{1}{\omega}\ctcnupm{C}{_{AB}} \quad,\\
				\ctcnupm{\tilde{D}}{_{A}}&\eqs\frac{1}{\omega^2}\ctcnupm{D}{_{A}}\quad,&\ctcnupm{\tilde{C}}{_{A}}&\eqs\frac{1}{\omega^2}\ctcnupm{C}{_{A}} \quad,\\
				\cspm{\tilde{D}}&\eqs\frac{1}{\omega^3}\cspm{D}\quad,&\cspm{\tilde{C}}&\eqs\frac{1}{\omega^3}\cspm{C} \quad.
				\end{align}	
	\section{CKVs on 2-dimensional Riemannian manifolds}\label{app:conformal-miscellaneous}	
								Let us first recall that a conformal Killing vector field (CKV) on an $n$-dimensional pseudo-Riemannian manifold with metric $g$ is a vector field $\vec{\xi}$ that satisfies
								\begin{equation*}
								\lied_{\vec{\xi}} g =2\Psi g \hspace{5mm} (\Longrightarrow \, \, \mbox{div} \vec{\xi} =n \Psi ).
								\end{equation*}
								From known results (see e.g. \cite{Yano1957}) one knows that, in the 2-dimensional case,
									\begin{equation}
									\lied_{\vec{\xi}} K = -2\Psi K -\Delta \Psi .\label{eq:LieK}
									\end{equation}
								The Lie derivative of the connection reads ($\Psi_A := \cdc{_A} \Psi$) 
									\begin{equation}
									\lied_{\vec{\xi}} \ctc{\Gamma}{^C_{AB}} = \cdc{_A}\cdc{_B} \xi^C -K(\delta^C_A\xi_B -\mc{_{AB}}\xi^C)=\delta^C_A \Psi_B +\delta^C_B \Psi_A - \mc{^{CE}} \Psi_E \mc{_{AB}} 
									\end{equation}
								and thus
									\begin{equation}
									\lied_{\vec{\xi}} \cdc{_A} z_B - \cdc{_A} \lied_{\vec{\xi}} z_B = -z_C \lied_{\vec{\xi}} \ctc{\Gamma}{^C_{AB}}=
									-z_A \Psi_B- z_B \Psi_A + \mc{^{CE}}z_C \Psi_E \, \mc{_{AB}}
									\end{equation}
								for any arbitrary one-form $z_A$. From this it is immediate to get, for any arbitrary function
								\begin{equation*}
								\mc{^{AB}} \lied_{\vec{\xi}} (\cdc{_A} \cdc{_B} f) -\mc{^{AB}} \cdc{_A} \cdc{_B} \lied_{\vec{\xi}} f=0
								\end{equation*}
								and therefore
									\begin{equation}\label{eq:LieDelta}
									\lied_{\vec{\xi}} \Delta f= \Delta \lied_{\vec{\xi}} f - 2\Psi \Delta f , \hspace{1cm} \forall f\in C^3(\Sc).
									\end{equation}
								
								The conformal Killing vector fields remain so under a conformal rescaling of the metric, (and thus, under a residual gauge change on our cuts). This easily follows from
								\begin{equation*}
								\lied_{\vec{\xi}} q'_{AB} = \lied_{\vec{\xi}} \left( \omega^2 \mc{_{AB}}\right)=2\left(\Psi
								+\frac{1}{\omega} \lied_{\vec{\xi}} \omega\right) q'_{AB} 
								\end{equation*}
								which provides the gauge behaviour of $\Psi$ (also deducible from $\cdc{_A} \xi^A =2\Psi$):
									\begin{equation}\label{eq:Psiprimed}
									\Psi' = \Psi +\lied_{\vec{\xi}} \ln \omega .
									\end{equation}
								Obviously, any fixed points of the CKV remain so under allowed conformal rescaling. However, if the conformal factor $\omega$ has global problems, leading to changes in the topology of the manifold, then some of the fixed points may disappear (or new ones arise).
								
								From the known transformation of the scalar curvature under conformal re-scalings one also knows the gauge behaviour of $K$:
									\begin{equation}\label{eq:Kprimed}
									K' =\frac{1}{\omega^2} \left(K -\frac{1}{\omega} \Delta \omega +\frac{1}{\omega^2} \omega_B \omega^B \right)=
									\frac{1}{\omega^2} \left(K- \Delta \ln \omega \right).
									\end{equation}
								On using this together with \cref{eq:LieDelta,eq:LieK,eq:Psiprimed} one can compute
									\begin{align}
									\lied_{\vec{\xi}} K' &= -2K'\lied_{\vec{\xi}} \ln\omega  +\frac{1}{\omega^2} \left(\lied_{\vec{\xi}} K-
									\lied_{\vec{\xi}} \Delta\ln\omega \right)\\
									&= -2K' \lied_{\vec{\xi}} \ln \omega  +\frac{1}{\omega^2} \left(-2\Psi K -\Delta \Psi -\Delta  \lied_{\vec{\xi}}\ln\omega +2\Psi \Delta \ln\omega\right)\\
									&= -2K' \lied_{\vec{\xi}} \ln \omega  -2\Psi K' -\frac{1}{\omega^2} \Delta \Psi'\\
									&= -2K'\Psi' - \Delta' \Psi'
									\end{align}
								rendering \eqref{eq:LieK} gauge invariant, as it must be. In the latter step we have used
									\begin{equation}\label{eq:Deltaprimed}
									\Delta' f =\frac{1}{\omega^2} \Delta f, \hspace{1cm} \forall f\in C^2(\Sc)
									\end{equation}
								which can be easily proven and allows one to rewrite \eqref{eq:Kprimed} as
								\begin{equation*}
								K'-\frac{K}{\omega^2} =- \Delta' \ln\omega .
								\end{equation*}
								From \cref{eq:LieK} and the previous result \eqref{eq:Deltaprimed} one obtains,  \emph{in arbitrary gauge} and for compact $\Sc$
									\begin{equation}\label{eq:LieK0}
									\int_\Sc (\lied_{\vec{\xi}} K +2 K \Psi ) \epsilon =0,
									\end{equation}
								as must be the case because $\lied_{\vec{\xi}} K +2 K \Psi =$ div$(K\vec \xi)$. However, {\em if the manifold $ \Sc $ is $\mathbb{S}^2$}, then each of the summands in \cref{eq:LieK0} integrates to zero by itself, as follows from 
								\cref{thm:colRhothm}.

								\subsection*{General CKV on 2-dimensional Riemannian manifolds}
								As it is known, any 2-dimensional Riemannian manifold is {\em locally} conformal to the flat Riemannian manifold. Hence, let us start by finding the general solution of CKVs in the flat Euclidean plane, $(\mathbb{R}^2,\delta_{AB})$. By using standard (global) Cartesian coordinates $\{x,y\}$ the metric reads
									\begin{equation}\label{eq:flatq}
									dx^2 + dy^2
									\end{equation}
								and the conformal Killing equations reduce to 
									\begin{equation*}
									\partial_x \xi^x =\Psi , \hspace{1cm} \partial_y \xi^y = \Psi, \hspace{1cm} \partial_x\xi^y +\partial_y \xi^x =0.
									\end{equation*}
								These are the Cauchy-Riemann equations for the real ($\xi^x$) and imaginary ($\xi^y$) parts of a complex analytic function on $\mathbb{C}$. Thus, the general solution for CKV on the plane is given by
									\begin{equation}
									\vec{\xi} = u\partial_x +v \partial_y, \hspace{1cm} f(z) = u(x,y) + i v(x,y) , 
									\end{equation}
								where $f(z)$ is any analytic function of the complex variable $z=x+i y$. This freedom is huge, and the algebra of CKV on the Eucliean plane is truly infinite dimensional. The function $\Psi$ is given by
									\begin{equation*}
									\Psi =\partial_x u =\partial_y v
									\end{equation*}
								and therefore Killing vectors are characterized by $u=a-w y$ and $v =b+w x$, with $a,b,w \in \mathbb{R}$ and, as is known there are three (linearly independent) of them.
								
								Some elementary examples of CKV for simple choices of $f(z)$ are
									\begin{itemize}
									\item $f=a+i b=$ constant. Then $\vec{\xi} = a \partial_x + b \partial_y$ represents the translations, which are KVF.
									\item $f(z)=(\nu+i w) z$. Then $\vec{\xi} = \nu (x\partial_x +y\partial_y) + w(x\partial_y-y\partial_x)$, representing a dilatation and a rotation, the latter being actually a KVF.
									\item $f(z)=(c-id) z^2 /2$. Then, $2\vec{\xi} = c [(x^2-y^2)\partial_x +2xy \partial_y]+d[2xy \partial_x+(y^2-x^2)\partial_y]$.
									\end{itemize}					
								One could continue with every power of $z$... As an example of a different kind, one can take for instance $f(z)=\ln (1+z)$. This is a multi-valued function with a singularity at $z=-1$. The corresponding Killing reads
									\begin{equation*}
									\vec{\xi} = \frac{1}{2} \ln (1+x^2+y^2) \partial_x + \arctan\left(\frac{y}{1+x}\right)\partial_y
									\end{equation*}
								and we see that this vector field has a problem at $(-1,0)$. Hence, the singularities of the functions $f(z)$ translate into problems for the associated CKV, and thus the function $f(z)$ must really be an entire function (analytic everywhere). 
								
								Furthermore, if the topology of the manifold is changed, even keeping the flatness of the metric, many of the CKV fields are not global and thus must be discarded. As a simple example, consider the case of $\Sc = \mathbb{S}^1\times \mathbb{R}$ by simply closing the coordinate $x$, that is, identifying $x$ with $x+2\pi$ for all $x$. Let us choose a new more graphical name for this periodic coordinate, say $x\rightarrow \sigma$. It is easily checked that all the previous CKV fail to be single valued except for the two translational KVFs. Not even the rotational Killing survives. In order to see how many CKV remain, one needs to consider functions $f(z)$ which are periodic. A couple of examples would be
									\begin{itemize}
									\item $f(z)= (\alpha +i \eta) \sin z$ which is periodic with period $2\pi$. For this function one derives the following CKV
									\begin{equation}
										\vec{\xi} = \alpha\left(\cosh y \sin \sigma \partial_\sigma +\sinh y \cos\sigma \partial_y \right)+
										\eta \left(-\sinh y \cos\sigma \partial_x + \cosh y\sin\sigma \partial_y \right).
									\end{equation}
								These are globally well defined. Notice that these CKV possess a fix point at $\sigma=0$ with $y=0$.
								\item $f(z)= (\alpha +i\eta) e^{iz}$, periodic with period $2\pi$. This leads to the CKV
									\begin{equation}
									\vec{\xi} = \alpha e^{-y} \left(\cos\sigma\partial_\sigma + \sin\sigma\partial_y \right)+\eta e^{-y} \left(-\sin\sigma\partial_\sigma +\cos\sigma\partial_y \right)
									\end{equation}
								again globally well defined, but now without fixed points. 
									\end{itemize}
								Had we closed the coordinate $y$, we obviously should require periodic functions with imaginary periods, such as $e^z$ with period $2\pi i$. This is, of course, equivalent to the previous by interchanging $x$ and $y$. 
								
								There are many other periodic functions with real period, such as $\sin^2 z$, $\cos z$, etc, and they lead to globally well defined CKV on $(\mathbb{S}^1\times \mathbb{R}, \delta_{AB})$. The zeroes of the generating function $f(z)$ translate into fixed points for the derived CKVs.
								
								There is also the possibility of changing the topology by closing both coordinates identifying $x \leftrightarrow x+2\pi $ for all $x$  (coordinate $\sigma$) and $y \leftrightarrow y+T$ for all $y$ (coordinate $\hat \sigma$, say). (We keep the $2\pi$ periodicity in the first case for simplicity, but the two periods can have any value). Then the globally defined CKV can only arise from doubly periodic $f(z)$, and these are known to be the {\em elliptic} functions. Apart from the constant function, any other doubly periodic $f(z)$ has singularities (poles or essential) ---as follows from the fact that, if it were bounded in the fundamental rectangle $[0,2\pi] \times [0,T]$, then it would be bounded in the whole complex plane and Liouville's theorem applies. Therefore, there are no globally defined CKV in the flat torus $\mathbb{S}^1\times \mathbb{S}^1$ apart from the translational KVFs.
								
								Let us pass now to consider conformal re-scalings but, in principle, keeping the topology. Starting in reverse order as before, consider the standard torus ($\mathbb{S}^1\times \mathbb{S}^1$ with the standard metric inherited from $\mathbb{R}^3$), whose metric reads in coordinates $\{\sigma,\phi\}$, with $\sigma\in (0,2\pi)$ and $\phi\in [-\pi,\pi]$
									\begin{equation*}
									\mc{_{\text{torus}}} = B^2 d\phi^2 +(A+B\cos\phi)^2 d\sigma^2, 
									\end{equation*}
								where $A$ and $B<A$ are constants that represent the two radia of the torus. The following change of coordinate
									\begin{equation*}
									\phi \rightarrow \hat\sigma, \hspace{1cm} \tan \left(\frac{\sqrt{A^2-B^2}}{2B} \hat\sigma\right)= \left(\frac{A-B}{A+B} \right)^{1/2}\tan (\phi/2)
									\end{equation*}
								where $\hat\sigma$ is a periodic coordinate with period $T=B2\pi/\sqrt{A^2-B^2}$, brings the metric to the conformally flat form
									\begin{equation*}
									\mc{_{\text{torus}}} =\omega_{tor}^2 \left(d\sigma^2 + d\hat\sigma^2 \right), 
									\end{equation*}
								with
									\begin{equation*}
									\omega_{\text{torus}}=\frac{A^2-B^2}{A-B\cos(2\pi \hat\sigma/T)} = A+B \cos\phi .
									\end{equation*}
								We see that $\partial_\sigma$ remains as a KVF, but the second `translational' Killing vector $\partial_{\hat\sigma}$ becomes now a CKV. For completeness, the Gaussian curvature is now
									\begin{equation*}
									K_{\text{torus}} = \frac{\cos \phi}{B(A+B\cos\phi)} .
									\end{equation*}
								
								Going back to the case of the cylinder, $\mathbb{S}^1\times \mathbb{R}$, an interesting case arises by using as conformal factor $\omega_f = e^y$, which is obviously positive everywhere (remember that $y$ runs over the whole real line). The new metric on $\mathbb{S}^1\times \mathbb{R}$ reads now
									\begin{equation*}
									q_f = e^{2y} (d\sigma^2 + dy^2) 
									\end{equation*}
								which is obviously flat (e.g., use \cref{eq:Kprimed}), however the Riemannian surface $(\mathbb{S}^1\times \mathbb{R},q_f)$ is not {\em complete}. This is more easily seen by the simple change of coordinate $\rho = e^y$, $0< \rho < \infty$, that transforms the metric to the explicitly flat expression
									\begin{equation*}
									q_f =d\rho^2 + \rho^2 d\sigma^2 .
									\end{equation*}
								Of course, this metric can be further transformed to the standard flat metric \eqref{eq:flatq} by means of the standard polar-coordinate change; {\em nevertheless, the point $\rho\rightarrow 0$  is missing} (this point corresponds to the origin in the new Cartesian coordinates). Observe also that the KVF $\partial_\sigma$ remains as a KVF, while $\partial_y =\rho\partial_\rho$ is now a homothetic Killing vector (with $\Psi =2$). Obviously one can {\em complete} the Riemannian manifold by attaching the missing point $\rho=0$, but {\em then we are changing the topology of the manifold} which now becomes $\mathbb{R}^2$. This is a perfectly acceptable procedure {\em unless} there are pre-defined reasons (such as impossibility of attaching the point if the surface is embedded in a larger space) which prohibit the completion. 
								
								A similar, but even more interesting situation arises with the metric of constant positive curvature ---and the 2-sphere---, that we consider in detail next.
								
								
								\subsection*{The round gauge}
								Consider now a conformal re-scaling of the flat metric \eqref{eq:flatq} which achieves \emph{ positive constant} Gaussian curvature $K$. The conformal factor $\omega$ is well known and the metric reads 
								
									\begin{equation}\label{eq:q-round}
									\mc{_{\text{round}}} = \left[1+\frac{K}{4} (x^2+y^2) \right]^{-2} \left(dx^2 + dy^2 \right).
									\end{equation}
								Observe that the conformal factor is positive everywhere. All the CKV of the flat Euclidean plane remain as such in principle. However, the above metric is not complete either. The manifold $\mathbb{R}^2$ is not compact, still the distances are all finite: the length of any inextendible curve is always bounded (take e.g. the length from the origin to $x=\infty$ along the curve $y=0$ as an example). Alternatively, one could consider the geodesics and see that their affine parameter is always finite, the geodesic ends, but there is no curvature obstacle nor anything like that. Moreover, the area of the surface is also finite. 
								
								Of course, one can complete the manifold by simply adding a point at `infinity', corresponding to $x^2+y^2\rightarrow \infty$. By doing this, as before, we are changing the topology of the manifold but this is licit if there are no other reasons forbidding the completion. In any case, this completes the manifold, which is now compact and topologically $\mathbb{S}^2$. 
								
								Nevertheless, this completion that changes the topology also changes drastically the quantity of CKV available. The reason is clear: the norm of any such vector field $\xi$ is now 
									\begin{equation}\label{eq:norm}
									\mc{_{\text{round}}}(\xi,\xi) =  \left[1+\frac{K}{4} (x^2+y^2) \right]^{-2}(u^2+v^2) =
									 \left[1+\frac{K}{4} |z|^2 \right]^{-2} |f(z)|^2
									\end{equation}
								where $f(z)$ is the analytic function generating the CKV. Therefore, the norm of $\vec{\xi}$ diverges at the added point for any $f(z)$ other than polynomials of degree less or equal than 2. All those $\vec{\xi}$ with divergent norm must be removed, they are not global CKV as they actually explode on the completed manifold.
								
								Hence, the most general conformal Killing vector field is just the combination of those presented above for $f(z)$ constant, linear or quadratic on $z$, so that keeping the same notations as above
									\begin{equation*}
									\vec{\xi}=\left(a +\nu x+(cx+dy) x -\frac{c}{2} (x^2+y^2) -w y \right)\partial_x +
									\left(b +\nu y+(cx+dy) y -\frac{d}{2} (x^2+y^2) +w x \right)\partial_y  
									\end{equation*}
								where $a,b,\nu,c,d$ and $w$ are constants. Of course, these are also CKVs of the flat Euclidean plane ---and of any other conformally flat 2-dimensional metric. There are 6 parameters. Equivalently, a basis of the conformal Lie algebra is given by
									\begin{align}
									\vec{\xi}_1 &=\partial_x, \quad &\vec{\xi}_2& = \partial_y,\quad &\vec{\xi}_3& =x\partial_x +y\partial_y , \\
									\vec{\xi}_4 &=\frac{1}{2}(x^2-y^2) \partial_x +xy \partial_y, \quad &\vec{\xi}_5& = xy\partial_x +\frac{1}{2} (y^2-x^2)\partial_y, \quad &\vec{\xi}_6& = y\partial_x -x\partial_y .
									\end{align}
								
								Furthermore, we know that these CKV have a fixed point at the `added' point at infinity if and only if the norm \eqref{eq:norm} vanishes there, implying that all such CKV with $c=d=0$ have a fixed point. In other words, $\vec{\xi}_1,\vec{\xi}_2,\vec{\xi}_3$ and $\vec{\xi}_6$ have a fixed point there. 
								
								Each of these generators satisfies $\lied_{\vec{\xi}_I} \mc{_{AB}} = 2\Psi_I \mc{_{AB}}$ ($I=1,\dots ,6$). Taking into account that $2\Psi_I =\cdc{_A} \xi^A_I$ a quick calculation provides
									\begin{align}
									2\Psi_1 &= - \left[1+\frac{K}{4} (x^2+y^2) \right]^{-1} Kx, \quad &2\Psi_2& = -\left[1+\frac{K}{4} (x^2+y^2) \right]^{-1} Ky,\\
									2\Psi_4&= \left[1+\frac{K}{4} (x^2+y^2) \right]^{-1} 2x, \quad &2\Psi_5 &=\left[1+\frac{K}{4} (x^2+y^2) \right]^{-1} 2y, \\
									2\Psi_3 &= 2\left[1+\frac{K}{4} (x^2+y^2) \right]^{-1} \left[1-\frac{K}{4} (x^2+y^2) \right], \quad &\Psi_6& =0.
									\end{align}
								It easily follows that a basis of the subset of Killing vectors is given by
									\begin{equation*}
									\{\vec{\xi}_6, 2\vec{\xi}_1+K\vec{\xi}_4, 2\vec{\xi}_2 +K\vec{\xi}_5\}.
									\end{equation*}
								
								Choosing canonical angular coordinates on $\mathbb{S}^2$ via the standard stereographic projection from the north pole (the new point attached in the completion, corresponding to $\theta =0$)
									\begin{align}
									x=\frac{2}{\sqrt{K}} \cot \frac{\theta}{2} \cos\varphi , \hspace{1cm} y=\frac{2}{\sqrt{K}} \cot \frac{\theta}{2} \sin\varphi ,\\
									\theta =2\arctan \frac{2}{\sqrt{K(x^2+y^2)}} , \hspace{1cm} \varphi =\arctan \frac{y}{x}
									\end{align}
								with $\theta\in (0,\pi]$ and $\varphi\in [0,2\pi)$, the metric becomes
						\begin{equation}
				 \mc{_{\text{round}}} =\frac{1}{K} \left(d\theta^2 +\sin^2\theta d\varphi^2 \right)\label{eq:round}
						\end{equation}
								and using 
									\begin{equation}
									\partial_x= -\sqrt{K}\left(\sin^2\frac{\theta}{2} \cos\varphi \partial_\theta +\frac{1}{2}\tan\frac{\theta}{2}\sin\varphi \partial_\varphi \right),\\
									\partial_y= \sqrt{K}\left(-\sin^2\frac{\theta}{2} \sin\varphi \partial_\theta +\frac{1}{2}\tan\frac{\theta}{2}\cos\varphi \partial_\varphi \right)
									\end{equation}
								the Killing vectors become
									\begin{align}
									\vec{\xi}_6 &=\partial_\varphi :=\vec{\tilde\xi}_3,\label{eq:Kil3}\\
									\frac{1}{\sqrt{K}}  \vec{\xi}_1+\frac{\sqrt{K}}{2} \vec{\xi}_4 &= -\cos\varphi \partial_\theta +\cot\theta \sin\varphi\partial_\varphi :=-\vec{\tilde\xi}_2,\label{eq:Kil2}\\
									\frac{1}{\sqrt{K}}   \vec{\xi}_2+\frac{\sqrt{K}}{2} \vec{\xi}_5 &= -\left(\sin\varphi \partial_\theta +\cot\theta \cos\varphi\partial_\varphi \right):=\vec{\tilde\xi}_1 \label{eq:Kil1}
									\end{align}
								(generating so(3)). Concerning the {\em proper} CKVs  we have analogously
									\begin{align}
									\vec{\xi}_3 &= -\sin\theta \partial_\theta:= \vec\eta_3, \label{eq:CKil3}\\
									\frac{1}{\sqrt{K}} \vec{\xi}_1-\frac{\sqrt{K}}{2}\vec{\xi}_4 &= \cos\theta\cos\varphi\partial_\theta -\frac{\sin\varphi}{\sin\theta}\partial_\varphi :=\vec\eta_1 , \label{eq:CKil1}\\
									\frac{1}{\sqrt{K}}   \vec{\xi}_2-\frac{\sqrt{K}}{2} \vec{\xi}_5 &= \cos\theta\sin\varphi\partial_\theta +\frac{\cos\varphi}{\sin\theta}\partial_\varphi :=\vec\eta_2.\label{eq:CKil2}
									\end{align}

								Observe that the three CKVs of \cref{eq:CKil1,eq:CKil2,eq:CKil3} are all closed as one-forms (and on the simply connected $\mathbb{S}^2$ actually exact):
									\begin{equation*}
									\boldmath{\eta}_3 =\frac{1}{K} d(\cos\theta), \hspace{6mm} \boldmath{\eta}_1 = \frac{1}{K} d(\sin\theta\cos\varphi), \hspace{6mm} \boldmath{\eta}_2= \frac{1}{K} d(\sin\theta\sin\varphi).
									\end{equation*}
								This, of course, agrees with the known result ($i=1,2,3$)
									\begin{equation}
									\cdc{_A} \cdc{_B} n^{(i)} =-K \mc{_{AB}} n^{(i)}, \label{eq:gradientCKV}
									\end{equation}
								where
									\begin{equation*}
									n^{(i)} =\left(\sin\theta\cos\varphi,\sin\theta\sin\varphi, \cos\theta\right)=-\left(\Psi_{\vec\eta_1},\Psi_{\vec\eta_2}, \Psi_{\vec\eta_3} \right)
									\end{equation*}
								because 
									\begin{equation*}
									\eta_{(i)}^B = \frac{1}{K} \cdc{^B} n^{(i)}.
									\end{equation*}
								Observe that 
									\begin{equation*}
									\epsilon^{AB}\eta_{(i)B} =\frac{1}{K} \epsilon^{AB} \cdc{_B} n^{(i)}
									\end{equation*}
								are the three Killing vector fields \cref{eq:Kil3,eq:Kil1,eq:Kil2}. 
								Notice that in particular $\Delta n^{(i)} = -2K n^{(i)}$, meaning that $n^{(i)}=-\Psi_{\vec\eta_i}$ are the three spherical harmonics $Y_1^i$, with $\ell =1$. More generally one has for $s\geq 2$
									\begin{equation}\label{eq:Deltans}
									\Delta (n^{(i_1)} \dots n^{(i_s)})= 2K {s\choose 2} \delta^{(i_1i_2} n^{(i_3)}\dots n^{(i_s))}-Ks(s+1) n^{(i_1)} \dots n^{(i_s)}
									\end{equation}

								Another interesting remark that will be useful later to analyze the condition \eqref{eq:conditionY} is that the Lie algebra of CKVs, as a whole, is preserved by the Hodge dual. Actually, in the basis provided for the round metric, the duals of the KVs become CKVs, and viceversa, as follows:
									\begin{equation*}
									\star\vec{\tilde\xi}_3 =-\vec\eta_3, \hspace{9mm} \star\vec{\tilde\xi}_1=-\vec\eta_1, \hspace{9mm} \star\vec{\tilde\xi}_2= -\vec\eta_2.
									\end{equation*}
								In particular this implies that if $\boldmath{X}$ satisfies \eqref{eq:conditionY}, then so does $\star\boldmath{X}$. 
								
								Their scalar products are known (or can be directly computed)
									\begin{align}
									\vec{\tilde\xi}_i\cdot\vec{\tilde\xi}_j&= \vec\eta_i\cdot \vec\eta_j =\mc{^{AB}} \cdc{_A} n^{(i)} \cdc{_B} n^{(j)} = \frac{1}{K} (\delta^{ij} -n^{(i)} n^{(j)}),\label{eq:etadoteta}\\
									\vec\eta_i\cdot \vec{\tilde\xi}_j &= \frac{1}{K} \epsilon_{ijk}n^{(k)}  \label{eq:etadotxi}.
									\end{align}
								A final identity that may be useful is
									\begin{equation}
									n^{(i)} \vec\eta_i =\vec 0. \hspace{1cm} n^{(i)} \vec{\tilde\xi}_i =\vec 0.\label{eq:neta}
									\end{equation}

\printbibliography
\end{document}